\documentclass[submission,Phys]{SciPost}

\hypersetup{
    colorlinks,
    linkcolor={red!50!black},
    citecolor={blue!50!black},
    urlcolor={blue!80!black}
}

\usepackage[bitstream-charter]{mathdesign}
  \usepackage[normalem]{ulem} 

  \usepackage{alltt}
  \usepackage{xspace}

  \usepackage{soul}  
  \definecolor{hlc}{rgb}{1,1,.7}
  \sethlcolor{hlc}

  \usepackage[mono]{inconsolata} 

  \usepackage[frozencache,cachedir=.]{minted}

  \definecolor{cemph}{rgb}{0,.447,.7410} 
  \colorlet{cidx}{cemph!66!black}
  \colorlet{cEmph}{cemph!90!white}

  \definecolor{mlgray}{rgb}{.95,.95,.95} 
  \definecolor{shgray}{rgb}{.96,.96,1.0} 

  \definecolor{csrc}{rgb}{.4,.4,.4}
  \definecolor{msrc}{rgb}{0.93 0.46 0.15}

  \colorlet{mSrc}{msrc!90!black}
  \colorlet{mIdx}{msrc!66!black}

  \definecolor{mcol}{rgb}{.05,.70,.20} 
  \definecolor{ccol}{rgb}{.50,.50,.50} 
  \definecolor{mccl}{rgb}{.04,.39,.11} 

  \definecolor{ccmt}{rgb}{0.25,0.47,0.47} 

  \definecolor{cOM}{rgb}{0.439,0.678,0.278}

  \setminted[matlab]{bgcolor=mlgray,breaklines,mathescape=true,%
     fontsize=\small,linenos,numbersep=1ex,%
     framesep=1ex}

  \setminted[text]{mathescape=true,
     fontsize=\small,linenos,breaklines=false,numbersep=1ex,%
     frame=leftline,framerule=1pt,rulecolor=lightgray,framesep=1ex}

  \setminted[tex]{mathescape=true,%
     fontsize=\small,linenos,breaklines=false,numbersep=1ex,%
     frame=leftline,framerule=1pt,rulecolor=lightgray,framesep=1ex}

  \setminted[bash]{bgcolor=shgray,mathescape=true,xleftmargin=2ex,%
     fontsize=\small,linenos,breaklines=false,numbersep=1ex,%
     frame=leftline,framerule=1pt,rulecolor=lightgray,framesep=1ex}

  \def\myempty{myemptytoken}

  \newcommand{\mlbl}[1]{\phantomsection\label{#1}}
  \newcommand{\mline}[1]{line~\ref{#1}}
  \newcommand{\mLine}[1]{Line~\ref{#1}}
  \newcommand{\mlines}[1]{lines~\ref{#1}}
  \newcommand{\mLines}[1]{Lines~\ref{#1}}

  \newcommand{\mintLabelA}[1]{
    \begin{eqnarray}\label{#1}\end{eqnarray}
    \phantom{x}\vspace{-.65in}%
  }

  \newcommand{\mintLabelB}[2][\myempty]{\phantom{x}%
    \ifx\myempty#1%
       \vspace{-0.71in}
	\else%
       \vspace{#1}
	\fi
    \begin{eqnarray}\label{#2}\end{eqnarray}
  }

  \newcommand{\mtt}[1]{\mintinline[breaklines=true]{text}{#1}}

  \def\qt{\texttt{\textquotesingle}}

  \hypersetup{linkcolor=cidx} 

  \titleformat{\subparagraph}[runin]{%
  \normalfont\normalsize\itshape}{\thesubparagraph}{0.1em}{\Emph}[\normalfont: ]

  \titlespacing*{\subparagraph}{0pt}{1.5ex plus 1ex minus 0.2ex}{1.5ex plus 0.2ex}

  \newcommand{\SecI}[1]{Sec.\,\ref{idx:#1}}
  \newcommand{\Sec}[1]{Sec.\,\ref{sec:#1}}
  \newcommand{\Secs}[1]{Secs.\,\ref{sec:#1}}
  \newcommand{\SEC}[1]{Section\,\ref{sec:#1}}

  \newcommand{\App}[1]{App.\,\ref{app:#1}}
  \newcommand{\Aps}[1]{App.\,\ref{sec:#1}}
  \newcommand{\Apps}[1]{Apps.\,\ref{app:#1}}
  \newcommand{\APP}[1]{Appendix\,\ref{app:#1}}

  \newcommand{\Eq}[1]{Eq.\,\eqref{eq:#1}}
  \newcommand{\Eqs}[1]{Eqs.\,\eqref{eq:#1}}
  \newcommand{\EQ}[1]{Equation\,\eqref{eq:#1}}

  \newcommand{\Fig}[1]{Fig.\,\ref{fig:#1}}

  \newcommand{\SRC}[2][]{{\hyperref[src:#1]{#2}}}

  \newcommand{\rsym}[1]{\ensuremath{%
  \mathfrak{r}_{\mathrm{\ifx\@empty{}\else{#1}\fi}}}}

  \newcommand{\idx}[2][]{{\hyperref[idx:#1]{#2}}}
  \newcommand{\Idx}[2][]{{\hyperref[idx:#1]{\textcolor{mIdx}{#2}}}}
  \newcommand{\Idt}[2][]{{\hyperref[idx:#1]{\textcolor{mIdx}{\texttt{#2}}}}}
  \newcommand{\HSec}[2][]{{\hyperref[sec:#1]{\textcolor{mIdx}{\texttt{#2}}}}}

  \newcommand{\heq}[2][]{{\hyperref[eq:#1]{#2}}}
  \newcommand{\hsec}[2][]{{\hyperref[sec:#1]{#2}}}
  \newcommand{\happ}[2][]{{\hyperref[app:#1]{#2}}}
  \newcommand{\hfig}[2][]{{\hyperref[fig:#1]{#2}}}
  \newcommand{\RCR}[1]{\hsec[RCStore]{#1}}

  \newcommand{\hcite}[2][]{{\hyperlink{cite.#1}{#2}}}

  \newcommand{\src}[1]{\textcolor{gray!90!black}{\texttt{#1}}} 

  \newcommand{\Srt}[1]{\textcolor{mSrc}{\texttt{#1}}} 
  \newcommand{\Src}[1]{\textcolor{mSrc}{{#1}}}

  \newcommand{\Var}[1]{\src{\$(#1)}}
  \newcommand{\var}[1]{\src{\$#1}}

  \def\Cpp{\texttt{C++}\xspace}

  \newcommand{\str}[1]{\src{\textquotesingle{#1}\textquotesingle}}

  \newcommand{\qdir}[1]{\mbox{\texttt{#1}}\xspace}

  \newcommand{\Emph}[1]{\textcolor{cEmph}{{#1}}}
 
  \def\QSpace{\mbox{\texttt{QSpace}}\xspace}
  \def\QSpaces{\mbox{\texttt{QSpace}s}\xspace}

  \def\atQSpace{\mbox{\texttt{@QSpace}}\xspace}
  \def\AtQSpace{\mbox{\textcolor{msrc}{\texttt{@QSpace}}}\xspace}
  \def\mQSpace{\mbox{\textcolor{msrc}{\texttt{QSpace}}}\xspace}

  \def\atSymStore{\mbox{\texttt{@SymStore}}\xspace}
  \def\atSymOp{\mbox{\texttt{@SymOp}}\xspace}
  \def\atHamilton{\mbox{\texttt{@Hamilton1D}}\xspace}

  \def\itags{\texttt{itag}s\xspace}
  \def\itag{\texttt{itag}\xspace}

  \def\Xsymbol{\texttt{X}-symbol\xspace}
  \def\Xsymbols{\texttt{X}-symbols\xspace}

  \def\onej{\ensuremath{1j}\xspace}

  \def\MLR{\mbox{\texttt{MATLAB\_ROOT}}}
  \def\srcMLR{\src{\MLR}\xspace}
  \def\MLRh{\idx[MLR]{\MLR}\xspace}

  \def\MYM{\mbox{\texttt{MYMATLAB}}}
  \def\srcMYM{\src{\MYM}\xspace}
  \def\srcMYMv{\src{\$\MYM}\xspace}
  \def\hMYM{\idx[MYM]{\MYM}\xspace}
  \def\hMYMv{\idx[MYM]{\$\MYM}\xspace}

  \def\RCS{\mbox{\texttt{RC\_STORE}}\xspace}
  \def\RCSy{\mbox{\src{RC\_SYNC}}\xspace}

  \def\iu{\ensuremath{\mathrm{i}\mkern1mu}}

  \def\irec{\texttt{i}\xspace}

  \def\Id{\ensuremath{1\hspace{-0.85ex}1}}

  \def\rsym{\ensuremath{{\mathfrak{r}}}\xspace}

  \DeclareMathAlphabet{\mathesstixfrak}{U}{esstixfrak}{m}{n}
  \def\Rsym{\ensuremath{\mathesstixfrak{r}}\xspace}

\DeclareSymbolFont{usualmathcal}{OMS}{cmsy}{m}{n}
\DeclareSymbolFontAlphabet{\mathcal}{usualmathcal}

\begin{document}

\begin{center}{\Large \textbf{
QSpace -- An open-source tensor library
for Abelian and non-Abelian symmetries \\
}}\end{center}

\begin{center}
Andreas Weichselbaum\textsuperscript{$\star$} 
\end{center}

\begin{center}
Department of Condensed Matter Physics and Materials Science, \\
Brookhaven National Laboratory, Upton, NY 11973-5000, USA
\\

${}^\star$ {\small \sf weichselbaum@bnl.gov}
\end{center}

\begin{center}
\today
\end{center}

\section*{Abstract}

{\bf
   This is the documentation for the tensor library \QSpace (v4.0),
   a toolbox to exploit `quantum symmetry spaces'
   in tensor network states in the quantum many-body context.
   \QSpace permits arbitrary combinations of symmetries
   including the abelian symmetries $\mathbb{Z}_n$ and $U(1)$,
   as well as all non-abelian symmetries based on the
   semisimple classical Lie algebras: $A_n$, $B_n$,
   $C_n$, and $D_n$, or respectively, the special unitary
   group SU($n$), the odd orthogonal group SO($2n+1$),
   the symplectic group Sp($2n$), and the even orthogonal group SO($2n$).
   The code (\Cpp embedded via the MEX interface into Matlab)
   is available \href{https://bitbucket.org/qspace4u/}{open
   source as of \QSpace v4.0 on bitbucket}
   \cite{qspace4u} under the Apache 2.0 license.
   
   \QSpace is designed as a bottom-up approach for non-abelian
   symmetries. It starts
   from the defining representation and the respective Lie algebra.
   By explicitly computing and tabulating
   generalized Clebsch-Gordan coefficient tensors,
   \QSpace is versatile in the type of operations
   that it can perform across all symmetries.
   At the level of an application, much of the
   symmetry-related details are hidden within the
   \QSpace \Cpp core libraries.
   Hence when developing tensor network algorithms with \QSpace,
   these can be coded (nearly) as if there are
   no symmetries at all, despite being able to fully exploit
   general non-abelian symmetries. 
}

\vspace{10pt}
\noindent\rule{\textwidth}{1pt}
\tableofcontents\thispagestyle{fancy}
\noindent\rule{\textwidth}{1pt}
\vspace{-2ex}

\section{Introduction}
\label{sec:intro}

The treatment of tensor network states in numerical
simulations requires a set of standardized
elementary routines: ways to initialize tensors, 
Kronecker products to build tensor product spaces,
contractions to compute matrix elements or expectation
values, etc. Together these routines form a tensor library
that is designed to provide a user-friendly environment.
While the individual tensor operations are straightforward
linear algebra operations per se, the complexity of a
tensor library increases significantly if one intends
to exploit symmetries. Yet symmetries are important,
both for the sake of numerical efficiency, but
also to gain detailed symmetry-resolved physical
insights into the fabric of particular tensor network state
realizations.
By now several mature open-source tensor libraries
are available. Yet the overwhelming majority of these
only permit one to exploit abelian symmetries. This includes,
for example, iTensor \cite{iTensor,Fishman22itensor},
Uni10 \cite{Kao15}, ALPS \cite{Bauer11alps}, or
TeNPy \cite{TeNPy18}. Implementations of non-abelian
symmetries in tensor network libraries are scarce,
with TensorKit \cite{TensorKit} a notable recent exception.

\QSpace also started from plain abelian symmetries
originally in \hsec[version]{version} 1 (v1).
Yet~with~\hcite[Wb12_SUN]{v2},
\QSpace implemented non-abelian symmetries
\cite{Wb12_SUN}. It was designed as a bottom-up approach,
based on the realization \cite{Singh10} that,
as a generalization of the Wigner-Eckart theorem
\cite{Sakurai94,Krishna80I,Toth08,Bulla08,Moca12},
Clebsch-Gordan coefficients of
an arbitrary-rank symmetric tensor can be {\it factorized}
in the form of a tensor product with 
reduced matrix elements. \QSpace explicitly computes and tabulates
Clebsch-Gordan coefficients for standard tensor
product decomposition (rank-3), as well as higher-rank
Clebsch-Gordan tensors (CGTs) obtained via contractions.
It is in this sense that \QSpace is considered a {\it bottom-up}
approach: by explicitly computing and utilizing CGTs,
the tensor representation is known 
\hsec[tensor:decomp]{in full numerical detail}
which then enables general tensor operations
in a self-contained fashion. Specifically, \QSpace does
not rely on or use \idx[6j]{$6j$-symbols} \cite{Singh10,Wb20},
bearing in mind that these are known analytically
for SU(2) only, in which case these are simple numbers.
For general non-abelian symmetry, however,
the $6j$ symbols become rank-4 tensors in outer
multiplicity (`\idx[4M]{$4$M-symbols}')
which introduces significant complications.

Tensor network simulations typically deal with
lattice models. These are comprised of sites whose
state space is considered small (microscopic).
By iteratively adding site after site, ever larger many-body Hilbert spaces
can be built. Eventually, these are truncated in tensor
network simulations in a controlled fashion based on 
the underlying entanglement structure.
This iterative build-up of a Hilbert space precisely
also underlies the design principle of \QSpace for non-abelian symmetries:
one starts from small local units where all symmetry-related
aspects can be generated, and thus are known
in a simple, transparent way. With these initial
building blocks then one can proceed to `play lego':
one builds ever larger more-complex tensor network structures. 
From the \QSpace perspective, this iterative nature automatically
generates the respective multiplet spaces {\it as they
occur on demand} at each step, based on well-defined elementary
steps such as tensor product decomposition,
contractions, etc. Newly generated 
symmetry related data is tabulated and stored much like
in a database (the \RCS in \QSpace). \QSpace thus automatically
adapts to what is needed in the simulations that are run,
bearing in mind that it is 
impossible to store all symmetry-related data for
continuous non-abelian symmetries, since that data set
is infinite.

\subsection{Version history}
\label{sec:version}

The \QSpace tensor library has a long-standing history
that started around 2006.
Since its inception, it has already been thoroughly
scrutinized, debugged, and optimized.
A distinctive feature of \QSpace has been to exploit
arbitrary symmetries
from the very beginning for the sake of numerical efficiency.
The initial motivation of \QSpace was 
at the interface between the density matrix renormalization
group (DMRG \cite{White92,Schollwoeck11}) and the numerical
renormalization group (NRG \cite{Wilson75,Bulla08}).
Yet since it was designed as a general tensor library
from the very beginning, \QSpace is also equally applicable
in other tensor network algorithms.
\QSpace went open-source as of version 4 \cite{qspace4u}.
\begin{itemize}
  \setlength{\itemsep}{-0.1\baselineskip}

  \item \QSpace v1.* (2006-2011) started out as a tensor
     library for arbitrary sets of U(1) abelian symmetries,
     with applications in the realm of
     NRG (fdm-NRG \cite{Wb07,Costi09,Tureci11}),
     DMRG \cite{Saberi09,Muender10,Holzner11},
     and their crossover \cite{Wb09,Saberi08}.

  \item \QSpace v2.* (2012-2015) newly introduced
     general continuous non-abelian symmetries to its
     capability \cite{Wb12_SUN,Hanl13,Hanl14b}, then including the
     special unitary group SU($N$) and the symplectic
     group Sp($2N$) where $N\geq 2$ is dealt with as a parameter.

  \item \QSpace v3.* (2015-2022)
     completed the set of semi-simple symmetries
     by also implementing the special orthogonal groups
     SO($N$), both for even and odd $N$, noting that these
     represent different Lie algebras $D_n$
     and $B_n$ with $N=2n$ or $N=2n+1$, respectively.
     \QSpace v3 also included significant performance
     upgrades, with full tabulation of
     generalized Clebsch-Gordan tensors (CGTs)
     as well as their contractions via the introduction
     of \Xsymbols \cite{Wb20}.
     
  \item \QSpace v4 (since 2022) finally was geared towards open
     source in response to frequent requests from the community.
     \QSpace v4  is mostly the state of \QSpace v3.2 (last version
     in v3.*), yet cleaned up in the sense of having completed
     and merged still open development branches.
     It went open-source in 2022 as a git-repository on 
     \href{https://bitbucket.org/qspace4u/}{bitbucket}
     \cite{qspace4u}
     under the Apache 2.0 license.

\end{itemize}

\subsection{Target audience and additional literature}

This documentation of \QSpace assumes basic familiarity
with tensor network states for correlated quantum
many-body states.
This includes general tensor network operations and
strategies, as well as their graphical description
in the form of tensor network diagrams.
Many excellent introductory papers, tutorials, and reviews
already exist in the literature in this regard,
e.g., see \cite{Eisert13,Orus14tns,TeNPy18,Orus19}
or the lecture notes in Ref. \cite{JvD-lectures}
amongst many others. This documentation does not intend
to replicate these. Instead, it focuses on the implementation
of general symmetries in tensor network states,
with a strong emphasis on the distinguishing feature
of \QSpace, namely non-abelian symmetries.
This documentation builds on the earlier publications
on \QSpace in Refs. \cite{Wb12_SUN,Wb20} which already also put
an emphasis on a more self-contained pedagogical presentation
(e.g., see the extensive appendices in \cite{Wb12_SUN}
including App.\,A on {\it Non-abelian symmetries 101}).
As such, these are highly recommended additional literature
that complement this documentation.

\subsection{Format conventions}
\label{sec:format}

The format for the \src{coding syntax} 
(in this font and color coding)
used throughout this documentation 
is in intuitive compact Matlab semantics.
The main reason is that this also 
reflects the \happ[ENV:ML+unix]{\QSpace environment}
that the \Cpp \QSpace core routines
are embedded into
(the many \hsec[format]{cross-references}
within this documentation via hyperlinks are formatted
in blue as shown;
non-hyperlinked \Emph{terms or highlighted text}
are differentiated in a somewhat lighter color).
A detailed discussion
of the structure of the vast low-level \Cpp
code of the core routines, on the other hand,
is beyond the scope of this documentation. 
The Matlab syntax used here
is basic and concerns, for example,
access to fields in structures, like \src{X.field},
cell arrays \src{Q\{i\}},  the representation of \str{strings}, 
or matrix notation such as \src{M(i,:)} or \src{M(:,j)}
for row $i$ or column $j$, etc.
The color format is changed when required
for the sake of differentiation or emphasis
of \Src{Matlab-specific} \QSpace topics, e.g.,
like the general (\Cpp) \QSpace implementation
vs. its Matlab \Src{\QSpace}, i.e., \atQSpace counterpart.
To visually indicate when the latter terms are hyperlinked,
the color changes to a \Idx[QS:QSpace]{darker shade}.
Optional arguments to functions are usually bracketed
as in \src{[$\ldots$]}.  At the level of the operating
system, (environmental) variables like \src{v} are also
referenced as \var{v} or \Var{v} as in \src{bash}
or \src{Makefile} semantic.

In this documentation, snippets of Matlab code
using \QSpace will be displayed
based on the \src{minted} latex package
that also enables syntax highlighting,
\begin{minted}[escapeinside=??]{matlab}
>> fprintf(1,'\n   Hello world from Matlab prompt!');  % some comment
>> [a,b]=helloworld('some','input',pi);                % helloworld MEX example
\end{minted}
Here \src{{>}>} indicates the Matlab command line prompt.
The second line, for example,
is a simple \idx[MEX]{MEX} test routine,
that assigns \src{a=\str{some}}, \src{b=\str{input}}, etc.
As with the above example, \QSpace always assumes
Matlab strings in single quotes (\str{...}).
Textual output from Matlab commands such
as the above is displayed without any further
syntax highlighting in the format
\begin{minted}[firstnumber=last]{text}
   Hello world from Matlab prompt!
   Hello world from MEX! (.mexmaci64: nargout=2, nargin=3)
\end{minted}
The line numbers to the left are included for ease
of reference in subsequent discussions.
These output displays may include further commands
shown with a leading Matlab prompt, together with their
respective output.
The output is frequently adapted for the purpose
of this documentation, e.g., by skipping empty
or trivial lines such as \str{ans = } indicating the 
answer to, i.e., result of a command.

\subsection{Outline}

This documentation makes use of a broad range of
semantics and definitions.
Essential terminology
and concepts are introduced in detail in \Sec{QSpace:1}.
This is complemented by \Apps{acronyms} and \ref{app:TNsym}
which provide glossaries on frequently used acronyms 
and general terminology for tensor networks and symmetries,
respectively. These short glossaries are grouped
by their meaning, rather than alphabetically,
such that quickly glossing through them
from top to bottom should be meaningful.
The bare-bones data structure of \QSpace tensors
is discussed in \Aps{QS:struct}.

The general \QSpace approach to implementing physical
model systems is summarized in \Sec{QSpace:2}.
Because \QSpace examples need \QSpace tensors
that need to be generated first, more detailed
examples for \Secs{QSpace:1} and \ref{sec:QSpace:2}
are collected into \Sec{examples} for clarity, with
cross-references as relevant.
\SEC{tutorials} provides a compilation
of more involved examples, intended as simple
tutorials already geared toward applications.
The general part of this documentation concludes
in \Sec{conclusion} which also comments
on feedback and support.

System requirements are detailed in \App{requirements}.
Download, compilation, and \QSpace environment
are discussed in \Apps{download} and \ref{app:QS:env}.
\APP{QS:apps} comments on state-of-the-art NRG and
DMRG applications that are already also present in \QSpace,
but whose detailed documentation is beyond the scope
of this documentation.
\App{Xdocs} comments on detailed usage information
(like `\src{man}' pages)
and further documentation in the \QSpace repository.
\APP{MEX:list} provides a listing of the most relevant
compiled binary \QSpace \idx[MEX]{MEX} routines.
\APP{QS:env-ML} finally gives an overview of the
Matlab-intrinsic environment that \QSpace is embedded into.

\section{General \QSpace Approach and Conventions}
\label{sec:QSpace:1}

The simple main idea underlying \QSpace is
that for any tensor $X$ with an arbitrary but fixed
number~$r$ of legs or indices, referred to as
\idx[rank]{(tensor) rank} $r$,
the presence of a global symmetry leads to a block
decomposition of the tensor.
Then for each block with well-defined symmetry sectors $q$
on all its legs, the symmetry-related aspects 
factorize \cite{Singh10,Wb12_SUN}, 
\begin{eqnarray}
   X = \bigoplus_{q}\ 
      \Vert X\Vert_{q} \otimes C_{q}
\label{eq:tensor:decomp:0}
\end{eqnarray}
which is generalized and discussed in significantly
more detail in \Sec{tensor:decomp} still. 
Here $\Vert X\Vert_{q}$ represents the reduced matrix
element tensors (RMTs), and $C_q$ the generalized
Clebsch-Gordan coefficient tensors (CGTs).
Both mimic the same structure of the original tensor $X$,
and thus, in particular, are also of the same rank $r$.
A scalar rank-2 tensor, like for example a Hamiltonian,
becomes block diagonal.
The listing of such non-zero blocks is denoted
by the direct sum ($\oplus$) over symmetry block
configurations that are permitted by symmetry.
These are labeled by $q$ here which is assumed collective
over all legs. Simply put, a \QSpace
tensor consists of a list of RMTs $\Vert X\Vert_{q}$
with references to CGTs $C_q$. \QSpace 
takes care of all the symmetry-related aspects
and the required book keeping. In this sense,
the tensor \src{X} is also referred to as \QSpace \src{X}.

The decomposition in \eqref{eq:tensor:decomp:0}
is also applicable to abelian symmetries.
This is useful from the coding point of view for the
sake of a shared consistent data structure
when having abelian and non-abelian symmetries
present at the same time. 
Fusing state spaces with abelian symmetries can
also be cast into the framework of CGTs, albeit trivially 
so by taking $C_q=1$ when the set of labels $q$
collected over all legs is permissible from a fusion
point of view, and $C_q=0$ otherwise.
These trivial factors can be created on the spot,
of course, with no need to tabulate.

In \QSpace, the CGTs $C_q$ for non-abelian symmetries
are explicitly constructed on demand once and for all, and
maintained in a database for later reference.
In the presence of multiple symmetries, the CGTs in
the decomposition in \eqref{eq:tensor:decomp:0}
further factorize, $C_q = \bigotimes_s C_{q_s}^{(s)}$,
with $C_{q_s}^{(s)}$ the CGT for symmetry $s$.
The symmetry labels for any symmetry sector on a
particular leg then also
become the collection of labels across all
$C_{q_s}^{(s)}$, i.e.,  $q \equiv \{q_s\}$ .
All bookkeeping for this additional layered structure
is automatically taken care of by \QSpace.

The remainder of this section summarizes frequently used
terminology, conventions, and acronyms that are extensively
used with \QSpace. For more elementary tensor-network
and symmetry-related definitions, it is advised to
briefly also glance over \Apps{acronyms} and \ref{app:TNsym}
before reading on.

\subsection{\QSpace stores tensors (not state spaces)}
\label{idx:nolegs}

\QSpace is a tensor library.
Therefore it represents tensors (blobs in a pictorial
representation), but not state spaces per se (individual
isolated legs or lines in tensor networks, i.e., indices).
\QSpace operates on
tensor objects that have indices (legs) attached.
From the point of view of symmetries then, e.g.,
when incoming charge needs to be preserved
into outgoing charge, this necessitates at least
two indices.
In this sense, \QSpace necessarily represents tensors of
\idx[rank]{rank} $r\geq 2$.
There are only two trivial exceptions that permit rank $r<2$.
These are:

\paragraph{\QSpace tensor with rank $r<2$}
\label{idx:rank01}

\begin{itemize}
\item rank 0:
when computing expectation values or overlaps,
the resulting object is fully contracted,
i.e., has no more open legs.
Hence this represents a tensor of rank $r=0$.
For an example, see \eqref{eq:rank0} where a tensor
is fully contracted with itself.

\item rank 1:
the only rank-1 tensor that is permitted from
a symmetry point of view is a tensor with 
a single leg in the vacuum symmetry sector
i.e., with \idx[qlabels]{symmetry labels} $q=0$
(if there is no charge coming in, there
is no need for it to leave). 
For an example, see~\eqref{eq:rank1}.
A \mbox{rank-1} tensor can
occur naturally out of contractions, e.g.,
when fully contracting a tensor of rank $r$ on 
all its $r$ indices with a tensor of rank $r+1$.
On general grounds, the result of such a contraction
can only have a non-zero contribution
in the $q=0$ symmetry sector on the single remaining
open leg.

\end{itemize}

\paragraph{Referencing state spaces}

To specify a particular state space
in \QSpace, this needs to be done by referring to
respective legs on an existing tensor
(e.g., see \idx[getIdentityQS]{\src{getIdentity}}).
Since objects are referenced
when handed over to routines (both Matlab
and MEX routines alike), this is efficient
in the sense that tensors do not 
need to be copied for this purpose
[cf. Matlab's just-in-time (JIT) concept
where copies are only generated once an
object is actually altered].
Nevertheless, if desired, the minimal setting
that gets closest to storing a state space
on leg $l$ of \QSpace \src{X} of some rank $r\geq l$,
is by defining the identity tensor on that leg
in compact diagonal form: \src{diag(getIdentity(X,l))}.
This also inherits the dimensionality 
of each symmetry sector, as well as the 
index tag (\idx[itags]{\itag}) for that leg.

\subsection{Composite index for state spaces}
\label{idx:states}

In the presence of symmetries, any state space, e.g.,
as associated with legs of tensors,
needs to be organized into symmetry sectors.
When exploiting non-abelian symmetries, any state~$s$
must be identifiable by answering the following questions:
(1) what symmetry sector does it belong to? Within that symmetry
sector, (2) which multiplet does it belong to? And in that
particular symmetry multiplet, (3) what state does it represent?
Hence for the description of any state space $s$
in the presence of non-abelian symmetries,
this  naturally acquires a composite index structure 
that contains the answers to the three questions above
\cite{Wb12_SUN},
\begin{eqnarray}
    |s\rangle  \equiv | q n ; q_z\rangle
\ .\label{eq:composite-index}
\end{eqnarray}
Here $q$ denotes the combined set of symmetry labels 
that specify (1) the symmetry sector for all symmetries included.
The index $n$ then specifies (2)
the multiplet in that symmetry sector,
and $q_z$ spans (3) the internal structure of 
the individual multiplets in this symmetry sector $q$.
To emphasize that the symmetry labels can refer to
any symmetry or combinations thereof,
\QSpace generally uses the letter $q$
for \idx[qlabels]{symmetry labels}, as in
\texttt{\Emph{Q}Space} for quantum symmetry spaces.
For the particular case
of a single \idx[SU2:qlabels]{SU(2)} symmetry,
for example, $(q,q_z) \equiv (2S,2S_z)$
reflects the spin quantum numbers.

The composite index structure in \eqref{eq:composite-index}
can also be trivially applied to abelian symmetries.
Since each abelian `multiplet' has only one state,
this also represents the `\idx[max-weight]{maximum weight}' state.
Hence the $q_z$ label may be promoted to 
the multiplet label, i.e., $q \equiv q_z$.
For the case of all-abelian symmetries, the redundant
$q_z$ labels together with their respective trivial
CGTs can be skipped, altogether.
In \QSpace this explicitly happens for the case
of all-U(1) symmetries which results in an empty \src{X.info.cgr} 
field in the \hsec[QS:struct]{\QSpace data structure}
for \QSpace \src{X}
(this reflects the original setting as of \QSpace v1).

Within the \hsec[QS:struct]{\QSpace data structure}
[see \Eq{F:struct} and subsequent lines for an explicit example],
the symmetry labels $q$ in \Eq{composite-index}
for a \QSpace tensor \src{X} are stored
as rows in the matrices \src{X.Q\{l\}}
where $l=1,\ldots,r$ indexes the leg of the tensor.
There is a one-to-one correspondence of the rows
\src{X.Q\{l\}(\irec,:)} with the respective
RMT \src{X.data\{\irec{}\}} which describes the \irec'th
non-zero block in \QSpace \src{X}.
The index $n$ in the decomposition \eqref{eq:composite-index}
is not explicitly stored per se, but is implicit
as the index within the blocks of the RMTs
stored in the cell array \src{X.data}, e.g., when taking
some matrix element \src{X.data\{i\}($\ldots,n_l,
\ldots$)} with respect to leg $l$
[for an example, see discussion of output to \eqref{eq:SdotS}].
Similarly, $q_z$ in \Eq{composite-index}
corresponds to an implicit index within the CGTs 
that also needs to deal with {\it inner multiplicity}
\cite{Wb12_SUN}. The CGTs are tabulated and stored in
\RCR{\RCS/CStore} since fixed by symmetry.
Hence an actual \QSpace tensor \src{X}
only stores the relevant {\it reduced} data based on 
the multiplet structure $(q,n)$. Concerning the symmetry
related aspects ($q_z$) it suffices to only {\it reference}
the underlying Clebsch-Gordan data, referred to as
\idx[CGR]{Clebsch-Gordan references} (CGRs)
and stored in \src{X.info.cgr}. Overall,
this switches the description of a tensor from a
state-based to a multiplet-based setting.

\subsection{Direction of legs for tensors}
\label{sec:qdir}

All lines in a tensor network diagram are directed,
i.e, carry an arrow (e.g., see \Fig{arrows}; \cite{Wb20}).
This binary scheme of
whether an index is `incoming' (\qdir{+}) or `outgoing'
(\qdir{-}) from a tensor derives from, and thus is 
equivalent to the respective binary concepts of raised vs.
lowered indices, or contra- vs. co-variant indices, but also
bra- vs. ket-states. This binary concept is intrinsic
to tensors. For this reason, the direction of legs is
meticulously enforced and tracked for all tensors
in \QSpace. 

The semantical preference on the notion of direction is
mainly due to pictorial representations where directed lines
are easier to visualize. 
Besides, this also gives a simple pictorial representation
of the Einstein summation convention: a raised index 
that is summed over with a paired-up lowered index
simply corresponds to a well-defined direction of the line
that depicts that contracted index. This directed
line necessarily leaves from one tensor and 
and enters another.
From the perspective of a tensor, the related notation of using
\qdir{+}/\qdir{-} for directions of its legs, as also
used in the code this way, is motivated from the point of view of
symmetries: for example, for U(1) charge conservation, the
total combined incoming charge must equal the total outgoing
charge, i.e., $\sum_{\rm in} q_{\rm in} = \sum_{\rm out}
q_{\rm out}$ which may be trivially rewritten as
$+\sum_{\rm in} q_{\rm in} - \sum_{\rm out} q_{\rm out} = 0$.
This motivates that from the point of view of a
tensor, an incoming (outgoing) index adds (removes) charge,
and hence is associated with a
\qdir{+} (\qdir{-}), respectively.

\begin{figure}[tbh]
\begin{center}
\includegraphics[width=0.85\linewidth]{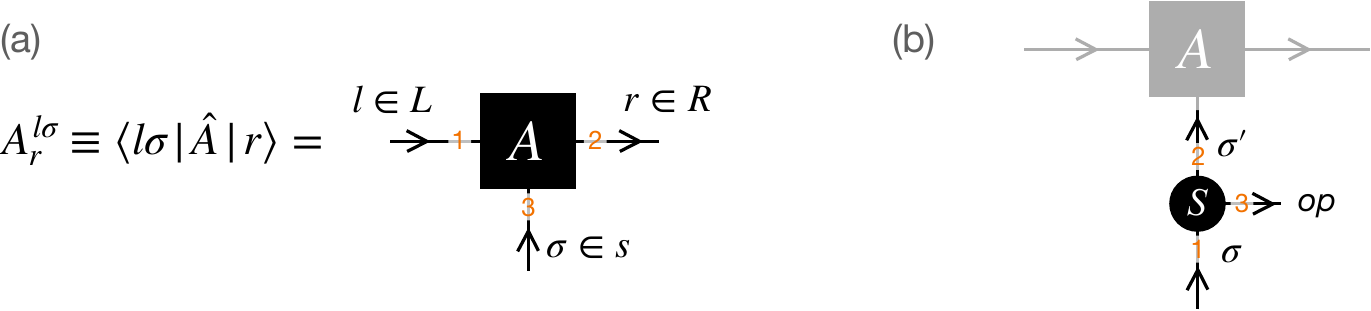}
\end{center}
\caption{
(a) Fusion of the local state space $\sigma_{(i)}$ by a tensor $A_i$ 
    as in an MPS.  Skipping the index $i$ for readability, the left
    state space $l\in L$, and the local state space
    $\sigma \in s$ are fused into the state space $r\in R$
    to the right of site $i$.
    Irrespective of leg directions on $l$ and $r$, 
    \idx[Atensor:LRs]{$A$-tensors} use an \src{\idx[Atensor:LRs]{LRs}}
    index order convention as indicated by the orange labels.
(b) (Irreducible) operator $S$
    like a spin operator as in \Eq{Sop}
    acting on local state space
     $\sigma,\sigma' \in s$. By convention, operators use
    ($\sigma$,$\sigma'$,op) index order convention
    as indicated by the orange labels. The \idx[irop]{\it operator index}
    (third leg) may be skipped and thus absent for scalar operators
    where this represents a trivial singleton dimension.
}\label{fig:Aop} 
\end{figure}

\subsubsection{Convention: Incoming leg to tensor is written
as superscript index} \label{idx:arrows}

Consider an \idx[Atensor]{$A$-tensor} that fuses a local state
space $s_{(i)}$, e.g., with a matrix product state (MPS)
$|A\rangle$ and local state
spaces $\sigma \in s_{(i)}$ for some site $i$ in mind
[\Fig{Aop}(a)]. For simplicity, 
the index $i$ is skipped as well as 
any other indices present on $A$ for the sake of the argument here.
The local identity operator is given by
\begin{eqnarray}
   \hat{\Id} &=& 
   \sum_{\sigma \in s}\ 
     \underbrace{|\sigma\rangle}_{\equiv | s _\sigma \rangle}
     \underbrace{\langle\sigma|}_{\equiv \langle s^\sigma|  } 
   \ \equiv \ | s _\sigma \rangle \langle s^\sigma|
   \ \ \equiv \ \ \ 
   \raisebox{-1.2ex}{\includegraphics[width=1.5in]{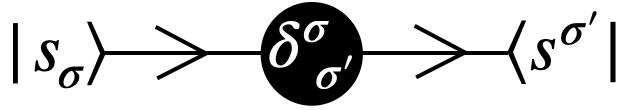}}
\label{eq:Id:sigma}
\end{eqnarray}
with implicit summation over pairs of raised
and lowered indices if not explicitly specified
(cf. Einstein summation convention).
The above starts with the convention that
(i) ket-states have lowered indices,
and hence are considered covariant.
Therefore bra-states are contra-variant and written
with a raised index.
Applying the identity in \Eq{Id:sigma} to $\hat{A}$,
one obtains
\begin{eqnarray}
  \hat{A} \ =\ 
   \sum_\sigma |s _\sigma \rangle
   \underbrace{\langle s^\sigma|\,\hat{A}}_{\equiv A^\sigma}
 \ \equiv A^\sigma |\sigma \rangle
\text{ .}\label{eq:Aop:sigma}
\end{eqnarray}
In a numerical context, one always has to use `matrix elements'
$A^\sigma$ that are obtained in a particular basis.
By convention, such tensor coefficient spaces are written
\idx[Op:hats]{without hats}. 
Now also with graphical depictions in mind, 
raised or lowered indices will be assigned directions.
Intuitively from the perspective of a tensor,
by convention, (ii) the index of the
local state space $\sigma$ is considered `incoming'
to the tensor $A$. Hence in \QSpace an {\it incoming} index
to a tensor coefficient space [as in $A^\sigma \equiv
\langle s^\sigma|\,\hat{A}$ in \Eq{Aop:sigma}]
corresponds to a projection onto a bra state, and is written as a raised index.
The identity in \Eq{Id:sigma} may thus be depicted
graphically as shown at its r.h.s.,
which also shows that a ket-state in itself is considered outgoing.
Eventually, via contractions, a lowered index needs to be
paired with a raised index, or equivalently,
a ket- with a bra-state, or
a co- with a contra-variant index,
or an outgoing with an incoming leg.

The incoming index in $A^\sigma$ in \Eq{Aop:sigma}
is contracted with the outgoing index of the
ket state $|\sigma\rangle$ [see also right term in \Eq{Id:sigma}].
As such this represents
the `operator' $\hat{A}$ with no open index,
thus written with a \idx[Op:hats]{hat}. It is more
of an abstract object that is independent of a particular basis.
In an actual numerical context, however, only matrix elements occur, 
such as $A^\sigma = \langle s^\sigma | \hat{A}$.
This way, the index $\sigma$ as in in $A^\sigma$ becomes an open,
i.e., uncontracted index.  Similarly, for example, 
the identity operator $\hat{\Id}$ as written in \Eq{Id:sigma},
has no open indices. However,
when expressed in a particular basis, the matrix elements are
$\langle s^\sigma | \hat{\Id} | s_{\sigma'}\rangle
= \delta^{\sigma}_{\ \sigma'}$. This `strips off' the ket
and bra from the r.h.s. in \Eq{Id:sigma},
resulting in \raisebox{-1.0ex}{%
\includegraphics[width=0.8in,trim = 5.2em 0ex 6em 0ex, clip=true]{fig_Id}} \ 
with open indices $\sigma$ and $\sigma'$.

Now when applying an operator $\hat{S}$ on the site
that is fused by the tensor $\hat{A}$, 
the \idx[matel]{matrix element} corresponds to the projection
onto the local state $\sigma$
[cf. \Fig{Aop}(b)],
\begin{eqnarray}
  \langle \sigma|\, \hat{S} \hat{A} &=&
   \sum_{\sigma'}
   \underbrace{
      \langle s^{\sigma}| \hat{S}|s _{\sigma'} \rangle}_{
      \equiv S_{\ \sigma'}^{\sigma}}
   \underbrace{\langle s^{\sigma'}|\hat{A}}_{\equiv A^{\sigma'}}
 \ \equiv\  
    S^{\sigma}_{\ \sigma'} \, A^{\sigma'}
\text{ ,}\label{eq:SA:indices}
\end{eqnarray}
having inserted the identity as in \Eq{Id:sigma}.
Since by convention the index $\sigma'$ is {\it incoming} to $A$,
it has to leave and thus be {\it outgoing} for the tensor
$S^{\sigma}_{\ \sigma'}$
via the projection to the ket-state $|s _{\sigma'} \rangle$.
Therefore $S^{\sigma}_{\ \sigma'} \equiv
\langle s^{\sigma}| \hat{S}|s _{\sigma'} \rangle$ has 
bra-state $\sigma$ as incoming, and the
ket-state $\sigma'$ as outgoing.
The order of the bra indices relative to the ket
indices is irrelevant, in that it can be trivially changed.
For physical operators, nevertheless,
by standard convention, bra indices are typically
listed before ket indices. Hence their \hsec[idx-order]{index order
convention} is $\langle 1 |\hat{S}|2\rangle
\equiv S^1_{\ \ 2} \equiv S^1_2$, and therefore 
\begin{eqnarray}
   S^{\sigma}_{\sigma'} \equiv
   S^{\sigma}_{\ \,\sigma'} \equiv
   \langle s^{\sigma}| \hat{S}|s _{\sigma'} \rangle
\text{ .}\label{eq:S:indices}
\end{eqnarray}
When following the direction of the indices
in a pictorial representation, \Eq{SA:indices} `flows'
from the left towards the right
(upwards in \Fig{arrows}). This is reverse to
how one would read such an expression, e.g., as part of an
expectation value from a `time-ordered' perspective, namely
right to left (downwards in \Fig{arrows}): one starts with the
ket-state $|A\rangle$ at the right,
and then applies particular operators in a well-defined
sequence, and in this sense `time order'.
This proceeds towards the left, until it meets a bra state at the
very left for the case of a matrix element or expectation value.

\subsubsection{Leg directions in \QSpace}
\label{idx:qdir}

The concept of raised or lowered indices applies to every
level of a \QSpace tensor. It applies to the tensor
as a whole, but also to all RMTs as well as CGRs individually.
The CGRs are references to sorted CGTs (these are tabulated
with \idx[CGT:sorted]{sorted $q$-labels} to avoid
proliferation of entries).
Bearing in mind that the relative index order of
raised relative to the lowered indices is irrelevant,
the raised and lowered $q$-labels in CGTs can be simply
grouped without paying attention to their relative order.
For example,
\idx[irop]{irops} have CGTs of the form $C^{q'}_{q q_{\rm
op}} \equiv \langle q' | C_{q_{\rm op}}| q\rangle$.
For the case of a scalar operator with a singleton
operator dimension having $q_{\rm op}=0$, this
can be reduced to $C_q^q$. Standard \idx[CGT]{CGCs} have $C^{q_1
q_2}_{q_3}$ as they fuse $(q_1,q_2) \to q_3$.
\hsec[1j]{\onej-symbols}
have CGTs $C^{q\bar{q}}$ or their conjugate $C_{q\bar{q}}$
which fuse $(q,\bar{q}) \to 0$ or vice versa,
with $\bar{q}$ the \idx[dual]{dual} irep to $q$.
Again since the relative index order of raised relative
to the lowered indices does not matter, this
documentation also uses the alternative notation
\begin{eqnarray}
   (q_1^{\,} \ldots q_{l}^{\,} q_{l+1}^\ast \ldots q_{r}^\ast)
   &\equiv& (q_1^{\,} \ldots q_{l}^{\,}|q_{l+1} \ldots q_{r})
   \ \equiv\ (q_1^{\,}, \ldots, q_{l}^{\,};q_{l+1}, \ldots, q_{r})
\label{eq:qlabels:tensor}
\end{eqnarray}
for a rank-$r$ CGT  $C^{q_1 \ldots q_{l}}_{q_{l+1} \ldots
q_{r}}$, where the asterisk \src{*} on the l.h.s.
explicitly denotes an outgoing index (see also \idx[conj]{tensor
conjugation}). By listing all $l$ incoming legs first,
followed by the remaining $r-l$ outgoing legs with $l \in [0,r]$,
this permits the notation on the r.h.s.
This index ordering is adopted in the \RCR{\RCS/CStore}
to avoid a proliferation of CGTs.
On top of this simple grouping of indices, the
\idx[CGT:sorted]{sorted CGTs} in the \RCS also have the
$q$-labels non-trivially sorted
in a lexicographical manner within the incoming
as well as the outgoing indices.

\QSpace permits an \hsec[idx-order]{arbitrary permutation}
of legs of a tensor, with an example shown in \Eq{F2:perm}.
For this the information on in- or outgoing cannot be
simply grouped as in \Eq{qlabels:tensor}, but needs
to be specified explicitly with each leg.
It is stored with the \QSpace tensor as a whole
with the tensor's \idx[itags]{\itags}. By convention,
a \idx[itag:markers]{trailing} \str{*}
in the string \src{X.info.itags\{$l$\}}
indicates an outgoing index for leg $l$. If there is 
no trailing \str{*}, the index is considered ingoing, instead.
\QSpace insists that directions are specified with the
\itags globally for a tensor, i.e., the field
\src{X.info.itags} must be set, at the very minimum
containing the conjugate flags for every leg.
Having the directions
specified with the \itags is important for the
purely abelian setting, since in that case
\src{X.info.cgr} is not required and can be set empty.
In the presence of non-abelian symmetries, however,
the leg directions specified with the \itags is
redundant since this information can be derived
from each individual \idx[CGR]{CGR}.
The $q$-directions specified with the \idx[itags]{\itags} 
via trailing asterisks for the tensor overall 
are thus fixed for a given tensor and must be
preserved when modifying its \itags.
To ensure this, \QSpace provides the function 
\idx[QS:setitags]{setitags()},
with an example shown in \Eq{itags:A}.

The \Emph{$q$-direction} for CGTs as referenced by CGRs 
are collected and encoded as plain strings
\src{qdir=}\str{$\eta_1 \eta_2 \ldots \eta_r$}, with $\eta_i \in
\{\qdir{+},\qdir{-}\}$.
For example, see \src{X.info.cgr(i,j).qdir}
for a \QSpace~\src{X}.
The $q$-directions may also be derived in string form
from the \itags of a \QSpace tensor
via \src{qdir=getqdir(X,\str{-s})}.
For example, this yields the strings
\begin{eqnarray}
  && \str{\qdir{+-}}      \qquad \text{for \idx[scalar:op]{scalar} operators}
\notag \\ 
  && \str{\qdir{+{-}-}} \quad\,\ \text{for \idx[irop]{irops}}
\label{eq:qdir:ops} \\
  && \str{\qdir{+{-}+}} \quad\,\ \text{for \idx[Atensor]{$A$-tensors}}
  \text{, etc.} \notag
\end{eqnarray}
\QSpace also uses this string notation to organize much
of the folder structure for each non-abelian symmetry
in the \RCR{\RCS}.
Having sorted CGTs there, this groups all \src{+}'s
to the front.

\subsection{Conventions in pictorial representations}
\label{sec:pictorial}

\begin{figure}[tbh]
\begin{center}
\includegraphics[width=0.9\linewidth]{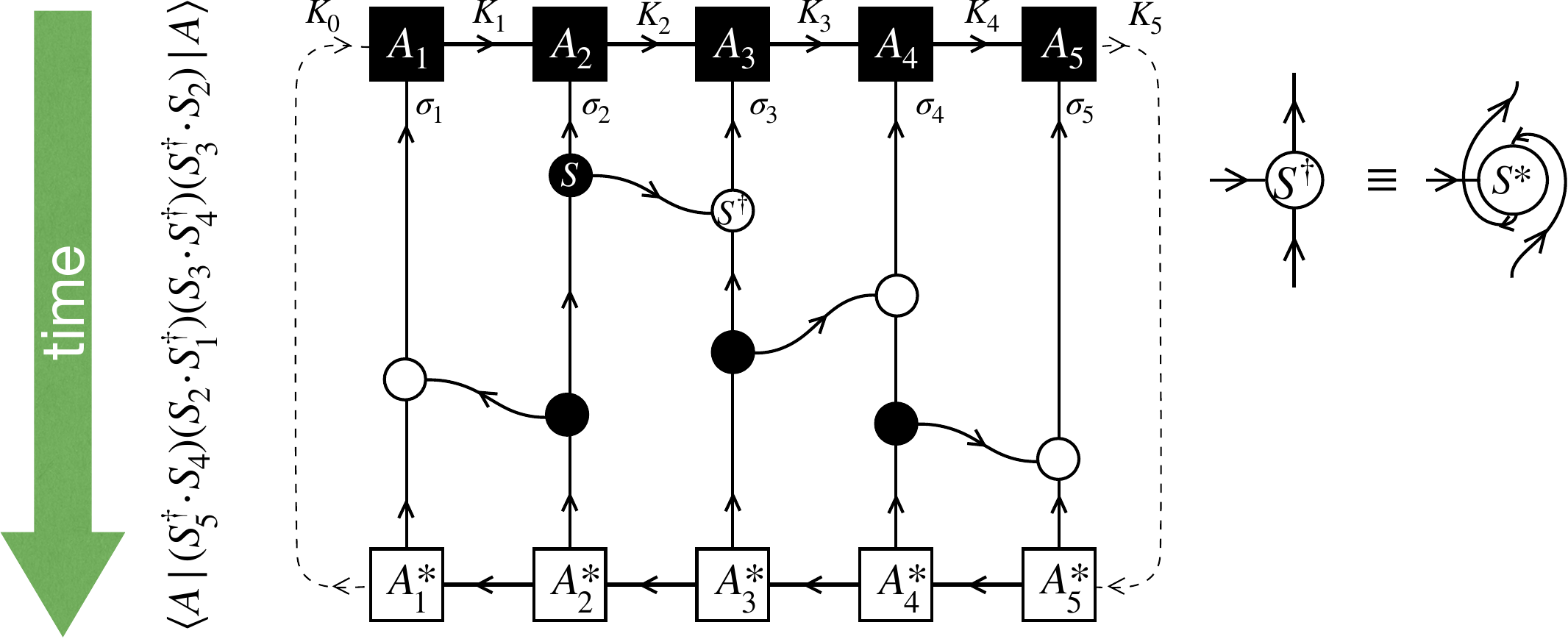}
\end{center}
\captionsetup{singlelinecheck=off}
\caption{
   Conventions in tensor network diagrams based
   on an MPS-based example for the expectation value
   of some product of interactions 
   $\hat{S}_i^\dagger\cdot \hat{S}_j$ (pairs of connected circles)
   like for Heisenberg spins.
   The corresponding mathematical expression is
   specified to the left and
   vertically aligned with the various terms in the pictorial
   representation.
   The definition of the operator $S^\dagger$
   is depicted to the right
   in terms of the \idx[conj]{conjugate tensor} $S^\ast$.
   It is the mirror image of $S$
   reflected horizontally and leg directions reverted,
   such that the operator index enters $S^\ast$
   (and thus also $S^\dagger$) from the left,
   but exits $S$ from the right
   (see also \Fig{F1F2} for a similar context).
   The state is given by $| A\rangle \equiv \prod_i^L A_i$
   (ket-state described by the MPS at the top)
   operating on the local state space $\sigma_i$ for sites
   $i=1,\ldots,L$, having $L=5$.
   Bond states, e.g., up to and including site $i$
   are denoted as $K_i$ (like `kept' many-body states
   up to iteration~$i$ in the presence of truncation).
   Here $K_0$ is assumed the left vacuum state
   with trivial \idx[dimension]{bond dimension} $D_0=1$.
   Similarly, $K_5$ describes the full MPS $|A\rangle$.
   Being a single many-body state, hence also $D_5=1$. 
   The outer contractions (dashed lines) are thus trivial.
   The direction of `time' (green arrow)
   indicates the order of operators
   applied to the initial ket-state $|A\rangle$ at the top.
}
\label{fig:arrows}
\end{figure}

Pictorial representations are very useful in describing
tensor network approaches and algorithms
\cite{Corboz10,Schollwoeck11,Orus14tns}.
For example, \Fig{arrows}
shows a tensor network diagram for
an expectation value of some product of
interactions $\hat{S}_i^\dagger\cdot \hat{S}_j$.
Based on this, general conventions in graphical
depictions of tensor networks with \QSpace
are as follows
\cite{Wb12_SUN,Wb20}:
\begin{itemize}
\item
   Tensors are denoted as some blob (box, circle, etc.).
   Each tensor has as many legs attached as it has indices,
   the number of which defines its \idx[rank]{rank}.
   The legs of tensors
   represent starting or end points of lines.
   
\item
   Every line in a tensor network carries an arrow:
   from the point of view of a tensor, arrows specify
   incoming/outgoing legs or, equivalently, raised/lowered indices. 
   A contracted line has no open ends: it starts from
   some tensor, and
   terminates in another (possibly also the same)
   tensor. Such a contracted line represents 
   a summed-over index where, by construction,
   a lowered index leaving some tensor needs to terminate
   as a raised index with some other tensor
   (cf. Einstein summation convention).

\item
   The tensor network diagram as in \Fig{arrows} is
   read top to bottom:
   starting from the ket-state $|A\rangle$ at the top,
   a set of operators, gates, or interactions are applied
   (round objects $S_i^\dagger\cdot S_j$).
   The order of the operators applied is relevant,
   up to trivial shifts along the vertical line they act upon,
   as long as they do not cross with other operators
   acting on the same line (state space).
   The diagram is terminated eventually by a
   bra-state $\langle A|$ at the bottom. Overall, this results
   in the desired matrix elements or expectation value.

\item
   By way of reading the diagram,
   this suggests an implicit `temporal' order
   top to bottom (big green arrow to the left),
   as defined by the particular order of operators
   applied onto $|A\rangle$. 
   With the intuitive convention that the local state space
   \hfig[Aop]{\it enters} a ket state, however,
   the `temporal' direction in \Fig{arrows} is opposite
   to the arrows shown with the lines that are associated
   with the local state space $\sigma_i$.

\item   
   When taking a ket-state $|A\rangle$,
   one implies its {\it coefficients} $\langle s^\sigma|A\rangle$ 
   with index $\sigma$. That index represents
   a projection onto a {\it bra}-state [cf. \Eq{Aop:sigma}].
   This reflects the general situation in numerical
   simulations that tensors are
   necessarily always cast into a particular basis.
   By attaching legs to a tensor in a tensor network
   diagram, one implies that it is cast
   into a coefficient space in some arbitrary but fixed basis
   that may be given a label (cf. \idx[itags]{\itags}).
   By convention the ket-state like $|\sigma\rangle$
   represents an outgoing index. Therefore
   the tensor coefficients projected
   onto a bra-state are reverse and thus ingoing.

\item 
   When a tensor is shown as a filled (empty) object
   such as a box, circle, etc.,
   its \idx[conj]{conjugate tensor}
   is shown as a mirrored empty (filled) shape, 
   respectively, for visual differentiation
   (e.g., see $A^\ast$, $S^\ast$ or $S^\dagger$ in \Fig{arrows}).

\end{itemize}

\subsection{Tensor conjugation (denoted by \str{*})}
\label{idx:conj}

The standard concept of {\it Hermitian conjugate},
or equivalently {\it conjugate transpose}, is well-defined
only for rank-2 tensors, i.e., matrices or operators.
For an \idx[irop]{irreducible operator} $S$, for example, 
its Hermitian conjugate \src{$S^\dagger$
= permute($S$,\str{21*})} \footnote{
The potentially missing \src{3} in the
\idx[perm]{permutation} \str{21*} $\to$ \str{213*}
is implied in \QSpace for trailing indices
that keep their position preserved,
i.e., act like an identity permutation.
This way, \src{S} may be a
rank-2 or rank-3 operator.}
consists of transposition on the first two indices
(\str{21}), and conjugation (the trailing \str{*}).
For a general tensor, however, transposition in 
itself is not well-defined, as it 
generalizes to a non-unique permutation
that is chosen depending on the context.
The situation is different, though, 
in pictorial representations where the generalization
of the transpose to arbitrary-rank tensors is
well-defined in that one simply takes the mirror image of
the tensor (cf. $A^\ast_i$ or $S^\ast$ in \Fig{arrows};
\cite{Schollwoeck11,Orus14tns,Bruognolo21}).
Nevertheless, this leaves the freedom to rotate
the tensor as a whole in the pictorial setting.
But then this holds for any tensor, conjugate or not.
More specifically, tensor conjugation thus
proceeds as follows \cite{Wb20}:
\begin{enumerate}
\renewcommand{\labelenumi}{(\arabic{enumi})}

\item draw the {\it mirror image} of the original tensor
in pictorial representations

\item complex conjugate the array entries 
   (relevant for complex RMTs only)

   \item revert arrows on all legs (i.e., swap raised/lowered
   indices)

   \item index spaces remain the {\it same}, 
   such that, e.g., also symmetry labels are left intact.
\end{enumerate}
Point (4) is added for emphasis only here concerning
symmetry labels.
For better visual differentiation of conjugate tensors
concerning point (1), one may furthermore switch
from filled tensor objects to empty (outlined)
ones, or vice versa (see conventions in \Sec{pictorial},
or $A^\ast_i$ or $S^\ast$ in \Fig{arrows} as explicit examples).

The concept of point (1) only applies to pictorial
representations, but not to the practical numerical context. 
Because the permutation underlying point (1)
is not well-defined, tensor conjugation in \QSpace adheres
to points (2-4) only. \QSpace preserves the index order
in tensor conjugations, unless specified otherwise
by the user (see \Sec{permute} for examples).
Point (1) reflects the generalization of the transposition
of matrices. Effectively, this fully reverts the order
of indices. But for tensors of arbitrary rank,
the starting point of what is
considered the first index, is not well-defined
unless one adheres to some arbitrary but fixed convention.
For example, there are many different ways 
where the mirror plane can be drawn.
In \Fig{arrows}, for example,  the $A_i^\ast$ (bottom)
are mirrored vs. their respective $A_i$ (top)
by a horizontal plane, whereas for the interactions,
the operator $S^\ast$ is mirrored 
by a vertical plane w.r.t. their paired up $S$ operator
(see definition of $S^\dagger$ to the right of \Fig{arrows}).

Due to (3), tensor conjugation
(or `Hermitian conjugate') where present must always
be explicitly specified and included in \QSpace,
even if the \idx[matel]{matrix elements} themselves in the RMTs
are not complex (point 2). This is important
since reverting directions is equivalent
to raising all lowered indices and vice versa.
This is an intrinsic concept of tensors that needs
to be respected for overall consistency.
From the symmetry point of view,
directions on legs frequently also indicate
types of orthonormalization. Hence by \QSpace insisting
on well-defined directions of legs throughout the tensor
network, point (3) also helps to avoid errors in the
coding of tensor network algorithms.

When conjugating a tensor, the action in steps
(3-4) will be also referred to as `conjugating' indices
or legs. For a given leg and symmetry sector $q$,
this is denoted by $q \to q^\ast$.
It needs to be differentiated from the
\idx[dual]{dual} representation denoted by $\bar{q}$.
The latter only comes into the picture when reverting
directions of individual legs by applying, i.e., contracting
\hsec[1j]{\onej-symbols}.
This then not only reverts the direction of the affected leg,
but also maps symmetry labels to their duals, $q \to \bar{q}$.
Hypothetically, if one were to explicitly revert all arrows
after tensor conjugation back to their original
direction by applying, i.e., contracting
the respective \onej-symbols onto every leg,
the resulting (\QSpace) tensor is generally different 
from the original one. Aside from having flipped symmetry labels
to their \idx[dual]{duals}, for the case of complex RMTs
the reduced matrix elements would still also
remain complex conjugated, bearing in mind
that \onej-symbols only have non-trivial structure
within the CGTs, but behave like an identity within the RMTs.

\subsection{\onej symbols (individual ireps)}
\label{sec:1j}

\begin{figure}[tb]
\begin{center}
\includegraphics[width=.95\linewidth]{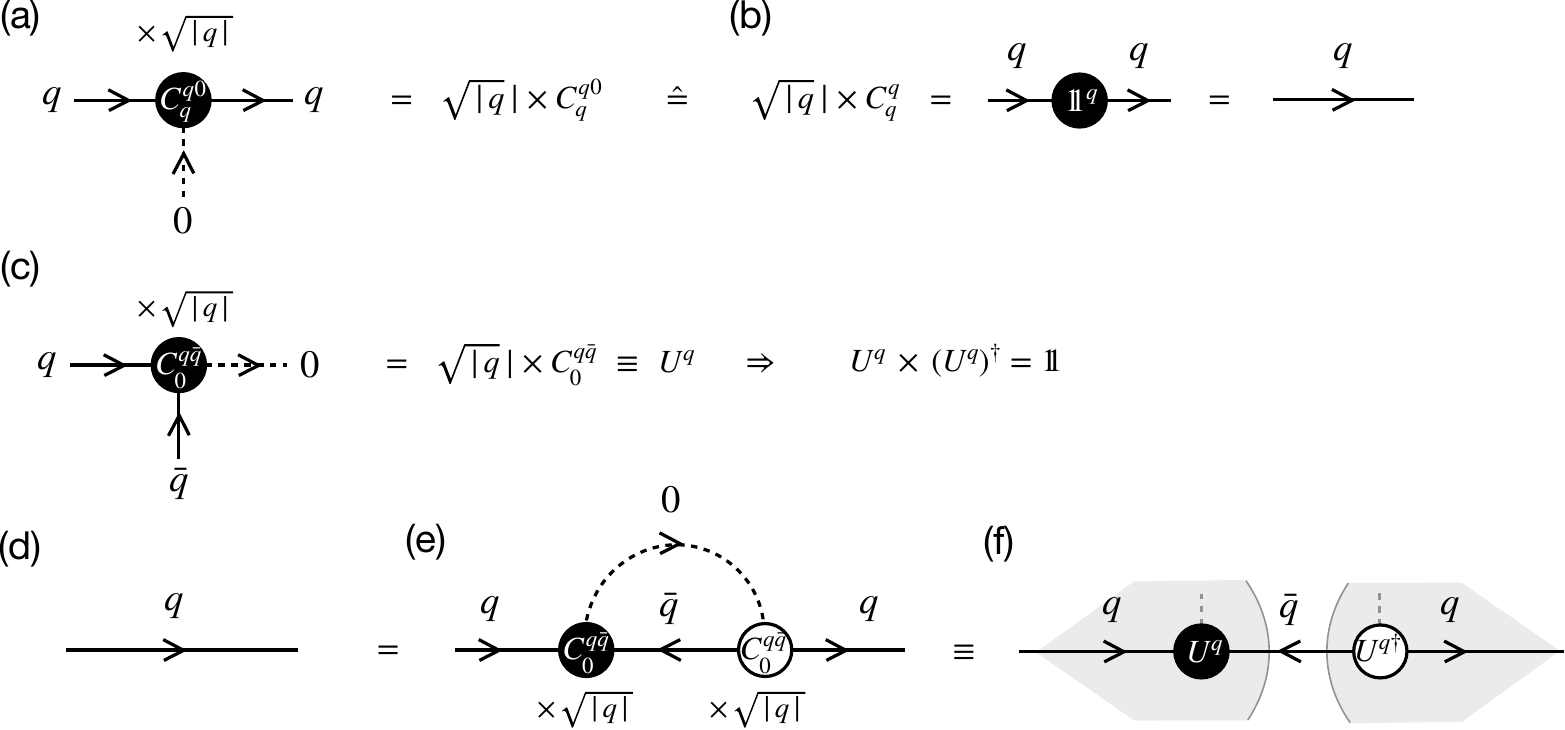}
\end{center}
\caption{Identity and \onej symbols --
(a) Fusing a multiplet $q$ with a scalar (singleton $0$)
    trivially leads to the same multiplet $q$. Hence the
    underlying CGT must be the identity $\Id^q/\sqrt{|q|}$ up to the
    \hsec[CGT:norm]{CGT (ortho)normalization convention}.
    Since the singleton index for the scalar irep ($0$)
    can be skipped, the same holds for rank-2 CGTs as in
(b). The identity $\sqrt{|q|}\, C^q_q = \Id^q$ can be 
    trivially inserted
     into or removed from any directed line.
(c) Construction of the \onej symbol $U^q$
    (`one $q \to j$' symbol \cite{Derome65,Butler75})
    from a standard rank-3 CGC $C^{q\bar{q}}_0$
    up to normalization
    involving the \idx[dual]{dual} representation $\bar{q}$.
    As apparent from the construction in (d-f), \onej symbols
    represent orthogonal matrices (unitaries). 
}\label{fig:1j} 
\end{figure}

In the literature of SU(2), $3j$ symbols refer
to standard CGCs  $(j_1 j_2|j_3)$ that fuse angular
momenta $(j_1,j_2) \to j_3$. 
Within \QSpace, symmetry labels are generally written as $j \to q$
to emphasize that these may represent
symmetry labels for a general (non-)abelian symmetry.
Hence a standard CGC is written as rank-3 \idx[CGR]{CGT}
$C^{q_1 q_2}_{q_3}$.
Now for the case that two multiplets $q_1$ and $q_2$
fuse into the scalar irep $q_3=0$ [\Fig{1j}(c)],
the respective CGT has a \idx[singleton]{singleton dimension}
as third index that may be skipped.
This case is possible if and only if $q_2$ is the dual
representation to $q_1$, i.e., $q \equiv q_1 = \bar{q}_{2}$. 
This results in the CGT $C^{q\bar{q}} \equiv C^{q\bar{q}}_0$
which exists and is unique for every irep $q$,
since no outer multiplicity can occur for CGTs of rank $\leq 2$,
the CGTs $C^{q\bar{q}}$ are unique up to the standard
CGC sign convention. In \QSpace, these particular
rank-2 CGTs $C^{q\bar{q}}$ are automatically computed
and stored for every new irep $q$ that is generated in a
tensor product decomposition for non-abelian symmetries
without necessarily performing the full decomposition
of $(q, \bar{q})$ (see Ref.\, \cite{Wb20} for details).
For abelian symmetries, the \onej symbols can be 
trivially generated on the fly with no need to store.

Importantly, every $C^{q\bar{q}}$ exactly also represents
an orthogonal transformation up to a 
normalization factor [\Fig{1j}(c)],
\begin{eqnarray}
   U^{q} \equiv \sqrt{|q|}\, C^{q\bar{q}}
   \qquad \text{ (\onej symbol)}
\text{ ,}\label{eq:1j}
\end{eqnarray}
where $|q|$ represents the state space dimension
of irep $q$, such that $U^q U^{q\dagger} = U^{q\dagger} U^q = \Id$
(bearing in mind that with all CGTs real,
the dagger is equivalent to a simple transpose).
These $U^{q}$ can be used to revert arrows
on lines in a tensor network state, 
as sketched in \Fig{1j}(d-f) \cite{Wb20}
from the symmetry perspective of CGTs [in practice,
reverting arrows typically also affects the RMTs,
e.g., when orthonormalizing state spaces in the process of
shifting the \idx[OC]{orthogonality center} (OC), etc.]. 
By inserting the identity $UU^\dagger$
into a fully contracted line for irep $q$ 
in between, say, tensors $A$ and $B$,
splitting this product, and contracting
$U$ and $U^\dagger$ onto $A$ and $B$, respectively
[indicated by the shaded gray arrows in the background
of \Fig{1j}(f)],
this results in a (i) reverted arrow on that line,
and also on the respective legs of
the tensors $A$ and $B$. Furthermore, by construction,
this simultaneously (ii) switches the symmetry sector
on that line to the dual representation, $q \to \bar{q}$.
Because of its usefulness, $U^{q}$ is given
a name. Since it only depends on a single symmetry label
$q$ [or `one $j$' symbol for SU(2)] with $\bar{q}$ simply inferred,
it is referred to as \onej-symbol with reference to the
literature of SU(2) \cite{Derome65,Butler75,Wb20}.

\paragraph{Caveat self-dual ireps}
\label{idx:caveat:1j}

In the construction of the \onej-symbol as rank-2 tensors,
the singleton dimension in the scalar representation $0$
[black dashed line in \Fig{1j}(e,f)] was skipped.
This results in a caveat, in that the truncated `stem'
of this singleton dimension
[remnant of the black dashed line in \Fig{1j}(f)]
needs to emerge on the  {\it same} side for both
$U^q$ and $U^{q\dagger}$, here
emerging at the top. Alternatively, one could have
closed the black dashed line in \Fig{1j}(e) at 
the bottom.

More specifically, the caveat concerns self-dual ireps
here, $q=\bar{q}$, like all ireps in SU(2).
In that case the \onej-symbol $U^q$
becomes indistinguishable from $U^{\bar{q}}$
in terms of symmetry labels and $q$-directions,
despite that they may differ by a sign \cite{Wb20},
since $U^q \equiv C^{q\bar{q}} = C^{qq}$,
yet $U^{\bar{q}} \equiv C^{\bar{q}q} \equiv (C^{q\bar{q}})^T = 
(C^{qq})^T = \pm U^{q} $. The last sign is negative,
for example, for all half-integer spins for SU(2).
Hence when inserting $U^{q} U^{q\dagger} = \Id$
into a contracted line of a tensor network
for the sake of reverting its arrow,
proper care must be taken that the {\it same} object
is used twice, once with, and once without the dagger.
Otherwise, this can give rise to erroneous signs.
As a safeguard in that regard, trailing primes are
introduced with \itags to \idx[markers]{mark dual}
state spaces, as in $C^{q\bar{q}}$ even if $q=\bar{q}$.

\subsection{Identity operator vs. \onej tensor and marker
for dual state space}
\label{sec:1j:tensor}

The \onej-symbol was introduced above
for a particular multiplet $q$ in some symmetry.
Yet it can be easily generalized to cover
full state spaces as implemented with
\idx[getIdentityQS]{\src{getIdentity}}.
Based on its origin,
this will be also referred to as \onej symbol
or \onej tensor, nevertheless.
The \onej tensor is closely related to the identity
operator [cf. \Fig{1j}]. Hence both of them 
refer to the state space associated with a particular
leg of a tensor and are also  obtained by the same
\QSpace routine \idx[getIdentityQS]{getIdentity}.
See \Sec{1j-markers} for detailed examples.
To start with, both objects are of rank $2$.
However, while the identity operator is a plain scalar
operator, the \onej tensor is not an operator
per se, as it has both of its indices incoming (or both
of them outgoing for its conjugate). Therefore it is
also not necessarily block-diagonal, but rather
anti-block-diagonal with respect to its symmetry sectors.

The identity operator $E$
as obtained by \src{\idx[getIdentityQS]{getIdentity}}
adheres to the index order convention of \idx[irop]{operators}
as in \Fig{Aop}(b). By having a scalar rank-2 operator here,
$E$ has one in- and one out-going index.
The \onej tensor is obtained by simply adding the
option \str{-0} (`fuse into $q=0$')
to the same call to \src{getIdentity} otherwise.
By contrast to the identity operator,
this returns a unitary \onej-tensor $U$ on the same state space,
yet with both indices ingoing. It is real and therefore
satisfies $UU^{T} = UU^{\dagger} = \Id$.
The \QSpace $U$  returned by \src{getIdentity}
has exactly the same RMTs as the identity operator.
However, instead of identity CGTs, it
references the respective \onej CGTs required
for a given state space, properly normalized via the respective CGRs. 
In this sense, the \onej-tensor $U$ defined over an 
entire state space of a leg of a tensor is also frequently
referred to as a \onej symbol itself.

\paragraph{Marking \itag of dual state space}
\label{idx:markers}

When calling \src{U=getIdentity(A,l,\str{-0})}
this returns the full state space as present in \QSpace A
on leg $l$ (first index in $U$) with its dual 
on the second index in $U$. To safeguard against
the \idx[caveat:1j]{above caveat}, the \itag 
of the \Emph{dual state space} on the second index is 
\idx[itag:markers]{marked with a trailing prime} (\src{'}).
After all, the dual state space is {\it different}
from the original state space, and hence its
\itag also should be differentiated.
The trailing marker character serves as a toggle:
reverting the arrow twice returns to the original \itag.
Therefore adding a prime to an existing trailing prime,
annihilates both of them,
i.e., \src{foo{'}{'} $\to$ foo}. So if by the history
of an algorithm, the input space already has
a mark on its \itag, then that mark is removed
for the dual space on the second index.
Detailed examples are given and discussed 
in \Sec{1j-markers}.

\subsection{Generating scalar (interacting) Hamiltonian terms from irops}
\label{sec:SdotSd}

A Hamiltonian represents a scalar operator.
Hence from the point of view of symmetries,
all terms constituting a Hamiltonian term must be scalars
themselves. With this in mind, general two-body interaction terms
in a Hamiltonian can always be written as or built
from scalar bilinears $\hat{S}_i^\dagger\cdot \hat{T}_j$
acting on sites $i$ and $j$, also permitting $i=j$.
Typically, $\hat{S} = \hat{T}$, with pictorial examples
shown in \Fig{arrows}, and explicit \QSpace contractions
to describe such pairwise interactions provided 
in \Eq{FdotF} or \Sec{2sites},
e.g., \Eqs{SdotS}--\eqref{eq:FdotF:T12} together with \Fig{F1F2}.

From a symmetry point of view, 
the bilinear $\hat{S}_i^\dagger\cdot \hat{T}_j$
requires (i) that the operators $\hat{S}$ and $\hat{T}$
are \idx[irop]{irops}, or more generally operators.
In the case of a pair of irops,  
(ii) these must transform according to the  {\it same}
operator irep $q_{\rm op}$, as this is summed over (contracted)
in the dot-product $\hat{S}_i^\dagger\cdot \hat{T}_j$.
In addition, for the contraction to be valid, the index
must be out-going from one irop, and incoming to the
other irop. Therefore the general natural structure includes the
\idx[conj]{conjugate tensor}, i.e., always makes explicitly use of
the dagger, as in  $\hat{S}_i^\dagger\cdot \hat{T}_j$. The dagger
may equally well be written with the second operator, instead.
For the example, consider the SU(2) spin operator written
as the irop that transforms like an $S=1$ multiplet
\begin{eqnarray}
  S &=& \begin{pmatrix}
  -\tfrac{1}{\sqrt{2}} S_+ \\ S_z \\ \tfrac{1}{\sqrt{2}} S_-
  \end{pmatrix}
\label{eq:Sop}
\end{eqnarray}
where the minus sign with the first component originates
from the application of raising or lowering operators
\cite{Wb12_SUN}.
The inverse square root factors 
ensure the simple dot-product structure for 
isotropic spin interactions,
having $\hat{S}_i^\dagger\cdot \hat{S}_j
= \frac{1}{2}( \hat{S}_{i+}\hat{S}_{j-} + {\rm H.c.})
+ \hat{S}_{iz}\hat{S}_{jz}
= \hat{S}_{ix} \hat{S}_{jx}
+ \hat{S}_{iy} \hat{S}_{jy}
+ \hat{S}_{iz} \hat{S}_{jz}
$.
The components of the spin operator in \Eq{Sop}
are evidently non-hermitian,
and thus do require the dagger with
$\hat{S}_i^\dagger\cdot \hat{S}_j$.
For general operators,
if the scalar result out of $\hat{S}_i^\dagger\cdot \hat{T}_j$
is non-hermitian [e.g., as is the case for
fermionic hopping described by $S \to F$
with $F$ some set of fermionic annihilation operators;
see \Eq{FdotF:T12} and subsequent discussion],
then the Hermitian conjugate
$(\hat{S}_i^\dagger\cdot \hat{T}_j)^\dagger
= \hat{T}_j^\dagger\cdot \hat{S}_i$ needs to be added
if part of a Hamiltonian.
Hence a typical Hamiltonian can be written
as a sum of scalar operator contributions,
\begin{eqnarray}
   \hat{H}
   = \sum_{i\alpha} \varepsilon^{\alpha}_{i} \hat{n}_{\alpha i}
   + \sum_{i j, \alpha} V_{ij}^\alpha\ 
   \hat{S}_{\alpha i}^\dagger\cdot \hat{T}_{\alpha j}
   \ \ [+\ {\rm H.c.}]
\label{eq:H:gen}
\end{eqnarray}
which may be further extended to also include products
of terms such as the above.
Here $\hat{n}_{\alpha} \equiv \hat{c}_\alpha^\dagger \cdot
\hat{c}_\alpha$ describes some set of local scalar operators,
and $\hat{S}_{\alpha}$ and $\hat{T}_{\alpha}$ some set of
local irops, typically having $\hat{S}=\hat{T}$
for a particular interaction. The Hermitian
conjugate is required for all interactions
whose operator irep $q_{\rm}$ is not self-dual.
For example, it is not required for SU(2) spin
interactions $\hat{S}_i^\dagger\cdot \hat{S}_j$.
It is required for fermionic hopping, though,
when using U(1) charge symmetry,
since the annihilation operators $F$ transform
like charge $q=-1$ (they reduce charge by $1$),
whose dual are creation operators with $\bar{q}=+1$.

Taking the irop $S$, its daggered version $S^\dagger$ 
could be explicitly constructed, in principle:
the tensor $S$ would have to be \idx[conj]{conjugated}
which reverts all legs, including the \idx[irop]{irop index}.
By subsequently applying a \onej symbol $U$ onto the irop
index 3 and permuting indices 1 and 2, this brings the irop
$\tilde{S}\equiv U S^\dagger$ back to canonical
\QSpace form for irops, with the crucial difference,
that the irop now transforms according to $\bar{q}$.
In practice, such a transformation to obtain
a canconical form of $S^\dagger$ by using \onej symbols etc.
is never required when computing $\hat{S}_i^\dagger\cdot \hat{T}_j$
[e.g., see \Fig{F1F2} and corresponding text].
Much to the contrary, it would be rather impractical
and prone to errors.
Having both, $S$ and $\tilde{S}\equiv S^\dagger$
in canonical irop form, the irop index is outgoing
in both tensors. This would prevent a simple contraction
as in $\tilde{S}\cdot S$ on the irop index,
because leg directions are incompatible.
Considering $\tilde{S}^\dagger\cdot S$, instead,
this would be contractible in principle for self-dual $q_{\rm op}$,
such as for the SU(2) spin operator, but wrong in most cases,
in the sense that it gives a different operator
$\tilde{S}^\dagger\cdot S =
(SU^\dagger)\cdot S \neq  S^\dagger \cdot S$.

Hence there shall never be a need to explicitly construct
$\tilde{S}\equiv S^\dagger$ as canonical irop based on \onej symbols,
etc., when computing interaction terms in a Hamiltonian.
What is solely required is to contract the \idx[conj]{conjugate}
tensor of $S_i$. This can be indicated on the fly as an option
when calling \src{contract} itself, while ensuring that
the `transposed' indices are correctly contracted.
The latter is also dictated by the directions of the legs
[e.g., see \Fig{F1F2} and corresponding text].
Hence much of this is taken care of automatically
when using auto-contraction based on \itags.

\subsection{Compact symmetry labels for non-abelian
symmetries} \label{sec:compact:qs}

The motivation of a `compact' notation for
\idx[qlabels]{symmetry labels ($q$-labels)}
of non-abelian symmetries is mainly for
readability purposes, e.g., to have well-aligned
displayed tables, but also a means for a
simple standardized notation in text.
The compact notation also reflects the 
situation in practice in the code, where
an $n$-tupel of labels is just a vector
of numbers without any white space or separators.
The contact notation permits a lean concise
specification of symmetry labels. It is specific
to non-abelian symmetries since ireps 
for any abelian symmetry are always described
by a single symmetry label only (see \idx[rsym]{symmetry rank}).
Therefore plain signed integer notation is used
for abelian symmetries. This is also the only setting
where negative symmetry labels $q<0$ can occur.

\paragraph{\QSpace label convention for SU(2): $q=2S$}
\label{idx:SU2:qlabels}

The concept of compact symmetry labels
for non-abelian symmetries relies on symmetry labels 
that are (i) non-negative and (ii) integer to start with.
This is the case for all
simple Lie algebras, except for the half-integer spins in SU(2)
for historical reasons.
The latter `exception', however, is only by convention.
Hence \QSpace adopts the alternative labeling convention
for SU(2)
which is fully aligned with the labeling scheme
for SU($N>2$) ireps more generally,
namely by choosing the SU(2) multiplet
label $q=2S \in \mathbb{N}_0$. This is consistent
with the symmetry labels for SU($N$) in general,
in that they specify Young tableaus
[e.g., see \Eq{YTableau}]: an SU(2) spin $S$ has
a Young tableau of a single row with $q=2S$ boxes.
This convention for SU(2) also permits the usage of plain
\hsec[QS:int]{integer labels} for symmetry labels, and thus avoids,
e.g., the need to print half-integers. Half-integer spins
$S$ then simply map to odd integers for $q=2S$.

As a corollary, if an SU(2) symmetry
gets broken down to U(1), the symmetry labels
become $q = 2S_z \in \{-2S,-2S+1,\ldots,2S\}$,
thus also ensuring integer labels in the U(1) 
context for the case of half-integer spins.
For example, a spin $S=1/2$ then acquires
the U(1) spin labels $q_S \equiv 2S_z \in \{-1,1\}$.

\paragraph{\QSpace label convention for U(1) charge}
\label{idx:U1:labels}

From a practical point of view, it is desirable
in tensor network simulations to keep symmetry labels
around $q\sim 0$. This way the low-energy symmetry
sector remains the same with increasing system size
also towards the thermodynamic limit. 
This avoids a `running' symmetry label 
that keeps growing proportional to block size.
For counting particle number (charge), it is therefore
desirable to count particles relative to the
average filling, i.e., $q \equiv n-n_0$ with $n_0$
the average filling per site for the entire system,
typically half-filling.
For example, for a single
spinful fermionic level, the total occupation 
relative to half-filling $n_0=1$ is given by the sectors
$q_C \equiv n-n_0 \in \{ -1, 0, 1\}$, corresponding
to empty, half-filled, and fully occupied,
respectively. This reflects the charge symmetry labels
as returned by \hsec[gls]{getLocalSpace}.

For the description of an odd number of flavors, however, 
such as a single spinless fermionic level,
the prescription above with a focus on `relative to half-filling'
would result in half-integer labels for charge.
Hence in the case of an odd number of flavors per site,
where $n_0$ for half-filling becomes a half-integer,
\QSpace adopts the convention $q_C \to 2(n-n_0)$,
which includes a factor of $2$
for the sake of having integer symmetry labels for charge.
For a single spinless fermionic level, therefore
\hsec[gls]{getLocalSpace} returns $q_C \in \{ -1, 1\}$.
As for the case of \idx[SU2:qlabels]{SU(2) labels} above,
this convention of applying a factor $2$ is mainly
for convenience and readability, yet also for consistency across
symmetries. For example, SU(2) particle/hole symmetry always takes
the viewpoint of particle number relative to half-filling.
In this sense, when breaking down an SU(2) particle/hole symmetry
to U(1) charge, $q_C \to 2(n-n_0)$ is consistent with 
the \QSpace convention $q_S \to 2S_z$ for spin.

On general grounds, U(1) labels may undergo an arbitrary
linear map $n \to \ q(n) = b n-a$ with $a,b \in \mathbb{N}$.
Thus if the desired target value for the average filling 
$n_0 = a/b > 0$ is rational, this ensures that the 
\hsec[QS:int]{integer} charge labels $q \in \mathbb{Z}$
remain centered around $q\sim 0$ when increasing block
or system size. Since this is a rather specialized
setting, though, it is not implemented in
\hsec[gls]{getLocalSpace} as is. Nevertheless, one
may simply \hsec[gls:x]{tweak its output} in this regard.

\paragraph{Compact alpha-numeric labeling scheme}

By \QSpace convention, all symmetry labels for non-abelian
symmetries are compact non-negative integers.
Therefore any set of $q$-labels
$q=(q_1,q_2,\ldots,q_\rsym)$ with $q_i\in \mathbb{N}_0$
for a particular \idx[rsym]{symmetry of rank \rsym}
is a meaningful irep. As already pointed out above,
for the example of SU($N$), these $q$-labels directly
specify the {\it Young tableau} in the sense
that $q_i\geq0$ specifies the offset (i.e., difference) of
boxes from rows $i$ to $i+1$,
\begin{eqnarray}
\includegraphics[width=0.7\linewidth]{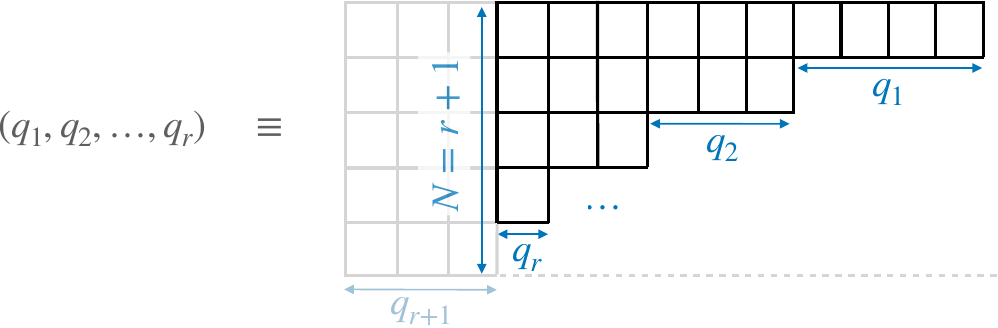}
\notag \\[-4ex]\label{eq:YTableau} 
\end{eqnarray}
This represents the Dynkin symmetry labels for SU($N$) ireps.
For a rank-$\rsym$ symmetry, 
there are at most $\rsym$ rows since any $q_{r+1}$ of
completely filled columns, i.e., with $\rsym+1$ boxes, if present,
can be split off as a scalar factor in a product-state like fashion:
a completely filled column represents the largest set of
$N$ particles that can be brought into a completely
antisymmetrized state which represents a singlet state.
Hence from an irep point of view, the label $q_{r+1=N}$ is
irrelevant and can be skipped, as indicated 
in \Eq{YTableau} by fading
out the leftmost $q_{r+1}$ completely filled columns.

When exploiting non-abelian symmetries, one usually assumes
that there is no spontaneous symmetry breaking on physical grounds.%
\footnote{
If a non-abelian symmetry of a Hamiltonian is broken
spontaneously in its low-energy regime,
non-abelian multiplets become macroscopically large.
In this case, however, there is no gain in exploiting
non-abelian symmetries. Rather, the symmetry should
be reduced, e.g., to an abelian U(1) symmetry in 
the simulation itself for the sake of numerical efficiency
based on the underlying physics. As a corollary, the tendency
of a system to generate ever larger multiplets during
initialization of a tensor network wave function,
e.g., via iterative diagonalization in an NRG-like fashion
even for a uniform system,
is a strong indicator for a spontaneously symmetry
broken ground state.}
Therefore assuming a `non-magnetic' state,
the symmetry labels $q_i$
do not grow macroscopically large with system size,
but are rather spread around $q_i \geq 0$.
Then if the values for $q_i\in \mathbb{N}_0$ 
do not grow macroscopically large,
\QSpace adopts the following extension to the hexadecimal 
format for the sake of a compact notation:
the alphabetic characters
\src{A-Z} are used do describe the range $q_i \in [10, 35]$.
If the operating system supports a case-sensitive file system,
the range is further extended 
to $q_i \in [36,61]$ which maps to the lower-case alphabet
\src{a-z}, respectively
(the reference to file system comes into play here
since $q$-labels are also inserted into file names
within the file-based database \RCR{\RCS}).
For example then, for SU(2) the irep \src{q=\str{Y}$\equiv 34$} 
corresponds to a multiplet with spin $S=34/2 = 17$. 
Therefore if a given a set of $q$-labels satisfies
$q_i\in \mathbb{N}_0 \leq q_1^{\rm max}$
for all $i=1,\ldots,\rsym$ with 
\begin{eqnarray}
   q_1^{\rm max}  = \left\{
     \begin{array}[c]{ll}
       61 & (= 9+2\times 26) \ 
          \text{ for a fully case-sensitive file system, e.g., Linux} \\
       35 & (= 9+1\times 26)\ \text{ otherwise, like MacOS}
     \end{array}
  \right.
\label{eq:compact:qs}
\end{eqnarray}
the above compact extended hexadecimal notation is adopted
for this particular set of symmetry labels in \QSpace.

The range in \Eq{compact:qs} covers the vast
majority of tensor network simulations.
For example for SU(2),
typical tensor network ground state calculations in the singlet
symmetry sector $q=0$ stay within $q\lesssim 10$
i.e., $S\lesssim 5$. This also generalizes to general
non-abelian symmetries, typically having $q_i\lesssim 10$,
yet with significantly more possible combinations due
to the presence of $i=1,\ldots,\rsym$ symmetry labels.
Hence, the compact symmetry labels overwhelmingly stay
within the `numeric' range $0\ldots 9$ for all $q_i$,
yet permit to exceed this range significantly
within the compact notation by including alphabetic
characters.
Overall this motivates the compact alpha-numeric
notation of symmetry labels
\begin{eqnarray}
   q &\equiv& (q_1,q_2,\ldots,q_\rsym)
     \ \equiv\ (q_1q_2\ldots q_\rsym)
     \ \equiv\ \qt q_1q_2\ldots q_\rsym \qt
\end{eqnarray}
for a particular irep $q$, with all spacing (commas)
in between the $q_i$ removed, and written with or without
brackets, like simple strings.
For the case that $q_i > q_1^{\rm max}$,
a more extended display with regular integers is chosen
with the $q_i$ separated by spaces.

For example, the scalar representation is given by
$q_{\rm scalar}=$ $(0\ldots 0_\rsym) \equiv 0$,
which is also conveniently denoted by a single $0$
in this documentation for any symmetry for readability.
The defining representation is given by $q_{\rm def}
= (10 \ldots 0_\rsym)$. For example, for SU($N$),
\begin{eqnarray}
   q_{\rm def}  &=& \left\{
     \begin{array}[c]{lll}
       (1)   & \text{equivalent to }S=q/2=1/2 & \text{SU(2)} \\
       (10)  & \equiv(1,0)   & \text{SU(3) } \\ 
       (100) & \equiv(1,0,0) & \text{SU(4) } \\ 
       \ \  \ldots & \quad\ \  \ldots & \text{etc.}
      \end{array}
  \right.
\label{eq:qdef}
\end{eqnarray}
The \idx[dual]{dual representation} for SU($N$) simply flips
the order of labels in the standardized scheme above,
i.e., $\bar{q} = (q_\rsym q_{\rsym-1}\ldots q_1)$.

\begin{figure}[b!]
\begin{center}
\includegraphics[width=0.8\linewidth]{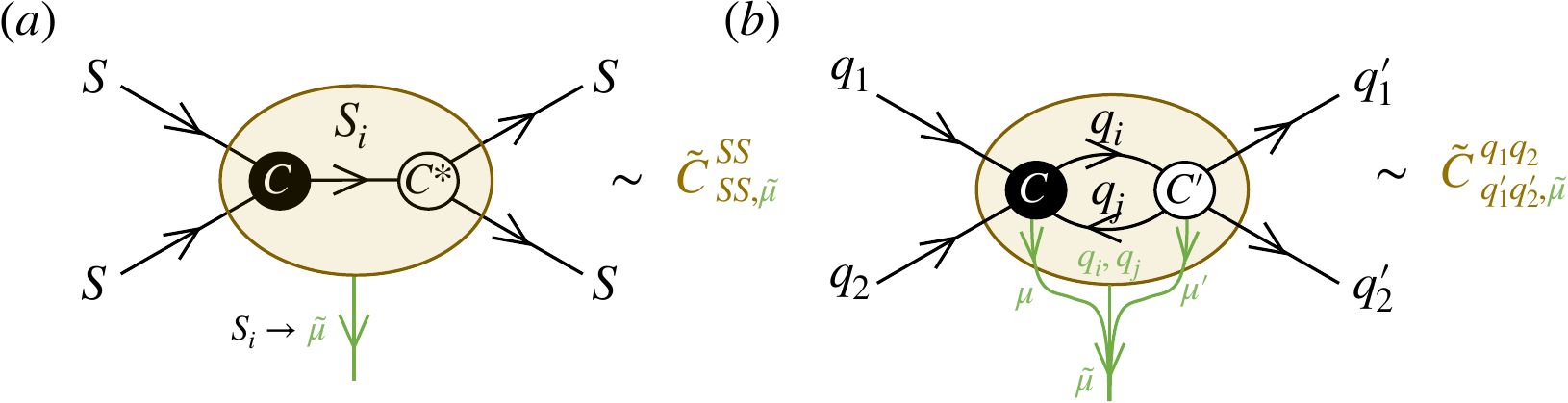}
\end{center}
\caption{
    Outer multiplicity (OM; figure adapted from \cite{Wb20})
    based on the simple specific example of a rank-4
    CGT of SU(2) in panel (a) where the CGT contraction 
    $(SS|S_i) \ast_i (SS|S_i)^\ast \equiv
    (SS|SS)_i$ on the intermediate index for fixed 
    intermediate multiplet $S_i \in 0,\ldots,2S$
    gives rise to
    outer multiplicity for the rank-4 CGT 
    $\tilde{C}^{qq}_{qq,S_i \to \tilde{\mu}}$
    with $q=2S$ and $\tilde{\mu} = 1,\ldots,M$ having $M=2S+1$.
(b) More generally,
    contracting two CGTs $C$ and $C'$ on two shared indices
    with fixed intermediate symmetry sectors
    $q_i$ and $q_j$. Each of the CGTs $C$ and
    $C'$ can have OM on their own, for general
    non-abelian symmetries already so at the
    level of rank-3 CGTs as in (a).
    The combined multiplicity
    is therefore determined by obtaining the
    resulting CGTs $\tilde{C}$ for all combinations
    $(\mu,\mu',q_i,q_j)$. If multiple intermediate
    indices are contracted simultaneously,
    like $q_i$ and $q_j$ above,
    then the plain contractions $\tilde{C}(\mu,\mu',q_i,q_j)$
    are generally
    not orthogonal and typically overcomplete.
    Hence these need to be orthonormalized into
    an arbitrary but fixed combined OM basis indexed
    by $\tilde{\mu}$ [cf. \Eq{Cnorm:2}].
} \label{fig:OM} 
\end{figure} 

\subsection{Outer multiplicity}
\label{idx:OM}

Outer multiplicity (OM) is ubiquitous when dealing with
non-abelian symmetries in tensor network algorithms.
It adds significant overhead in the book-keeping
when dealing with symmetries, yet is fully taken
care of by \QSpace in its MEX core routines.
The discussion here therefore serves as background 
information only. While it is important to be aware of
on general grounds when interpreting \QSpace
tensors, the actual book-keeping
is hidden from, and in this sense of no further
concern for the typical \QSpace user.

To be specific, consider a rank-4 tensor having SU(2)
symmetry with two incoming legs and two outgoing legs
which, for simplicity, all carry the same spin $S$.  This
is described by $(S,S|S,S)$ or, equivalently, the CGT
$C^{qq}_{qq}$ with $q=2S$ in \QSpace [cf. \Fig{OM}].
Now since $(S,S)$ can be fused into $M=2S+1$ intermediate
spins $S_{i} = 0,1,\ldots,2S$, each of these can again be
`unfused' symmetrically into the outgoing indices. Hence
one may contract $(SS|S_i) \ast_i (SS|S_i)^\ast \equiv
(SS|SS)_i$ on the intermediate index for any fixed
$S_i$ (this represents a projector onto the intermediate
multiplet $S_i$ of dimension $2 S_i+1$).  Then $i\to \mu
=1,\ldots, M$ takes the role of an outer multiplicity
index: for given CGT $C^{qq}_{qq}$ with fixed ireps on all
legs, there exist {\it multiple} orthogonal CGT's
permissible by symmetry.  Hence the CGT $C_{qq}^{qq,\mu}$
naturally acquires an additional index, the
\Emph{OM~index} $\mu$ which spans the OM space of dimension $M$.
The CGT $C_{qq}^{qq}$ is unique then, up to an arbitrary
orthogonal rotation in OM space [this generalizes the sign
convention of standard rank-3 CGCs for SU(2)]. The choice
of basis within this OM space in \QSpace is discussed with the
\hsec[CGT:norm]{CGT (ortho)normalization convention}.
Overall then, one has $M>1$ in the {\it presence} of OM,
$M\leq1$ in the {\it absence} of OM, and $M=0$ if a
particular CGT does not exist at all, i.e., is not
permissible from a symmetry point of view.

For abelian symmetries, one always has $M\leq1$,
i.e., outer multiplicity does not occur. It is clear
from the above example, however, that OM is already ubiquitous even
for SU(2) for tensors of rank $r\geq 4$. For general non-abelian
symmetries of symmetry rank $\rsym>1$, OM already occurs
at the fundamental level of standard CGCs, i.e, CGTs of rank $r=3$
\cite{Elliott79,Alex11,Wb12_SUN}
(there can never be OM for $r\leq 2$, like scalar operators).

\paragraph{OM index is specific to tensor (not particular leg)}
\label{idx:OM:tensor}

Standard CGC decomposition $C^{q_1 q_2}_{q_3}$ may 
tempt one to assign the OM index $\mu=1,\ldots,M$
with the fused multiplet $q_3$, as there are $M$ of these
occurring in the decomposition.
However, by applying \onej-symbols,
any other raised index in $C^{q_1 q_2}_{q_3}$ can be
lowered, while raising $q_3$. While this changes
symmetry labels to their duals, this nevertheless also
demonstrates that, by construction, precisely
the same multiplicity will occur
in all these other instances, too.
Therefore the OM index may be associated
with any of the legs, depending on the context.
In this sense, the OM index really belongs to a CGT
as a whole, and not to any particular leg of it.
This disregard for the direction of the legs
is also reflected in the CGT norm convention
in \Eq{Cnorm:1}.

\paragraph{OM can become arbitrarily large}
\label{idx:OM:large}

For non-abelian symmetries, OM typically
\idx[high-rank]{grows exponentially}
with the number of legs, i.e., the tensor rank \cite{Wb20}.
Yet even at the level of standard rank-3 CGCs,
OM can become arbitrarily large. For example, for SU(3)
the CGT $C^{qq}_q$ with $q=(nn)$ with $n\in \mathbb{N}_0$
has an outer multiplicity $M=n+1$. 
That is, fusing two SU(3) multiplets $(nn)$
results in $n+1$ distinct multiplets with the {\it same}
fused multiplet label $(nn)$ amongst others,  i.e.,
$(nn)\otimes (nn) = (00) \oplus (nn)^{n+1} \oplus \ldots$\,,
where the exponent indicates OM.
The multiplet $(nn)$ has dimension $|q| = (n+1)^3$,
and is self-dual, such that the scalar $(00)$
also occurs in the decomposition above.
To be specific, fusing
two ireps in the \idx[qadj]{adjoint} representation (11), i.e.,
$n=1$ above, one obtains the full decomposition
$(11)\otimes (11) = 
(00) \oplus 
(03) \oplus 
(11)^{2} \oplus
(22) \oplus 
(30)$,
this gives the adjoint (11) {\it twice}. In terms of multiplet
\idx[qlabel-dim]{dimensionality}
this reads ${\bf 8} \otimes {\bf 8} 
= {\bf 1} + {\bf 10} + (2\times{\bf 8}) + {\bf 27} + {\bf 10}$,
respectively, which adds up to the expected total
of $8^2 = 64$ states.
Based on the unboundedness of OM even for rank-3 CGTs,
there appears no simple way to fix a canonical OM basis, e.g.,
w.r.t. to permutations of legs, etc., since that symmetry is
necessarily finite given the finite number
of legs. To the best of our
knowledge, there is no superior systematic way to deal
with arbitrarily large OM for general non-abelian symmetries.
Hence a bottom-up constructive approach is adopted 
in \QSpace that is based on demand, and thus arbitrary
but fixed, at the expense of a \idx[RCS:hist]{history
dependent \RCS} as discussed with \Sec{CGT:norm}.
\vspace{-1ex}

\subsection{General tensor decomposition}
\label{sec:tensor:decomp}

\begin{figure}[tbh]
\begin{center}
\includegraphics[width=0.45\linewidth]{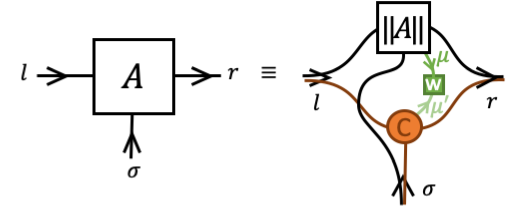}
\end{center}
\caption{Pictorial representation of tensor 
   decomposition in
   \Eq{tensor:decomp} for the case of an
   \idx[Atensor]{$A$-tensor}
   for fixed symmetry sectors and hence a particular
   contribution $q$ for simplicity 
   (figure adapted from \cite{Wb20}). The bare
   indices $(l,r,\sigma)$ in the absence of
   symmetries become composite indices
   [cf. \Eq{composite-index}]. The RMTs $\Vert A\Vert_q$
   are stored with a \QSpace tensor, whereas the CGTs $C_q$
   are only referenced via \idx[CGR]{CGRs} which 
   point into the \RCS database. 
   The CGRs also store the matrices $w_q$ with the tensor. 
}\label{fig:cgw} 
\end{figure}

The general decomposition of a \QSpace X 
in the presence of symmetries
was already briefly touched upon
in the \hsec[QSpace:1]{introduction}
[cf. \Eq{tensor:decomp:0}]
in the absence of OM. In the presence of OM,
this becomes \cite{Wb12_SUN,Werner19,Werner20,Wb20},
\begin{eqnarray}
   X = \bigoplus_{q}
   \Bigl[
   \Vert X\Vert_{q}^{\textcolor{cOM}{\bf\mu}} \otimes \bigl(
      w^{ \textcolor{cOM}{\bf\mu'}}
       _{ \textcolor{cOM}{\bf\mu }}
      C_{q \textcolor{cOM}{\bf\mu'}}
      \bigr)
   \Bigr]
\text{ ,}\label{eq:tensor:decomp}
\end{eqnarray}
where the OM indices \textcolor{cOM}{$\mu$} and
\textcolor{cOM}{$\mu'$}
are implicitly summed over pairwise in a plain sum
($\sum_{\textcolor{cOM}{\mu \mu'}}$).
With this, the OM indices are fully contracted
in \Eq{tensor:decomp} as visualized in \Fig{cgw}).
In practice, however,
the CGTs $C_{q \textcolor{cOM}{\bf\mu'}}$ are
referenced, and the summation over multiplicity
is postponed depending on the context.
For pairwise contractions, for example, this summation
is translated to generating \Xsymbols \cite{Wb20}
from the CGTs and contracting these
onto the RMTs $\Vert X\Vert_{q}^{\textcolor{cOM}{\bf\mu}}$.
That is, the tensor-product in 
\Eq{tensor:decomp} across RMTs and CGTs is never
explicitly expanded per se, as this would defeat
the purpose of exploiting symmetries.

The leading direct sum ($\oplus$) over $q$ 
in \Eq{tensor:decomp} implies block structure,
which translates to the \hsec[QS:struct]{record index} \irec
in the \QSpace tensor structure.
Via the Wigner Eckart theorem
one obtains the tensor product ($\otimes$)
of the RMTs $\Vert X\Vert_{q \mu}$ with the CGTs
$C_{q\mu}$ (assuming $w=\Id$ for simplicity).
The RMTs and CGTs can be dealt with separately,
and hence are stored separately.
The RMTs can be chosen freely depending on the
physics or the algorithm, whereas the CGTs are fixed
by symmetry, and hence referenced via the \idx[CGR]{CGRs}.
The CGRs thus store the matrices $w^{ \mu'}_{\mu}$,
while the RMTs $\Vert X\Vert_{q}^{\mu}$
carry trailing open OM indices $\mu$ for $M>1$.
\vspace{-1ex}

\paragraph{$w$-matrix}
\label{idx:cgw}

An important element in the tensor decomposition
\eqref{eq:tensor:decomp} is the direct link across
the tensor product structure of the RMT and CGT
via a plain regular sum on the OM indices.
It simply states that for every OM instance
there can be a different RMT (the additional space within
the RMT due to OM only needs to be explicitly allocated
when fusing state spaces \cite{Wb12_SUN} which is
taken care of by the \idx[getIdentityQS]{getIdentity}
routine).
When summing over the OM index
denoted in green in \Eq{tensor:decomp},
one may insert an additional matrix, referred to as the
$w$-matrix, which lives directly on the OM link
that connects the RMT to the CGT [cf. \Fig{cgw}].
The $w$-matrix can always be absorbed, i.e., contracted
onto the RMT by carrying out the summation over $\mu$.%
\footnote{
The $w$-matrix cannot be absorbed onto the CGT, since CGTs
are defined once and for all. Besides, the data format is
different: CGTs are stored in \idx[MPFR]{higher precision}, whereas
the RMTs, same as the $w$-matrices, have plain double precision
arithmetic.}
However, there is practical value in keeping the
$w$-matrix nevertheless:
it permits one to efficiently absorb
rotations in OM space, e.g., as they occur out of permutations
of the legs of the tensor, but also to absorb scale factors. 
For this reason the $w$-matrix is \hsec[QS:struct]{stored
with each \QSpace tensor via the CGR}.
It is a real matrix with the default normalization convention
\begin{eqnarray}
   w^\dagger w = \Id
\label{eq:wnorm:1} \text{ ,}
\end{eqnarray}
typically taking the first non-zero matrix element
in $w$ to be positive.
When $w$ is contracted onto the RMT,
it can be simply replaced by $w \to \Id$ afterwards. 
By the convention that the 
\heq[S:indices]{superscript index
is the first index}, i.e., 
$w^{\mu'}_{ \mu} \equiv
 w^{\mu'}_{\ \ \mu}$,
the index $\mu'$ specifies its rows.
Then by \Eq{tensor:decomp}
the row index of $w$ links to the CGT,
whereas the column index links to the RMT.
Typically, the OM dimension on the RMT is the same
as for the CGT, but it may also be smaller.
At any given time, the maximal OM dimension is clearly 
set by the CGT (OM may grow dynamically during simulations,
until the full OM is exhausted). Yet the RMT of a given tensor may
only relate to a particular subset of OM directions,
hence may have lower OM.
For this reason, $w$ can have fewer columns than rows,
i.e., it is a matrix of dimensions $M\times M'$ with $M'\leq M$.
If $M'<M$, $w w^\dagger \neq \Id$,
whereas \Eq{wnorm:1} holds in any case.

\phantomsection\label{sec:cgw:exceptions}
There are only \Emph{two exceptions} to the normalization
convention in \Eq{wnorm:1}. They are intended for
the sake of readability on the Matlab interface only
(by contrast, the \QSpace core routines always adhere
to \Eq{wnorm:1} internally).
In these cases the normalization in \Eq{wnorm:1}
is only proportional to the identity, 
  $w^\dagger w = |w|^2 \Id$.
For \QSpace records $q$ where $|w_q|\neq 1$, these normalization
factors are printed
to the right when displaying a \QSpace tensor (see display
in \Fig{QS:disp} for a specific example):
\begin{itemize}
\item rank-2 tensors (e.g., scalar operators):
   for the sake of the argument consider an identity operator $X$.
   In this case, for the sake of readability
   and also to avoid confusion,
   one would like the RMTs themselves to be identities,
   i.e., $\Vert X_q\Vert  = \Id$
   an identity operator acting within symmetry sector $q$.
   But the CGTs have fixed
   \hsec[CGT:norm]{CGT normalization convention}
   themselves. In particular, for a scalar operator like the identity,
   they are trivially given by
   $C_q^q = \frac{1}{\sqrt{|q|}} \Id^{|q|}$
   with $|q|$ the multiplet dimension.
   Therefore in order to reconcile the desired behavior
   of the RMTs with the CGT normalization convention,
   the remaining scale factor can be absorbed at the
   level of the CGR with the $w$ matrix by setting
   \begin{eqnarray}
      w_q = \sqrt{|q|}
      \text{ .}\label{eq:wnorm:x2}
   \end{eqnarray}
   This cancels the normalization factor of the CGT,
   thus resulting in a combined identity operator, $w C = \Id$.
   Since there is no OM for rank-2 tensors, the $w$
   in \Eq{wnorm:x2} is always a plain number.
   Aside from identity operators,
   the altered normalization convention on the $w$-matrix
   for the case of rank-2 tensors 
   intuitively shows the `actual' value of matrix elements
   in the RMTs for scalar operators such as Hamiltonians
   or density matrices.

\item rank-3 tensors with only one in- or outgoing index:
   similar to the previous case for \mbox{rank-2}
   consider here for the sake of the argument the
   identity $A$-tensor \cite{Wb12_SUN}
   that describes the fusion of two incoming state spaces
   without truncation into their combined state space.
   Intuitively, here too, one would like to see
   parts of sliced up identity matrix in the RMTs.
   That is, the entries in the RMTs should be either 0 or 1,
   as this indicates a map of input multiplets to fused multiplets.
   For this, however, the rank-3 CGTs would have to be 
   normalized like standard CGCs, thus desiring
   $\sqrt{|q_3|} C^{q_1q_2}_{q_3}$.
   Again this can be simply achieved from the CGR point of view
   by setting
   \begin{eqnarray}
      w_q = \sqrt{|q_3|} \ \Id
   \text{ .}\label{eq:wnorm:x3}
   \end{eqnarray}
   Now an $A$-tensor is a rank-3 tensor with one outgoing
   index only. Conversely, \idx[irop]{irops} are rank-3 tensors,
   with only one incoming index. In this sense,
   for generality, the normalization in \Eq{wnorm:x3}
   is adopted for all rank-3 tensors that have one
   incoming or one outgoing index only. The irep
   on this leg then
   defines the dimensional factor in \Eq{wnorm:x3}.   
   In the absence of OM for these rank-3 tensors
   [such as for SU(2) throughout],
   $w$ is again just a plain number.
   Otherwise, the $\Id$ in \Eq{wnorm:x3} is the
   identity matrix in the OM space of the underlying
   $C^{q_1q_2}_{q_3}$ or $C_{q_1q_2}^{q_3}$.
\end{itemize}

\paragraph{Avoiding high-rank tensors}
\label{idx:high-rank}

The block decomposition
via $q$ in \Eq{tensor:decomp}
only includes permissible symmetry combinations
for the given number of legs. However,
it is important to bear in mind in the context of tensor
networks that this number of possible
combinations $q$ (or equivalently, the number
of \hsec[QS:struct]{records} in a \QSpace)
can become quickly very large (exponentially so)
if tensors have many open legs, i.e., are high-rank.

This proliferation of non-zero blocks
for large-rank tensors is already present for
plain abelian symmetries. Consider,
for example, a tensor of even rank $r$, with $r/2$ incoming
and $r/2$ outgoing indices and where for the sake
of the argument, all legs have the same state space,
say, with $n_{s}$ symmetry sectors each.
Then {\it all} $n_s^{r/2}$ symmetry combinations
of the $r/2$ incoming legs are permissible,
since they can again be `unfused' into the outgoing legs. 
Therefore the number of blocks (RMTs) grows
exponentially with the rank.
In practice, there will be typically
many more combinations still that satisfy
$\sum_{\rm in} q_{\rm in} = \sum_{\rm out} q_{\rm out}$.
The proliferation of non-zero blocks with tensor rank
eventually holds generally. It is also irrespective
of the direction of legs since these can be easily flipped
by contracting \hsec[1j]{\onej-symbols} which does not
affect the number of non-zero blocks.

For non-abelian symmetries, the situation is further
exacerbated: the fusion of multiplets typically
gives rise to several combined ireps, which thus further
increases the number of internal fusion channels
and hence non-zero blocks
(e.g., hundreds of thousands for rank-6 tensors
\cite{Bruognolo21}).
Furthermore, when building larger-rank tensors by contraction,
also the OM of the contracted tensors themselves will
grow exponentially with increasing tensor rank \cite{Wb20}.
If then one tries, for example, to contract two such
large-rank tensors, this quickly translates
into extremely long lists of elementary pairwise block
contractions 
(billions when contracting rank-6 tensors \cite{Bruognolo21}).
While not necessarily memory consuming, this may
take far too long.

Hence reducing the tensor rank of large-rank tensors
becomes mandatory. This can be achieved in two ways:
(i) simply fusing legs which in itself has a significant 
positive impact
on the overall performance \cite{Bruognolo21}.
Alternatively, if the situation permits, one may also
(ii) split up large-rank tensors into a tensor network of
several lower-rank tensors such as tensor trains
\cite{Hackbusch12}. The latter has strong links
to {\it data compression} which is of central importance 
for tensor network algorithms such as DMRG/MPS
\cite{PerezGarcia07,Schollwoeck11,Orus14}
all along, even in the absence of any symmetry.

\subsection{CGT orthonormalization convention}
\label{sec:CGT:norm}

All CGTs in \QSpace adhere to the Frobenius-like
normalization convention,
\begin{eqnarray}
   \Vert C_q \Vert ^2
   \equiv {\rm tr}(C_q^\dagger C_q)
   \equiv {\rm tr}(C^q C_q)
   = 1
\text{ ,}\label{eq:Cnorm:1}
\end{eqnarray}
where the tensor trace is to be interpreted such that
it fully contracts $C_q$ with the conjugate of itself
on all legs, with only the OM index left open if present.
To start with, \Eq{Cnorm:1} assumes no OM, i.e., $M=1$.
The tensor $C_q$ denotes a general CGT with symmetry
labels $q$ combined over all legs with certain directions which are of no
further interest for the argument here. For simplicity,
this collective $q$ is written as a subscript here,
where nevertheless its
conjugate is symbolically written as $C^q$ with raised $q$.

Importantly,
the norm in \Eq{Cnorm:1} is invariant under an arbitrary
orthogonal transformation on any of the tensor's legs.
As this includes reverting directions
of legs based on \hsec[1j]{\onej-symbols}, the normalization
in \Eq{Cnorm:1} is therefore {\it independent} of the directions
of the legs (while also switching to dual ireps), fully analogous
to Wigner-$3j$ symbols. This bears significant practical advantages.

In the presence of \idx[OM]{outer multiplicity},
CGTs acquire an additional OM index $\mu = 1,\ldots, M$.
By convention, this is taken as trailing and thus slowest index
at position $r+1$ with $r$ the rank of the tensor,%
\footnote{\QSpace uses generalized column-major storage
for all tensors, throughout,
RMTs as well as CGTs. While RMTs are stored as full tensors,
CGTs are stored in \idx[CGT:sparse]{sparse} format generalized
to arbitrary rank tensors \cite{Wb12_SUN}. In either case,
listing the OM index as trailing index implies that
data is blocked in storage w.r.t. the OM index since slowest.}
thus having $C_q \to C_{q\mu}$.
In the presence of OM then, i.e., $M>1$, the r.h.s. of \Eq{Cnorm:1}
becomes an identity matrix in OM space. That is,
with $1 \to \Id^{(M)}$, the individual OM components
can be fully orthonormalized \cite{Wb20},
\begin{eqnarray}
   \mathrm{Tr}\bigl( C_{q\mu}^\dagger C_{q\mu'} \bigr) \equiv
   \mathrm{Tr}\bigl( C^{q\mu} C_{q\mu'} \bigr)  = 
   \delta^{\mu}_{\mu'} 
\text{ .}\label{eq:Cnorm:2}
\end{eqnarray}
As apparent from the above, CGTs are only
unique up to an arbitrary but fixed orthogonal
transformation in OM space. By contrast, the basis within the state 
space of any leg is defined and thus fixed with the respective
multiplet decomposition which may have to deal with
inner multiplicity \cite{Wb12_SUN}. 
In the absence of OM [such as in standard CGCs for SU(2)],
the normalization convention in  \Eq{Cnorm:1}
reduces to a sign-convention.
This sign convention is still adopted in any case
in the standard fashion of CGCs,
namely that the first non-zero entry in the CGT $C_{q\mu}$ 
for any $\mu$ is positive.

Aside from the sign-convention on the basis
vectors in OM space, this still leaves considerable
arbitrariness in the presence of OM, i.e., $M>1$.
This arbitrariness is fixed within \QSpace
by its operational procedure: CGTs are constructed
once and for all with iteratively increasing $M'$
{\it as they occur} (first come first serve) and
then stored in the \RCR{\RCS/CStore} \cite{Wb20},
until eventually at $M'=M$ the full OM space is exhausted,
and thus complete. This procedure is motivated by the fact,
that for larger-rank tensors, depending on the type
of contractions performed, the full OM dimension $M$
space may never be explored and thus is not required. In this sense,
\QSpace does not preemptively generate the full OM space 
for any CGT $C_q$ the first time it occurs.
This is different for standard rank-3 CGCs, 
which are derived from an explicit numerical tensor
product decomposition. By construction, their
OM space is always complete from the first time
they are encountered.
For other rank-3 CGTs that derive out of contractions
or higher-rank tensors, the OM space is built
as it occurs out of contractions.
If a rank-3 CGT was first generated via a contraction,
and later on is completed via a tensor product
decomposition, the existing CGT always has preference
to ensure consistency within the existing \RCR{\RCS}.
In this case, any additional OM space derived from
the tensor product decomposition is properly
projected and renormalized.
\vspace{-1.5ex}

\paragraph{History dependent \RCS}
\label{idx:RCS:hist}

By the operational procedure above, namely the deliberate decision 
to not generate the full OM space in a unique well-defined
manner for any CGT encountered, the \RCS becomes history dependent.
As such, the \RCS cannot be simply switched across different tensor
network simulations. This makes access to a global
centrally maintained \RCR{\RCS} essential.
By meticulously applying and checking IDs together with
high-resolution time stamps with CGTs \cite{Wb20},
\QSpace insists on a particular \RCS database within
a given tensor network simulation in order
to ensure overall consistency.
\vspace{-1.5ex}

\paragraph{Norm of a \QSpace tensor}
\label{idx:QS:norm}

As a corollary, the norm of a \QSpace \src{X} is defined 
by the full tensor trace of $X$ with the conjugate
of itself, i.e., the Frobenius norm
\begin{eqnarray}
   \Vert X\Vert ^2 \equiv \Srt{norm(X)}\verb|^2| 
   \equiv {\rm tr}(X^\dagger X)
\text{ ,}\label{eq:QS:norm}
\end{eqnarray}
With the CGTs \hsec[CGT:norm]{normalized} as in \Eq{Cnorm:2},
when contracted with themselves, they result in simple
identity matrix in OM space. Then if the \idx[cgw]{$w$ matrices}
on the OM links also follow their own orthonormalzation convention,
$w^\dagger w=\Id$ [\Eq{wnorm:1}], all that remains for obtaining
the norm of a \QSpace tensor is simply
to add up the square of the Frobenius norm for all RMTs
(for the case that $w^\dagger w\neq \Id$ [e.g., see \Eqs{wnorm:x2}
or \eqref{eq:wnorm:x3}],
this needs to be contracted (or multiplied if a simple number
$\propto \Id$)
onto the RMTs when computing their norm squared).
The normalization conventions therefore considerably simplify
obtaining the norm of a \QSpace tensor.
In any case, the norm can be simply obtained
using the \idx[normQS]{class routine \src{norm}}
which automatically takes care of all the above.
\vspace{-1.5ex}

\paragraph{Representation of OM in \QSpace tensors}
\label{idx:OMindex}

The tensor decomposition in \Eq{tensor:decomp}
has an RMT $\Vert X\Vert_{q}^{\textcolor{cOM}{\bf\mu}}$
for every symmetry configuration $q$, i.e.,
with symmetry labels fixed for every leg.
The combined set of labels for all legs
as represented by $q$ 
thus is unique in the direct sum ($\oplus$).
Consequently, also for a \QSpace \src{X} the
combined, i.e, catenated symmetry labels
\src{Q=X.Q; Q=[Q\{:\}];} (cf. \QSpace \hsec[QS:struct]{data
structure} and also \idx[QS:subsref]{subsref})
are {\it unique} for any \QSpace \src{X}.
That is, \QSpace records as of \QSpace v4
have unique symmetry labels for each record
(this is in contrast to v2 \cite{Wb12_SUN} or v3
which had OM split up as multiple records in a
\QSpace tensor with `degenerate' $q$-labels).
In the presence of outer multiplicity, i.e., $M>1$,
the OM index $\mu$ is carried as additional open
trailing index in the RMTs (single index fused
over the combined OM for all non-abelian symmetries
present). Hence it occurs at index position $r+1$
with $r$ the rank of the tensor.
In the \QSpace display, it is indicated as
\src{@M} with the RMT dimensions
[see \Eq{itags:A} and subsequent display for an example].
This OM index eventually is connected via
the intermediary $w$-matrix to the referenced
CGTs which then also carry trailing OM indices.

\subsection{CGT data storage and precision}
\label{idx:MPFR}

All \idx[CGR]{CGTs} are real, highly sparse, and computed
starting from better than quad-precision
by linking \QSpace with the \idx[MPFR:lib]{GMP/MPFR} multi-precision
library \cite{Fousse07,mpfr} during MEX compilation.
The CGTs are also stored
in a higher precision format in \RCR{\RCS/CStore}.
The major reason for higher precision is the
iterative nature of \QSpace \cite{Wb20} that mimics
the iterative nature of any tensor network simulation:
starting out from the defining representation,
an initial set of smaller ireps is generated which usually
more than suffices for the description of a physical lattice site.
Starting from this local state space (cf. \hsec[gls]{\Srt{getLocalSpace}}),
ever more complex tensors can be built by adding
sites to the physical Hilbert space and thus to the
tensor network. The process of tensor
product decomposition is fully numerical \cite{Wb12_SUN},
and hence subject to numerical precision error.
The generalized raising/lowering operators (in the literature
also referred to as {\it simple roots}
in the \idx[roots]{root space} of a Lie algebra, \cite{Cahn84,Wb12_SUN})
are also computed fully numerically within \QSpace [cf. \RCR{\RCS/RStore}]
subsequent to a tensor product decompositions via their
projection into the space of the newly encountered ireps
[the generalized raising/lowering operators could be generated
directly, e.g., based on Gelfand-Tsetlin pattern calculus
for SU($N$) \cite{Alex11}; but with general non-abelian
symmetries also beyond SU($N$) in mind, this approach
is not adopted in \QSpace{}].
The numerical approach in \QSpace also automatically ensures
consistency w.r.t. {\it inner multiplicity} (which describes
the situation ubiquitous in rank $\rsym>1$ symmetries, namely 
where within a given multiplet the same $q_z$-labels
can occur multiple times \cite{Elliott79,Alex11,Wb12_SUN}).
Therefore, by the design choice of \QSpace as a true
bottom-up approach for non-abelian symmetries,
any inaccuracy at an earlier stage is inherited
to any later stage. In this sense, the accuracy of CGTs
deteriorates as the \RCS becomes ever larger.
Empirically, however, by starting
better than quad precision (about 40 digits), all CGTs obtained
in the history of \QSpace so far have been accurate at least
up to 24 digits by monitoring numerical noise. In this sense
all CGTs are exact within double precision accuracy $10^{-16}$.
This is also important when exploiting the sparseness
of CGTs since for that purpose, one needs to be able
to reliably distinguish small CGT coefficients from
numerical noise
[for example, for SU(4) CGT coefficients as small as $10^{-9}$
have been encountered].

The sparse CGT data is stored in GMP/MPFR format.
The MPFR library does not support binary storage per se.
Hence the numerical data needs to be converted to \src{int8}
strings in a specified base\cite{Fousse07,mpfr}.
In \QSpace, for compactness, this numerical base is
chosen as \src{2*3*5 = 30} \cite{Wb12_SUN}
(for comparison, hexadecimal has base $16$).
The reason for this choice is that it permits exact
representation of fractions based on the prime numbers
$2$, $3$, and $5$, which thus includes binary as well
as decimal fractions.
Within the mat-binary format, these strings are automatically
compressed, such that eventually the storage overhead of the
string representation is expected to be minor.
\vspace{-1.5ex}

\paragraph{Sparse CGTs of arbitrary rank}
\label{idx:CGT:sparse}

All CGTs are stored in sparse format.
For rank $r>2$, the tensors are reshaped by grouping
leading vs. trailing indices,
such that the resulting `matrix' is closest to square.
Thereafter the sparse data
can be stored as a standard sparse matrix.
By also storing the actual dimensions for all legs
of the original tensor, this permits one to restore the entire
sparse tensor of arbitrary rank by returning to the
original rank and recomputing sparse index positions.

\subsection{Order of legs (index order conventions)}
\label{sec:idx-order}

\QSpace supports arbitrary order of legs in a tensor.
When permuting the index order (legs) of a \QSpace \src{X},
this applies the relevant permutation onto each of the
RMTs \src{X.data\{\irec{}\}}, and permutes the symmetry labels
for the legs in \src{X.Q\{:\}}, as well as the \itags.
The general index order is also
\idx[CGT-CGR]{supported by the CGRs} stored with a \QSpace.
\vspace{-1.5ex}

\paragraph{Index order and CGRs}
\label{idx:CGT:perm}
\label{idx:CGT:qdeg}

The CGRs store references to \idx[CGT:sorted]{sorted 
CGTs} in the \RCS. These CGTs have sorted
$q$-labels to avoid the proliferation of entries.
The required permutation (and possible conjugation)
to match the current index order
of a \QSpace \src{X} is auto-determined 
and stored with the CGRs {\it internally}
in the MEX core routines. When handed over to the 
user space in Matlab,
however, this permutation is applied to the CGRs
in \src{X.info.cgr} and hence not directly visible.
An arbitrary permutation of the legs on a \QSpace \src{X}
is translated then to the respective permutation
relative to the underlying sorted CGT.
The order of raised indices relative to the lowered indices
is irrelevant and can be trivially permuted.
Subtleties arise, however, for \QSpace records with 
the same $q$-labels and directions on multiple legs,
referred to as \idx[qdeg]{degenerate $q$-labels},
as this may lead to rotations in OM space
(cf. \idx[2M]{2M-symbols}).
In any case, all of this is taken care of
in the \QSpace core routine \src{permute[QS]}
(MEX routine \src{permuteQS} as called by the
Matlab class wrapper routine \Src{permute}).
\vspace{-1.5ex}

\paragraph{Compact notation for permutations
(with optional conjugation)}
\label{idx:perm}

The order of legs can be changed using permutations.
A permutation of length $n\leq r$
permutes the first $n$ legs (or all legs for $n=r$).
In this sense, the specified permutation can be shorter
than the rank of the tensor.
At the same time, one can also \idx[conj]{conjugate}
the tensor as a whole by specifying a trailing \str{*}
with the permutation.
For permutations of length $n\leq 9$, which is nearly
always the case, \QSpace accepts a compact
string notation for specifying permutations.
For example, \str{21*}
denotes a transposition of the first two legs, i.e.,
the permutation
\src{p=[2\,1]}  together with overall conjugation \str{*}.
In \QSpace syntax, for example, the Hermitian 
conjugate of an irop $S$ can be obtained then via
\src{permute(S,\str{21*})} which is equivalent
to the non-compact semantic
\src{permute(S,[2\,1],\str{conj})}.
\vspace{-1.5ex}

\paragraph{Compact notation for index sets
(with optional conjugation)}
\label{idx:compact:idx}

Specifying a subset of legs permits the same compact notation
as the \idx[perm]{permutations} of legs above, including an
optional trailing conjugation flag \str{*}.
This is relevant, for example, for contractions
when several indices are contracted simultaneously.
For example, in \Eq{nloc} 
\str{13*} specifies the simultaneous contraction of
the preceding tensor on indices $1$ and $3$
while also taking its \idx[conj]{conjugate}.
\vspace{-1.5ex}

\paragraph{Index order for operators and leg directions}
\label{idx:irop}

An irreducible operator (irop) consists of an irreducible
set of operators that transform like a multiplet,
yet with symmetry operations applied via a commutator.
The set of operators thus acquires an additional
index (referred to as \Emph{irop index}) which,
by convention, is always listed last, i.e., third.
The natural representation of an operator
in the presence of symmetries therefore is a
rank-3 tensor [cf. \Fig{Aop}(b)].
If the operator transforms like a \idx[scalar:op]{scalar},
i.e., with symmetry labels $q=0$ on the irop index,
this index represents a \idx[singleton]{singleton} dimension
which can be skipped. Thus the irop for scalar operators,
such as a Hamiltonian term or density matrix,
can be reduced to rank $2$.

The action of a particular operator is characterized
by the symmetry labels on its irop index.
Consider, for example, the matrix elements of 
an annihilation operator, e.g.,
$\langle s |\hat{c}|s'\rangle$. Ignoring spin for simplicity,
this operator {\it reduces} charge $q$ by 1,
i.e., $q_s = q_{s'} - 1$.
Based on this intuitive notion, namely that the action of the
operator is to {\it lower} the charge, this implies 
the symmetry label of the irop $q_{\rm op}=-1$.
For this to hold, however, the irop index must be
grouped with the \idx[arrows]{ket state}, and therefore is also
outgoing. Together with the index order convention
$S^1_{23} \equiv \langle 1 | S_{[3]} |2\rangle$
for an operator $S$,
the leg directions of an irop are thus \qdir{+{-}-}.

Not all rank-3 tensors with leg directions \qdir{+{-}-}
are necessarily operators that act within a particular
state space. Therefore to explicitly
declare an arbitrary \QSpace tensor \src{X}
of rank $3$ with leg directions \qdir{+{-}-}
an irop or operator more generally,
\QSpace recommends (and in certain applications insists on)
the additional flag \str{operator} with rank-3 operators
(see field \src{X.info.otype} in \Sec{QS:struct}).
\vspace{-1.5ex}

\paragraph{Index order convention with $A$-tensors (MPS)}
\label{idx:Atensor:LRs}

The \idx[Atensor]{$A$-tensors} as with an MPS adhere to the 
\idx[LRs]{LRs} index order convention [cf. \Fig{Aop}(a)],
having $\langle l | A^{[\sigma]} |r\rangle$
where $l\in L$, $r \in R$ refer to left and right
block states, respectively, with $\sigma \in s$
the local state space. When L$\to$R orthonormalized,
$A^{l\sigma}_r$ has leg directions \qdir{+-+},
when L$\leftarrow$R orthonormalized, $A^{r\sigma}_l$
has leg directions \qdir{-++},
and when at the \idx[OC]{orthogonality center} (OC), 
$A^{lr\sigma}$  has leg directions \qdir{+++} (all in).
\vspace{-1.5ex}

\paragraph{Index order out of contractions}
\label{idx:ctr:index-order}

The contraction of a pair of tensors, symbolically written as $A*B$,
always preserves the index order of the non-contracted indices.
The ones of the first tensor ($A$) are collected first,
followed by the non-contracted indices of the second tensor ($B$).
Hence the index order out of contractions is fixed
by the order of arguments handed over to \src{contract[QS]}.

\subsection{Fully contracted CGT networks and $M$-symbols}
\label{idx:Msym}

When projecting a CGT network onto a particular OM
basis of the resulting CGT, this corresponds to the 
evaluation of a fully contracted tensor network.
It is relevant, for example, for pairwise contractions
of tensors [cf. \idx[Xsym]{\Xsymbols} further below]. 
With this in mind, consider then such a fully contracted
tensor network of $n$ CGTs.
Each participating CGT may carry an OM index $\mu$ that
needs to be kept open since it links to an RMT
[cf. \Eq{tensor:decomp}] that is outside the discussion here.
In this sense, a fully contracted tensor network
of $n$ CGTs has $n$ open OM indices, and thus is
referred to as an \Emph{$n$M-symbol} with $n\in \mathbb{N}$.
The $q$-labels for all contracted legs are
considered additional metadata.
\vspace{-2ex}

\paragraph{1M and 2M-symbols}
\label{idx:2M}

The normalization convention
in \Eq{Cnorm:2} represents a 2M-symbol
$\delta^{\mu}_{\mu'}$.
However, since the multiplicity
indices always refer to the OM space
of the same CGT, 
in the spirit of \hsec[1j]{\onej}-symbols, one may also 
refer to this as a 1M-symbol. It is always
simply a square identity matrix in OM space.
More non-trivially, in the presence of
\idx[qdeg]{degenerate $q$-labels} on multiple legs
with the same directions, a given CGT may be contractable with a 
\idx[CGT:perm]{permuted} version of itself.
This can result in an orthogonal matrix and
thus an actual 2M-symbol.
\vspace{-2ex}

\paragraph{3M-symbols}
\label{idx:3M}

When contracting a pair of CGTs and projecting
the result onto the target CGT, this corresponds to a
3M-symbol. In the specific context of tensor contractions,
this is equivalently referred to as \idx[Xsym]{\Xsymbols}.
Note that by construction ($n=3$)M-symbols with $n$
an odd number require the presence
of at least some non-rank-3 CGTs in its network, i.e.,
the presence of non-CGCs.
\vspace{-2ex}

\paragraph{4M = $6j$-symbols, 6M = $9j$-symbols, etc}
\label{idx:6j}\label{idx:4M}

In the context of SU(2), a fully contracted
CGC network consists of CGTs of rank $3$ only.
For this to be fully contractable, the total number~$n$
of CGCs must be necessarily even. It has a total of
$3n/2 \equiv 3\tilde{n}$ contracted legs,
each of which carries a multiplet label.
Using $j$ with reference to SU(2), therefore a $(3\tilde{n})$-$j$
symbol is a fully contracted network
of $n=2\tilde{n}$ CGCs. For example,
a $6j$-symbol ($\tilde{n}=2$) contracts 
$n=4$ CGCs, and thus represents a \Emph{4M-symbol}.
Similarly, a $9j$-symbol ($\tilde{n}=3$)
represents a 6M-symbol, etc.
In general, a $(3\tilde{n})$-$j$
symbol is a $(n=2\tilde{n})$-M symbol,
with all $j$-labels on the contracted legs
considered additional metadata \cite{Wb20}.
Therefore fully contracted networks of rank-3 CGTs
correspond to $n$M-symbols with $n$ an even number.

For the simple case of SU(2) where CGTs for rank $r\leq 3$
do not have OM, $(3\tilde{n})$-$j$ symbols
are plain numbers.
However, for a more general non-abelian
symmetry of symmetry rank $\rsym >1$ like SU($N>2$),
for example, the $6j$-symbols
are rank-4 tensors with open OM indices $\mu$ on
all its legs, hence 4M-symbols.
Since there is no simple canonical
basis in OM space, besides that the OM space even for
CGCs (rank 3) can become \idx[OM:large]{arbitrarily large}
already for SU(3), this makes it considerably
more cumbersome to generalize $6j$-symbols, etc., 
to arbitrary non-abelian symmetries.
Therefore analytical results for $6j$
symbols are only available for SU(2),
but not for rank $\rsym>1$ symmetries.
For this very reason, \QSpace
does not use $6j$-symbols at all.
It rather uses 3M-symbols (\Xsymbols), instead.

\subsection{Pairwise tensor contractions and \Xsymbols}
\label{idx:Xsym}

When contracting a set of tensors, 
in practice, this is always carried out
as a sequence of pairwise contractions.
For a particular pairwise contraction then,
respective contractions occur in parallel at the level
of the RMTs, but also at the level of the CGTs for each
non-abelian symmetry included in the tensor
representation \cite{Wb12_SUN}.

\begin{figure}[tbh]
\begin{center}
\includegraphics[width=0.8\linewidth]{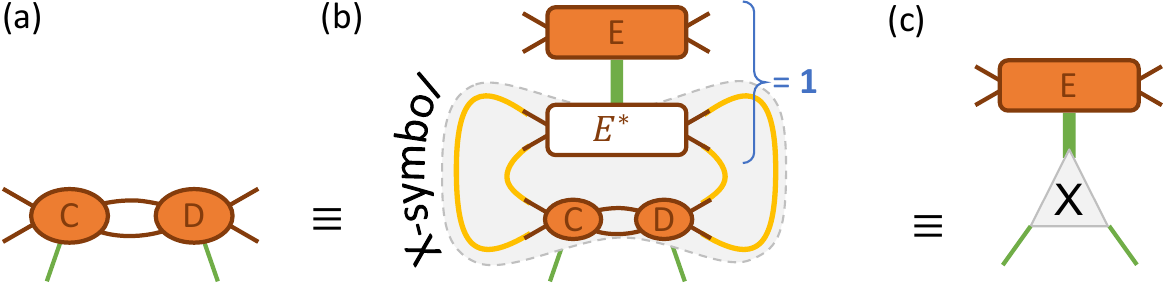}
\end{center}
\caption{\Xsymbols (figure adapted from \cite{Wb20}) --
   When contracting a pair of CGTs $C$ and $D$ 
   on some subset of legs [panel (a)]
   with OM indices indicated in green, 
   this gives rise to a new CGT $(C D) \equiv C\ast D$.
   The $q$-labels and
   $q$-directions are assumed arbitrary but fixed
   for all non-green lines, but not shown for readability.
   The resulting contracted tensor is expected to have
   a single OM index, hence the two green lines
   in (a) need to be fused, as indicated in the remainder
   of the panels.
   If a CGT $E$ with the same rank and $q$-labels
   as $(C D)$ already exists in the \RCS 
   (up to \idx[CGT:sorted]{sorting})
   the contraction in (a) may be simply projected
   into the existing OM space of CGT $E$ by inserting the
   identity $\Id = E^\ast \cdot E$, with the sum
   in the product performed
   over the OM space of $E$ [thick green line in (b)].
   If the CGT $E$ does not yet exist or is incomplete,
   the additional OM components are generated via
   Schmidt decomposition out of $(C D)$.
   The lower gray-shaded area in (b) now encompasses 
   a fully contracted network for three CGTs,
   i.e., a 3M symbol. For the explicit purpose of
   pairwise contractions, these are referred to
   as \idx[Xsym]{\Xsymbols}. \vspace{-2ex}
} 
\label{fig:xsym} 
\end{figure}

The contraction of a pair of CGTs $C \ast D$ 
on a subset of indices results in another CGT~$E$
[cf. \Fig{xsym}].
By projecting the outcome of $C \ast D$
onto $E$ (while completing the OM space of $E$ 
whenever relevant), the fully contracted
${\rm tr}( E^\ast (C D))$ 
corresponds to a 3M-symbol [\Fig{xsym}(b)].
In the context of pairwise contractions (`\texttt{X}'),
\QSpace refers to these as \Emph{\Xsymbols} \cite{Wb20}.

In practice, the CGTs $C$ and $D$ in \Fig{xsym}(a) as well as
$E$ in \Fig{xsym}(c) are referenced in \QSpace tensors and
stored in the \RCS once and for all once generated.  Hence all
that is required eventually for a pairwise contraction of
\QSpace tensors are the \Xsymbols in \Fig{xsym}(b,c).
These are contracted onto the OM indices of the input pair of
RMTs \cite{Wb20}, and in this sense {\it fuse} their OM indices.

\subsection{Index tags (\itags)} 
\label{idx:itags}

\QSpace supports user-defined names on tensor legs
with strings up to a length of 8 characters.
Since they represent tags that
are assigned to indices, they are referred to as \itags.
The usage of \itags makes \QSpace tensors significantly
more user-friendly. For one, they allow one to 
identify individual \QSpace tensors more easily
in a simulation. Hence the \itags
are also shown when \idx[QS:display]{displaying}
a \QSpace tensor.
Moreover, \QSpace supports auto-contraction based
on \itags, in that a matching pair of \itags with opposite
directions (i.e., one in- and one outgoing)
are auto-contracted unless indicated otherwise.
See help on \idx[contractQS]{contractQS}
for more on this.

Within the \QSpace core routines, \itags are
converted to \src{unsigned long} integers,
which thus sets the size limit of 8 bytes.
If an \itag is shorter than 8 characters, it is padded by zeros.
For a \QSpace tensor \src{X} of rank $r$, its \itags are stored
in the mandatory field \src{X.info.itags\{$l$\}}
for leg $l=1,\ldots,r$.

Itags are expected to be printable, mostly alpha-numeric
characters in ASCII code range 33-126, excluding \itag separators \src{[,;|]}
(these characters are sometimes used to
separate multiple \itags combined into a single string).
It is recommended to also avoid the trailing marker
characters \src{['*]} (see below) {\it within} an itag
string, in order to avoid confusion.  The above ASCII range
is case-sensitive and excludes white space.
The sign bit for each character
is exploited in \QSpace core routines
for other internally purposes, such as flagging
itags that \idx[ctr:semantics]{shall not be contracted}.
With this then, for example, 
the following itags are all different from
each other: \str{s1}, \str{s01}, \str{S01},
\str{s:1}, \str{s-1}, etc. \vspace{-1ex}

\paragraph{Trailing marker characters in \itags}
\label{idx:itag:markers}

The \itags are also used to store the
\idx[qdir]{direction of legs}
for a given \QSpace tensor:
a trailing `conjugate' character \str{*} indicates
outgoing legs, whereas the absence of a trailing \str{*}
indicates incoming.
These directions must be maintained and present
even if the \itags are empty otherwise.
In case of inconsistency 
of the leg directions specified with the \itags
of a \QSpace tensor
as compared to the $q$-directions in the
\hsec[QS:struct]{CGRs} (if present),
\QSpace will throw an error.

Moreover, \QSpace uses trailing primes to mark 
dual state spaces.  When reverting the direction of a leg,
this changes the interpretation of the associated
block state space. Therefore an automated modification of
the \itag is natural.
It is useful for better interpretability, but also as a safety
measure that guards against contracting wrong 
indices that would carry the same \itag otherwise
(see also \hsec[1j]{\onej symbols}).

In summary, \QSpace thus assigns special meaning
to the following trailing marker characters in \itags:
\begin{itemize}

\item A trailing asterisk (\src{*}) indicates a conjugate, i.e.,
outgoing index. Internally within the \Cpp core routines,
this is split off as
$q$-directions for an entire tensor. Therefore they do not 
count towards the 8-char limit of an \itag.

\item
A trailing prime (\src{'})
indicates dual state space which occurs when a leg direction
is reverted. The prime is considered integral part of an \itag,
and hence counts towards the 8-character length limit.
For this reason, a trailing prime also occurs {\it before}
a trailing conjugate flag \str{*} if present.

\end{itemize}
Both of the trailing marker characters above
act as a toggle. 
Since the dual of the dual of a state space
is the original state space,
marking an already marked itag removes the
trailing prime.
Similarly, taking the conjugate of an already
conjugate index reverts its direction
twice, and therefore results in the original
direction of that leg. Hence in this case,
a trailing conjugate character \str{*} is removed.
\\[1ex]
{\it Remark on notation} -- 
In this documentation, an itag \str{foo}
with a trailing prime is frequently printed as \str{foo'}.
The outer single quotes \str{...} denote a string,
and differ in font from the prime (\src{'}),
for the sake of the discussion.
They are considered the same, though, on the coding level.
Specifically, with Matlab syntax in mind, \str{foo'}
$\equiv$ \str{foo\textquotesingle\textquotesingle}
$\equiv$ \src{"foo\textquotesingle"},
yet noting that as of \QSpace v4, Matlab strings are always
assumed to use single quotes.

\paragraph{Color coding}
\label{idx:itag:color}

When using \QSpace in \idx[ml:term]{interactive Matlab terminal mode},
trailing marker characters in \itags are replaced by color coding
for the sake of readability:
outgoing indices become gray-ish
(while skipping the trailing \src{*} character),
and \itags marked by the trailing prime (\src{'}),
acquire a color other than the default text color
of the terminal (darker green as of \QSpace v4,
thus replacing the trailing prime; see \Sec{1j-markers}
for examples).
This way, for example, within an MPS, bond \itags for $A$-tensors
associated with the left block (left relative to the OC)
acquire a different color
as compared to the \itags for $A$ tensors associated
with the right block.

The color coding of \itags, of course, is applicable to
non-empty \itags only (i.e., \itags that contain actual
printable string after removing trailing marker characters).
If all \itags are empty, then the \idx[qdir]{\qdir{+/-}} notation
for $q$-direction is used as strings when displaying \itags
(e.g., see \Fig{QS:disp}) with color coding applied.
When only some \itags are empty, no color substitution
is performed, with all trailing markers shown as is.
By default, no color coding is applied throughout for
\idx[ml:term]{non-interactive} batch or cluster jobs
where log output is expected to be redirected to log files.
The behavior can be changed via the environmental
variable \src{\idx[QS:color]{QS\_LOG\_COLOR}}.

\section{QSpace Essentials}
\label{sec:QSpace:2}

\subsection{Starting point: {\tt getLocalSpace()}}
\label{sec:gls}

The starting point of any tensor network setup in \QSpace
is \Srt{getLocalSpace}. This permits one to define
a lattice site via a set of \idx[nolegs]{operators}
that act within that state space.
However, because these operators
are expressed as matrix elements in a particular symmetry
eigenbasis of the site, this implicitly also represents
its full state space, e.g., via the identity operator.
The way \src{getLocalSpace} operates is that it builds
the local state space in Fock space without symmetries
first, since everything can be simply constructed there
including the symmetry operators for all requested symmetries.
Based on this, \src{getLocalSpace}
rotates and groups the basis into symmetry multiplet spaces,
and then `compresses' the data via \src{\idx[compactQS]{compactQS}}
based on the Wigner-Eckart theorem to obtain the RMTs
while splitting off the CGCs as tensor product factors
[cf. \Eq{tensor:decomp}].

The overall argument structure of {\tt getLocalSpace()}
is as follows (for more detailed usage information,
see \src{\,help getLocalSpace}):
\begin{flalign}
  \src{[op1,...,opN,Isym] = getLocalSpace(type [, symmetries] [, options])}
\label{eq:gls}
\end{flalign}
The input specifies the \src{type} of a site,
together with its \src{symmetries}, and further \src{options}.
It returns a set of representative operators 
\src{op1,...,opN} for the specified site (therefore \src{N} 
depends on the input), together with an info structure \src{Isym}.
The \src{type} of a site is a string that can be any of
the following:
\begin{flalign*}
\begin{tabular}{p{1in}p{4.2in}}
 \str{FermionS} & spinful fermions \\
 \str{Fermion}  & spinless fermions \\
 \str{Spin}     & spin models \\
 \ \ etc.       & other experimental setups of spin models \\
\end{tabular} 
\end{flalign*}
The argument \src{symmetries} consists of a comma-separated
catenated string that specifies the type and also order of
symmetries to use.
\vspace{-1ex}

\paragraph{Spinfull fermions}

For spinful fermionic models, i.e., \src{type=\str{FermionS}},
the following symmetry entries are supported,
\begin{flalign*}
\begin{tabular}{p{1in}p{4.2in}}
  \str{Acharge}      & abelian total charge, i.e., U(1)  \\
  \str{ZNcharge}      & abelian total charge,
        modulo \src{N} $\geq 2 \in \mathbb{N}$, e.g.,
        \str{Z2charge} for Z$_2$ \\[1ex]
  \str{SU2charge}    & SU(2) total particle-hole \\
  \str{Aspin}        & abelian total spin, i.e, U(1) \\
  \str{SU2spin}      & SU(2) total spin ($S$) \\
  \str{SU2spinJ}     & SU(2) total spin ($J=L+S$) \\
  \str{AspinJ}       & U(1) total spin $(J=L+S)_z$ \\[1ex]
  \str{SUNchannel}   & SU($N$) channel symmetry \\
  \str{SONchannel}   & SO($N$) channel symmetry \\
  \str{SpNchannel}   & Sp($N$) symmetry with $N$
  even for $N/2$ channels
  combined with SU(2) particle-hole charge symmetry \\
\end{tabular}
\end{flalign*}
The order in which these symmetry labels are specified also
determines the \hsec[QS:struct]{order of symmetries} used,
and hence the order of the symmetry labels for any state space.
The last three symmetries only apply in the presence
of multiple channels (or flavors). The number of channels
can be specified via the optional argument \str{NC}, 
or directly with the symmetry itself. For example,
\str{SUNchannel} with \src{\str{NC},3} is equivalent
to \str{SU3channel}.
For the case of $\src{NC}>1$ channels, 
abelian (non-abelian) charge in all channels individually
can be requested via \str{Acharge(:)} or \str{SU2charge(:)},
respectively.

\paragraph{Spinless fermions}

Spinless fermionic models, i.e., \src{type=\str{Fermion}},
support the symmetries
\str{Acharge} and \str{SUNchannel} similar to the
spinful case above. A subtlety concerns the
case of an odd number of channels \src{NC}.
There the particle number relative to
half-filling becomes half-integer for the state
space of a single site or, more generally,
an odd number of sites.
In order to avoid this
(also bearing in mind that \QSpace insists on integer
symmetry labels), a factor $2$ is applied to the charge
symmetry labels. See \SecI{U1:labels} for more on this.

The fermionic models require that fermionic signs
are explicitly taken care of in the applications
via the fermionic parity operator \src{Z}.
That is, {\it graded tensors}
\cite{Bultinck17mps,Mortier24} are not implemented 
in \QSpace yet which eventually may allow one to
take care of fermionic signs in an automated fashion.
Bosonic degrees of freedom with a finite local
state space may be implemented as \hsec[gls:tweaks]{tweaks} of the
U(1) fermionic models while stripping any fermionic
signs based on the parity operator \src{Z}.
Bosonic degrees of freedom with infinite state
space such as harmonic oscillators need to be
truncated. This can be done efficiently, e.g., via
a shifted optimal boson basis \cite{Guo12}.

\paragraph{Spin models}

Plain spin models such as SU(2) spin-$S$ Heisenberg
models can be setup with \src{type=\str{Spin}}.
This typically only allows a single symmetry
that is specified as an additional argument,
with the default being \str{SU2}
(see \Sec{spin} below for examples).
To obtain other more
complex spin models based on several symmetries,
the fermionic models may be \hsec[gls:x]{tweaked}
e.g., via projections (see \Sec{tweakS} for an example).

\paragraph{Special case: No symmetries}
\label{idx:nosym}

\QSpace also permits simulations in the absence of
any symmetry. Yet since this is a rather atypical
case when using \QSpace, this proceeds via a tweak.
For example, no symmetries can be implemented via
an auxiliary trivial abelian symmetry such as U(1),
where only the scalar symmetry label $q=0$ 
of the vacuum is present on any leg.
With this, every tensor contains a single \QSpace
record only (thus a single full block as RMT).
An example is provided with the spin setting
in line \eqref{eq:S3:irop:nosym} below,
which returns objects of the above type.
As shown in \idx[nosym:ex]{\Sec{spinhalf}},
for spin $S=1/2$, \src{S(i).data\{1\}} contains
the Pauli matrices $[
 \sigma_z ,\, 
 \frac{1}{\sqrt{2}} \sigma_-,\,
-\frac{1}{\sqrt{2}} \sigma_+
]$ for $i=1,2,3$, respectively.
The order and normalization of operators can always
be simply double-checked by inspection.

\subsubsection{Example: Single fermionic site}

The following call to \src{getLocalSpace}
returns \QSpace operators for a spinful
fermionic site with a single level (i.e., $d=4$ states)
\begin{flalign}
\begin{tabular}{p{4.8in}}
   \src{[F,Z,S,IS] = 
   getLocalSpace(\str{FermionS},\str{Acharge,SU2spin});}
\end{tabular}
\label{eq:gls:ferm}
\end{flalign}
The detailed structure of the returned \QSpace tensors
as displayed within Matlab is discussed by the example of \src{F}
in \Sec{QS:disp} (e.g., see \Fig{QS:disp}) and also
with \eqref{eq:gls:NC1}.
The operators returned by \eqref{eq:gls:ferm} are
the fermionic annihilation operators \src{F}
(irop with \hyperref[F1:start]{two components},
one for spin up, and one for spin down),
the fermionic \hyperref[Z:start]{parity operator \src{Z}}, and
the \idx[Sop:SU2]{total spin operator \src{S}}
(hence an irop with three components).

The last argument \src{IS} returned by \eqref{eq:gls:ferm}
is an \idx[NC1:IS]{info structure} that contains additional
detailed information concerning the chosen symmetry setup.
As seen in \eqref{eq:IS:struct},
it contains the identity operator \src{IS.E},
or \src{IS.sym} as a more verbose description
of the symmetries as specified with the input
to \src{getLocalSpace}. This permits one to look up the order of 
symmetries later, e.g., to reconfirm whether the charge
or spin symmetry is listed first, etc., since the symmetries
specified with a \QSpace tensor, such as \src{A}
or \src{SU2}, no longer contain that type of information.
Other fields in \src{IS} also contain low-level
information relevant to the internal workings of
\src{getLocalSpace} which may be safely ignored.

The symmetries of a model hold throughout a tensor
network simulation. In this spirit,
the call to \src{getLocalSpace} defines the symmetry
once and for all during the setup of a model.
Hence symmetries can only be switched during the setup
for different simulations. For example,
for the same local state space of a single fermionic site,
it may be changed then across the following settings
depending on the symmetries of the model Hamiltonian
based on the underlying model parameters:
\begin{flalign*}
\begin{tabular}{llll}
   \str{Acharge,Aspin}
   & $\text{U(1)}_{\rm charge} \otimes \text{U(1)}_{\rm spin}$
   & $d^\ast = 4 = d$ & (fully abelian)
\\
   \str{Acharge,SU2spin}
   & $\text{U(1)}_{\rm charge} \otimes \text{SU(2)}_{\rm spin}$
   & $d^\ast = 3$ 
\\
   \str{SU2charge,Aspin}
   & $\text{SU(2)}_{\rm charge} \otimes \text{U(1)}_{\rm spin}$
   & $d^\ast = 3$ 
\\
   \str{SU2charge,SU2spin}
   & $\text{SU(2)}_{\rm charge} \otimes \text{SU(2)}_{\rm spin}$
   & $d^\ast = 2$ & (fully non-abelian)
\\
\end{tabular}
\end{flalign*}
All describe the same local state space of $d=4$ states.
With increasing symmetry from top to bottom, however,
the local state space dimension is \idx[Deff]{effectively reduced}
from $d^\ast = 4 \to 2$ multiplets.
For comparison, the more involved example for
\idx[gls:Sp6]{\src{NC=3} channels with
symplectic} symmetry,
\begin{flalign*}
   \src{[F,Z,S,IS]=getLocalSpace(\str{FermionS},\str{SU2spin,SpNchannel},\str{NC},3);}
\end{flalign*}
contains \src{4*NC=12} components in the irop \src{F} 
(creation and annihilation operators for each spin and channel).
The total spin irop \src{S} still has three components.
The scalar parity operator \src{Z} shares much of the structure
of an identity matrix except that its diagonal entries
are $\pm 1$. As such it also reflects the full state
space of a site. By contrast, the spin operator
only acts within partially occupied sectors
since the fully occupied or empty states are singlets
with spin matrix elements all-zero.
Yet zero blocks are not stored. 
In the present example of \src{NC=3} spinful levels,
the largest non-abelian symmetry  
is given by the symplectic \src{Sp(2*NC=6)} symmetry.
Within this setting, the local space dimension
$d=4^3=64$ states is reduced to an effective
dimension of a mere $d^\ast = 4$ multiplets!

\subsubsection{Example: Spin models}
\label{sec:spin}

For spin-models, i.e., \src{type=\str{Spin}} in \Eq{gls},
the argument \src{symmetries} is only allowed to contain
a single symmetry, assuming  SU(2) if not specified.
The options can specify the particular irep to use,
assuming the defining irep, by default.
For example, the following two lines are equivalent for SU(2),
\begin{flalign}
\begin{tabular}{p{4.8in}}
   \src{[S,IS] = getLocalSpace(\str{Spin},S);} 
\end{tabular}
\label{eq:gls:spin-S:1} \\
\begin{tabular}{p{4.8in}}
   \src{[S,IS] = getLocalSpace(\str{Spin},\str{SU2},2*S);}
\end{tabular}
\label{eq:gls:spin-S:2}
\end{flalign}
These return the spin operator for a spin $S$
multiplet in the local state space with 
\idx[SU2:qlabels]{$q$-label} $q=2S$.
The usage in line \eqref{eq:gls:spin-S:2}
permits to switch to other non-abelian symmetries.
For example,
\begin{flalign}
\begin{tabular}{p{4.8in}}
   \src{[S,IS] = getLocalSpace(\str{Spin},\str{SU3},\str{02});}
\end{tabular}
\label{eq:gls:spin}
\end{flalign}
returns the `spin' operator [synonymous for the irop
in the \idx[qadj]{adjoint} representation]
for SU(3), with the local state space in the 6-dimensional
representation \src{q=\str{02}}.
Here the input supports both,
\hsec[compact:qs]{compact string} (\str{02}) and
non-compact numeric notation (\src{[0\,2]}).
\vspace{-1ex}

\paragraph{Switch to abelian or no symmetry}

In the case of SU(2), the usage in line \eqref{eq:gls:spin-S:1}
also accepts the following switches to abelian symmetries,
\begin{flalign}
\begin{tabular}{p{4.8in}}
   \src{[S,IS] = getLocalSpace(\str{Spin},S,\str{-A});}
\end{tabular}
\label{eq:S3:irop:A} \\
\begin{tabular}{p{4.8in}}
   \src{[S,IS] = getLocalSpace(\str{Spin},S,\str{{-}-Z2});}
\end{tabular}
\label{eq:S3:irop:Z2} \\
\begin{tabular}{p{4.8in}}
   \src{[S,IS] = getLocalSpace(\str{Spin},S,\str{{-}-nosym});} 
\end{tabular}
\label{eq:S3:irop:nosym}
\end{flalign}
where the \heq[gls:spinhalf]{option \str{\heq[gls:spinhalf]{-A}}}
returns the spin operator in abelian U(1) symmetries
(equivalent to $Z_\infty$),
\str{{-}-Z2} returns it with \hyperref[Sop:Z2]{$Z_2$ symmetry},
and \str{{-}-nosym} with \hyperref[Sop:nosym]{no symmetry} at all
(effectively implemented as trivial `U(1)' with $q=0$ only).
In all cases, the returned spin operator \src{S} 
has three entries corresponding to
\src{S}\,$= [\hat{S}_z,\, \tfrac{1}{\sqrt{2}} \hat{S}_-,\,
   -\tfrac{1}{\sqrt{2}} \hat{S}_+]$.
See \Sec{spinhalf} for a more detailed discussion.
\vspace{-1ex}

\paragraph{Normalization of spin operator}
\label{idx:S:norm}

The spin operator \src{S} is a specific irop
that transforms in the \idx[qadj]{adjoint} representation.
It has as many components as there
are generators in the Lie algebra.
The spin operators out of \src{getLocalSpace}
[e.g., \Eq{gls:spin}] adhere to 
standard normalization conventions in the
literature. This can be easily
double-checked, e.g., via \src{normQS()}.
For example, for an \idx[Sop:SU2]{SU(2) spin}
in the defining irep,
$\Vert S \Vert^2 \equiv {\rm tr}(S^\dagger\cdot S)
= (3/4) \,{\rm tr}(\Id^{(2)}) = 3/2$.
Therefore it must hold for the respective \QSpace 
spin operator that {\tt norm(S)}\verb|^2| {\tt = 3/2},
seen as $\Vert S\Vert = \sqrt{3/2} = 1.225$
in \mline{idx:Sop:SU2:hdr} below.

The normalization and sign of the spin operator for all ireps
is fully determined by the definition of the spin operator
in the defining representation.
After all, the spin operators in any other irep
can be constructed by combining defining ireps 
and then compute the matrix elements of $S_{\rm tot}$
in the combined Hilbert space. This construction
can be used to reach any other irep. With this
the normalization {\it and sign}
of $S_{\rm tot}$ is fixed by specifying these for
the defining irep. Hence one needs to be careful
when manually constructing spin operators 
directly for an irep other than the defining
representation, e.g., based on existing
CGTs.

Based on the above constructive approach,
the spin operator is also well-defined 
in the presence of OM for larger ireps:
while the spin operator is unique
in the defining irep (this particular case never has OM),
OM can arise for larger local multiplets.
For example, consider the SU(3) spin operator in the
local irep \src{q=\str{11}} also being in the
\idx[qadj]{adjoint} representation. This has $M=2$. Yet from the
constructive approach above, there is also only {\it one}
well-defined spin operator in this setting, since it can be
expressed as $S_{\rm tot}$ in the tensor product space
of three defining ireps (while there will be two ireps
in \str{11} in this construction, both replicate
the {\it same} spin operators, since for both the RMT
links the same way to their shared CGR).

\subsubsection{Beyond models implemented in \src{getLocalSpace}}
\label{sec:gls:x}

The implemented settings for fermionic or spin models
cover a wide range of applications. Yet, in practice,
situations will arise where the desired
symmetry setting is not yet implemented in \src{getLocalSpace}.
In these cases, it is often sufficient to tweak
the \QSpace output of \src{getLocalSpace} that captures
the closest most relevant symmetry setting.
For this, however, it is important that one fully understands
in detail the general \hsec[QS:struct]{\QSpace data structure}.
See \Sec{gls:tweaks} for detailed examples.
As a last resort, one may \idx[contact]{contact}
\QSpace support for help.
Typical strategies of tweaking the output of \src{getLocalSpace}
may include:
\begin{itemize}
  \setlength{\itemsep}{-0.1\baselineskip}

\item appending additional symmetries
[see \src{\@QSpace/addSymmetry.m}].

\item projecting fermionic models to a fixed charge sector,
thus effectively obtaining a spin model. If the charge
label (relative to half-filling) is non-zero, the
respective symmetry labels (columns in \src{\hsec[QS:struct]{X.Q}})
may be simply set to zero because they are all the same
due to the projection.

\item converting abelian symmetries, like U(1) labels to $Z_N$
labels by changing the symmetry name
(cf. \src{\hsec[QS:struct]{X.info.qtype}}) and
converting the symmetry labels in \src{\hsec[QS:struct]{X.Q}})
accordingly.

\item the absence of any symmetries was already
\idx[nosym]{discussed above}.

\end{itemize}

\begin{figure}[p]
\begin{center}
\includegraphics[width=1\linewidth]{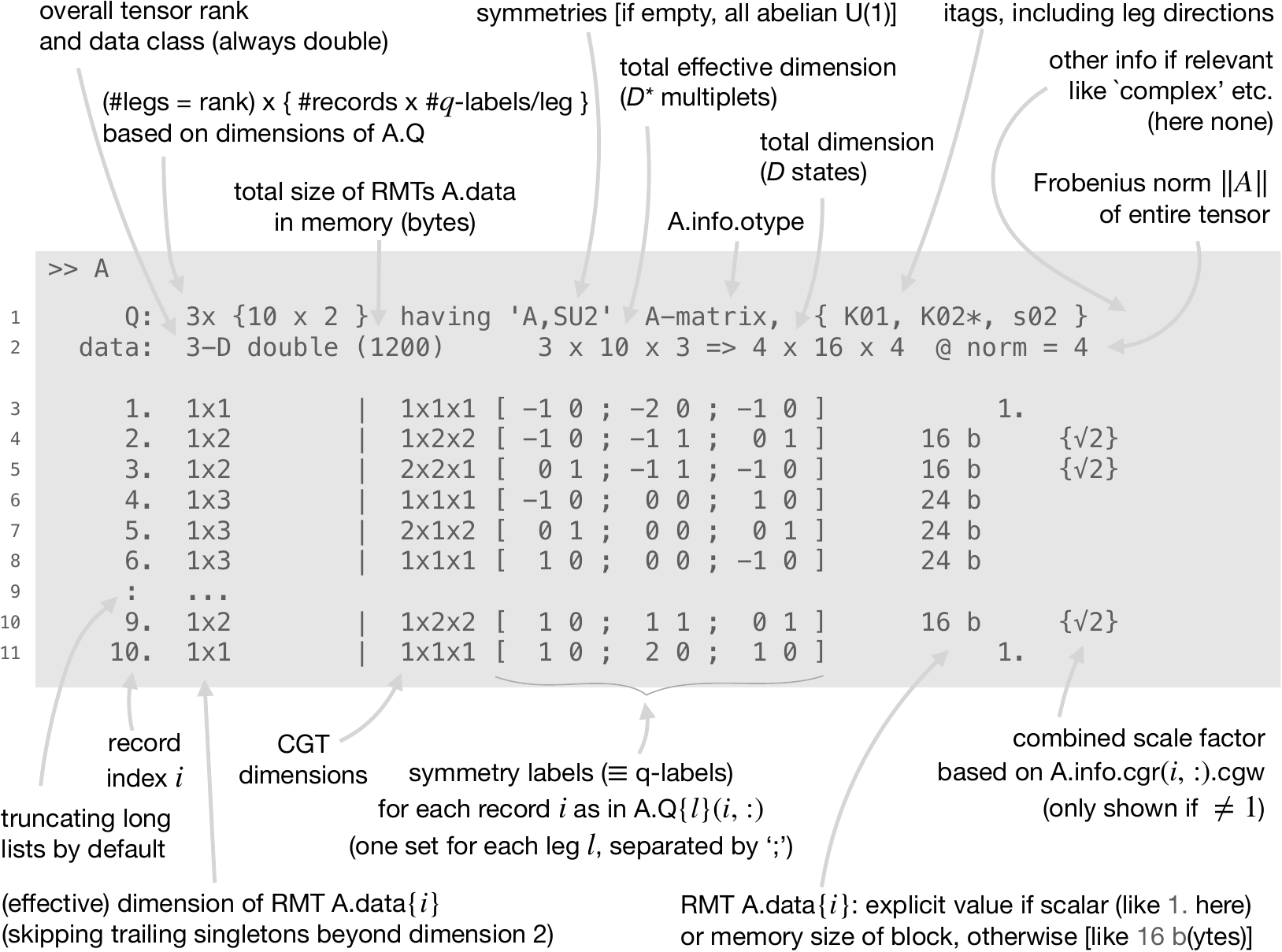}
\end{center}
\caption{ \label{fig:QS:disp}
  Typical \QSpace \Idx[QS:display]{display}, e.g.,
  as printed when typing the name of a \QSpace tensor
  on the Matlab prompt (\src{{>}>}) in the user interface.
  This display summarizes the tensor's \hsec[QS:struct]{data structure} 
  into a more readable form.
  The example here is based on the \idx[Atensor:LRs]{$A$-tensor}
  as obtained in \eqref{eq:A2}.
  It fuses the state spaces of two spinful 
  fermionic sites with $D = 4\times 4 \to 16$ states
  [\src{4 x 16 x 4} in line~2 after permutation
  to standard \idx[Atensor:LRs]{LRs} index order;
  cf. \eqref{eq:A2}].
  The symmetry setting of U(1) charge
  (abelian~\str{A}) with SU(2) spin is shown as \str{A,SU2}
  in line~1. Hence all multiplets carry two symmetry labels
  $q=(q_1\ q_2)$,
  as reflected in the dimensions of \src{Q} in line~1,
  and the columns of $q$-labels listed for each leg.
  The charge $q_1$ represents particle number relative to
  half-filling and hence can become negative, whereas
  $\idx[SU2:qlabels]{q_2=2S} \geq 0$
  reflects the SU(2) spin multiplet.
  Due to the presence of SU(2) spin,
  switching to the multiplet-based
  description reduced the Hilbert space to an effective
  dimension $D^\ast = 3 \otimes 3 \to 10$ multiplets
  (\src{3 x 10 x 3} in line~2).
  Since each fused multiplet carries unique symmetry labels here,
  this also translates to a total of 10 \QSpace records.
  The record index is shown to the left for reference.
  The \idx[RMT]{RMTs} \src{A.data\{\irec{}\}} represent reduced
  matrix elements. Hence they are tensors with {\it effective} dimensions
  as shown in the second column (this adheres to Matlab
  conventions where trailing \idx[singleton]{singleton} dimensions
  beyond dimension 2 are skipped).
  The CGT dimensions are displayed in the third column
  (shown including trailing singletons for non-abelian symmetries
  only; if a \QSpace has all-abelian
  symmetries, the CGT dimensions are skipped entirely).
  Long listings are truncated, by default, as in line~9
  by \str{:   ...} for the skipped records~7-8.
  For a full listing of all \QSpace records, one 
  needs to explicitly call the underlying function
  \src{display(A,\str{-f})},
  since this is the only way to specify options
  (like \str{-f} for full display).
  The combined scale factor out of the CGRs
  [cf. \Eqs{wnorm:x2} and \eqref{eq:wnorm:x3}]
  is shown to the right. 
  Since it always represents the square root of some (product of)
  multiplet dimensions $d \in \mathbb{N}$, it is printed
  for the sake of readability only for those records
  where $d\neq1$ as $\surd d$.
}
\end{figure}

\subsection{QSpace display}
\label{sec:QS:disp}

A summary of the content of a \QSpace tensor is displayed
by simply typing its name at the Matlab command line prompt
followed by enter (technically speaking, this is realized
by overloading the \idx[QS:display]{display}
function in \AtQSpace). 

For example, \Fig{QS:disp} displays
the \idx[Atensor:LRs]{$A$-tensor}
as generated in \eqref{eq:A2},
with detailed additional explanations.
The symmetry setting is a combination of U(1) charge
with SU(2) spin, with the order fixed by the call
to \src{getLocalSpace} in \eqref{eq:gls:NC1}.
The display consists a header info section, followed
by a listing of the symmetry blocks including symmetry labels
and dimensional information of the underlying
RMTs and CGRs. The pair of header lines
show global tensor information, including \itags
for all legs together with their directions.
In case of \itags that are empty up to trailing
markers, the \idx[qdir]{$q$-directions}
are shown in the \heq[qdir:ops]{\src{+/-} format}
as this enables \idx[itag:color]{color coding}.
See also \SRC[F]{applications} for more examples.

The square-root dimensional factors to the
right in \Fig{QS:disp} are shown only  if 
the \idx[cgw]{$w$-matrices} provided with the CGRs
have \hsec[cgw:exceptions]{anomalous} normalization, i.e.,
$w^\dagger w= d\Id$ with $d\neq 1$
(combined as a product across all symmetries).
In this case ``\src{\{$\surd d$\}}'' is included
in the display as a reminder of the altered normalization
affecting the RMTs. From \Eqs{wnorm:x2} and \eqref{eq:wnorm:x3},
it is generally expected that $d$ reflects some multiplet
dimensions, and therefore is an integer.
This motivates the notation with the surd ($\surd$\,)
for the sake of readability and compactness.

\section{Collection of Examples}
\label{sec:examples}

\QSpace uses a compact syntax for
tensor network commands with
usability and readability in mind. This compact notation,
while it may be self-explanatory to some extent,
takes some time to digest and get used to,
though. Hence it is discussed in detail with the examples
in this documentation.
This section provides a combined collection of examples
referenced earlier in this documentation.
Essential syntax-highlighted Matlab code 
is shown in a gray box.
This is followed by a detailed discussion
of output objects and sanity checks on their structure.
The source and output lines are given line numbers
to the left for ease of reference.

To provide \QSpace examples,
the first step is defining a set of exemplary \QSpace tensors.
Hence to be specific,
the examples below use the \QSpace operators
that act within the local state space of a
single spinful fermionic site with U(1) charge
and SU(2) spin symmetry, \\
\begin{minipage}{1\linewidth}
\begin{minted}[escapeinside=??]{matlab}
>> [F,Z,S,IS]=getLocalSpace('FermionS','Acharge,SU2spin');
\end{minted}
\mintLabelB{eq:gls:NC1}
\\[-5ex]
\end{minipage}
The local state space $s \in \{
|0\rangle, |{\uparrow}\rangle, 
|{\downarrow}\rangle, |{\uparrow\downarrow}\rangle \}$
represents empty, singly up/down, and doubly occupied states,
respectively (in contrast to earlier, e.g., \Fig{Aop},
in what follows $\sigma \in \{\uparrow,\downarrow\}
\equiv \{1,2\}$ specifies the
spin flavor index that underlies the local many-body
state space $s$).
By not specifying the number of flavors/channels above, by default,
this assumes a single flavor (for examples with \src{NC=3}
flavors, \idx[gls:NC3]{see further below}).
The above call to \src{getLocalSpace} thus
returns the fermionic annihilation operators
$\hat{F} \equiv (\hat{f}_\uparrow,\hat{f}_\downarrow)^T$
in \idx[irop]{irop} index order convention,
\\[-3ex]\phantomsection\label{src:F}
\begin{minted}[escapeinside=??,firstnumber=last,samepage]{text}
F = ?\mlbl{F1:start}?
     Q:  3x { 2 x 2 }  having 'A,SU2'  operator,  { +-- }         ?\mlbl{F1:hdr1}?
  data:  3-D double (224)      2 x 2 x 1 => 3 x 3 x 2  @ norm = 2 ?\mlbl{F1:hdr2}?

     1.  1x1        |  1x2x2 [ -1 0 ;  0 1 ; -1 1 ]     -1.41421 ?\mlbl{F1:rec1}?
     2.  1x1        |  2x1x2 [  0 1 ;  1 0 ; -1 1 ]          -1. ?\mlbl{F1:rec2}? {?$\surd$?2}
\end{minted}
\begin{samepage}
the fermionic parity operator $\hat{Z} = (-1)^{\hat{n}}$,
with $\hat{n}$ the particle number operator [cf. \Eq{nloc}], 
\begin{minted}[escapeinside=??,firstnumber=last,samepage]{text}
Z = ?\mlbl{Z:start}?
     Q:  2x { 3 x 2 }  having 'A,SU2'  { +- }
  data:  2-D double (336)      3 x 3 => 4 x 4  @ norm = 2

     1.  1x1        |    1x1 [ -1 0 ; -1 0 ]           1.
     2.  1x1        |    2x2 [  0 1 ;  0 1 ]          -1.  {?$\surd$?2}
     3.  1x1        |    1x1 [  1 0 ;  1 0 ]           1.
\end{minted}
\end{samepage}
the spin \idx[irop]{irop} as in \Eq{Sop}
with $\hat{S}_a = \sum_{\sigma\sigma'}
\hat{f}_\sigma^{\dagger} S_a^{\sigma\sigma'}
\hat{f}_\sigma^{\,}$
and $S_\pm \equiv S_x \pm \iu S_y$,
\begin{minted}[escapeinside=??,firstnumber=last]{text}
S = ?\mlbl{idx:Sop:SU2}?
     Q:  3x { 1 x 2 }  having 'A,SU2'  operator,  { +-- }
  data:  3-D double (112)      1 x 1 x 1 => 2 x 2 x 3  @ norm = 1.225   ?\mlbl{idx:Sop:SU2:hdr}?

     1.  1x1        |  2x2x3 [  0 1 ;  0 1 ;  0 2 ]    -0.866025  {?$\surd$?2}  ?\mlbl{Sop:SU2:1}?
\end{minted}
and the structure \src{IS} with additional information,
\mintLabelA{eq:IS:struct}
\begin{minted}[escapeinside=??,firstnumber=last]{text}
IS = ?\mlbl{idx:NC1:IS}?
     NC: 1
    sym: 'Acharge,SU2spin'
      E: [1x1 QSpace]
    ...
\end{minted}
The latter contains the identity operator (\src{IS.E})
as a reference for the complete local state space,
a reminder of the number of flavors (\src{I.NC}),
as well as more verbose specification
of the symmetries used (\src{IS.sym}) for later idenfication.
The remaining fields in \src{IS} are mostly for debugging purposes,
partly also duplicate data such as the parity \src{Z},
and hence can be safely ignored.
The above \QSpace displays for \src{F}, \src{Z}, and \src{S}
are simply obtained by typing the respective name followed by enter,
e.g., \ \src{{>}> F}, where the leading \str{{>}>}
indicates the Matlab prompt.
The generated output conveniently summarizes
the content of the specified \QSpace tensor
(technically, by overloading the
\Idt[QS:display]{display} routine).
The printed output
is discussed in detail with \Fig{QS:disp}.

In the case of the fermionic annihilation operator \src{F},
\mline{F1:hdr1}
identifies the symmetries used in their specified order
[\src{A} for abelian U(1), and \src{SU2} for SU(2) spin].
It also shows that \src{F} is an \idx[irop]{\src{operator}}
(see \hsec[QS:struct]{info.otype})
with standard index order \src{+{-}-}. 
The symmetry labels in \mlines{F1:rec1}--\ref{F1:rec2}
consist of two numbers
$q=[q_1 \ q_2]$ specified for each leg, and separated by 
a semicolon. The first label $q_1$
encodes the particle number (filling) relative to half-filling.
Here $q_1 \in [-1,0,1]$ on the first two legs
which corresponds to $[0,1,2]$ particles
occupying the fermionic level. The second label specifies
the SU(2) symmetry labels, given by the
\idx[SU2:qlabels]{integers $q_2=2S$}.
The irop \src{F} transforms in the labels of the third leg,
namely $q^{\rm op}{=}$\,\src{[-1 1]}: that is, the operator
\src{F} reduces particle
number by one (hence $q_1=$\,\src{-1}), and represents a spinor
of two components (with $q_2=2S=$\,\src{1} it transforms like
a spin $S=1/2$ representing up and down spin). 
All of this is as expected for the set of fermionic
annihilation operators
$\hat{F} \equiv (\hat{f}_\uparrow,\hat{f}_\downarrow)^T$.
Since the irop index represents a single multiplet
(\mline{F1:hdr2}), the operator $q$-labels
are the same for all records. Hence the operator
is irreducible, i.e., it is an irop, indeed.
For comparison, a pair-annihilation operator
$\hat{f}_\uparrow \hat{f}_\downarrow$
would have $q^{\rm op}_1 = -2$.
Therefore by simply looking at the action of an operator
based on its irop symmetry labels
allows one to (roughly) identify an operator. 

By inspection, the first record for \src{F} (\mline{F1:rec1})
describes the matrix elements
$\langle 0|\hat{f}_\sigma |\sigma\rangle$
from singly occupied to empty, with the initial state
on leg 2 and the final state on leg 1
(ket and bra state for matrix element, respectively).
The corresponding CGT has the printed symmetry labels $(0|1,1)$,
bearing in mind $q=2S=1$ which happens to correspond
to a \onej symbol for the first record. It is of size \src{1x2x2}.
Conversely, the second record in \mline{F1:rec2}
describes the matrix elements
from doubly to singly occupied state, 
$\langle -\sigma|\hat{f}_\sigma |{\uparrow\downarrow} \rangle$ where
$-\sigma$ denotes the spin opposite to
$\sigma \in \{\uparrow, \downarrow \}$.
Its CGT hast the $q$-labels $(1|0,1)$ and dimensions
\src{2x1x2}. Therefore the CGT for the second record
corresponds to an identity matrix,
when ignoring the \idx[singleton]{singleton} on the second index.
The respective reduced
matrix elements (RMTs) for the operator \src{F}
turn out to be plain numbers,
thus of size \src{1x1}$\,[\src{x1}]$ (left column),
skipping the last trailing singleton since beyond rank 2).
Their value is directly displayed to the right.
The signs (similar to the sign in \mline{Sop:SU2:1} for~\src{S})
are determined by the sign convention of the underlying
CGT for the irop. They are of no further concern,
otherwise. What is important, though, is that the
operator \src{F} is defined once and for all,
so that the {\it same} object is used for all sites
that are identical.

The way \src{getLocalSpace} proceeds to obtain 
the \QSpace representations above, is to start from the
Fock space description. For a single spinless
fermionic level, in the basis (0,1) for empty
and occupied, the annihilation operator $c$,
fermionic parity operator $z$,
and identity operator $e$ are given by 
\begin{eqnarray}
   c &=& \begin{pmatrix}
      0 & 1 \\
      0 & 0
   \end{pmatrix}
\ , \qquad
   z = [c,c^\dagger] = \begin{pmatrix}
      1 &  0 \\
      0 & -1
   \end{pmatrix}
\ , \qquad
   e = \{c,c^\dagger\} = \begin{pmatrix}
      1 &  0 \\
      0 & 1
   \end{pmatrix}
\end{eqnarray}
with the sole non-zero matrix element of the annihilation
operator being $\langle 0| \hat{c} |1\rangle = 1$.
Hence when combining two fermionic levels with different spin,
for example, the matrix representation for the
fermionic annihilation operators 
can be written by the $4\times 4$ matrices
(assuming the first factor in the tensor product to be
the fast index, consistent with column major),
\begin{eqnarray}
   f_\uparrow = c \otimes e = \begin{pmatrix}
      0 & 1 & 0 & 0 \\
      0 & 0 & 0 & 0 \\
      0 & 0 & 0 & 1 \\
      0 & 0 & 0 & 0
   \end{pmatrix}
   \ ,\qquad
   f_\downarrow = z \otimes c = \begin{pmatrix}
      0 & 0 & 1 & 0 \\
      0 & 0 & 0 &-1 \\
      0 & 0 & 0 & 0 \\
      0 & 0 & 0 & 0
   \end{pmatrix}
\label{eq:fops}
\end{eqnarray}
having assumed the fermionic order
$|{\uparrow\downarrow}\rangle \equiv
 \hat{c}^\dagger_\uparrow \hat{c}^\dagger_\downarrow |\rangle $
thus already leading to the fermionic sign via $z$ 
for $f_\downarrow$ above. It is important for the
fermionic anticommutator relations which must be
already built-in in the above matrix elements, e.g.,
\begin{eqnarray*}
  \{ f_\uparrow, f_\downarrow \} 
  &=& (c \otimes e)(z \otimes c) + (z \otimes c)(c \otimes e)
  =  \{c,z\} \otimes c
  = 0.
\end{eqnarray*}
since $cz = -zc$ by the very property of $c$ that it
changes the particle number by one.
Combining the annihilation operators into the
set $F = (f_\uparrow,f_\downarrow)^T$,
then for example 
\begin{eqnarray*}
   N = F^\dagger\cdot F
 \ =\ f_\uparrow^\dagger f_\uparrow^{\,}
    + f_\downarrow^\dagger f_\downarrow^{\,}
 \ =\ (c^\dagger c \otimes e ) 
    + (c^\dagger c \otimes z^\dagger z)
 \ =\ (n \otimes e) + (e \otimes n)
 \ \equiv\ N_\uparrow + N_\downarrow
\end{eqnarray*}
as expected, with $n = c^\dagger c$ the occupation operator.
\QSpace takes the combined
set \src{F=(f$_\uparrow$,f$_\downarrow$)$^T$} above,
organizes the state space based on symmetry operators
(\hsec[SymOp]{Class/@SymOp}),
and then `compresses' it via \idx[compactQS]{compactQS}.
This diligently factorizes the CGTs as present
in the \RCS, with the remainder being in the RMTs.

\paragraph{Explicit data structure}

By typing \src{F} at the Matlab prompt,
this invokes \Srt{\atQSpace/display} to conveniently summarize
the specified \QSpace data, e.g., as shown for \src{F}
in \mlines{F1:start}--\ref{F1:rec2}.
The bare \QSpace data structure as in \Aps{QS:struct}
is obtained by stripping the tensor \src{F} of
its \QSpace class assignment via \\
\begin{minipage}{1\linewidth}
\begin{minted}[escapeinside=??,firstnumber=last]{matlab}
>> struct(F)
\end{minted}
\mintLabelB{eq:F:struct}
\end{minipage}
\begin{minted}[escapeinside=??,firstnumber=last]{text}
  struct with fields:

       Q: {[2x2 double]  [2x2 double]  [2x2 double]}
    data: {2x1 cell}
    info: [1x1 struct]
\end{minted}
which is considerably less readable and informative
as compared to \mlines{F1:start}--\ref{F1:rec2}
above, despite referring to the same bare data.
The two RMTs $\irec=1,2$ in \mlines{F1:rec1}--\ref{F1:rec2}
are stored in \src{F.data\{\irec{}\}}
\begin{minted}[escapeinside=??,firstnumber=last]{text}
>> F.data'

  1x2 cell array

    {[-1.4142]}    {[-1.0000]}
\end{minted}
with the respective symmetry labels
stored in \src{F.Q\{l\}\{\irec,:\}} for leg $l$,
\begin{minted}[escapeinside=??,firstnumber=last]{text}
>> F.Q{1}
    -1     0
     0     1
     
>> F.Q{2}
     0     1
     1     0
     
>> F.Q{3}
    -1     1
    -1     1
\end{minted}
These are displayed  as \idx[QS:recs]{records} \irec
with \mlines{F1:rec1}--\ref{F1:rec2}
above\footnote{As a \idx[QS:subsref]{Matlab technicality}, note that
\src{F.Q\{:\}} would not work here to display all entries;
rather one would have to use an intermediate assignment,
as in \src{Q=F.Q; Q\{:\}}.}.
The info structure
\begin{minted}[escapeinside=??,firstnumber=last]{text}
>> F.info  ?\mlbl{F.info}?
    qtype: 'A,SU2'
    otype: 'operator'
    itags: {''  '*'  '*'}
    ctime: 0
      cgr: [2x2 struct]
\end{minted}
is described in more detail with the data structure
in \Aps{QS:struct}. It contains symmetry-related content,
such as the symmetry type \src{F.info.qtype} 
or the CGRs in \src{cgr}.
The field \src{itags} specifies the \itag for each leg.
Hence the number of entries has to match 
\src{numel(F.Q)}, or equivalently the rank of the tensor.

The field \src{info.cgr} is another structure array
with one column per symmetry,
and as many rows as there are records in the \QSpace{}.
In case of the operator \src{F}, for example, the entry
\src{F.info.cgr(1,2)} contains the CGR for 
the \QSpace record $\irec=1$ and symmetry $2$,
i.e., the SU(2) spin symmetry as per the present setup
in \eqref{eq:gls:NC1},
\begin{minted}[escapeinside=??,firstnumber=last]{tex}
>> F.info.cgr(1,2)

    type: 'SU2'                            % symmetry type
    qset: [0 1 1]        ?\mlbl{F:cgr:qset}?                  % symmetry labels ($q$-labels, here $\idx[SU2:qlabels]{q=2S}$)
    qdir: '+--'                            % $\idx[qdir]{q\texttt{-directions}}$
     cid: [1.50e+09 1.50e+09 690057 0 8]   % $\RCS$ specific IDs incl. time stamps
    size: [1 2 2]         ?\mlbl{F:cgr:size}?                 % CGT size (dimensions, uint64 data type)
     nnz: 2                                % number of non-zero entries in CGT
     cgw: 1                                % $w$-matrix
     cgt: [1x0 double]                     % auxiliary internal field for trace

\end{minted}
The symmetry labels \src{qset} in \mline{F:cgr:qset}
are those from the referenced CGT and reflect the ones
in \src{X.Q\{l\}(1,2)} for leg $l$
as listed with the \QSpace display 
of this record $\irec=\src{1.}$ in \mline{F1:rec1}
earlier. The dimensions in 
\mline{F:cgr:size} are also displayed there.

\noindent 
By contrast, the CGRs \src{X.info.cgr(:,1)}
for the leading abelian U(1) symmetry
are all trivial. The underlying CGTs are simply $C=1.$,
bearing in mind that only non-zero CGTs are present in a \QSpace
tensor, i.e., only ones that are permissible
from a symmetry point of view.
For example, the trivial CGR from the first record
reads
\begin{minted}[escapeinside=??,firstnumber=last]{text}
>> F.info.cgr(1,1)

    type: 'A'
    qset: [-1 0 -1]
    qdir: '+--'
     cid: [0 0 0 0 1]
    size: [1 1 1]
     nnz: 0
     cgw: 1
     cgt: [1x0 double]
\end{minted}
It reflects the plain number $1.$ by having
\src{size = 1}, \src{cgw = 1}, \src{nnz} is ignored, 
no internal IDs \src{cid}, etc.
The symmetry labels \src{qset = [-1 0 -1]} are again
consistent with those
in \src{X.Q\{l\}(1,1)} for leg $l$
as displayed in \mline{F1:rec1} earlier.

\paragraph{State space from operators}

The annihilation operator \src{F} is non-hermitian
and contains only off-diagonal symmetry blocks.
Its full dimensions are shown as \src{3 x 3 x 2}
in \mline{F1:hdr2} which only references 3 out of the
total of 4 states with either of the first two indices.
The reason being, that an annihilation operator
destroys an empty ket state $|0\rangle$, yet also has no
matrix elements that include the fully occupied
bra state $\langle {\uparrow\downarrow}|$.
However their union
of symmetry sectors as obtained by \src{getIdentity}
when called on an operator like \src{F} here, \\
\begin{minipage}{1\linewidth}
\begin{minted}[escapeinside=??,firstnumber=last]{matlab}
>> E=getIdentity(F)  ?\mlbl{E1:F1}?
\end{minted}
\mintLabelB{eq:Id:F}
\end{minipage}
\begin{minted}[escapeinside=??,firstnumber=last]{text}
     Q:  2x { 3 x 2 }  having 'A,SU2'  { +- }
  data:  2-D double (336)      3 x 3 => 4 x 4  @ norm = 2  ?\mlbl{E1:hdr2}?

     1.  1x1        |    1x1 [ -1 0 ; -1 0 ]           1.
     2.  1x1        |    2x2 [  0 1 ;  0 1 ]           1.  {?$\surd$?2}
     3.  1x1        |    1x1 [  1 0 ;  1 0 ]           1.

\end{minted}
recovers the full \src{4 x 4} dimensional local
state space (\mline{E1:hdr2})
of the single spinful fermionic level
under consideration.
By not having specified an index with the operator
\src{F} in \mline{E1:F1},
the default behavior of \src{getIdentity}
assumes an \src{\idx[irop]{operator}}
as input with standard index order, 
and hence takes the union of the state spaces of
the first two indices of \src{F}.
The fact that the first two indices of the annihilation
operator must span the
full state space of a site is also apparent
from the realization that its fermionic anticommutator 
relations must yield the identity [for an explicit
sanity check in this regard, see \eqref{eq:acomm}].

\begin{figure}[tbh]
\begin{center}
\includegraphics[width=.8\linewidth]{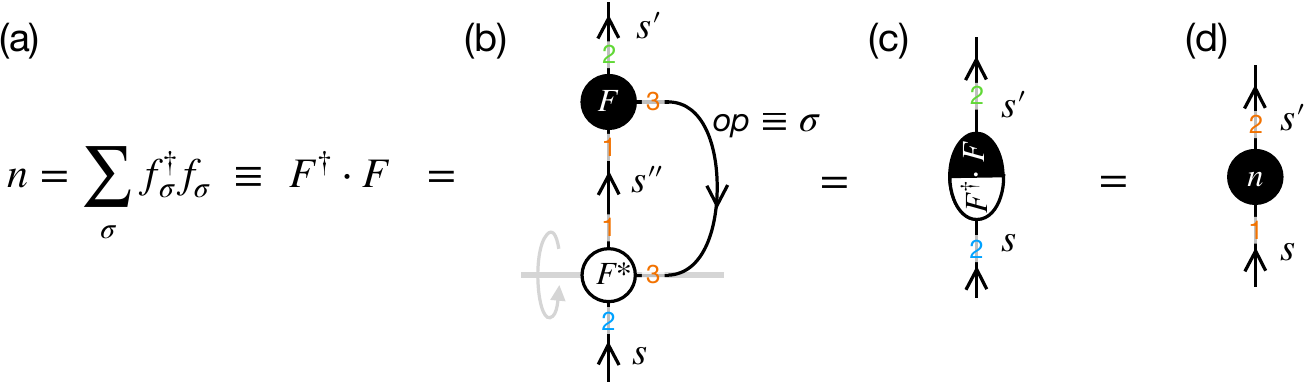}
\end{center}\vspace{-2ex}
\caption{Contraction $n=F^\dagger\cdot F$ to obtain the
occupation operator in some (local) state space $s$,
having spin flavor $\sigma \in\{\uparrow,\downarrow\}$.
(b) The contraction in \eqref{eq:nloc}
takes the conjugate tensor of \src{F}
reverts all arrows and also \idx[conj]{reflects}
the tensor, here vertically as indicated with
the gray line and rotating arrow.
Hence the index order in $1$ and $2$ is already
flipped up/down with~$F^\ast$.
By contracting indices~\str{13*}
on the first, i.e., lower tensor, the~\src{1} takes the transpose,
thus completing the Hermitian conjugate. 
The remaining open indices are collected (c), thus giving
rise to the new index order in (d) for the
occupation operator. Uncontracted, and thus
open indices are collected 
in the order of the input \QSpace tensors to
\src{contract}. Therefore the chosen order of input
arguments in the command \Eq{nloc}
already returns the \idx[irop]{standard index order for
operators}, as indicated by the orange labels in~(d).
\vspace{-2ex}
}
\label{fig:nctr} 
\end{figure} 

\paragraph{Occupation number operator}
\label{idx:nloc}

The operators returned by \src{getLocalSpace} in \eqref{eq:gls:NC1}
are considered elementary for the description of the specified site.
Any other local operator can be constructed from these,
usually by simple one-liners.
By construction, these operators are then already
consistent with the chosen symmetries.
For example, the local occupation number operator
is obtained by \\
\begin{minipage}{1\linewidth}
\begin{minted}[escapeinside=??,firstnumber=last]{matlab}
>> nloc = contract(F,'13*',F,'13')
\end{minted}
\mintLabelB{eq:nloc}
\end{minipage}
\begin{minted}[escapeinside=??,firstnumber=last]{text}
     Q:  2x { 2 x 2 }  having 'A,SU2'  { +- }
  data:  2-D double (224)      2 x 2 => 3 x 3  @ norm = ?$\surd$?6

     1.  1x1        |    2x2 [  0 1 ;  0 1 ]           1. ?\mlbl{nloc:rec1}? {?$\surd$?2}
     2.  1x1        |    1x1 [  1 0 ;  1 0 ]           2. ?\mlbl{nloc:rec2}?
\end{minted}
with a pictorial representation presented in \Fig{nctr}.
In addition to the contraction of index \src{1}
from the matrix product, the sum over spin corresponds
to the simultaneous contraction of the irop index \src{3}
as specified by the strings \str{13*} or \str{13}
in \idx[compact:idx]{compact} notation for
the first and second operator, respectively.
The above display for \src{nloc} shows two block entries
corresponding to RMTs with \src{1.} or \src{2.} particles
as printed to the right in
\mlines{nloc:rec1} and \ref{nloc:rec2},
respectively.
These values reflect \src{nloc.data\{\irec{}\}}.
There is no block record for \src{0.} particles,
as this would represent a zero-block
(more precisely, since block-diagonal zero
blocks are usually \idx[QS:zblocks]{kept in rank-2 operators},
in the present case this block simply 
does not arise out of the contraction in the first place).
Note that for the value
\src{1.} to appear in \mline{nloc:rec1}, 
the normalization of the \idx[cgw]{$w$-matrix}
had been tweaked, as indicated by the $\{\surd2\}$ factor
to the very right [cf. discussion with \Eq{wnorm:x2}].

The definition of \src{nloc} permits one to relate its trace
${\rm tr}(n_{\rm loc}) = {\rm tr}(F^\dagger
\cdot F) = \Vert F\Vert^2$
with the norm of $\Vert \src{F}\Vert=2$ in \mline{F1:hdr2}.
This can be quickly double-checked as follows,
\begin{minted}[escapeinside=??,firstnumber=last]{text}
>> [ trace(nloc), norm(F)^2 ]

    4.0000    4.0000

\end{minted}
The trace above includes a factor of $2$ for the sector \src{q=[0 1]}
in \mline{nloc:rec1}, since by representing a spin half ($2S=1$),
this contains two states. Hence
${\rm tr} (n_{\rm loc}) = 0 + 2\times 1 + 2 = 4$.
The fermionic anti-commutator relations can be checked by  \\
\begin{minipage}{1\linewidth}
\begin{minted}[escapeinside=??,firstnumber=last]{matlab}
>> E=acomm(F,F')/2           %  $=\frac{1}{2} \sum_\sigma \{ \hat{f}_\sigma, \hat{f}_\sigma^\dagger \} \ \equiv\ \frac{1}{2} (F \cdot F^\dagger + F^\dagger \cdot F)\ =\ \Id \mlbl{E1:acomm}$ 
\end{minted}
\mintLabelB{eq:acomm}
\end{minipage}
\begin{minted}[escapeinside=??,firstnumber=last]{text}
     Q:  2x { 3 x 2 }  having 'A,SU2'  { +- }
  data:  2-D double (336)      3 x 3 => 4 x 4  @ norm = 2

     1.  1x1        |    1x1 [ -1 0 ; -1 0 ]           1.
     2.  1x1        |    2x2 [  0 1 ;  0 1 ]           1.  {?$\surd$?2}
     3.  1x1        |    1x1 [  1 0 ;  1 0 ]           1.

>> isIdentityQS(E) ?\mlbl{E:isId}?

     1
\end{minted}
This needs to sum over $\sigma \in \{\uparrow,\downarrow\}$
since \src{F} exists as an irop, where one cannot single out a
particular component $\sigma$, as this would break the symmetry.
\mLine{E:isId} checks whether \src{E} represents an
identity operator (solely within the symmetry space
present in the \QSpace~\src{E}, i.e., without knowledge
of any potentially `absent' \idx[QS:zblocks]{zero blocks};
this is no issue here, though).
The call to the \QSpace routine \src{acomm()} above for the
anti-commutator is equivalent to performing the
contraction in \eqref{eq:nloc} plus its respective 
counterpart to complete the anti-commutator
[see comment with \eqref{eq:acomm}],
\begin{minted}[escapeinside=??,firstnumber=last]{text}

>> E_ = ( contract(F,'13*',F,'13' ) + ...
          contract(F,'23', F,'23*') ) / 2;
>> norm(E-E_) ?\mlbl{E1:diff}?

     0

>> E==E_   ?\mlbl{E1:same}?

     1
\end{minted}
\mLines{E1:diff} and \ref{E1:same}
verify that \src{E} and \src{E\_} are the
same tensors, indeed (\mline{E1:diff} is somewhat
more flexible as compared to the overload in \src{\atQSpace/eq.m},
as one may use the former to accept differences
on the order of double precision noise in a general context).
Similar to the fermionic occupation operator in \Eq{nloc},
one can construct the local Casimir operator from the spin
operator, \\
\begin{minipage}{1\linewidth}
\begin{minted}[escapeinside=??,firstnumber=last]{matlab}
>> S2=contract(S,'13*',S,'13')                 %  $\hat{S}^2 = \hat{S}^\dagger \cdot \hat{S}\ =\ \tfrac{3}{4}\, \hat{\Id}$ 
\end{minted}
\mintLabelB{eq:S2}
\end{minipage}
\begin{minted}[escapeinside=??,firstnumber=last]{text}
     Q:  2x { 1 x 2 }  having 'A,SU2'  { +- }
  data:  2-D double (112)      1 x 1 => 2 x 2  @ norm = 1.061   ?\mlbl{S2:hdr2}?

     1.  1x1        |    2x2 [  0 1 ;  0 1 ]         0.75  {?$\surd$?2}
\end{minted}
Since the spin operator \eqref{eq:Sop} is
non-hermitian, the dagger is relevant.
The above display shows a single entry for the half-filled
symmetry sector \src{[0 1]} only, which being  spin-half
($2S=1$), is the only one with a non-zero Casimir, indeed,
having the value \src{0.75} shown to the right.
All other symmetry sectors, namely empty and doubly
occupied, have $S^2=0$. Thus being \idx[QS:zblocks]{zero-blocks},
these are absent. Matter of fact, these are already absent in 
the spin operator \src{S} that entered the contraction.
The Casimir is thus already also reflected
in the RMTs of the spin operator, e.g., having the RMT
$\sqrt{3/4} = 0.866025$ in \mline{Sop:SU2:1} up to a sign.
The norm $\Vert S^2 \Vert$ is
given by $\frac{3}{4} \Vert \ \Id^{(2)} \Vert
= \sqrt{9/8} = 1.061$ as seen in \mline{S2:hdr2}.
This is in contrast to $\Vert S\Vert = \sqrt{3/4}\
\Vert \Id^{(2)} \Vert = \sqrt{3/2} = 1.225$
in \mline{idx:Sop:SU2:hdr} above.
\\[2ex]
Based on the set of \QSpace operators introduced above,
the following proceeds with explicit examples
for the main text earlier.
This starts with examples for \SecI{nolegs}.

\subsection{\QSpace tensors of rank \texorpdfstring{$r<2$}{r<2}}

\paragraph{Rank-0 tensor}
\label{idx:rank0}

A tensor of \idx[rank01]{rank $r=0$} has no open legs, 
and hence can only represent a scalar number.
It naturally occurs when all indices have been contracted,
e.g., when computing expectation values or overlaps. 
For example, computing the Frobenius norm of a tensor
translates to contracting a tensor fully with itself,
\begin{eqnarray*}
   \Vert \hat{f}\Vert^2 &\equiv&
    {\rm tr}(\hat{f^\dagger} \cdot \hat{f})
  = \sum_\sigma
    {\rm tr}(\hat{f}_\sigma^\dagger \cdot \hat{f}_\sigma)
\end{eqnarray*}
the result has no more open legs. In \QSpace, this becomes

\begin{samepage}
\noindent
\begin{minipage}{1\linewidth}
\begin{minted}[escapeinside=??,firstnumber=last]{matlab}
>> X=contract(F,'123*',F,'123')
\end{minted}
\mintLabelB{eq:rank0}
\end{minipage}
\begin{minted}[escapeinside=??,firstnumber=last]{text}
     Q:  []  having 'A,SU2'  double scalar ?\mlbl{rank0:1}?
  data:  { 4 }  ?\mlbl{rank0:2}?

>> getscalar(X) ?\mlbl{X:scalar}?

    4.0000
\end{minted}
Therefore it has empty
$q$-labels \src{X.Q}, and  only a single RMT \src{X.data\{1\}}
[cf. \Aps{QS:struct}] with value
\src{X.data\{1\}} $= \Vert F\Vert^2 = 4$ (cf. \mline{F1:hdr2}).
For such rank-0 tensors, the \QSpace display switches to the
simplified format in \mlines{rank0:1}--\ref{rank0:2} above.
The scalar value can be extracted by \src{getscalar} (\mline{X:scalar})
which also performs consistency checks (e.g., if \src{X}
does not represent a scalar, it throws an error).
\end{samepage}

\paragraph{Rank-1 tensor}
\label{idx:rank1}

The only permissible \idx[rank01]{rank-1 tensor} is a tensor
that carries the symmetry sector of the vacuum
state, namely $q=0$.
For a general non-empty \QSpace \src{X} of rank $r\geq2$,
the rank-1 tensor representing the vacuum state
can be obtained by
 \\
\begin{minipage}{1\linewidth}
\begin{minted}[escapeinside=??,firstnumber=last]{matlab}
>> V=getvac(F,'-1d')
\end{minted}
\mintLabelB{eq:rank1}
\end{minipage}
\begin{minted}[escapeinside=??,firstnumber=last]{text}
     Q:  1x { 1 x 2 }  having 'A,SU2'  { + }
  data:  1-D double (112)      1 x 1  @ norm = 1

     1.  1          |      1 [  0 0 ]           1.   ?\mlbl{V:rec1}?
\end{minted}
With the option \str{-1d}, this returns a vector of
length $d=1$ (\mline{V:rec1}).
Rank-1 tensors in $q=0$ with dimension $d>1$ can
occur naturally out of contractions, e.g.,
when fully contracting a tensor of rank $r$ on 
all its $r$ indices with a tensor of rank $r+1$.
Without the \str{-1d} option, \src{getvac} in \Eq{rank1}
returns a rank-2 tensor that represents the identity operator
in the vacuum state, and thus is of dimension $1\times 1$.

\subsection{Permutations and index order}
\label{sec:permute}

\QSpace tensors support an arbitrary index order
that can be permuted as desired.
While some tensors such as \idx[irop]{operators} or
\idx[Atensor]{$A$-tensors} assume certain index order
conventions for coding purposes, these are relevant
for applications only, but not for \QSpace tensors per se.
For the sake of the argument, consider
an operator that describes fermionic hopping
from site \src{s02} and \src{s01}, i.e.,
$\hat{f}_1^\dagger \cdot \hat{f}_2\ =\ \sum_\sigma\hat{f}_{1\sigma}^\dagger \cdot \hat{f}_{2\sigma}$
with $\sigma \in \{\uparrow,\downarrow\}$, \\
\begin{minipage}{1\linewidth}
\begin{minted}[escapeinside=??,firstnumber=last]{matlab}
>> F2=contract(F,'-op:s01','*', Z*F,'-op:s02', [2 3 1 4]) ?\mlbl{F2:op}?
\end{minted}
\mintLabelB{eq:FdotF}
\end{minipage}
\begin{minted}[escapeinside=??,firstnumber=last]{text}
     Q:  4x { 4 x 2 }  having 'A,SU2'  { s01, s02, s01*, s02* }  ?\mlbl{F2:hdr1}?
  data:  4-D double (448)      2 x 2 x 2 x 2 => 3 x 3 x 3 x 3  @ norm = ?$\surd$?8

     1.  1x1          | 2x1x1x2 [  0 1 ; -1 0 ; -1 0 ;  0 1 ]      1.41421
     2.  1x1          | 2x2x1x1 [  0 1 ;  0 1 ; -1 0 ;  1 0 ]     -1.41421
     3.  1x1          | 1x1x2x2 [  1 0 ; -1 0 ;  0 1 ;  0 1 ]     -1.41421
     4.  1x1          | 1x2x2x1 [  1 0 ;  0 1 ;  0 1 ;  1 0 ]     -1.41421
\end{minted}
written as rank-4 tensor here by only contracting the operator
index via the dot product in the operators, thus having
$\langle s_1 s_2|\hat{f}_1^\dagger \cdot \hat{f}_2
|s'_1 s'_2\rangle$.
By specifying a conjugate flag \str{*} with the first
operator \src{F}, this takes its Hermitian conjugate.
The option \str{-op:s01} assigns
\idx[irop]{operator} \itags for site \str{s01} on the fly.
Specifically, this assigns the temporary \itags\ 
\src{\{\str{s01},\str{s01*},\str{op*}\}}\ 
to the preceding input operator [cf. \Fig{Aop}(b)].
Similarly, the second operator \src{Z*F} gets assigned
the operators \itags for site \str{s02}
(because in a simulation typically more than 10 sites
are included, \QSpace frequently uses a 2-digit 
format for site or bond indices in \itags, like
\str{s01} and \str{s02} above for the sake of
aligned displays).

Fermionic signs are applied by contracting the parity
operator \src{Z}$\equiv (-1)^{\hat{n}}$
from the left onto the operator \src{F} acting 
on site $2$. This assumes a fermionic order
where site $1$ comes before site~$2$ when building
a Fock state space, e.g., $\hat{f}^\dagger_2 \hat{f}_1^\dagger
|\rangle$ similar to \Eq{fops}.
Hence the matrix elements for $\hat{f}_1$ acquire
fermionic signs depending on the state of site $2$
which thus gets the parity operator applied
(see also appendix in \cite{Wb12} for more discussion
on fermionic signs).
Since both, \src{Z} and \src{F}, are operators,
one can simply use the operator \src{*} for matrix multiplication
in \src{Z*F}.%
\footnote{The matrix multiplication operator \src{*}
is overloaded via \Srt{\atQSpace/mtimes.m}
to perform a \QSpace contraction;
this also deals with potential rank-3 operators as input,
like \src{F} here. As a safety measure,
this requires the explicit marking of the \QSpace
as an `operator' via the flag \src{F.info.otype=\str{operator}}.
Here this is automatically set via \src{getLocalSpace},
e.g., see \mline{F1:hdr1} above.}
The default operator \itag assigned above
by the option \str{-op:..} is \str{op}.
In the present example, this is the only matching \itag
in the auto-contraction. Therefore the
contraction results in a rank-4 tensor that explicitly
acts on the two separate sites \src{s01} and \src{s02}.

Command \eqref{eq:FdotF} 
also includes a trailing permutation that is applied
to the overall result. Therefore it is equivalent to
\\[2ex]
\begin{minipage}{1\linewidth}
\begin{minted}[escapeinside=??,firstnumber=last]{text}
>> F2=contract(F,'-op:s01','*',Z*F,'-op:s02'); ?\mlbl{F2:Z}?
>> F2=permute(F2,'2314')     % may also write permutation as [2 3 1 4] ?\mlbl{F2:perm}?
\end{minted}
\mintLabelB{eq:F2:perm}
\end{minipage}
\\[3ex]
The contraction in \mline{F2:Z}
collects the uncontracted legs of
the first operator \src{conj(F)} [note the \str{*} flag
in the contraction], followed by the uncontracted ones
from the second operator, maintaining their original order.
The bra and ket indices of the rank-4 operator \src{F2}
out of the contraction can therefore be properly grouped
by the permutation \src{[2 3 1 4]} (\mline{F2:perm},
or trailing option in \mline{F2:op})
with the effect
$(s^\ast_1,s_1,s_2,s^\ast_2)
\to (s_1,s_2, s^\ast_1 s^\ast_2)$.
This finalizes the transpose for the first operator,
which was conjugated by the flag \str{*} in the contraction.
With this, the \itags of the final output for \src{F2}
have $q$-directions \qdir{++{-}-} as seen from the trailing
conjugate flags \str{*} in the \itags in \mline{F2:hdr1}.

The permutation in \mline{F2:perm}
accepts \idx[perm]{compact string} notation aside from the
plain numerical representation. By contrast,
\mline{F2:op}
insists on a numerical representation
of the permutation, in order to differentiate it from 
other string options meant for contractions.
\mLine{F2:perm}
accepts any valid permutation to be applied
to \src{F2}, including an optional trailing conjugation flag \str{*}.
For example, the following swaps bra with ket indices
and also applies overall tensor \idx[conj]{conjugation},
i.e., obtains the Hermitian conjugate
$\langle s_1 s_2|\hat{f}_2^\dagger \cdot \hat{f}_1
|s'_1 s'_2\rangle$ for given 
rank-4 tensor, 
\begin{minted}[escapeinside=??,firstnumber=last]{text}
>> F2_ = permute(F2,'3412*')

     Q:  4x { 4 x 2 }  having 'A,SU2'  { s01, s02, s01*, s02* }
  data:  4-D double (448)      2 x 2 x 2 x 2 => 3 x 3 x 3 x 3  @ norm = ?$\surd$?8

     1.  1x1          | 1x2x2x1 [ -1 0 ;  0 1 ;  0 1 ; -1 0 ]      1.41421
     2.  1x1          | 1x1x2x2 [ -1 0 ;  1 0 ;  0 1 ;  0 1 ]     -1.41421
     3.  1x1          | 2x2x1x1 [  0 1 ;  0 1 ;  1 0 ; -1 0 ]     -1.41421
     4.  1x1          | 2x1x1x2 [  0 1 ;  1 0 ;  1 0 ;  0 1 ]     -1.41421

\end{minted}
While the \itags of the resulting \src{F2\_}
are identical to the ones from \src{F2},
their content is clearly different, as can be simply checked
by \src{norm(F2-F2\_)} which results in 4.

\newpage

\subsection{\onej tensor and trailing marker characters}
\label{sec:1j-markers}

The identity operator in \Eq{Id:F} is a regular
\idx[irop]{operator} with one in- and one out-going index.
Instead of the identity, one may request
the \hsec[1j:tensor]{\onej tensor}
by the same call
to \src{getIdentity}, yet adding the option \str{-0}
(`fusing to $q=0$'),
\\[1ex]
\begin{minipage}{1\linewidth}
\begin{minted}[escapeinside=??,firstnumber=last]{text}
>> U=getIdentity(F,'-0') ?\mlbl{U:1j}?
\end{minted}
\mintLabelB{eq:1j-example}
\end{minipage}
\begin{minted}[escapeinside=??,firstnumber=last]{text}

     Q:  2x { 3 x 2 }  having 'A,SU2'  { +?\textcolor{mcol}{+}? } ?\mlbl{U:1j:hdr1}?
  data:  2-D double (336)      3 x 3 => 4 x 4  @ norm = 2         ?\mlbl{U:1j:hdr2}?

     1.  1x1        |    1x1 [ -1 0 ;  1 0 ]           1.
     2.  1x1        |    2x2 [  0 1 ;  0 1 ]           1.  {?$\surd$?2}
     3.  1x1        |    1x1 [  1 0 ; -1 0 ]           1.  ?\mlbl{U:1j:end}?
\end{minted}
This shows a colored \itag in \mline{U:1j:hdr1}
when using the \idx[ml:term]{Matlab terminal}
mode, for readability. The actual \str{U.info.itags} are
\src{\{\str{},\str{'}\}}, with the first itag empty,
and the second \itag just a \idx[itag:markers]{prime (\src{'})}.
Since no string delimiters are shown with \itags in the \QSpace
display, empty strings would not display at all.
Hence, in case of empty strings, the \idx[qdir]{$q$-directions}
\src{+/-} are displayed, instead. In the presence of a trailing
prime marker character, the \itag is colored as above.
The color coding is understood by terminals only,
but not the Matlab desktop environment.
For the latter by default, \mline{U:1j:hdr1}
above would display as
\begin{minted}[escapeinside=??,firstnumber=last]{text}
     Q:  2x { 3 x 2 }  having 'A,SU2'  { ++' }
\end{minted}
instead, which shows the marker character with the second \itag,
\src{++\textquotesingle} $\equiv$
\src{\{\src{+},\src{+\textquotesingle}\}}.
The same can also be enforced by setting
\src{setenv QS\_LOG\_COLOR 0} in the \idx[QS:color]{environment},
e.g., on the Matlab prompt itself.
If \src{getIdentity} is called with reference to a tensor
leg that carries an \itag, then that \itag is inherited,
\begin{minted}[escapeinside=??,firstnumber=last]{text}
>> E=setitags(getIdentity(F),{'s01','s01'});
>> U=getIdentity(E,'-0')

     Q:  2x { 3 x 2 }  having 'A,SU2'  { s01, ?\textcolor{mcol}{s01}? } ?\mlbl{U:1j:hdr1b}?
     ...

\end{minted}
again, with colored output, as in Matlab terminal mode.
The remainder (\src{...}) is the same as in
\mlines{U:1j:hdr2}--\ref{U:1j:end}.
The actual \itags \src{U.info.itags} are \{\str{s01},
\str{s01'}\}, as displayed in the Matlab desktop,
where \mline{U:1j:hdr1b}
would read
\begin{minted}[escapeinside=??,firstnumber=last]{text}
     Q:  2x { 3 x 2 }  having 'A,SU2'  { s01, s01' }  ?\mlbl{U:1j:hdr1c}?
\end{minted}
If one were to take the conjugate tensor of $U$,
this results in all outgoing legs, with the second leg
still also marked,
which in terminal mode would translate to the color coding
for the sake of readability, 
\begin{minted}[escapeinside=??,firstnumber=last]{text}
>> conj(U)
     ...
     Q:  2x { 3 x 2 }  having 'A,SU2'  { ?\textcolor{ccol}{s02}?, ?\textcolor{mccl}{s02}? }
     ...
\end{minted}
where the an outgoing \itag
is printed in gray, and if marked, the original brighter green
in \mline{U:1j:hdr1b}
becomes a darker green.
In the Matlab desktop, the \itags are fully spelled out 
\begin{minted}[escapeinside=??,firstnumber=last]{text}
>> conj(U)
     ...
     Q:  2x { 3 x 2 }  having 'A,SU2'  { s02*, s02'* }
     ...
\end{minted}
The trailing marker toggle inserted by \src{getIdentity}
with the option \str{-0} may be disabled by using the
option \str{-z}, instead (\src{z} as in zero in \str{-0}
for historical reasons; not recommended),
\begin{minted}[escapeinside=??,firstnumber=last]{text}
>> Uz=getIdentity(E,'-z')

     Q:  2x { 3 x 2 }  having 'A,SU2'  { s01, s01 } ?\mlbl{Uz:hdr1}?
  data:  2-D double (336)      3 x 3 => 4 x 4  @ norm = 2

     1.  1x1        |    1x1 [ -1 0 ;  1 0 ]           1.
     2.  1x1        |    2x2 [  0 1 ;  0 1 ]           1.  {?$\surd$?2}
     3.  1x1        |    1x1 [  1 0 ; -1 0 ]           1.

\end{minted}
In this case, the \itags (\mline{Uz:hdr1})
are the same for both legs and become indistinguishable 
from the ones for the transpose of the tensor.
However, taking the transpose of \src{Uz},
\begin{minted}[escapeinside=??,firstnumber=last]{text}
>> Uz_=permute(Uz,'21')

     Q:  2x { 3 x 2 }  having 'A,SU2'  { s01, s01 }
  data:  2-D double (336)      3 x 3 => 4 x 4  @ norm = 2

     1.  1x1        |    1x1 [  1 0 ; -1 0 ]           1.
     2.  1x1        |    2x2 [  0 1 ;  0 1 ]           1.  {-?$\surd$?2}
     3.  1x1        |    1x1 [ -1 0 ;  1 0 ]           1.

\end{minted}
give rise to a minus sign with the second entry
(with the permutation absorbed into the \idx[cgw]{$w$-matrix},
and hence showing up with the curly bracket to the very right).
Therefore \src{Uz} is clearly
different from \src{Uz\_}, having \src{norm(Uz-Uz\_) = $\surd 8$},
even though indistinguishable, say, pictorially when including \itags.
Both are unitaries, though, and therefore can act as \onej symbols.
However, mixing them up when reverting directions of legs, e.g.,
by inserting \src{Uz * Uz\_}$^\dagger \neq \Id$ into a line
in a tensor network, this is likely to give wrong results
because of sign errors, even though the contraction is 
possible on principle grounds. To avoid such 
sources of errors that may be difficult to spot, the trailing prime
as a \idx[itag:markers]{marker character} was introduced
with \QSpace~v4.
After all, the \idx[dual]{dual state space}
is different from its original one, 
even if the considered state space may happen to be
self-dual in the sense that it maps onto itself as a whole,
as in the present example.
Trying to contract a marked with
a non-marked \itag then, will result in an error
because such \itags are considered different.

\subsection{Spin-half spin operators}
\label{sec:spinhalf}

The spin operators for a single spin $S=1/2$
are obtained as in \Eq{S3:irop:A}, \\
\begin{minipage}{1\linewidth}
\begin{minted}[escapeinside=??]{matlab}
>> [S,IS] = getLocalSpace('Spin',1/2,'-A');
\end{minted}
\mintLabelB{eq:gls:spinhalf}
\\[-5ex]
\end{minipage}
resulting in
\begin{minted}[escapeinside=??,firstnumber=last]{text}
>> S  % equivalent to display(S)

 1.     (U1)   { +-- }        3D double      0.7071         224  2x2x1
 2.     (U1)   { +-- }        3D double      0.7071         112  1x1x1      operator
 3.     (U1)   { +-- }        3D double     -0.7071         112  1x1x1      operator
\end{minted}
Since \src{S} is a \QSpace array of more than two entries,
the \QSpace display switches to a list view with
one-liners for each entry in \src{S} that further summarize
the header lines in the detailed display for each \src{S(i)}
with $i=1,2,3$.  To see the full \QSpace display of all entries
in the \QSpace array \src{S},
one needs to call the underlying \src{display}
routine together with the option~\str{-v}
that stands for more `verbose',
\begin{minted}[escapeinside=??,firstnumber=last]{text}
>> display(S,'-v')

S(1) =
     Q:  3x { 2 x 1 }  abelian U(1)  { +-- }
  data:  3-D double (224)      2 x 2 x 1  @ norm = 0.7071

     1.  1x1      [  1 ;  1 ;  0 ]          0.5 ?\mlbl{Sop:1}?
     2.  1x1      [ -1 ; -1 ;  0 ]         -0.5

S(2) =
     Q:  3x { 1 x 1 }  abelian U(1)  operator,  { +-- }
  data:  3-D double (112)      1 x 1 x 1  @ norm = 0.7071

     1.  1x1      [ -1 ;  1 ; -2 ]     0.707107 ?\mlbl{Sop:2}?

S(3) =
     Q:  3x { 1 x 1 }  abelian U(1)  operator,  { +-- }
  data:  3-D double (112)      1 x 1 x 1  @ norm = 0.7071

     1.  1x1      [  1 ; -1 ;  2 ]    -0.707107 ?\mlbl{Sop:3}?

\end{minted}
By inspection, this shows that the returned spin operator \src{S} 
has three entries corresponding to
\begin{flalign}
   \src{S}\,= [\hat{S}_z, \tfrac{1}{\sqrt{2}} \hat{S}_-,
   -\tfrac{1}{\sqrt{2}} \hat{S}_+]
\text{ .}\label{eq:S3:irop}
\end{flalign}
For algorithmic convenience,
this keeps $\hat{S}_z$ to the front
by permuting $\hat{S}_+$ to the end w.r.t. \Eq{Sop},
such that \src{S(1)} is either
$\hat{S}_z$ or the full spin irop in case of SU(2) spin.
The spin operator \src{S} above is derived from the
representation as $S=1$ irop in \Eq{Sop},
here corresponding to symmetry labels
$S_z \in \{0,-1,+1\}$ or
$\idx[SU2:qlabels]{q_{\rm op}=2S} \in \{0,-2,+2\}$
for the irop $q$-label with the third index
in lines \ref{Sop:1}f, \ref{Sop:2}, and \ref{Sop:3},
respectively [hats are used with the operators in \eqref{eq:S3:irop}
to differentiate them from the symmetry
labels $|S{=}1;\, S_z\in -1,0,1\rangle$ here for the 
$S=1$ multiplet underlying the irop].
This explains the normalization and signs in the components
of \src{S} up to the permutation of the entries.
Note that the inverse square root factors with
the lowering and raising operator lead to the
same norm $|\src{S(i)}| = \frac{1}{\sqrt{2}} = 0.7071$ ($i=2,3$)
as compared to the $S_z$ operator ($i=1$).
As such, they also allow for a convenient $S^\dagger \cdot S$
contraction for isotropic spin interactions.

Having U(1) spin, all three operators in \src{S} 
are distinguishable by their irop symmetry labels.
Therefore the three spin operators may
be combined into a single spin operator by simply `adding'
them,
\begin{minted}[escapeinside=??,firstnumber=last,samepage]{text}
>> Sop=sum(S) ?\mlbl{Sop:sum}?

     Q:  3x { 4 x 1 }  abelian U(1)  { +-- }
  data:  3-D double (448)      2 x 2 x 3  @ norm = 1.225 ?\mlbl{Sop:hdr}?

     1.  1x1      [ -1 ; -1 ;  0 ]         -0.5
     2.  1x1      [ -1 ;  1 ; -2 ]     0.707107
     3.  1x1      [  1 ; -1 ;  2 ]    -0.707107
     4.  1x1      [  1 ;  1 ;  0 ]          0.5

\end{minted}
Because all irop symmetry sectors in \src{S} are different,
addition translates into a direct sum $\oplus$ here.
This simply {\it catenates} all \QSpace \idx[QS:recs]{records}
into a single list,
since the entries are still differentiated by their U(1)
irop labels, thus preserving the RMT block structure.
The resulting \src{Sop} is no longer irreducible from
a symmetry perspective, though. Therefore in contrast
to the individual entries \src{S(i)}, the combined \src{Sop}
no longer qualifies
as an `irop', since it has multiple operator symmetry
labels listed with its third index. It is still a perfectly
well-defined operator, nevertheless.
The norm of the spin operator \src{Sop},
\begin{eqnarray*}
  \Vert S\Vert^2 &\equiv& {\rm tr}(S^\dagger \cdot S)
  = \sum_{i=1,2,3} {\rm tr}\,(S_i^\dagger S_i)
  = \sum_{a=x,y,z} {\rm tr}\,(S_a^\dagger S_a)
  = 3 \times (2\,\tfrac{1}{2^2}) \
  = \ \tfrac{3}{2}
\end{eqnarray*}
agrees with \mline{Sop:hdr}, having $|S| = \sqrt{3/2} = 1.225$.
The combined spin operator \src{Sop}
also permits one to easily obtain the Casimir
$\hat{S}^2 \equiv \hat{S}^\dagger \cdot \hat{S}$,
\begin{minted}[escapeinside=??,firstnumber=last]{text}
>> S2=contract(Sop,'13*',Sop,'13')  ?\mlbl{Sop:sum:S2}?

     Q:  2x { 2 x 1 }  abelian U(1)  { +- }
  data:  2-D double (224)      2 x 2  @ norm = 1.061

     1.  1x1  [ -1 ; -1 ]         0.75
     2.  1x1  [  1 ;  1 ]         0.75

\end{minted}
resulting in $\frac{3}{4} \Id$ as expected for the
underlying spin-half.
The contraction on the third index above implements
the dot product in $\hat{S}^\dagger \cdot \hat{S}$.
\vspace{-1ex}

\paragraph{Switch to $Z_2$ symmetry}

The option \str{-A} in the setup \eqref{eq:gls:spinhalf} 
requested abelian U(1) symmetry.
By specifying \str{-{-}Z2}, instead,
this switches to a $Z_2$ symmetry representation
[cf. \Eq{S3:irop:Z2}],
\begin{minted}[escapeinside=??,firstnumber=last]{matlab}
>> [S,IS] = getLocalSpace('Spin',1/2,'--Z2') ?\mlbl{Sop:Z2}?
\end{minted}
\vspace{-4ex}
\begin{minted}[escapeinside=??,firstnumber=last]{text}

S(1) = ?\mlbl{Sop:Z2:1}?
     Q:  3x { 2 x 1 }  having 'Z2'  operator,  { +-- }
  data:  3-D double (224)      2 x 2 x 1  @ norm = 0.7071

     1.  1x1      [  1 ;  1 ;  0 ]         -0.5
     2.  1x1      [  0 ;  0 ;  0 ]          0.5

S(2) = ?\mlbl{Sop:Z2:2}?
     Q:  3x { 1 x 1 }  having 'Z2'  operator,  { +-- }
  data:  3-D double (112)      1 x 1 x 1  @ norm = 0.7071

     1.  1x1      [  0 ;  1 ;  1 ]     0.707107

S(3) = ?\mlbl{Sop:Z2:3}?
     Q:  3x { 1 x 1 }  having 'Z2'  operator,  { +-- }
  data:  3-D double (112)      1 x 1 x 1  @ norm = 0.7071

     1.  1x1      [  1 ;  0 ;  1 ]    -0.707107

\end{minted}
By comparison to the earlier U(1) spin operator,
the setup of the model changed from \idx[SU2:qlabels]{$q=2S$}
for U(1) spin to a $Z_2$ setting with $q\in\{0,1\}
\equiv \{\uparrow,\downarrow\}$,
as seen from $S_z$ in \mlines{Sop:Z2:1}ff.
The spin operator \src{S(1)}\,$=S_z$ carries
$q_{\rm op} = 0$, whereas the operators
$\hat{S}_+$ and $\hat{S}_-$ become indistinguishable
in terms of irop symmetry labels, both having
$q_{\rm op} = 1$ due to the $Z_2$ symmetry.
However, when including the symmetry labels
for all legs, all entries for the spin operator
above are still distinguishable from a symmetry
sector point of view.
Hence the earlier U(1) constructions in \mlines{Sop:sum}
and \ref{Sop:sum:S2} still also work.
\vspace{-1ex}

\paragraph{Switch to no symmetry}
\label{idx:nosym:ex}

By switching the symmetry option in \eqref{eq:gls:spinhalf}
from \str{-A} to \str{-{-}nosym} (`no symmetries'), 
this effectively turns off symmetries
[cf. \Eq{S3:irop:nosym}],
\begin{minted}[escapeinside=??,firstnumber=last]{matlab}
>> [S,IS] = getLocalSpace('Spin',1/2,'--nosym') ?\mlbl{Sop:nosym}?
\end{minted}
\vspace{-4ex}
\begin{minted}[escapeinside=??,firstnumber=last]{text}

S(1) =
     Q:  2x { 1 x 1 }  having 'A'  { +- }
  data:  2-D double (136)      2 x 2  @ norm = 0.7071

     1.  2x2  [  0 ;  0 ]      32 b

S(2) =
     Q:  2x { 1 x 1 }  having 'A'  { +- }
  data:  2-D double (136)      2 x 2  @ norm = 0.7071

     1.  2x2  [  0 ;  0 ]      32 b

S(3) =
     Q:  2x { 1 x 1 }  having 'A'  { +- }
  data:  2-D double (136)      2 x 2  @ norm = 0.7071

     1.  2x2  [  0 ;  0 ]      32 b

\end{minted}
This is a \hsec[gls:tweaks]{tweak} to abelian U(1),
seen as \str{A} in the header lines,
that only uses symmetry labels $q=0$. This trivial
setting is equivalent to having no symmetry at all.
The RMTs are $2\times 2$ matrizes,
\begin{minted}[escapeinside=??,firstnumber=last]{text}
>> S(1).data{1}

    0.5000         0
         0   -0.5000

>> S(2).data{1}

         0         0
    0.7071         0

>> S(3).data{1}

         0   -0.7071
         0         0
\end{minted}
now reflecting in detail the spin operators 
in terms of the Pauli matrices, 
$[ \frac{1}{2}\sigma_z, \, \frac{1}{\sqrt{2}} \sigma_-, \,
-\frac{1}{\sqrt{2}} \sigma_+ ]$,
respectively.
In the present case of no symmetry, 
there are no symmetry labels that can describe the action
of the operators.
Hence all operators are rank-2 \idx[scalar:op]{scalars}.
This way, all spin operators become
indistinguishable in terms of irop $q$-label.
These spin operators can no longer
be combined into a single spin operator as in \mline{Sop:sum},
since the RMTs are no longer distinguishable
by symmetry labels. Hence \src{sum(S)} would no longer
perform the direct sum ($\oplus$) by catenating
the \QSpace records, but actually add the RMTs
which is not meaningful.

\newpage


\section{Simple Tutorials and Applications}
\label{sec:tutorials}

This section extends on the simple examples in \Sec{examples}
which supplemented the earlier sections in the main text.
Here the focus shifts more toward tensor
network applications.
For this purpose, it should be noted that
the public \QSpace repository already
contains significantly more than the bare tensor
library documented here, e.g., see the listing of (MEX)
applications in \App{QS:apps}. The full documentation
of this is left for the future. The standard help
explaining the purpose and usage of any function is available
as part of the repository, nevertheless.
Yet their discussion as well as the description of the
related setup scripts is beyond the scope of this documentation 
which rather focuses on \QSpace as a tensor library.
Still, the interested user may find it rewarding
to explore. Since fdm-NRG \cite{Wb07} has been one
of the very first applications of \QSpace, it is discussed
in more detailed in \Sec{NRG}.

Sanity checks for tensors that have a simple
interpretation and structure are also frequently included.
These are helpful in practice quite generally,
as they ensure that one has a good understanding
of all the operators and tensors that
one is dealing with.
Much of the same spirit also underlies 
the earlier examples given, and \QSpace as a whole. There
are many consistency and plausibility checks internally in the
\QSpace MEX core routines, as well as in the Matlab environment.
At negligible numerical overhead, these
\idx[safeguards]{assertions} are
safeguards to ensure overall consistency.

\subsection{Iterative build-up of many-body state spaces:
Two fermionic sites}
\label{sec:2sites}

An elementary step to building many-body state spaces
is the iterative addition of a new site. The
example here shows how one can start this within \QSpace
by adding a second site. To be specific,
suppose one wants to describe two interacting
spinful fermionic sites with respective
local state spaces $\sigma_i$ and site index $i=1,2$
in a combined state space.
The description of a single site
again starts with \eqref{eq:gls:NC1} above.
With this one can go ahead and start taking
iterative tensor products of that state space
with copies of itself to build a many-body basis.

Reminding oneself that \QSpace is a tensor library
that \idx[nolegs]{encodes tensors only},
state spaces are specified by pointing to particular
legs of a given tensor. Hence when expanding a
Hilbert space via a tensor product,
one needs to ensure to pick operators
that have complete state space, i.e.,
\idx[QS:zblocks]{non-zero blocks} in every symmetry
sector of the target state space. Natural
choices are the identity stored
with \src{IS.E} [cf. \Eq{gls:NC1}], or the parity \src{Z}.
When building a tensor product of state spaces,
the routine \src{getIdentity} takes two \QSpace{s} as input,
both representing a state space
of their own (here by having identical sites,
the operator \src{Z} is listed twice),
\\
\begin{minipage}{1\linewidth}
\begin{minted}[escapeinside=??]{matlab}
 A=getIdentity(Z,Z,[1 3 2]);        ?\mlbl{A2:Z}?
 A=setitags(A,{'K01','K02','s02'}); ?\mlbl{A2:itags}?
\end{minted}
\mintLabelB{eq:A2}
\end{minipage}
\\[2ex]
In \mline{A2:Z}, one may use Z $\to$ IS.E, instead.
One even could have used \src{F} instead of \src{Z},
since if no explicit leg index is provided with the input,
\src{getIdentity} assumes an \idx[irop]{operator},
and hence combines the state spaces
from the first two legs [same argument as with 
\Eq{Id:F} above].
Either way, \src{A} encodes the tensor-product fusion
of the state spaces of two sites.
The optional last argument \src{[1 3 2]} in \mline{A2:Z}
specifies a permutation to be applied to the returned
output $A$-tensor. By default, the output of
\src{getIdentity} in \mline{A2:Z}
has an index order
that lists the two input spaces first (in the order
specified), and lists the combined state space
last (third index). By permuting this order as
specified in \mline{A2:Z},
this returns the rank-3
\idx[Atensor]{$A$-tensor} with index order that adheres
to the \idx[Atensor:LRs]{LRs} index order convention
[cf. \Fig{Aop}].
As an implementational detail, the permutation in line~1
requires numeric format, since a \idx[perm]{(compact) string}
representation such as \str{132}
would be interpreted as an \itag for the fused index, instead.

From an MPS perspective, the above tensor product
of sites $\sigma_1$ and $\sigma_2$
may represent the start of a physical chain.
Virtual bond indices specify the many-body state spaces
that are {\it kept} from one iteration to the next.
Hence $\sigma_1 \to K_1 \equiv$ \str{K01}
may be considered the state space that is kept
from iteration 1 (cf. \Fig{arrows}). \QSpace,
and hence also this documentation, frequently
uses a 2-digit format for site or bond indices
for the sake of aligning display output,
bearing in mind that typically more than 10
sites are present in a simulation.
Hence \str{K01} instead of, e.g., \str{K1}. 
By adding site $\sigma_2\to$\str{s02}, one arrives
at the combined state space \str{K02}.
The \idx[itags]{itags} in \mline{A2:itags}
are thus chosen as in \Fig{arrows} for $A_2$
at the start of an MPS.
With this, the \idx[Atensor]{$A$-tensor} returned
with \mline{A2:itags}
reads (see \Fig{QS:disp} for detailed explanations of
this very example) \\[-3ex]
\begin{minted}[escapeinside=??,firstnumber=last]{text}
>> A     ?\mlbl{A2:disp}?
     Q:  3x {10 x 2 }  having 'A,SU2'  A-matrix,  { K01, K02*, s02 }  ?\mlbl{A2:hdr1}?
  data:  3-D double (1200)      3 x 10 x 3 => 4 x 16 x 4  @ norm = 4  ?\mlbl{A2:hdr2}?

     1.  1x1        |  1x1x1 [ -1 0 ; -2 0 ; -1 0 ]           1. ?\mlbl{A2:rec1}?
     2.  1x2        |  1x2x2 [ -1 0 ; -1 1 ;  0 1 ]      16 b     {?$\surd$?2} ?\mlbl{A2:rec2}?
     3.  1x2        |  2x2x1 [  0 1 ; -1 1 ; -1 0 ]      16 b     {?$\surd$?2}
     4.  1x3        |  1x1x1 [ -1 0 ;  0 0 ;  1 0 ]      24 b
     5.  1x3        |  2x1x2 [  0 1 ;  0 0 ;  0 1 ]      24 b
     6.  1x3        |  1x1x1 [  1 0 ;  0 0 ; -1 0 ]      24 b
     7.  1x1        |  2x3x2 [  0 1 ;  0 2 ;  0 1 ]           1.  {?$\surd$?3} ?\mlbl{A2:rec7}?
     8.  1x2        |  2x2x1 [  0 1 ;  1 1 ;  1 0 ]      16 b     {?$\surd$?2}
     9.  1x2        |  1x2x2 [  1 0 ;  1 1 ;  0 1 ]      16 b     {?$\surd$?2}
    10.  1x1        |  1x1x1 [  1 0 ;  2 0 ;  1 0 ]           1.  ?\mlbl{A2:recl}?   
\end{minted}
The \itags are shown in the header \mline{A2:hdr1}.
With \src{K02} the outgoing index, it has the trailing
conjugate flag~\str{*}.
Since $q$-directions need to be preserved, 
when setting itags via \src{setitags}
there is no need to specify
the trailing asterisk with \src{K02} in \mline{A2:itags}
(if trailing asterisks had been specified,
these would be ignored).

The norm in \mline{A2:hdr2} shows $\Vert A\Vert = 4$, 
i.e., $\Vert A \Vert^2 = {\rm tr}(\Id) = 16$,
which is consistent with the full Hilbert space dimension reflected in the dimensions \src{[4 16 4]} of $A$ in the same line.
The square root factors to the very right, as always,
are just reminders of the tweaked normalization with the \idx[cgw]{$w$-matrix} [cf. \Eq{wnorm:x3}],
so that the reduced matrix elements read \src{1.},
as one may expect for a normalized mapping
of state space, represented as a sliced-up {\it identity} matrix in multiplet space.
For example, in \mline{A2:rec7},
the factor $\sqrt{3}$ [which is due to having the fused multiplet $q=2S=2$ of dimension $3$] is split off into the $w$-matrix,
such that the RMT can read \src{1}.
When the RMT has dimensions $>1$, this shows the size of that RMT in bytes instead of its value.
For example, for the \src{1x2[x1]} RMT in \mline{A2:rec2},
2 * (8 bytes for \src{double}) = \src{16 b}.
The trailing \idx[singleton]{singleton} dimensions in the RMT on index 3 are not shown (second column).

\begin{figure}[tbh]
\begin{center}
\includegraphics[width=1\linewidth]{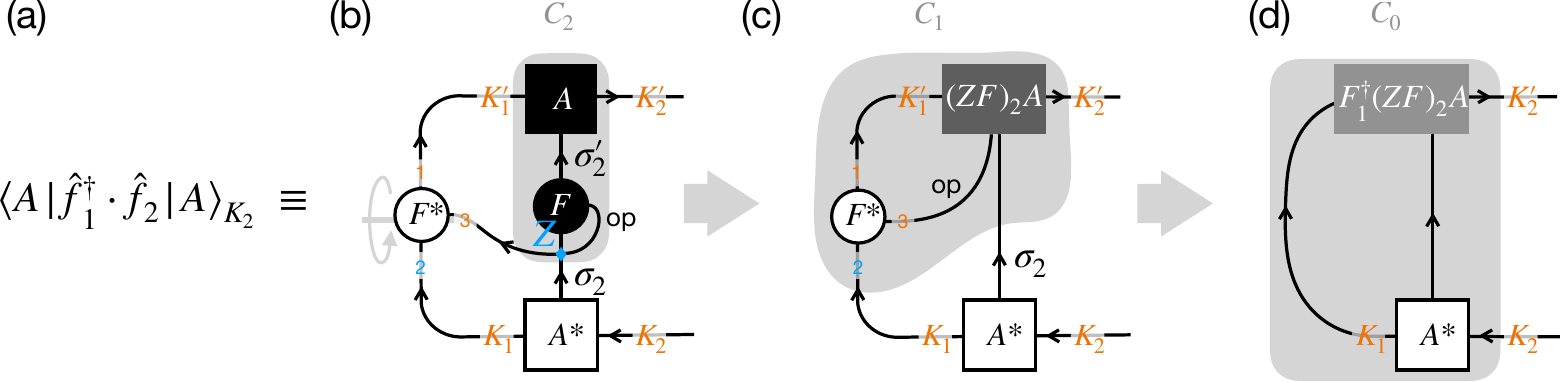}
\end{center}
\caption{
   Computing the matrix elements for the fermionic hopping
   $\hat{f}_1^\dagger \cdot \hat{f}_2$
   in the combined basis $K_2$ corresponding to the nested cell
   contraction in \eqref{eq:FdotF:T12}.
   The contraction includes four tensors as shown in (b).
   By taking the conjugate of the tensor $F_{1}$, i.e., $F$
   acting on $K_1 \equiv \sigma_1$, similar to \Fig{nctr} again
   the up/down \idx[conj]{reflected} mirror image is drawn for
   $F^\ast$. This is emphasized by the gray line with a rotating arrow
   and the reverted index order 1 and 2.  Auto-contraction based
   on \itags automatically links the indices correctly to take
   the transpose for the Hermitian conjugate $F_1^\dagger$ overall. 
   Since $F_2$, i.e., $F$ applied to $\sigma_2$, acts on
   $|A\rangle$ first [panel (b)], the operator line \src{op}
   leads to a crossing of the line $\sigma_2$ {\it below}
   $F_{2}$, indicated by the blue symbol. Because the
   operator line \src{op} carries an odd charge parity, this crossing
   gives rise to a fermionic swap gate \cite{Corboz10}.
   Here this translates into the parity operator $Z$ (blue)
   applied to $\sigma_2$ at the location of the blue
   symbol. This is equivalent to taking
   \src{F} $\to$ \src{ZF} for site~2.
   If spin operators $S_1^\dagger \cdot S_2$ had been 
   considered instead of the fermionic operators 
   $\hat{f}_1^\dagger \cdot \hat{f}_2$ here,
   the precise location of the line crossing
   (above or below $F_2 \to S_2$) would be of no further
   importance because the spin operators carry an
   even charge parity on \src{op}
   (hence they also commute, $S_1^\dagger \cdot S_2
   = S_2 \cdot S_1^\dagger$).
   The nested cell contraction in \eqref{eq:FdotF:T12}
   gives the sequence of pairwise contraction indicated
   in panels (b-d). The contractions  $C_2$, $C_1$, and $C_0$
   highlighted by the gray shaded area are
   with reference to \Eq{S1S2}.
   The last contraction $C_0$ in panel (d)
   gives the resulting scalar operator in the state space~$K_2$.
}
\label{fig:F1F2} 
\end{figure}

\begin{figure}[bh!]
\begin{center}
\includegraphics[width=0.9\linewidth]{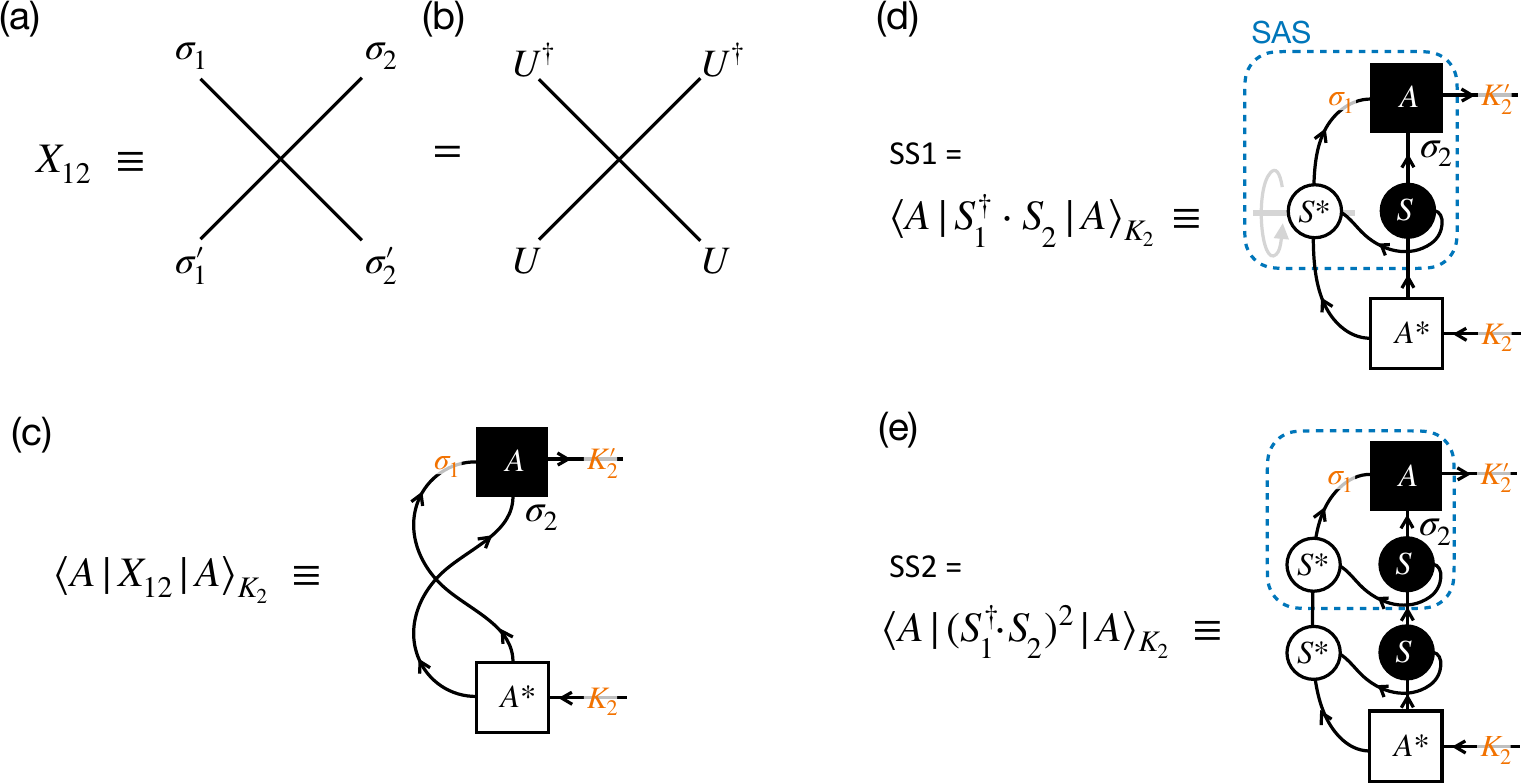}
\end{center}
\caption{
(a) The swap operator $X_{12}$ 
acting on two sites $i\in \{1,2\}$ with 
respective local state spaces $\sigma_i$.
(b) The swap operator is invariant under arbitrary
unitary transformations on the local state spaces
since $UU^\dagger=1$ annihilate each other along
the diagonal lines.
(c)~The matrix elements of the swap operator
expressed in the combined state space
as described by an $A$-tensor.
(d-e) The matrix elements of
$(S_1^\dagger\cdot S_2{\,})^n$ for $n=1$ and $2$,
respectively. The contraction in (d) is the same
as in \Fig{F1F2}, except that spin operators
are applied here.
Similarly, also the \idx[conj]{conjugate} operator
for the first site is drawn as the vertical mirror
image as indicated by the gray circular arrow and line.
The intermediate object in the contraction
indicated by the blue dashed box in (d), referred
to as \src{SAS} in the text (\mline{SAS}),
can be reused in (e).
}\label{fig:Xop} 
\end{figure}

With the 2-site Hilbert space described by $A$, one can now
proceed to compute matrix elements in the combined 
many-body state space.
Consider, for example, the spin interaction
$S_{12} \equiv A^\dagger (S_1^\dagger \cdot S_2^{\,}) A$
when cast into the basis $A$
[see \Fig{Xop}(d) or also \Fig{F1F2} with $(Z)F \to S$
for pictorial representations]
\\
\begin{minipage}{1\linewidth}
\begin{minted}[escapeinside=??,firstnumber=last]{matlab}
 S12 = contract(A,'!2*',{S,'-op:K01','*',{A,S,'-op:^s'}}) ?\mlbl{S12:op}?
\end{minted}
\mintLabelB{eq:SdotS}
\end{minipage}
\begin{minted}[escapeinside=??,firstnumber=last]{text}
     Q:  2x { 2 x 2 }  having 'A,SU2'  { K02, K02* }            ?\mlbl{S12:hdr1}?
  data:  2-D double (288)      4 x 4 => 6 x 6  @ norm = 0.866   ?\mlbl{S12:hdr2}?
 
     1.  3x3        |    1x1 [  0 0 ;  0 0 ]      72 b                  ?\mlbl{S12:rec1}?
     2.  1x1        |    3x3 [  0 2 ;  0 2 ]         0.25  {?$\surd$?3} ?\mlbl{S12:rec2}?

>> S12.data{1}

         0         0         0   ?\mlbl{S12:data1:1}?
         0   -0.7500         0   ?\mlbl{S12:data1:2}?
         0         0         0   ?\mlbl{S12:data1:3}?
\end{minted}
Before explaining the compact semantics of the contraction
in \Eq{SdotS}, the output in \QSpace \src{S12} is examined.
As expected for the operator $S_1^\dagger \cdot S_2^{\,}$,
there are only matrix elements at half-filling in the above display 
for \src{S12}, having $q_1=0$ for the first symmetry label.
More restrictive still, non-zero matrix elements
can only derive from the half-filled space on either site.
The respective tensor product of two spin-halfs gives one
singlet ($q_2=0$) and one triplet state ($\idx[SU2:qlabels]{q_2=2S}=2$).
The triplet state
is represented by the second record in \mline{S12:rec2}
which has $\langle S_1^\dagger \cdot S_2^{\,} \rangle
= $\,\src{0.25} as expected.
The singlet across both sites is buried within the first
record in \mline{S12:rec1}
where it appears as the second entry (\mline{S12:data1:2})
and carries the value
$\langle S_1^\dagger \cdot S_2^{\,} \rangle = $\,\src{-0.7500}, as expected for the singlet.
In the global half-filled sector,
there are two other states in the scalar sector \src{q=[0 0]},
namely the combination of completely empty at site~1
with completely filled at site~2, or vice versa.
Their expectation value $\langle S_1^\dagger \cdot S_2^{\,} \rangle = 0$,
thus corresponding to the other two zero entries along the diagonal
in \mlines{S12:data1:1} and \ref{S12:data1:3}.
Because they are part of the same global symmetry sector, they
also need to appear here within the same RMT indexed by $n$ as
in \Eq{composite-index}.
Here this results in the RMT of dimension $3\times 3$ 
in \mlines{S12:data1:1}--\ref{S12:data1:3}.
Since the Hamiltonian preserves spin symmetry, the matrix
elements of \src{S12} must already be in a diagonal representation, as seen above, indeed.
\vspace{-1ex}

\paragraph{Contraction semantics}
\label{idx:ctr:semantics}

The nested contraction in \Eq{SdotS}
consists of three levels,
based on the pairwise contraction pattern
fixed by adding brackets to
$S_{12} = A^\dagger\, (S_1^\dagger \cdot (S_2^{\,}\, A))$,
\begin{eqnarray}
    S_{12} &=& {\rm contract}(\underbrace{
    A,\str{!2*},\underbrace{
      \{S,\str{-op:K01},\str{*},\underbrace{
         \{A,S,\str{-op:\textasciicircum s}\}}_{
           \equiv C_2
         }\}}_{
         \equiv C_1
      }}_{\equiv C_0})
\label{eq:S1S2}
\end{eqnarray}
[see \Fig{Xop}(d) or also \Fig{F1F2} with $(Z)F \to S$
for pictorial representations]. The underlying
MEX routine \src{contractQS} supports
nested cell structures based on pairwise contractions,
also referred to as contraction pattern.
With this, one may contract
any arbitrary number of tensors in a single call
to \src{contract} (the Matlab wrapper routine \src{contractQS}).
For any included pairwise contraction,
the two involved tensors can be either an existing \QSpace
specified as argument, or another cell $\{\ldots\}$
thus adding another recursive `level' to the contraction.
The latter indicates a nested contraction that needs
to be carried out first, so that its result can be
used. If further cells are encountered,
\src{contractQS} proceeds recursively through the
nested calls as they are encountered.
Eventually, at the deepest level of any set of
nested calls must be two \QSpace objects to be
contracted [like with $C_2$ in \Eq{S1S2}]
which then is carried out first.
Each \QSpace input with empty itags may have an
option that assigns itags on the fly, typically via
\str{-op:..} as above. Each \QSpace or cell
also accepts an option that specifies what
indices (not) to contract, and whether to
take the conjugate. This option is specified
via a compact string with the format \src{[[!]cidx][*]}
where terms in square brackets are optional.
The term \src{cidx} represents a
\idx[compact:idx]{compact} string that
explicitly specifies which indices to contract
(or {\it not} to contract in the presence
of the leading \str{!}). If \src{cidx}
is empty, an implicit auto-contract searches
for all matching indices based on their \itags.
The trailing \src{*} takes the
\idx[conj]{conjugate} of the preceding
\QSpace or cell.
For an explanation of all options available
for contractions, see  \src{help contractQS},
or equivalently, \src{contractQS -h}.

The nested cell structure as provided by the
input to \src{contractQS}
fully determines the order of contractions.
Specifically, \QSpace does not optimize or
restructure the order of contractions.
The optimal order of contractions is thus
left for the user to determine. In any contraction
like $T_{12} \equiv T_1\ast T_2$,
the non-contracted, i.e., kept indices of the first tensor
$T_1$ are collected {\it before} the ones in the
paired up $T_2$, while maintaining their original order.
This holds for any pairwise construction in the
nested structure provided to \src{contractQS}.
This way, the structure of the input
to \src{contractQS} also fully determines
the index order of the final result.

Returning to the example in \Eq{S1S2},
the inner most cell $C_2$ occurs
at nested {\it contraction level} 2.
It is contracted first.
Since \src{S} does not have any
itags assigned yet (after all, it can be applied to any site),
they are assigned {\it on the fly}
by the provided `hint' whose semantics
is along the lines of a simplified {\it regular expression}:
by specifying \str{-op:..},
the tensor \src{S} is assumed
to be an \idx[irop]{operator}, i.e., of rank 2 or 3,
that acts on a particular leg on the {\it other}
paired up tensor in the contraction (here \src{A}
in $C_2$).
By looking for an \itag that {\it starts with \src{s}}
(denoted by \src{{\textasciicircum}s};
see \Cpp \src{std::regex} for more on this syntax),
in the present case,
this is sufficient to uniquely identify leg \src{s02} on \src{A}.
In this sense, the syntax in the contraction internally
assigns on the fly the \itags \mtt{{'s02','s02*','op'}}
to the operator \src{S} for this particular contraction.
Based on these \itags then, \QSpace auto-contracts
this spin operator onto \src{s02} of \src{A}.
The temporary \itags in \src{S} that are not contracted,
are inherited by and kept with the temporary tensor $C_2$.

Given $C_2$ as a temporary object in memory,
the next cell to be
contracted is $C_1$, at nested contraction level 1.
Analogous to $C_2$,
this now contracts \src{S} onto \src{K01}.
More precisely, because of the presence of the
\str{*} option, this applies the conjugate,
and hence effectively the Hermitian conjugate
and thus the dagger of \src{S} onto \src{K01}.
The subsequent auto-contraction in $C_1$ also
contracts the assigned default operator \itag \str{op} for both
instances of \str{S}. Overall,
therefore this contracts $S_1^\dagger \cdot S_2$.

The last contraction $C_0$ at {\it base level} 0
uses the temporary \QSpace tensor $C_1$.
The result out of $C_0$ is returned
and assigned to \src{S12} (technically,
$C_0$ is already generated in the memory space of 
the output \src{S12} to avoid copying objects).
For this last contraction, however,
all indices can be paired up, and hence auto-contraction
would contract them all. To avoid this,
\QSpace introduces the \Emph{`not' semantics \str{!...}}.
For example, here by specifying \str{!2..} after
the first operator, this indicates that index 2
of that operator (here referring to the combined state space \src{K02})
{\it shall not} be contracted and thus left open, even though
the \itags could be paired up, in principle.
The trailing \src{*}, finally,
indicates to also take the conjugate
tensor of \src{A}. This corresponds to bra states,
and therefore concludes the calculation of the
matrix elements in the combined state space \src{K02}
of sites 1 and 2.

Completely analogous to the spin interaction above,
one can also compute the matrix elements for the fermionic hopping
across sites 1 and 2, $T_{12} \equiv f_1^\dagger \cdot f_2^{\,}
\ \to\ \src{F}_1^\ast \cdot (\src{ZF})_2$, \\
\begin{minipage}{1\linewidth}
\begin{minted}[escapeinside=??,firstnumber=last]{matlab}
>> T12 = contract(A,'!2*',{F,'-op:K01','*',{A,(Z*F),'-op:^s'}});
\end{minted}
\mintLabelB{eq:FdotF:T12}
\end{minipage}
\\[2ex]
This contraction is the same as in \eqref{eq:SdotS},
except for the operators acting on
the two sites, substituting (\src{S},\src{S}) $\to$ 
(\src{F}, \src{Z*F}).
Here by having fermionic operators,
one needs to incorporate fermionic signs
\cite{Wb12,Corboz10}. As a short remnant from
a Jordan Wigner string for the nearest-neighbor
hopping here, the fermionic parity operator \src{Z}
is applied onto the second \src{F} operator.
In terms of fermionic order, this
adopts the convention that the first site
acts on the vacuum state {\it first}, followed by the
second site, etc. Hence similar to \eqref{eq:FdotF},
as $f_1$ needs to be
pulled past any $f_2^\dagger$ when computing matrix elements in Fock space, it leaves a trail of parity operators (here to be applied onto $f_2$ from the left).
In the pictorial representation of \Fig{F1F2},
the crossing \cite{Wb12,Corboz10} of the operator line
\src{op} with $\sigma_2$ is precisely the location where
the parity gate \src{Z} needs to be applied
if the local operator has a fermionic character.
This results in \src{S} $\to$ \src{ZF}
for the operator acting on site $\sigma_2$.
Adding the Hermitian conjugate to $T_{12}$,
one obtains a typical Hamiltonian term
\begin{minted}[escapeinside=??,firstnumber=last]{text}
>> H12 = T12 + T12'

     Q:  2x { 3 x 2 }  having 'A,SU2'  { K02, K02* }          ?\mlbl{H12:hdr1}?
  data:  2-D double (448)      7 x 7 => 11 x 11  @ norm = 4   ?\mlbl{H12:hdr2}?

     1.  2x2        |    2x2 [ -1 1 ; -1 1 ]      32 b     {?$\surd$?2} ?\mlbl{H12:rec1}?
     2.  3x3        |    1x1 [  0 0 ;  0 0 ]      72 b                  ?\mlbl{H12:rec2}?
     3.  2x2        |    2x2 [  1 1 ;  1 1 ]      32 b     {?$\surd$?2} ?\mlbl{H12:rec3}?

\end{minted}
It is hermitian, as can be explicitly checked by computing
\src{norm(H12-H12')} which yields \src{0}.
When looking at the overall dimensions of \src{H12},
one realizes a subtlety:
the combined Hilbert space for two spinful fermionic levels
has a dimension of 16, yet \src{H12}
shows a dimension of \src{11 x 11} in \mline{H12:hdr2}.
Apparently, 5 out of 16 states went `missing'.
The only way this can occur is that these 
belong to symmetry sectors of \src{H12} that
represent \idx[QS:zblocks]{zero blocks}, i.e., have Frobenius 
norm below or comparable to numerical double
precision noise (it is not possible for individual states to be missing within otherwise present symmetry sectors).
For the efficiency of a tensor library, such zero-blocks are not stored, by default. 
They are kept nevertheless, however, in \QSpace for scalar operators (cf. \src{\idx[skipZerosQS]{skipZerosQS}}).
With diagonal zero blocks absent here nevertheless, these sectors simply did not arise out of the contraction itself. Indeed, this is a consequence of the symmetry sectors that were present in the input tensors \src{F} and the contracted performed.

From the one-particle picture, there are two
spin-degenerate single-particle levels for the Hamiltonian \src{H12} above at energies $\varepsilon_\pm = \pm 1$.
When building the many-body state space for this,
the ground state represents the doubly-filled level
$\varepsilon_-$ at energy $E_0=-2$.
It corresponds to a half-filled state and hence 
belongs to the symmetry sector $q=\src{[0 0]}$ (\mline{H12:rec2}).
Adding or removing particles then gives the many-body
excitation spectrum $E \in \{-2,-1,0,1,2\}$
(e.g., see first column in 
\mlines{I12.ee:1}--\ref{I12.ee:end} below).
Here the completely empty or filled sectors
($[\pm 2\ 0]$, both 1-dimensional)
are missing in \src{H12}, since
indeed, these have hopping matrix elements equal to zero.
By comparing to the symmetry sectors on \src{A}
in \mlines{A2:rec1}--\ref{A2:recl}
[or equivalently, but more readable, by comparing
to the output of \src{E2} in \mline{E2:A:2} below],
one finds that the remaining symmetry sector
that is missing is \src([0 2]).
One also quickly realizes why: this is the triplet state
at half-filling (having $S=2/2=1$). Indeed,
say with an up-spin on each site,
there cannot be spin-preserving hopping.
This adds up to the total of $1+1+3 = 5$ `missing' states,
or correspondingly, 3 symmetry sectors.
They may be added to \src{H12} by an infinitesimal
(with the value zero also permitted)
to also include all-zero eigenvalue blocks.
Since \src{H12} has non-diagonal RMTs,
eigenvalue decomposition yields,
\begin{minted}[escapeinside=??,firstnumber=last]{matlab}
>> E2=getIdentity(A,2);                ?\mlbl{E2:A:2}?
   [U12,E12,I12] = eig( H12 + 0*E2 );  ?\mlbl{H12:eig}?
\end{minted}
Here \src{eig} is a wrapper to the MEX routine \src{eigQS}.
\mLine{H12:eig}
adds the identity operator of the full state space 
on leg 2 in \src{A}, i.e., \src{K02}, with weight zero. 
Being a scalar operator, the resulting
diagonal zero-blocks \idx[skipZerosQS]{are kept}.
The reference to the $A$-tensor ensures a complete
state space, by construction.
The output of the above can be inspected, as usual,
by simply typing the object's name,
\begin{minted}[escapeinside=??,firstnumber=last]{text}
>> U12
     Q:  2x { 6 x 2 }  having 'A,SU2'  { K02, K02* }
  data:  2-D double (784)      10 x 10 => 16 x 16  @ norm = 4

     1.  1x1        |    1x1 [ -2 0 ; -2 0 ]           1.
     2.  2x2        |    2x2 [ -1 1 ; -1 1 ]      32 b     {?$\surd$?2}
     3.  3x3        |    1x1 [  0 0 ;  0 0 ]      72 b
     4.  1x1        |    3x3 [  0 2 ;  0 2 ]           1.  {?$\surd$?3}
     5.  2x2        |    2x2 [  1 1 ;  1 1 ]      32 b     {?$\surd$?2}
     6.  1x1        |    1x1 [  2 0 ;  2 0 ]           1.

>> E12
     Q:  2x { 6 x 2 }  having 'A,SU2'  { K02, K02* }
  data:  2-D double (704)      6 x 10 => 10 x 16  @ norm = 4

     1.  1x1        |    1x1 [ -2 0 ; -2 0 ]           0.
     2.  1x2        |    2x2 [ -1 1 ; -1 1 ]      16 b     {?$\surd$?2}
     3.  1x3        |    1x1 [  0 0 ;  0 0 ]      24 b
     4.  1x1        |    3x3 [  0 2 ;  0 2 ]           0.  {?$\surd$?3}
     5.  1x2        |    2x2 [  1 1 ;  1 1 ]      16 b     {?$\surd$?2}
     6.  1x1        |    1x1 [  2 0 ;  2 0 ]           0.

>> I12.ee  ?\mlbl{I12.ee}?

   -2.0000    1.0000  ?\mlbl{I12.ee:1}?
   -1.0000    2.0000
   -1.0000    2.0000
    0.0000    1.0000
    0.0000    1.0000
    0.0000    3.0000
    0.0000    1.0000
    1.0000    2.0000
    1.0000    2.0000
    2.0000    1.0000  ?\mlbl{I12.ee:end}?
    
\end{minted}
The columns of \src{U12} encode the eigenstates that bring \src{H12} into diagonal form \src{E12}.
The latter is stored in \idx[diagQS]{compact diagonal format} (storing the diagonal only for each RMT as
a row vector, as seen by the RMT dimensions
in the second column).
The compact diagonal format can be expanded to a
regular diagonal matrix
via \src{\idx[diagQS]{diag}(E12)}. As seen from \src{E12}, there
are three symmetry sectors that have zero blocks
(all of them containing a single multiplet which
thus prints the \src{0.} to the right).

The full eigenspectrum is provided in plain numeric format
with the info structure \src{IS} returned as third argument.
The first column in \src{I12.ee}
in \mlines{I12.ee:1}--\ref{I12.ee:end}
shows the many-body spectrum $E \in \{-2,-1,0,1,2\}$ 
as discussed above.
Each row represents a particular multiplet.
The combined multiplet dimension over all symmetries,
i.e., the eigenenergy's degeneracy is specified
in the second column.
In the absence of non-abelian symmetries, the second column is omitted
since in that case it would contain only trivial \src{1}'s.
In the example here, the second column adds up to 16.
This confirms the full Hilbert space dimension.

\subsubsection{Changing local state space or symmetries}
\label{sec:gls:NC3}

In the above discussion, the overall symmetry was specified
exactly once, namely in \Eq{gls:NC1} which defines the type
of `site' to use. Since symmetries are global,
all symmetries to be included need to be 
defined there once and for all.
If one were to switch to a different symmetry combination, 
line \eqref{eq:gls:NC1} is the only one that needs to be changed for the entire subsequent discussion.
For example, one may use the same fermionic model, yet with
\src{NC=1} $\to$ \src{3} fermionic channels (flavors),
\begin{minted}[escapeinside=??]{matlab}
 [F,Z,S,IS]=getLocalSpace('FermionS','Acharge,SU2spin,SUNchannel','NC',3); ?\mlbl{idx:gls:NC3}?
 [F,Z,S,IS]=getLocalSpace('FermionS','Acharge,SU2spin,SU3channel');    % equivalent
\end{minted}
The local state space of a site now consists of 
$d_{\rm loc}=4^3 = 64$ states which can be reduced to an effective $d_{\rm loc}^\ast=10$ multiplets, e.g., as seen 
by inspecting the identity operator [cf. \eqref{eq:IS:struct}],
\begin{minted}[escapeinside=??,firstnumber=last]{text}
>> IS.E ?\mlbl{IS.E:NC3}?

     Q:  2x {10 x 4 }  having 'A,SU2,SU3'  { +- }                 ?\mlbl{IS.E:hdr1}?
  data:  2-D double (1120)      10 x 10 => 64 x 64  @ norm = 8    ?\mlbl{IS.E:hdr2}?

     1.  1x1        |    1x1    1x1 [ -3 0 00 ; -3 0 00 ]           1.
     2.  1x1        |    2x2    3x3 [ -2 1 10 ; -2 1 10 ]           1.  {?$\surd$?6}
     3.  1x1        |    1x1    6x6 [ -1 0 20 ; -1 0 20 ]           1.  {?$\surd$?6}
     4.  1x1        |    3x3    3x3 [ -1 2 01 ; -1 2 01 ]           1.  {?$\surd$?9}
     5.  1x1        |    2x2    8x8 [  0 1 11 ;  0 1 11 ]           1.  {?$\surd$?16} ?\mlbl{IS.E:rec5}?
     6.  1x1        |    1x1    6x6 [  1 0 02 ;  1 0 02 ]           1.  {?$\surd$?6}
     7.  1x1        |    4x4    1x1 [  0 3 00 ;  0 3 00 ]           1.  {?$\surd$?4} ?\mlbl{IS.E:rec7}?
     8.  1x1        |    3x3    3x3 [  1 2 10 ;  1 2 10 ]           1.  {?$\surd$?9}
     9.  1x1        |    2x2    3x3 [  2 1 01 ;  2 1 01 ]           1.  {?$\surd$?6}
    10.  1x1        |    1x1    1x1 [  3 0 00 ;  3 0 00 ]           1.  ?\mlbl{IS.E:NC3:end}?

\end{minted}
Here every symmetry sector 
on each of the \src{2} legs in 
the total of \src{10} records carries
\src{4} symmetry labels $q \equiv (q_1,q_2,q_3,q_4)
\equiv (q_1\ q_2\ q_3 q_4)$, as also indicated with the
dimensions \mbox{\src{2x \{10 x 4\}}} in \mline{IS.E:hdr1}.
Based on the order of symmetries requested with
\src{getLocalSpace},
the first symmetry label describes
the filling relative to half-filling.
Given \src{NC=3} flavors,
this has the range $q_1 \in \{-3,-2,\ldots,3\}$
from completely empty to completely filled,
respectively.
The second symmetry label describes SU(2) spin,
\idx[SU2:qlabels]{$q_1=2S$},
with the largest spin multiplet given by $q_2=3$
(\mline{IS.E:rec7}), 
i.e., spin $S=3/2$. It derives from the spin-half
for each half-filled level, and hence has
combined symmetry labels $[0\ 3\ 00]$.
The last two symmetry labels $(q_3 q_4)$ describe the
SU(3) channel or flavor symmetry.
The largest SU(3) multiplet above 
(11)\,$\equiv{\bf 8}$ in \mline{IS.E:rec5}
contains 8 states (hence referred to as octet;
for SU(3) this is also the \idx[qadj]{adjoint} representation).
Having combined symmetry labels $[0\ 1\ 11]$, again this describes states at half-filling.
With \src{NC=3} being odd,
half-filling necessarily has a non-zero spin, here $2S=1$.
Hence the combined multiplet dimension in $[0\ 1\ 11]$
is $2\times 8 = 16$ states (see CGT dimensions in \mline{IS.E:rec5}).

The \QSpace displays as the ones above are just for information. Their understanding requires at least a rudimentary understanding of the symmetries used, including their labeling structure. The display is generated automatically by \QSpace. Looking more closely at the individual entries permits simple consistency checks like the ones above. This ensures that one has a good understanding of the tensors under consideration.

Once a single site is defined together with the
symmetries employed, one can proceed identically to the earlier case of a single flavor to build a many-body state space. For example, two sites can be combined identically  as in \eqref{eq:A2} \\
\begin{minipage}{1\linewidth}
\begin{minted}[escapeinside=??,firstnumber=last]{matlab}
>> A=getIdentity(Z,Z,[1 3 2]);
>> A=setitags(A,{'K01','K02','s02'})
\end{minted}
\mintLabelB{eq:itags:A}
\end{minipage}
\\[2ex]
In the present case, however, this results in
\begin{minted}[escapeinside=??,firstnumber=last]{text}
A = ?\mlbl{A3:disp}?

    Q:  3x {258 x 4 }  having 'A,SU2,SU3'  A-matrix,  { K01, K02*, s02 }       ?\mlbl{A3:hdr1}?
 data:  3-D double (38.08k)      10 x 260 x 10 => 64 x 4,096 x 64  @ norm = 64 ?\mlbl{A3:hdr2}?

    1.  1x1       |    1x1x1  1x1x1 [ -3 0 00 ; -6 0 00 ; -3 0 00 ]        1.
    2.  1x2       |    1x2x2  1x3x3 [ -3 0 00 ; -5 1 10 ; -2 1 10 ]   16 b    {?$\surd$?6}
    3.  1x2       |    2x2x1  3x3x1 [ -2 1 10 ; -5 1 10 ; -3 0 00 ]   16 b    {?$\surd$?6}
    4.  1x1       |    2x1x2  3x3x3 [ -2 1 10 ; -4 0 01 ; -2 1 10 ]        1. {?$\surd$?3}
    5.  1x3       |    1x1x1  1x6x6 [ -3 0 00 ; -4 0 20 ; -1 0 20 ]   24 b    {?$\surd$?6}
    :   ...  ?\mlbl{A3:rec:i}?
  117.  1x8x1 @2  | 2x1x2x1 8x8x8x2 [  0 1 11 ;  0 0 11 ;  0 1 11 ]  128 b    {?$\surd$?8}?\mlbl{A3:rec117}?
  138.  1x12x1 @2 | 2x3x2x1 8x8x8x2 [  0 1 11 ;  0 2 11 ;  0 1 11 ]  192 b    {?$\surd$?24}?\mlbl{A3:rec138}?
    :   ...  ?\mlbl{A3:rec:j}?          
  257.  1x2       |    1x2x2  1x3x3 [  3 0 00 ;  5 1 01 ;  2 1 01 ]   16 b    {?$\surd$?6}
  258.  1x1       |    1x1x1  1x1x1 [  3 0 00 ;  6 0 00 ;  3 0 00 ]        1.

\end{minted}
As compared to the $A$-tensor for \src{NC=1} in \eqref{eq:A2}
and its output following \mline{A2:disp},
the fusion of two sites here leads to considerably
more entries (258 vs. 10 earlier).
By default, the listing is truncated, strongly
so in the present case, skipping
records 6--256 in \mlines{A3:rec:i}--\ref{A3:rec:j},
except for two largest entries, either
in size or \idx[OMindex]{outer multiplicity}.
\mLines{A3:rec117}--\ref{A3:rec138}
show records 117 and 138, both of which are entries with outer multiplicity $M=2$, denoted
via ``\src{@2}'' with the dimensions of the RMTs.
This outer multiplicity
is also reflected  in the trailing dimensions
of the respective CGTs [e.g., having 
a trailing ``\src{x2}'' for the CGTs for SU(3),
as in \src{8x8x8x2},
while having no OM with the CGTs for SU(2), hence showing
a trailing ``\src{x1}'' with their dimensions,
as in \src{2x1x2x1}].
To display all records in \src{A},
one can type \src{display(A,\str{-f})}.
The overall tensor dimensions in the header (\mline{A3:hdr2})
are as expected, fusing $64 \times 64 \to 4,096$ states.
In terms of multiplets, this fuses $10 \times 10 \to 260$
multiplets, total which is about an order of magnitude smaller.
From the header in \mline{A3:hdr2} still,
the norm is given by $\Vert A\Vert =\sqrt{4,096}=64$ also reflecting
the fused state space dimension.
While from the display of the identity for a single site 
in \mlines{IS.E:NC3}--\ref{IS.E:NC3:end}
all input multiplets have unique symmetry labels 
showing an RMT size of \src{1x ...},
record 138 (\mline{A3:rec138}) here reports
12 multiplets in the fused sector $[0\ 2\ 11]$.
This combines a spin $S=1$ with an SU(3) multiplet
(11), such that these multiplets contain
$3\times 8 = 24$ states each (cf. also
\heq[wnorm:x3]{$w$-factor} to the very right).

The other commands earlier with \src{NC=1}, such as the
spin interaction or fermionic hopping, can be used
here also for \src{NC=3} {\it identically} without any change.
Hence with the symmetries only specified at the very 
beginning of a setup with \src{getLocalSpace},
the actual tensor network commands 
can be implemented the same way irrespective of the
symmetry setting, whether abelian or non-abelian symmetries
were used, whether multiple symmetries are used in parallel, or just individual ones, or \idx[nosym]{no symmetry} at all. Hence once the local state space with
the desired symmetries is defined, the subsequent
code can proceed nearly as if there had been no
symmetries at all. This allows one to focus on
tensor network algorithms without having
to worry much about symmetries, while fully exploiting
complex symmetry settings, nevertheless. 

Further increasing the symmetry for \src{NC=3}
in \mline{idx:gls:NC3} above to symplectic
(e.g., by having particle/hole symmetry on a
bipartite lattice as well as SU(\src{NC}) 
channel symmetry \cite{Wb12_SUN}), \\
\begin{minipage}{1\linewidth}
\begin{minted}[escapeinside=??,firstnumber=last]{matlab}
  [F,Z,S,IS]=getLocalSpace('FermionS','SU2spin,SpNchannel','NC',3); ?\mlbl{idx:gls:Sp6}?
  [F,Z,S,IS]=getLocalSpace('FermionS','SU2spin,Sp6channel');    % equivalent
\end{minted}
\mintLabelB{eq:Sp6}
\end{minipage}
\\[2ex]
As seen from the fermionic parity (or equally from the 
identity in \src{IS.E}),
\begin{minted}[escapeinside=??,firstnumber=last]{text}
Z =
     Q:  2x { 4 x 4 }  having 'SU2,Sp6'  { +- }
  data:  2-D double (448)      4 x 4 => 64 x 64  @ norm = 8

     1.  1x1        |    4x4    1x1 [ 3 000 ; 3 000 ]          -1.  {?$\surd$?4} ?\mlbl{Sp6:Z:1}?
     2.  1x1        |    3x3    6x6 [ 2 100 ; 2 100 ]           1.  {?$\surd$?18} ?\mlbl{Sp6:Z:2}?
     3.  1x1        |    2x2  14x14 [ 1 010 ; 1 010 ]          -1.  {?$\surd$?28} ?\mlbl{Sp6:Z:3}?
     4.  1x1        |    1x1  14x14 [ 0 001 ; 0 001 ]           1.  {?$\surd$?14} ?\mlbl{Sp6:Z:4}?
\end{minted}
in the symplectic case the local state space
of $d=64$ states has been reduced to an effective $d^\ast =4$
multiplets. There are still a total of four symmetry
labels here to identify a multiplet, $q=(q_1\ q_2 q_3 q_4)$,
but the split up changed.
The first label \idx[SU2:qlabels]{$q_1 = 2S$}
now specifies the SU(2) spin multiplet.
Since Sp(6) is a rank $\rsym=3$ symmetry,
it carries the remaining three labels $(q_2 q_3 q_4)$.
The scalar \src{000}\,$\equiv {\bf 1}$ is
shown with \mline{Sp6:Z:1}, the defining
\src{100}\,$\equiv {\bf 6}$ in \mline{Sp6:Z:2}.
The largest Sp(6) muliplet occurs with \mline{Sp6:Z:4},
namely \src{001}\,$\equiv {\bf 14}$. Note that
Sp(6) like the leading spin SU(2) is self-dual.
The annihilation operator 
\begin{minted}[escapeinside=??,firstnumber=last]{text}
F =  ?\mlbl{NC3:F}?
     Q:  3x { 6 x 4 }  having 'SU2,Sp6'  operator,  { +-- }
  data:  3-D double (672)      4 x 4 x 1 => 64 x 64 x 12  @ norm = ?$\surd$?384 ?\mlbl{NC3:F:hdr2}?

     1.  1x1        |  3x4x2   6x1x6 [ 2 100 ; 3 000 ; 1 100 ]      1.15470  {?$\surd$?18}
     2.  1x1        |  4x3x2   1x6x6 [ 3 000 ; 2 100 ; 1 100 ]      2.44949  {?$\surd$?4}
     3.  1x1        |  2x3x2  14x6x6 [ 1 010 ; 2 100 ; 1 100 ]     -1.73205  {?$\surd$?28}
     4.  1x1        |  3x2x2  6x14x6 [ 2 100 ; 1 010 ; 1 100 ]     -2.16025  {?$\surd$?18}
     5.  1x1        |  1x2x2 14x14x6 [ 0 001 ; 1 010 ; 1 100 ]     -2.44949  {?$\surd$?14}
     6.  1x1        |  2x1x2 14x14x6 [ 1 010 ; 0 001 ; 1 100 ]      1.73205  {?$\surd$?28}
\end{minted}
now has become an irop with $(\src{NC=3})\times 4 = 12$
components (\mline{NC3:F:hdr2})  that transforms like a single
multiplet in the symmetry sector \src{[1 100]},
consistently also of dimension $2\times 6 = 12$.
Each of these components contributes
equally to the norm $\Vert F\Vert^2 = {\rm tr}(F^\dagger F)
= 12 * 2^{6-1} = 384$ (\mline{NC3:F:hdr2}).
The SU(2) spin operator acts trivially in the symplectic
sector having \src{000},
\begin{minted}[escapeinside=??,firstnumber=last]{text}
S =  ?\mlbl{Sp6:S}?
     Q:  3x { 3 x 4 }  having 'SU2,Sp6'  operator,  { +-- }
  data:  3-D double (336)      3 x 3 x 1 => 50 x 50 x 3  @ norm = ?$\surd$?72

     1.  1x1        |  4x4x3   1x1x1 [ 3 000 ; 3 000 ; 2 000 ]     1.936491  {?$\surd$?4} ?\mlbl{Sp6:S:rec1}?
     2.  1x1        |  3x3x3   6x6x1 [ 2 100 ; 2 100 ; 2 000 ]    -1.414214  {?$\surd$?18} ?\mlbl{Sp6:S:rec2}?
     3.  1x1        |  2x2x3 14x14x1 [ 1 010 ; 1 010 ; 2 000 ]    -0.866025  {?$\surd$?28} ?\mlbl{Sp6:S:rec3}?
\end{minted}
The spin sectors $S=0$ are absent
since these correspond to \idx[QS:zblocks]{zero blocks}.
The norm of the individual RMTs 
in \mlines{Sp6:S:rec1}--\ref{Sp6:S:rec3}
already reflect
the Casimir for the respective local spin, having values
$(\sqrt{\frac{15}{4}}, -\sqrt{2}, \sqrt{\frac{3}{4}}$),
and thus $\pm \sqrt{S(S+1)}$ for $q_1 = 2S \in \{3,2,1\}$.
This emphasizes the importance of the
relative weight across the RMTs. The relative signs
are also equally important.

The tensor product space of two sites can proceed
the same way as in the initial setting
in \eqref{eq:A2}. The $A$-tensor now becomes
\begin{minted}[escapeinside=??,firstnumber=last]{text}
A = ?\mlbl{A:Sp6}?
     Q:  3x {61 x 4 }  having 'SU2,Sp6'  A-matrix,  { K01, K02*, s02 }
  data:  3-D double (7.828k)      4 x 61 x 4 => 64 x 4,096 x 64  @ norm = 64

     1.  1x4      |  1x1x1  14x1x14 [ 0 001 ; 0 000 ; 0 001 ]      32 b
     2.  1x4      |  2x1x2  14x1x14 [ 1 010 ; 0 000 ; 1 010 ]      32 b
     3.  1x4      |  3x1x3    6x1x6 [ 2 100 ; 0 000 ; 2 100 ]      32 b
     4.  1x4      |  4x1x4    1x1x1 [ 3 000 ; 0 000 ; 3 000 ]      32 b
     5.  1x1      |  1x1x1 14x84x14 [ 0 001 ; 0 002 ; 0 001 ]           1.  {?$\surd$?84}
     6.  1x2      |  2x1x2 14x14x14 [ 1 010 ; 0 010 ; 1 010 ]      16 b     {?$\surd$?14}
     :   ...
    35.  1x6      |  3x3x3   6x14x6 [ 2 100 ; 2 010 ; 2 100 ]      48 b     {?$\surd$?42}?\mlbl{A:Sp6:intermediate:recs}?
    36.  1x6      |  4x3x2  1x14x14 [ 3 000 ; 2 010 ; 1 010 ]      48 b     {?$\surd$?42}
     :   ...
    60.  1x2      |  4x6x3    1x6x6 [ 3 000 ; 5 100 ; 2 100 ]      16 b     {?$\surd$?36}
    61.  1x1      |  4x7x4    1x1x1 [ 3 000 ; 6 000 ; 3 000 ]           1.  {?$\surd$?7}
\end{minted}
This reduces the full Hilbert space for the two sites
from $D=64^2 = 4096$ states to an effective $D^\ast = 
61$ multiplets across 23 symmetry sectors
[the latter can be seen from \src{getIdentity(A,2)}].
The matrix elements
in \eqref{eq:SdotS} or \eqref{eq:FdotF:T12} can be 
computed as previously without any change.

\subsection{Swap operator for two spins \texorpdfstring{$S=1$}{S=1}}
\label{sec:swap}

The following proceeds similar to the previous
two-site examples [e.g., see \eqref{eq:A2}],
yet switches from a fermionic state space
for a site to a spin $S=1$ in SU(2),
\begin{minted}[escapeinside=??]{matlab}
  [S,IS]=getLocalSpace('Spin','SU2',2); ?\mlbl{S1:op}? % 2S=2 => S=1
  A=getIdentity(S,S,[1 3 2]);           ?\mlbl{A2:S}? % may also use S $\to$ IS.E instead
  A=setitags(A,{'K01','K02','s02'});
\end{minted}
\vspace{-4ex}
\begin{minted}[escapeinside=??,firstnumber=last]{text}
>> A

     Q:  3x { 3 x 1 }  having 'SU2'  A-matrix,  { K01, K02*, s02 }
  data:  3-D double (336)      1 x 3 x 1 => 3 x 9 x 3  @ norm = 3

     1.  1x1        |  3x1x3 [ 2 ; 0 ; 2 ]           1.
     2.  1x1        |  3x3x3 [ 2 ; 2 ; 2 ]           1.  {?$\surd$?3}
     3.  1x1        |  3x5x3 [ 2 ; 4 ; 2 ]           1.  {?$\surd$?5}
\end{minted}
The state space of two sites as described by \src{A}
results in a total spin $S\in \{0,2,4\}/2 = \{0,1,2\}$,
as expected. The following then computes the `interaction' $X$
that describes a plain swap of the state space of two sites,
also known as the permutation operator.
This operator $X$ is Hermitian 
with eigenvalues $\pm 1$ since $X^2=1$.
Its matrix elements can be obtained as follows,
\begin{minted}[escapeinside=??,firstnumber=last]{matlab}
 A_= untag(A);      ?\mlbl{Xop:A_}?              % required for the next line
 X = contract(A_,'13*',A_,'31'); ?\mlbl{Xop:ctr}? % swap operator
\end{minted}
\vspace{-4ex}
\begin{minted}[escapeinside=??,firstnumber=last]{text}
>> X ?\mlbl{X:swp}?
     Q:  2x { 3 x 1 }  having 'SU2'  { +- }
  data:  2-D double (336)      3 x 3 => 9 x 9  @ norm = 3

     1.  1x1        |    1x1 [ 0 ; 0 ]           1.                ?\mlbl{X:swp:rec1}?
     2.  1x1        |    3x3 [ 2 ; 2 ]          -1.  {?$\surd$?3}  ?\mlbl{X:swp:rec2}?
     3.  1x1        |    5x5 [ 4 ; 4 ]           1.  {?$\surd$?5}  ?\mlbl{X:swp:recl}?
\end{minted}
This swap operator $X$ is already diagonal with eigenvalues
$\pm 1$ as shown with the RMTs.
The \itags in the contracted local indices were
temporarily removed (\src{untag} in \mline{Xop:A_}),
since the swap operator contracts legs
with otherwise {\it different} \itags
[see \Fig{Xop}(a)].
Leaving the \itags in place would result
in an error due to \itag mismatch [cf. \Fig{Xop}(c)].
The contraction in \mline{Xop:ctr} uses
\idx[compact:idx]{compact index} notation.
It contracts legs $1$ and $3$,
i.e., $K_1 \equiv \sigma_1$ and $\sigma_2$ of the conjugate
of the first tensor with legs $3$ and $1$,
i.e., $\sigma_2$ and $\sigma_1$
and thus reverse order, on the second tensor
(tensor \idx[conj]{conjugation}
leaves the order of legs intact).

Now an interaction of two spin-$S$ sites can be decomposed
into powers $(S_1^\dagger\cdot S_2^{\,})^n$ 
of the spin-spin interaction
which forms a (non-orthonormalized) operator basis.
Clearly, by only having
$2S+1$ RMTs, each of which is given by a single number,
one reaches a complete basis for $n\leq 2S$.
Hence powers $n\geq 2S+1$ can be expressed
in terms of smaller powers.
For example for spin-half sites
the identity and the spin operator
already exhaust the complete set of local operators,
since $2^2 = 1+3$.
For the swap operator here for $S=1$ sites, this implies
\begin{eqnarray}
   X = \sum_{n=0}^2 a_n \  (S_1^{\dagger}\cdot S_2^{\,})^n
\text{ .}\label{eq:X:SSn}
\end{eqnarray}
with coefficients $a_n$ to be determined.
The operator for $n=0$ refers to the identity
which has the same \QSpace structure as displayed with \src{X}
in \mline{X:swp}, 
except that it has all \src{+1}'s in the right column
for the RMTs.
The case $n=1$ gives the Heisenberg interaction
\src{SS1} [cf. \Fig{Xop}(d)],
\begin{minted}[escapeinside=??,firstnumber=last]{text}
>> SAS = contract(S,'-op:K01','*',{A,S,'-op:^s'}); ?\mlbl{SAS}?
>> SS1 = contract(A,'!2*',SAS)                     ?\mlbl{SS1}?

     Q:  2x { 3 x 1 }  having 'SU2'  { K02, K02* }
  data:  2-D double (336)      3 x 3 => 9 x 9  @ norm = ?$\surd$?12

     1.  1x1        |    1x1 [ 0 ; 0 ]          -2.
     2.  1x1        |    3x3 [ 2 ; 2 ]          -1.  {?$\surd$?3}
     3.  1x1        |    5x5 [ 4 ; 4 ]           1.  {?$\surd$?5}
\end{minted}
and $n=2$ the quadrupolar term \src{SS2} [cf. \Fig{Xop}(e)],
\begin{minted}[escapeinside=??,firstnumber=last]{text}
>> SS2 = contract(A,'!2*',{S,'-op:K01','*',{SAS,S,'-op:^s'}})

     Q:  2x { 3 x 1 }  having 'SU2'  { K02, K02* }
  data:  2-D double (336)      3 x 3 => 9 x 9  @ norm = ?$\surd$?24

     1.  1x1        |    1x1 [ 0 ; 0 ]           4.
     2.  1x1        |    3x3 [ 2 ; 2 ]           1.  {?$\surd$?3}
     3.  1x1        |    5x5 [ 4 ; 4 ]           1.  {?$\surd$?5}
\end{minted}
Now by collecting the RMTs for $n=0,1,2$ ,
as well as for \src{X}
in \mlines{X:swp:rec1}--\ref{X:swp:recl}
into columns, 
one can determine the coefficients $a_n$ in \Eq{X:SSn},
\begin{eqnarray}
   \begin{pmatrix} 1 \\ -1 \\ 1 \end{pmatrix}
   \overset{!}{=}
   \begin{pmatrix}
     1 & -2 & 4 \\
     1 & -1 & 1 \\
     1 &  1 & 1 
   \end{pmatrix}
   \begin{pmatrix} a_0 \\ a_1 \\ a_2 \end{pmatrix}
\quad\Rightarrow\quad
   \begin{pmatrix} a_0 \\ a_1 \\ a_2 \end{pmatrix}
 = \begin{pmatrix} -1 \\ 1 \\ 1 \end{pmatrix}
\end{eqnarray}
and therefore
\begin{eqnarray}
   X = -1 
    +  S_1^{\dagger}\cdot S_2^{\,}
    + (S_1^{\dagger}\cdot S_2^{\,})^2 
\text{ .}\label{eq:X:SSn-2}
\end{eqnarray}
A simple countercheck confirms the coefficients
in \Eq{X:SSn-2}, indeed,
\begin{minted}[escapeinside=??,firstnumber=last]{text}
>> E2 = getIdentity(A,2);
>> X_ = -E2 + SS1 + SS2

    Q:  2x { 3 x 1 }  having 'SU2'  { +- }
  data:  2-D double (336)      3 x 3 => 9 x 9  @ norm = 3

     1.  1x1        |    1x1 [ 0 ; 0 ]           1.
     2.  1x1        |    3x3 [ 2 ; 2 ]          -1.  {?$\surd$?3}
     3.  1x1        |    5x5 [ 4 ; 4 ]           1.  {?$\surd$?5}
\end{minted}
with \src{norm(X\_-X)} returning numerical double
precision noise $<10^{-14}$. This not being exactly
zero is due to the \idx[cgw]{square root factors} shown
to the right. For this reason, \src{X==X\_} returns false.
The \QSpace routine \src{\idx[QS:equal]{sameas}(X,X\_)} based
on \src{norm(X\_-X)} returns true.

\paragraph{Physical interpretation}

By using a simple swap operator on local state spaces
of dimension $d_{\rm loc}=3$, an arbitrary 3-dimensional
unitary $U$ can be inserted as $UU^\dagger$
into any of the two lines in \Fig{Xop}(a).
This takes the interpretation of transforming
the local state space for each site.
However, with $X_{12} \to
  (U\otimes U) \, X_{12} (U^\dagger\otimes U^\dagger)
= (U U^\dagger\otimes 1) (1\otimes U U^\dagger) X_{12}
= X_{12}$ being identical to $X$ [most easily seen by 
considering the pictorial representation in \Fig{Xop}(a)],
this implies that the swap operation in the above Hamiltonian
actually has an enlarged symmetry, namely SU($d_{\rm loc}$).
For the case of the spin $S=1$ here, this is the well-known
SU(3) symmetric point of the bilinear-biquadratic
Heisenberg model for $S=1$ at the phase boundary
of the Haldane phase \cite{Haldane83,Tu08}.

\subsection{Tweaking local state space}
\label{sec:gls:tweaks}

The local state space of a site is set up via a call
to \src{getLocalSpace} which has many
\hsec[gls]{standard symmetry setups} already
implemented. Nevertheless, situations can arise
that may go beyond to what is already provided by
\src{getLocalSpace}. The typical way to proceed then
is to tweak its output starting from a symmetry
setup that may be considered closest to target setup.

To be specific, consider for example SU(2) spin-halfs
with an additional two symmetric orbital flavors, i.e.,
${\rm SU}(2)_{\rm spin}\otimes {\rm SU}(2)_{\rm orbital}$.
This does not exist ready-made in \src{getLocalSpace}.
Two ways to deal with this
situation are discussed in what follows.

\subsubsection{Project fermionic setup to spin model
and complete operator basis}
\label{sec:tweakS}

Starting from a spinfull fermionic model with 
\src{NC=2} orbital flavors \\
\begin{minipage}{1\linewidth}
\begin{minted}[escapeinside=??,firstnumber=last]{matlab}
   [F,Z,S,IS]=getLocalSpace('FermionS','Acharge,SU2spin,SU2channel');
\end{minted}
\mintLabelB{eq:gls:SU2x2}
\end{minipage} \\[2ex]
the state space can be inspected by \src{Z} or \src{IS.E},
\begin{minted}[escapeinside=??]{text}
>> E=IS.E

     Q:  2x { 6 x 3 }  having 'A,SU2,SU2'  { +- }
  data:  2-D double (672)      6 x 6 => 16 x 16  @ norm = 4

     1.  1x1        |    1x1    1x1 [ -2 0 0 ; -2 0 0 ]           1.
     2.  1x1        |    2x2    2x2 [ -1 1 1 ; -1 1 1 ]           1.  {?$\surd$?4} ?\mlbl{EP:rec2}?
     3.  1x1        |    1x1    3x3 [  0 0 2 ;  0 0 2 ]           1.  {?$\surd$?3}
     4.  1x1        |    3x3    1x1 [  0 2 0 ;  0 2 0 ]           1.  {?$\surd$?3}
     5.  1x1        |    2x2    2x2 [  1 1 1 ;  1 1 1 ]           1.  {?$\surd$?4}
     6.  1x1        |    1x1    1x1 [  2 0 0 ;  2 0 0 ]           1.
\end{minted}
Looking for the spin $S=1/2$ sector at a filling of $n=1$
particle on the site, this corresponds to $q_1=-1$
relative to half-filling. This contains a single multiplet
in \mline{EP:rec2} 
which is thus the one of interest.
It can be projected by just picking this record
from the identity using \idx[QS:getsub]{getsub},
\begin{minted}[escapeinside=??,firstnumber=last]{text}
>> q0=-1; i=find(Z.Q{1}(:,1)==q0); ?\mlbl{gls:tweak:q0}? % finds record i=2. (?\mline{EP:rec2}? above)
>> E=getsub(E,i)                    % reduces E to record(s) i

     Q:  2x { 1 x 3 }  having 'A,SU2,SU2'  { +- }
  data:  2-D double (112)      1 x 1 => 4 x 4  @ norm = 2

     1.  1x1        |    2x2    2x2 [ -1 1 1 ; -1 1 1 ]           1.  {?$\surd$?4}

\end{minted}
Since now only a trivial single abelian symmetry sector is present in terms of charge, it can be skipped (removed) as follows,
\begin{minted}[escapeinside=??,firstnumber=last]{text}
>> E=rmAbelian(E,1) ?\mlbl{E:rmAbelian}?

E =
     Q:  2x { 1 x 2 }  having 'SU2,SU2'  { +- }
  data:  2-D double (112)      1 x 1 => 4 x 4  @ norm = 2

     1.  1x1        |    2x2    2x2 [ 1 1 ; 1 1 ]           1.  {?$\surd$?4}

>> clear F Z ?\mlbl{clearFZ}?
\end{minted}
The syntax in \mline{E:rmAbelian}
removes the symmetry at position \src{1}
from \src{E}, i.e., the abelian U(1) charge.
Projecting to a single charge sector renders the fermionic operators irrelevant (\mline{clearFZ}).
The spin operator, finally, can be projected the same way
as the identity above,
\begin{minted}[escapeinside=??,firstnumber=last]{text}
>> S=getsub(find(S.Q{1}(:,1)==q0));   % having q0=-1 from ?\mline{gls:tweak:q0}? above
>> S=rmAbelian(S,1) ?\mlbl{Sop:tweak}?

S =
     Q:  3x { 1 x 2 }  having 'SU2,SU2'  operator,  { +-- }
  data:  3-D double (112)      1 x 1 x 1 => 4 x 4 x 3  @ norm = ?$\surd$?3

     1.  1x1        |  2x2x3  2x2x1 [ 1 1 ; 1 1 ; 2 0 ]    -0.866025  {?$\surd$?4} ?\mlbl{Sop:tweak:rec1}?
\end{minted}
which transforms like a \idx[SU2:qlabels]{$S=q/2$}\,$=1$ irop.
This is the same spin operator as for a single fermionic
flavor [cf. \eqref{eq:gls:NC1}, and subsequent display
in \mlines{idx:Sop:SU2}ff]
that acts trivially, i.e., like an identity
in the orbital space.

The pseudo-spin operator in the orbital space
can be obtained from a \idx[op-basis]{complete operator
basis} for the local state space.
For its construction,
one needs to obtain the tensor product space
in \src{E} with the dual of its state space
as obtained by the \hsec[1j]{\onej}-symbol
in \mline{Xb:U},
\begin{minted}[escapeinside=??,firstnumber=last]{text}
>> U=getIdentity(E,'-0');  ?\mlbl{Xb:U}?
>> X=getIdentity(U,1,U,2);   ?\mlbl{Xb:1}?
>> X=contract(U,'!1*',X,[2 1 3]) ?\mlbl{Xb:2}?

     Q:  3x { 4 x 2 }  having 'SU2,SU2'  { +-- }                      ?\mlbl{Xb:hdr1}?
  data:  3-D double (448)      1 x 1 x 4 => 4 x 4 x 16  @ norm = 4    ?\mlbl{Xb:hdr2}?

     1.  1x1        |  2x2x1  2x2x1 [ 1 1 ; 1 1 ; 0 0 ]          0.5  {?$\surd$?4} ?\mlbl{Xb:rec1}?
     2.  1x1        |  2x2x1  2x2x3 [ 1 1 ; 1 1 ; 0 2 ]     0.866025  {?$\surd$?4} ?\mlbl{Xb:rec2}?
     3.  1x1        |  2x2x3  2x2x1 [ 1 1 ; 1 1 ; 2 0 ]     0.866025  {?$\surd$?4} ?\mlbl{Xb:rec3}?
     4.  1x1        |  2x2x3  2x2x3 [ 1 1 ; 1 1 ; 2 2 ]          1.5  {?$\surd$?4} ?\mlbl{Xb:recl}?
\end{minted}
The tensor product in \mline{Xb:1}
derives from the fact that the full
operator space of a state space of dimension $d$
is described by $d^2 \equiv d\otimes d$ operators.
However, the incoming dual state space,
i.e., \idx[markers]{primed} index in \src{X}
still needs to be reverted 
by contracting the same \onej-symbol 
in \mline{Xb:2},
so that \src{X} represents a standard operator
[cf. \Fig{Aop}].
The trailing permutation \src{[2 1 3]} restores
the \idx[irop]{operator index order} \qdir{+{-}-},
as seen in \mline{Xb:hdr1}.
Having a dimension of \src{1x1x4} in terms of
multiplets (\mline{Xb:hdr2}),
the object \src{X} is clearly
not an irop, though. Rather it contains a (direct)
sum of four irops, which spans the full operator
space of a single site. This consists of the four
records in above display, each of which represents
an irop with a unique operator $q$-label:
\begin{itemize}
  \setlength{\itemsep}{-0.1\baselineskip}

\item Record \src{1.} (\mline{Xb:rec1})
transforms like a scalar
$q_{\rm op}=[0\  0]$. Since this is the only operator
of this type and bearing in mind that \src{X} represents
a complete operator basis, it must be the identity operator
up to the scale factor given by RMT value \src{0.5} 
shown to the right.
Via the tensor-product in \src{getIdentity}
in \mline{Xb:1}
each operator has Frobenius norm 1 which explains
this factor, since $\Vert \frac{1}{2} \Id^{(4)} \Vert = 1$.

\item Record \src{3.} (\mline{Xb:rec3})
represents an irop that
transforms according to $q_{\rm op}=[2\  0]$.
This identifies it as the spin operator like
in \mline{Sop:tweak:rec1} above 
up to a sign
(note that the sign is irrelevant in \hsec[SdotSd]{bilinear
products}, as long as the {\it same} operator \src{S} is used
consistently throughout).
Incidentally, in the present setting the value of its RMT 
\src{0.866025}$=\frac{\sqrt{3}}{2}$ is consistent
already with both, the normalization of a spin-half
operator ($\hat{S}^2 = \frac{3}{4}\,\Id^{(4)}$),
as well as the normalization out of the tensor-product in \src{getIdentity} in \mline{Xb:1}, 
having $\Vert \hat{S}\Vert^2
= {\rm tr}(\frac{3}{4} \Id^{(4)}) = 3$ 
representing three normalized operators, indeed.

\item Record \src{2.} (\mline{Xb:rec2})
represents an irop that
transforms according to $q_{\rm op}=[0\  2]$.
Therefore this entry describes the pseudo-spin operator
in orbital space.

\item Record \src{4.} (\mline{Xb:recl}),
finally, represents an irop that
transforms according to $q_{\rm op}=[2\ 2]$.
As such this is a spin-orbit interaction \src{ST}
up to a factor $2$ when comparing to RMT value
of~\src{1.5} to the standard
SU(2) spin normalization $\sqrt{(\hat{S}\hat{T})^2}
= \sqrt{\hat{S}^2 \hat{T}^2} = \frac{3}{4}\,\Id^{(4)}$.

\end{itemize}
By construction, the above full operator basis in \src{X}
permits one to extract
all (other) irreducible operators within the
present symmetry setting. Here this includes
the pseudo-spin operator \src{T} in orbital space
(choosing the same sign convention as in \src{S}),
\begin{minted}[escapeinside=??,firstnumber=last]{text}
>> T=-getsub(X,3) ?\mlbl{gls:X:T}?

     Q:  3x { 1 x 2 }  having 'SU2,SU2'  { +-- }
  data:  3-D double (112)      1 x 1 x 1 => 4 x 4 x 3  @ norm = ?$\surd$?3

     1.  1x1        |  2x2x3  2x2x1 [ 1 1 ; 1 1 ; 2 0 ]    -0.866025  {?$\surd$?4}

\end{minted}
The remaining last operator within the present symmetry
setting is the spin-orbit term
\begin{minted}[escapeinside=??,firstnumber=last]{text}
>> ST= getsub(X,4)/2;  ?\mlbl{gls:X:ST}?

     Q:  3x { 1 x 2 }  having 'SU2,SU2'  { +-- }
  data:  3-D double (112)      1 x 1 x 1 => 4 x 4 x 9  @ norm = 1.5

     1.  1x1        |  2x2x3  2x2x3 [ 1 1 ; 1 1 ; 2 2 ]         0.75  {?$\surd$?4}
\end{minted}
The combination of these four irops that describe
spinors of dimension $1 + 3 + 3+ 3\times 3 =
16 = 4^2$ thus, indeed, exhausts the operator basis
of given local state space.

\subsubsection{Appending additional symmetries}

Alternative to the previous projective approach,
one may tweak the output of \src{getLocalSpace}
by adding other symmetries still.
Starting with a plain SU(2) spin-half,
one can manually add an additional SU(2) symmetry,
here to the identity operator returned by \src{getLocalSpace}.
For the example above, this starts here
with the simpler spin model,
\begin{minted}[escapeinside=??]{matlab}
  [S,IS]=getLocalSpace('Spin','SU2',1);   % initial spin-half (q=2S=1)
  E=addSymmetry(IS.E,'SU2','q',1)   ?\mlbl{E:addSym}?
\end{minted}
\vspace{-4ex}
\begin{minted}[escapeinside=??,firstnumber=last]{text}
E =
     Q:  2x { 1 x 2 }  having 'SU2,SU2'  { +- }
  data:  2-D double (112)      1 x 1 => 4 x 4  @ norm = 2 ?\mlbl{gls:X:E:hdr2}?

     1.  1x1        |    2x2    2x2 [ 1 1 ; 1 1 ]           1.  {?$\surd$?4}
\end{minted}
By default, \src{\idx[QS:addSymmetry]{addSymmetry}}
appends the specified
symmetry in the scalar irep $q=0$. By explicitly
specifying $q=1$ in \mline{E:addSym},
this adds the second SU(2) symmetry in the form of an
$S=1/2$ multiplet.
It is interpreted here as SU(2) symmetric orbital.
The local state space therefore becomes $2\times 2 = 4$
dimensional as seen in \mline{gls:X:E:hdr2}.
Since \src{addSymmetry} always applies a square-root
factor of the multiplet dimension added,
the identity \src{E} already has the correct normalization
(which matter of fact, is the motivation as to why
\src{addSymmetry} behaves this way).

Next the spin operators are tweaked. First, this generates
the orbital spin operator \src{T} in \mline{T:addSym}
based on \src{S}, before adapting the spin operator \src{S}
itself in \mline{S:addSym},
\begin{minted}[escapeinside=??,firstnumber=last]{matlab}
  T=addSymmetry(S,'SU2','pos',1,'q',1); ?\mlbl{T:addSym}? % `spin' operator in orbital sector
  S=addSymmetry(S,'SU2','q',1);         ?\mlbl{S:addSym}? %  adapts existing spin operator
\end{minted}
\mLine{T:addSym} 
uses the option \src{\str{pos},1}
which inserts the new symmetry at location 1,
i.e., {\it prepends} the symmetry rather than appending it
(see \src{help addSymmetry} for more detailed usage information in this regard).
The display of the operators above reads
\begin{minted}[escapeinside=??,firstnumber=last]{text}

T =
     Q:  3x { 1 x 2 }  having 'SU2,SU2'  operator,  { +-- }
  data:  3-D double (112)      1 x 1 x 1 => 4 x 4 x 3  @ norm = ?$\surd$?3 ?\mlbl{T:hdr2}?

     1.  1x1        |  2x2x1  2x2x3 [ 1 1 ; 1 1 ; 0 2 ]    -0.866025  {?$\surd$?4}

S =
     Q:  3x { 1 x 2 }  having 'SU2,SU2'  operator,  { +-- }
  data:  3-D double (112)      1 x 1 x 1 => 4 x 4 x 3  @ norm = ?$\surd$?3 ?\mlbl{S:hdr2}?

     1.  1x1        |  2x2x3  2x2x1 [ 1 1 ; 1 1 ; 2 0 ]    -0.866025  {?$\surd$?4}
\end{minted}
which is consistent with the alternative approach earlier.
With respect to the \idx[irop]{operator index} (index 3)
the above shows that $S$ ($T$) acts like a spin operator
in the first (second) symmetry, respectively, thus
confirming ${\rm SU}(2)_{\rm spin}\otimes
{\rm SU}(2)_{\rm orbital}$ in this order.
Similarly to \src{E} above, the operators
\src{S} and \src{T} are again also already correctly
normalized ($\surd3$ in header \mlines{T:hdr2} and \ref{S:hdr2}),
since, e.g., $\Vert S\Vert^2 = {\rm tr}(\ 1^{(2)}
\otimes \frac{3}{4} 1^{(2)}\,) = 3$.

\subsection{Local operators beyond \texttt{getLocalSpace}}

The routine \src{\hsec[gls]{getLocalSpace}} returns a set of
elementary operators that are representative for the
chosen site. Many operators
can be derived in simple one-liners as already shown earlier.
However, there are exceptions as already partly discussed
in the previous section, e.g., when generating the orbital
pseudy spin operator \src{T} in \mline{gls:X:T}
or the spin-orbit operator \src{TS} in \mline{gls:X:ST}
from a complete operator basis.
This approach is reviewed once more in a more elaborate example
for orbital spin operators below,
together with an alternative strategy for obtaining
local spin operators based on a Schrieffer-Wolff-like
construction. Both represent instructive approaches
for generating additional local operators.

The following discussion is based on the fermionic
state space for \src{NC=2} orbitals in \eqref{eq:gls:SU2x2}
with a local Hilbert space of $d_{\rm loc}=4^2=16$ states.
The spin operator \src{S} for the entire fermionic
state space as returned by \eqref{eq:gls:SU2x2} reads
\begin{minted}[escapeinside=??,firstnumber=last]{text}
>> S =  ?\mlbl{gls:NC2:S}?
     Q:  3x { 3 x 3 }  having 'A,SU2,SU2'  operator,  { +-- }
  data:  3-D double (336)      3 x 3 x 1 => 11 x 11 x 3  @ norm = ?$\surd$?12

     1.  1x1        |  2x2x3  2x2x1 [ -1 1 1 ; -1 1 1 ;  0 2 0 ]    -0.866025  {?$\surd$?4}?\mlbl{gls:NC2:S:rec1}?
     2.  1x1        |  3x3x3  1x1x1 [  0 2 0 ;  0 2 0 ;  0 2 0 ]     -1.41421  {?$\surd$?3}?\mlbl{gls:NC2:S:rec2}?
     3.  1x1        |  2x2x3  2x2x1 [  1 1 1 ;  1 1 1 ;  0 2 0 ]    -0.866025  {?$\surd$?4}?\mlbl{gls:NC2:S:rec3}?

\end{minted}
As with the \hyperref[{Sp6:S}]{earlier symplectic example}
in the symplectic case, the norm of the individual RMTs
already reflect
the Casimir for the respective local spin, having values
$(\sqrt{\frac{3}{4}}, \sqrt{2}, \sqrt{\frac{3}{4}}$),
and thus $\sqrt{S(S+1)}$ for $q_1 = 2S \in \{1,2,1\}$.
The \idx[S:norm]{normalization} as well as the
\idx[S:norm]{(relative) signs} of the records in the
irop \src{S} are important.
While the signs here all the same,
this is not always the case, e.g.,
as seen in the \hyperref[{Sp6:S}]{earlier example}.
The global sign of any
spin operator is \idx[S:norm]{fixed} after defining the sign of
the spin operator in the defining representation.
Furthermore, since the spin operator is well-defined
in the defining representation, so it must be
in any other many-body state space derived from it.
Hence the spin operator, representing `total spin',
is well-defined and thus unique in any state space.

The spin operator above has irop labels $q_{S}=$\src{[0 2 0]}.
Based on \eqref{eq:gls:SU2x2} with $q=(q_1,q_2,q_3)$,
these symmetry labels are $q_1$ for abelian U(1) charge,
$\idx[SU2:qlabels]{q_2 = 2S}$ for the spin SU(2), and
$\idx[SU2:qlabels]{q_3 = 2S}$ for the orbital SU(2).
As seen from the irop labels $q_S$,
the spin operator acts non-trivially only in the SU(2) spin sector.
There it reflects the \idx[qadj]{adjoint}
representation ($S=1)$ that incorporates the full set
of generators for that symmetry.

\paragraph{Complete operator basis}
\label{idx:op-basis}

Any local operator lives within the complete
set of local operators. While constructing the latter
may appear overkill for specific
local operators, this approach is instructive, nevertheless.
The complete operator basis \src{Eop} is built starting
from the tensor product of the local state space with
its dual via a \idx[1j]{\onej-tensor} (\mline{SU2x2:U}),
followed by a contraction of the same \onej-tensor
in the line below
[using the local operators out of \src{getLocalSpace}
as in \eqref{eq:gls:SU2x2};
same construction as in \mlines{Xb:U}--\ref{Xb:2}]
\begin{minted}[escapeinside=??,firstnumber=last]{matlab}
>> U = getIdentity(IS.E,'-0');   ?\mlbl{SU2x2:U}?
>> Eop = contract(getIdentity(U,1,U,2),2,U,'2*',[1 3 2])  ?\mlbl{SU2x2:Eop}?
\end{minted}
\vspace{-4ex}
\begin{minted}[escapeinside=??,firstnumber=last]{text}
     Q:  3x {60 x 3 }  having 'A,SU2,SU2'  { +-- }
  data:  3-D double (7.734k)      6 x 6 x 60 => 16 x 16 x 256  @ norm = 16

     1.  1x1x6      |  1x1x1  1x1x1 [ -2 0 0 ; -2 0 0 ;  0 0 0 ]      48 b
     2.  1x1x6      |  2x1x2  2x1x2 [ -1 1 1 ; -2 0 0 ;  1 1 1 ]      48 b     {?$\surd$?4}
     3.  1x1x3      |  1x1x1  3x1x3 [  0 0 2 ; -2 0 0 ;  2 0 2 ]      24 b     {?$\surd$?3}
     4.  1x1x3      |  3x1x3  1x1x1 [  0 2 0 ; -2 0 0 ;  2 2 0 ]      24 b     {?$\surd$?3}
     5.  1x1x2      |  2x1x2  2x1x2 [  1 1 1 ; -2 0 0 ;  3 1 1 ]      16 b     {?$\surd$?4}
     6.  1x1        |  1x1x1  1x1x1 [  2 0 0 ; -2 0 0 ;  4 0 0 ]           1.
     :   ...
    59.  1x1x6      |  2x1x2  2x1x2 [  1 1 1 ;  2 0 0 ; -1 1 1 ]      48 b     {?$\surd$?4}
    60.  1x1x6      |  1x1x1  1x1x1 [  2 0 0 ;  2 0 0 ;  0 0 0 ]      48 b
\end{minted}
By construction, \src{Eop} encodes all $(d_{\rm loc}=16)^2 = 256$
local operators. 
It represents 60 irops that transform in a total of
24 different ireps, as seen from \src{getIdentity(Eop,3)}.
If one were to project the local state space
to a particular single multiplet, this would generate
an effective spin model. Only a global
normalization of irops has to be fixed then.

For the full local state space, the spin operator as returned
by \src{getLocalSpace} must also be contained in \src{Eop} .
A search for all records in \src{Eop} that have the
symmetry labels $q_S=$\src{[0 2 0]}
on the operator index $3$
via \src{\idx[QS:getsub]{getsub}} results in,
\begin{minted}[escapeinside=??,firstnumber=last]{text}
>> S_ = getsub(Eop,[0 2 0],3)

     Q:  3x { 3 x 3 }  having 'A,SU2,SU2'  { +-- }
  data:  3-D double (384)      3 x 3 x 3 => 11 x 11 x 9  @ norm = 3

     1.  1x1x3      |  2x2x3  2x2x1 [ -1 1 1 ; -1 1 1 ;  0 2 0 ]      24 b     {?$\surd$?4}
     2.  1x1x3      |  3x3x3  1x1x1 [  0 2 0 ;  0 2 0 ;  0 2 0 ]      24 b     {?$\surd$?3}
     3.  1x1x3      |  2x2x3  2x2x1 [  1 1 1 ;  1 1 1 ;  0 2 0 ]      24 b     {?$\surd$?4}

\end{minted}
It describes an operator subspace of three entries
(having dimensions \src{1x1x3} for all RMTs) which
is solely due to having three \QSpace records here.
The spin operator can be built as some linear
superposition within that subspace.
However, a priori, it is unclear how to obtain
the correct coefficients for this linear superposition
as in \mlines{gls:NC2:S:rec1}--\ref{gls:NC2:S:rec3}
from the present setting.

The spin operator for 
the SU(2) spin symmetry is known, of course,
because it is explicitly returned by \src{getLocalSpace}.
This has the benefit that 
one has a reference to compare various approaches to.
If one needs the pseudo-spin operator
in the full local state space above
in the orbital space, or the intertwined spin-orbit
operator, these are no longer available from \src{getLocalSpace}.
For these, one needs to address the problem
of how to make the spin operator unique, at least
up to a global scale factor.
For this purpose, one may target the spin operators
more concretely as discussed next.

\paragraph{Spin operators from second order fermionic hopping}

An alternative route to generalized `spin operators'
is physically motivated by Schrieff-Wolff transformations,
where a second-order perturbation in the tunneling
maps fermionic cotunneling to an effective `spin model'.
There with $\hat{f}_{i}$
the complete set of local fermionic operators
with $i=1,\ldots,N_f$ indexing all flavors
or flavor combinations, one needs to decompose
the pool of operators
$\{ \hat{f}_{i}^\dagger \hat{f}_{i'}^{\,} \}$ into irops.
Hence rather than building the full operator space
for the local state space, one can constrain
oneself to the operators obtained out of the
bilinear \src{F$^\dagger\otimes$F}, here with a Kronecker
rather than a dot-product on the irop indices.

Since \src{F$^\dagger$} as a \idx[conj]{conjugate} operator
has the irop index ingoing, this procedure is
equivalent to the construction of the complete
operator basis in \mlines{SU2x2:U}--\ref{SU2x2:Eop},
except that one just takes the irop space of \src{F},
instead. 
Replacing \src{IS.E} in \mline{SU2x2:U} by \src{F} on 
leg 3, one obtains,
\begin{minted}[escapeinside=??,firstnumber=last]{matlab}
>> U_ = getIdentity(F,3,'-0');   ?\mlbl{SU2x2:Eop:F}?
   Eop_ = contract(getIdentity(U_,1,U_,2),2,U_,'2*',[1 3 2])  ?\mlbl{SU2x2:Eop_}?
\end{minted}
\vspace{-5ex}
\begin{minted}[escapeinside=??,firstnumber=last]{text}
     Q:  3x { 4 x 3 }  having 'A,SU2,SU2'  { +-- }
  data:  3-D double (448)      1 x 1 x 4 => 4 x 4 x 16  @ norm = 4 ?\mlbl{SU2x2:Eop_:hdr2}?

     1.  1x1        |  2x2x1  2x2x1 [ -1 1 1 ; -1 1 1 ;  0 0 0 ]          0.5  {?$\surd$?4}?\mlbl{SU2x2:Eop_:rec1}?
     2.  1x1        |  2x2x1  2x2x3 [ -1 1 1 ; -1 1 1 ;  0 0 2 ]     0.866025  {?$\surd$?4}?\mlbl{SU2x2:Eop_:rec2}?
     3.  1x1        |  2x2x3  2x2x1 [ -1 1 1 ; -1 1 1 ;  0 2 0 ]     0.866025  {?$\surd$?4}?\mlbl{SU2x2:Eop_:rec3}?
     4.  1x1        |  2x2x3  2x2x3 [ -1 1 1 ; -1 1 1 ;  0 2 2 ]          1.5  {?$\surd$?4}?\mlbl{SU2x2:Eop_:rec4}?

\end{minted}
The spinfull \src{NC=2} model here based
on \eqref{eq:gls:SU2x2} has a combined total
of $N_f = 2\times2 =4$ flavors. Hence the pool of operators
$\{ \hat{f}_{i}^\dagger \hat{f}_{i'}^{\,} \}$
has $4\times 4 = 16$ entries. These are cast into
three irops here that span the full space of the
operator pool, as seen from the overall dimensions
\mbox{\src{4 x 4 x 16}} in \mline{SU2x2:Eop_:hdr2}.
By having RMTs of dimension \src{1x1}\,[\src{x1}] only,
each record now corresponds to an irop
that is unique up to normalization.

The tensor \src{Eop\_} still needs to be contracted
onto the two irop indices in \src{F$^\dagger$F}.
By choosing the SU(2) spin irop in \mline{SU2x2:Eop_:rec2},
one obtains the spin operator
\begin{minted}[escapeinside=??,firstnumber=last]{matlab}
>> x = getsub(Eop_,[0 2 0],3);    % select spin operator having q=[0 2 0]
   Sop = contract({F,'1*',F,1},'24',x,'21')
\end{minted}
\vspace{-5ex}
\begin{minted}[escapeinside=??,firstnumber=last]{text}
     Q:  3x { 3 x 3 }  having 'A,SU2,SU2'  { +-- }
  data:  3-D double (336)      3 x 3 x 1 => 11 x 11 x 3  @ norm = ?$\surd$?12 ?\mlbl{SW:Sop}?

     1.  1x1        |  2x2x3  2x2x1 [ -1 1 1 ; -1 1 1 ;  0 2 0 ]    -0.866025  {?$\surd$?4}
     2.  1x1        |  3x3x3  1x1x1 [  0 2 0 ;  0 2 0 ;  0 2 0 ]     -1.41421  {?$\surd$?3}
     3.  1x1        |  2x2x3  2x2x1 [  1 1 1 ;  1 1 1 ;  0 2 0 ]    -0.866025  {?$\surd$?4}

\end{minted}
This is precisely the spin operator returned by 
\src{getLocalSpace} in \mlines{gls:NC2:S}--\ref{gls:NC2:S:rec3},
already also with the correct normalization,
as confirmed by \src{norm(S-Sop)}$\leq 10^{-14}$.
The normalization can be easily double checked,
bearing in mind that the spin operator returned
by \src{getLocalSpace} describes the total spin of the local
state space. For \src{NC=2} fermionic channels
with $S_i$ the spin operator for channel (orbital) $i$,
this leads to $\Vert S \Vert^2 = \Vert S_1 + S_2 \Vert^2
= 2 \,\Vert S_1 \Vert^2\, {\rm tr}(\Id^{(4)}_2) 
= 2 \cdot (\frac{3}{4} 2)\cdot 4 = 12$,
in agreement with \mline{SW:Sop}.

The present approach now also allows to extend to 
other generalized spin operators. For example,
one may choose the SU(2) pseudo-spin irop 
in the orbital sector (\mline{SU2x2:Eop_:rec3}),
\begin{minted}[escapeinside=??,firstnumber=last]{matlab}
>> x = getsub(Eop_,[0 0 2],3);    % select pseudo-spin operator
   Top = contract({F,'1*',F,1},'24',x,'21')
\end{minted}
\vspace{-5ex}
\begin{minted}[escapeinside=??,firstnumber=last]{text}
     Q:  3x { 3 x 3 }  having 'A,SU2,SU2'  { +-- }
  data:  3-D double (336)      3 x 3 x 1 => 11 x 11 x 3  @ norm = ?$\surd$?12

     1.  1x1        |  2x2x1  2x2x3 [ -1 1 1 ; -1 1 1 ;  0 0 2 ]    -0.866025  {?$\surd$?4}
     2.  1x1        |  1x1x1  3x3x3 [  0 0 2 ;  0 0 2 ;  0 0 2 ]     -1.41421  {?$\surd$?3}
     3.  1x1        |  2x2x1  2x2x3 [  1 1 1 ;  1 1 1 ;  0 0 2 ]    -0.866025  {?$\surd$?4}
\end{minted}
which shows the same RMT values as the previous spin operator
\str{Sop}, or the spin-orbital sector in \mline{SU2x2:Eop_:rec4},
\begin{minted}[escapeinside=??,firstnumber=last]{matlab}
>> x = getsub(Eop_,[0 2 2],3);    % select spin-orbit operator
   STop = contract({F,'1*',F,1},'24',x,'21') / 2
\end{minted}
\vspace{-4ex}
\begin{minted}[escapeinside=??,firstnumber=last]{text}
     Q:  3x { 4 x 3 }  having 'A,SU2,SU2'  { +-- }
  data:  3-D double (448)      4 x 4 x 1 => 14 x 14 x 9  @ norm = 3

     1.  1x1        |  2x2x3  2x2x3 [ -1 1 1 ; -1 1 1 ;  0 2 2 ]         0.75  {?$\surd$?4}
     2.  1x1        |  1x3x3  3x1x3 [  0 0 2 ;  0 2 0 ;  0 2 2 ]     0.866025  {?$\surd$?3}
     3.  1x1        |  3x1x3  1x3x3 [  0 2 0 ;  0 0 2 ;  0 2 2 ]     0.866025  {?$\surd$?3}
     4.  1x1        |  2x2x3  2x2x3 [  1 1 1 ;  1 1 1 ;  0 2 2 ]        -0.75  {?$\surd$?4}
\end{minted}
In the last case the norm was adapted to yield the expected value
$\Vert \src{STop}\Vert^2 = 2\cdot2\cdot (\frac{3}{4} 2)^2 = 9$,
having $S = S_1 + S_2$, $T=T_1+T_2$, where $S_i$ and $T_i$
are spin and orbital operators, respectively
with $i=1,2$ indexing either orbital or spin.
The same factor $1/2$ was already also encountered
\hyperref[gls:X:ST]{earlier} in a similar context
with the operator \src{ST} on p.~\pageref{gls:X:ST}.

\subsection{Building rank-5 PEPS tensor}
\label{sec:PEPS}

\begin{figure}[tbh]
\begin{center}
\includegraphics[width=0.6\linewidth]{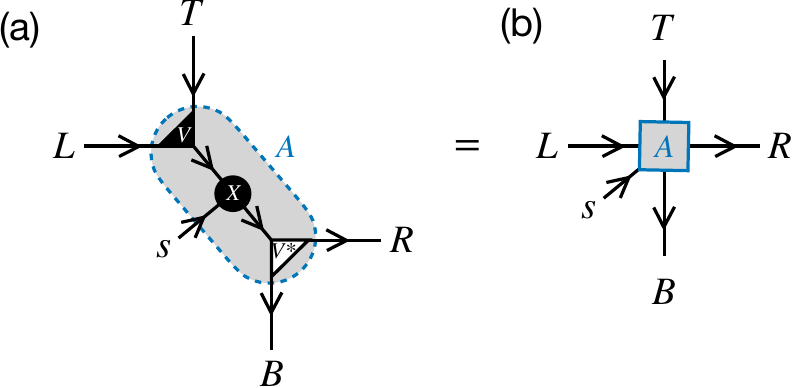}
\end{center}
\caption{Example for building rank-5 tensor for 2D PEPS,
   which ties the virtual state space at the $L$(eft),
   $T$(op), $R$(ight), and $B$(ottom) with the local state 
   space of a site ($s$). Contracting the tensor
   decomposition in (a)
   based on rank-3 tensors yields the rank-5 tensor in (b).
   Assuming periodic tiling of a 2D-PEPS
   based on a unit cell comprised of a single tensor $A$,
   the state spaces $L$
   and $R$ must match, yet with reverse direction,
   and ditto for $T$ and $B$.
} 
\label{fig:PEPS-A5} 
\end{figure}

Projected entangled pair states
(PEPS \cite{Verstraete04_peps,Cirac09,Orus14tns,Bruognolo21})
can be used to tile a tensor network state in
one or higher dimension. If a set of tensors is used to tile
the lattice to infinity, this is referred to as iPEPS
(or iTEBD in 1D \cite{Vidal07}).
As with any lattice model, one needs to 
(i)~define the local state space of a site 
(in \QSpace in terms of \src{getLocalSpace}),
followed by (ii) an initialization of
the tensor network state as a whole. Assuming
a 2D square lattice, an $A$-tensor in iPEPS becomes
a rank-5 tensor that combines the virtual indices
left (L), top (T), right (R), and bottom (B)
with the local state space of a site (s)
[cf. \Fig{PEPS-A5}].

\noindent
For example, starting with a local state space
described by an SU(3) spin in the defining representation
(10)\,$\equiv {\bf 3}$ [cf. \Eq{qdef}],
\begin{minted}[escapeinside=??]{matlab}
>> [S,IS]=getLocalSpace('Spin','SU3','10');
\end{minted}
one may proceed to initialize the rank-5 tensor
as sketched in \Fig{PEPS-A5}(a).
Assuming, for simplicity, that
the state spaces on all virtual indices are the same,
one nevertheless needs to make an initial choice
for this state space. To this end, one may build
some arbitrary but fixed combination of symmetry
sectors derived from the local state space \src{E},
\begin{minted}[escapeinside=??,firstnumber=last]{matlab}
>> E=IS.E; U=getIdentity(E,'-0');
   A2 = getIdentity(U,1,U,2);
   E2 = getIdentity(U,1) + getIdentity(U,2) + getIdentity(A2,3); ?\mlbl{PEPS:E2:1}?
   E2.data=repmat({eye(2)},size(E2.data));  ?\mlbl{PEPS:E2:2}?% 2 multiplets / symmetry sector
\end{minted}
\vspace{-4ex}
\begin{minted}[escapeinside=??,firstnumber=last]{text}
E2 =
     Q:  2x { 4 x 2 }  having 'SU3'  { +- }                             ?\mlbl{E2:hdr1}?
  data:  2-D double (544)      8 x 8 => 30 x 30  @ norm = ?$\surd$?30   ?\mlbl{E2:hdr2}?
 
     1.  2x2        |    1x1 [ 00 ; 00 ]      32 b
     2.  2x2        |    3x3 [ 01 ; 01 ]      32 b     {?$\surd$?3}
     3.  2x2        |    3x3 [ 10 ; 10 ]      32 b     {?$\surd$?3}
     4.  2x2        |    8x8 [ 11 ; 11 ]      32 b     {?$\surd$?8}
\end{minted}
\mLine{PEPS:E2:1} combines the local state space,
its dual via  the \hsec[1j]{\onej}-symbol \src{U} on index 2,
and their fused state space via \src{A2} into an
identity tensor \src{E2}.
Since there is no overlap of symmetry sectors of
the terms added here, one
can combine these spaces as a direct sum ($\oplus$)
simply by adding them up.  \mLine{PEPS:E2:2} 
then sets the dimension to two multiplets
per symmetry sector by setting all 
RMTs in \src{E2} to \src{eye(2)} $=\Id^{(2)}$.

With the state space \src{E2} for all virtual indices assumed
the same, one can combine two of these, like $L$ and $T$,
via the tensor $V$ as depicted in \Fig{PEPS-A5}(a),
\begin{minted}[escapeinside=??,firstnumber=last]{matlab}
>> V=getIdentity(E2,E2);
   X=getIdentity(V,3,E);   % combine with local state space E
\end{minted}
Since the fused state space in \src{X} as in 
\Fig{PEPS-A5}(a) shall be contracted to the conjugate
of \src{V} (which then links to $B$ and $R$),
their dimensions need to be matched,
\begin{minted}[escapeinside=??,firstnumber=last]{matlab}
>> [Q3,dd]=getQDimQS(V,3);       ?\mlbl{Q3:qdim}?
   [I,J]=matchIndex(X.Q{3},Q3);  ?\mlbl{Q3:match}?
   for i=1:numel(I)
      s=size(X.data{I(i)});
      X.data{I(i)}=randn([s(1:2), dd(J(i))]); ?\mlbl{X:randn}?
   end
\end{minted}
Here \src{getQDimQS} in \mline{Q3:qdim}
returns the symmetry
sectors (rows in \src{Q3}) and their respective number of
multiplets (integer vector \src{dd}) for leg $3$ in \src{V}.
\mLine{Q3:match}
finds all matching sectors in \src{X} on the third leg
with the entries in \src{Q3}, returned as \src{I} 
and \src{J}, respectively.
The loop then matches the 
dimension on the third leg of the respective RMT
to the one in \src{V} via \src{dd}, while
randomizing their content using \src{randn}.
With this, all ingredients in \Fig{PEPS-A5}(a)
are defined and contractable.
For transparency, one may add
\itags now,
\begin{minted}[escapeinside=??,firstnumber=last]{matlab}
>> LT=setitags(V,{'L','T','LT'}); ?\mlbl{LT:def}?                 % L,T $\to$ LT  
   BR=setitags(V,{'B','R','BR'}); ?\mlbl{BR:def}?                 % B,R $\to$ BR  
   X =setitags(getsub(X,I),{'LT','s','BR'});  ?\mlbl{X:tags}? % getsub is optional here
\end{minted}
The \QSpace \src{LT} describes the
upper tensor \src{V} in \Fig{PEPS-A5}(a)
that combines $L$ and $T$.
Similarly, \src{BR} describes the lower tensor \src{V}
that combines $B$ and $R$ (up to tensor conjugation
which will reverse arrows in the contraction).
\mLine{X:tags}
selects the subset of
records in \src{X} that matched \src{BR}
(this step is optional since non-matching symmetry sectors
drop out in the contraction anyway),
and also assigns \itags.
Based on the assigned \itags, the contraction into
the rank-5 tensor \src{A5}  in \Fig{PEPS-A5}(b)
can be carried out making use of 
\idx[ctr:semantics]{auto-contraction},
\begin{minted}[escapeinside=??,firstnumber=last]{matlab}
>> A5=contract({LT,X3},BR,'*',[1 2 4 5 3]); ?\mlbl{A5:ctr}? % final permution to LTBRs index order
   d2=getDimQS(E2);           ?\mlbl{A5:norm:d2}?
   A5=A5*(d2(end)/norm(A5));  ?\mlbl{A5:norm}?
\end{minted}
\vspace{-5ex}
\begin{minted}[escapeinside=??,firstnumber=last]{text}
A5 = ?\mlbl{A5:disp}?
    Q: 5x {85 x 2 }  having 'SU3'  { L, T, B*, R*, s } ?\mlbl{A5:hdr1}?
 data: 5-D double (39.26k)   8 x 8 x 8 x 8 x 1 => 30 x 30 x 30 x 30 x 3  @ norm = 30?\mlbl{A5:hdr2}?

     1.  2x2x2x2         | 1x1x1x3x3 [ 00 ; 00 ; 00 ; 10 ; 10 ]     128 b
     2.  2x2x2x2         | 1x1x3x3x3 [ 00 ; 00 ; 01 ; 01 ; 10 ]     128 b
     3.  2x2x2x2         | 1x1x3x1x3 [ 00 ; 00 ; 10 ; 00 ; 10 ]     128 b
     4.  2x2x2x2         | 1x1x3x8x3 [ 00 ; 00 ; 10 ; 11 ; 10 ]     128 b
     5.  2x2x2x2         | 1x1x8x3x3 [ 00 ; 00 ; 11 ; 10 ; 10 ]     128 b
     6.  2x2x2x2         | 1x3x1x1x3 [ 00 ; 01 ; 00 ; 00 ; 10 ]     128 b
     :   ... ?\mlbl{A5:dots}?
    84.  2x2x2x2x1 @10   | 8x8x3x8x3x10 [ 11 ; 11 ; 10 ; 11 ; 10 ]     1.2 k  ?\mlbl{A5:rec84}?
    85.  2x2x2x2x1 @10   | 8x8x8x3x3x10 [ 11 ; 11 ; 11 ; 10 ; 10 ]     1.2 k  ?\mlbl{A5:recl}?
\end{minted}
The trailing permutation in \mline{A5:ctr}
sets the index order \src{LTBRs},
as also seen in the display in \mline{A5:hdr1}.
\mLines{A5:norm:d2}--\ref{A5:norm}
apply some normalization to \src{A5}
given the earlier randomization in \mline{X:randn}
[\src{\idx[getDimQS]{getDimQS}} in \mline{A5:norm:d2} returns
the multiplet and state space dimensions
for all legs in \src{E2}, with \src{d2(end)}$=30$
the state space dimension of the last leg;
this stands for the total number of states on the
bond indices \src{[LTRB]} as described by \src{E2};
cf. \mline{E2:hdr2}].

The tensor \src{A5} has
multiplet dimensions \src{8 x 8 x 8 x 8 x 1}  (\mline{A5:hdr2})
that correspond to a respective \src{30 x 30 x 30 x 30 x 3} states.
The \src{1} multiplet
\src{q=(10)} on the last leg 
describes the local state space $\sigma\in s$ with \src{3} states.
It is the same throughout all 85 records
which reflect permissible symmetry combinations
across all five legs.
By default, the full listing
is truncated for readability (dots in \mline{A5:dots}).
The display also shows
two entries with largest outer multiplicity [$M=10$,
indicated by the trailing \src{@10} with the RMT dimensions
in \mbox{\mlines{A5:rec84}--\ref{A5:recl}},
as well as the trailing 6-th dimension \src{x10}
with the respective CGTs].
They also happen to be the largest and last two records,
otherwise these would be shown as
\hyperref[{A:Sp6}]{intermediate records}.



\subsection{fdm-NRG: Iterative diagonalization
of Wilson chain and spectral functions}
\label{sec:NRG}

The following more elaborate example
demonstrates how to use \QSpace for the
iterative diagonalization in the context of
the numerical renormalization
group (NRG \cite{Wilson75,Bulla08}).
It is included with this documentation
since the NRG was one of the first
applications of \QSpace \cite{Wb07,Wb12,Wb12_SUN}.
The example here is based on the standard
single impurity Anderson model (SIAM).

The following code represents the script \src{NRG/xnrgQS.m}
in the repository, yet with additional comments
and with a focus on auto-contractions based on \itags.
It reciprocates the hard-coded MEX
routine \src{NRGWilsonQS}.
For transparency and the sake of the discussion here,
however, the iterative diagonalization can also
be simply coded in Matlab itself, while nevertheless
utilizing the \QSpace toolbox via the MEX
routines for tensor product, contraction,
diagonalization, etc.
It also uses \QSpace Matlab utility routines in the
\src{lib/} directory of the repository, such as
\src{setdef}, \src{isset}, \src{add2struct}, \src{cto}, etc.,
whose purpose is simple and intuitive as will be indicated.
A full explanation can be found in their respective
\idx[ML:help]{help}. 
Since the following source mimics the algorithm behind
\src{NRGWilsonQS}, the format of the output data
is also chosen in a compatible way. This then
permits, for example, subsequent usage of the MEX routine
\src{fdmNRG\_QS} to compute correlation functions
(see \idx[fdmNRG-0]{fdm-NRG} below).

\paragraph{NRG basics}

The NRG is a long-standing powerful impurity solver
that tackles $0+1$ dimensional physical problems
\cite{Wilson75,Bulla08}.
For this, the impurity Hamiltonian is mapped onto
the so-called Wilson chain. It is described by the
general Hamiltonian
\begin{eqnarray}
   \hat{H} &=&
   \hat{H}_0(\hat{\mathbf{d}},\hat{\mathbf{f}}_{0}) + 
   \sum_{n=0}^\infty \bigl(
     t_n \hat{\mathbf{f}}_{n}^\dagger\cdot \hat{\mathbf{f}}_{n+1}
     + {\rm H.c.}
  \bigr)
\label{eq:Himp}
\end{eqnarray}
where $\hat{d}_\sigma \equiv (\hat{\mathbf{d}})_\sigma$
are the fermionic operators for the impurity
(`$d$-level' for historical reasons), where $\sigma$ denotes spin.
The fermionic operators for the bath
$(\hat{\mathbf{f}}_n)_{\sigma} \equiv \hat{f}_{n\sigma}$
are organized into a semi-infinite chain where the
impurity couples to the first Wilson site $\hat{\mathbf{f}}_0$
only. From a symmetry point of view,
any term in the Hamiltonian needs to represent a 
scalar. Hence the hopping term on the right 
naturally includes the \hsec[SdotSd]{scalar dot product}
$\hat{\mathbf{f}}_{n}^\dagger\cdot \hat{\mathbf{f}}_{n+1}
\equiv \sum_\sigma \hat{f}_{n\sigma}^\dagger \hat{f}_{n+1,\sigma}$
implemented via the contraction in \mline{Hcpl:ctr} below.
The hopping amplitudes $t_n \sim \Lambda^{-n/2}$ 
in \Eq{Himp} derive from the
NRG coarse-graining with its dimensionless discretization
parameter $\Lambda \gtrsim 2$. The resulting exponential
decay with increasing $n$ is essential to the NRG,
as this justifies the iterative diagonalization.
In this sense, the chain structure in NRG does not
represent real space, but rather energy scales which,
nevertheless, may be interpreted as inverse length scale
\cite{Wb11_aoc,Muender12}.
Due to the {\it effective} one-dimensional character
of the Hamiltonian in \Eq{Himp},
NRG is naturally suited to an MPS description.
As a characteristic of NRG, the MPS is not obtained
variationally, though, but rather via iterative
diagonalization \cite{Wb11_rho,Wb12}. Nevertheless,
it is arguably in close proximity to a variational
approach \cite{Wb09,Saberi08,Holzner11}.
\vspace{-2ex}

\paragraph{Setup of model (SIAM)}

The starting point of any NRG is the Hamiltonian
\src{H0} [$=\hat{H}_0$ in \eqref{eq:Himp}] that defines the impurity
($\hat{\bf d}$) together with its interaction
with the bath site at the
location of the impurity ($\hat{\bf f}_0$).
The matrix elements of \src{H0} are written
in the basis encoded in the $A$-tensor \src{A0}.
With index order \src{\idx[Atensor:LRs]{LRs}} in mind,
it fuses the impurity ($d\to$\,L) with 
the first Wilson site ($f_0 \to$\,s)
where \src{H0} is expressed in \src{R}.
Together this sets the stage for
the iterative diagonalization, fully defined by the fermionic
operators \src{F} (with \src{Z}) for a Wilson site
and the hopping amplitudes \src{ff}.

For the SIAM, the impurity is described by a single
spinfull fermionic level that already shares the same
structure as a Wilson site $\hat{\bf f}_n$.
Moreover, for the SIAM all interaction is contained
within the impurity itself via the Hubbard $U$,
whereas the `interaction' with the bath becomes
a plain fermionic hopping via the impurity's
hybridization function. Therefore for the case
of the SIAM, the NRG setup can be simplified
in that \src{H0} describes the impurity ($\hat{\bf d}$)
only. The bath-site $\hat{\bf f}_0$ can be already 
taken care of as an iterative NRG step. With this
the sum in \Eq{Himp} can be extended to start from
$n=-1$,
with the additional hopping term taking care of the
hybridization of the impurity via $t_{-1} \equiv \sqrt{\frac{2D\Gamma}{\pi}}$,
with the $D:=1$ the half-bandwidth of the bath.
The latter sets the unit of energy, unless specified 
otherwise. In any case, the iterative NRG diagonalization
only requires the intial \src{A0} and \src{H0},
the fermionic operators \src{F} (with \src{Z}), and the couplings \src{ff}
in the same units as \src{H0}.
This does not care about the precise starting
point of the iterative diagonalization
which is also in the spirit of the
MEX routine \src{NRGWilsonQS}.

For the SIAM, therefore the `left'
state space of \src{A0} can be taken as the vacuum state
($|\rangle \to $\,L),
and the local state space as the impurity itself
($d\to$\,s).
Then \src{H0} describes the impurity Hamiltonian only.
The hopping amplitude $\Gamma$ is prepended to the couplings
as \src{ff(1)}$=t_{-1}$. Since \src{getNRGcoupling}
takes $\Gamma$ as input parameter, the hoppings are already
returned this way (\mline{nrg:ff}). The setup for the SIAM thus chooses
the following adapted version of the Hamiltonian in \Eq{Himp},
\begin{eqnarray}
   \hat{H} &=&
   \underbrace{
   \varepsilon_d \hat{n}_d + B \hat{S}_z
   + \tfrac{U}{2} \hat{n}_d (\hat{n}_d-1)}_{ \quad \  \ 
     \equiv H_{\rm imp} \ \to\ \src{H0} \ \ 
     \text{(\mlines{SIAM:H0:hdr}--\ref{SIAM:H0:2})}
   }
   \quad + \hspace{-2ex}
   \underset{
     \text{(\mlines{NRG:iter:hdr}--\ref{NRG:iter:end})}}
   {\sum_{k=1}^N}\hspace{-2ex}
   \underbrace{\bigl(
     \src{ff(k)}\, \hat{\mathbf{f}}_{k-2}^\dagger\cdot \hat{\mathbf{f}}_{k-1}
     + {\rm H.c.}
  \bigr)}_{
     \text{(\mlines{Hcpl:hdr}--\ref{Hcpl:end})}
  }
\label{eq:Himp:ex}
\end{eqnarray}
with \src{ff(k)}$\equiv t_{k-2}$ and
$\mathbf{f}_{-1} \equiv \mathbf{d}$.\\

\noindent In the script \src{NRG/xnrgQS.m},
the initialization of the SIAM reads,%
\begin{minted}[escapeinside=??]{matlab}
% SIAM impurity parameters (all energies in units of the half bandwidth D:=1)
%
%   U        onsite interaction
%   epsd     impurity level position
%   Gamma    hybridization strength of impurity
%   B        magnetic field acting on impurity (Zeeman splitting; applied as $-B S_z$)

  setdef('U',0.12,'Gamma',0.01,'B',0);  ?\mlbl{nrg:sdef:1}? % set default values
  setdef('epsd',-U/2);                  ?\mlbl{nrg:sdef:2}? % half-filling by default

% NRG discretization parameters
%
%   Lambda   coarse graining strength ($\gtrsim 2$, dimensionless)
%   N        length of Wilson chain
%   Nkeep    number of multiplets to keep (or states if all-abelian)
%   Etrunc   energy truncation threshold (complementary to Nkeep)
%   z        shift of logarithmic discretization, as in $\Lambda^{(z-n)/2}$ with $z\in[0,1[$
%
% By setting Nkeep large, by default, preference is given to
% truncation by energy; using Nkeep, nevertheless, as safeguard.

  setdef('Lambda',2,'N',55,'Nkeep',1024,'Etrunc',7,'z',0); ?\mlbl{nrg:sdef}?

% Collect model and NRG parameters (global info structure for reference)
  global param
  param=add2struct('-',U,epsd,Gamma,B,N,Lambda,z); ?\mlbl{nrg:param}?

% NRG hopping parameters
  ff=getNRGcoupling(Gamma,Lambda,N,'z',z); ?\mlbl{nrg:ff}? % default z=0 if not specified

% Local state space: single spinfull fermionic level
% For the SIAM this is the same for the impurity as well as the Wilson sites.
% The option '-v' enables more verbose output.
  if ~isset('B')
       SYM='Acharge,SU2spin';
  else SYM='Acharge,Aspin'; ?\mlbl{SIAM:SYM}? % results in SS = $[S_+,S_-,S_z]$ below => SS(end)=$S_z$
  end 
  [FF,Z,SS,IS]=getLocalSpace('FermionS',SYM,'-v');

% Construct local occupation operator
% depending on SYM above, FF may contain different number of operators (nF)
  N0=QSpace(size(FF)); nF=numel(FF);
     for i=1:nF                                    % $i\equiv\sigma$ for nF>1
     N0(i)=contract(FF(i),'13*',FF(i),'13'); end   % $f_\sigma^\dagger \cdot f_\sigma$
  N0=sum(N0);                                      % $\sum_\sigma n_\sigma$
\end{minted}
The command \src{setdef} in \mlines{nrg:sdef:1},
\ref{nrg:sdef:2}, or \ref{nrg:sdef}
sets default values for the specified variables
if not yet set, i.e., if they are not defined
or empty.
The command \src{add2struct} in \mline{nrg:param}
adds specified variables as fields to the structure
\src{param} where the leading input argument \str{-} 
indicates to start from an empty structure. 
This collects physical model
and discretization parameters for general
information purposes later.
The call to \src{getNRGcoupling} in
\mline{nrg:ff}
returns the Wilson chain hopping parameters \src{ff},
starting with \src{ff(1)}$=\sqrt{2\Gamma/\pi}$.
The above also sets the stage for the definition of
\src{A0} and \src{H0},
\begin{minted}[escapeinside=??,firstnumber=last]{matlab}
% A-tensor for H0 (LRs index order convention)
  A0 = getIdentity(getvac(IS.E),IS.E,[1 3 2]); ?\mlbl{nrg:getvac}?

?\mlbl{SIAM:H0:hdr}?% Impurity Hamiltonian [in R-basis of A0, as in LRs]
  H0 = epsd*N0 + (U/2)*N0*(N0-IS.E) + 0*IS.E; ?\mlbl{SIAM:H0:1}?
  if isset('B'), H0 = H0 - B*SS(end); end  ?\mlbl{SIAM:H0:2}? % SS(3=end) contains $S_z \mlbl{nrg:S3}$
\end{minted}
The command \src{\idx[QS:getvac]{getvac}} 
in \mline{nrg:getvac}
returns the vacuum state space
for the symmetry setting of the input.
If a finite value for the magnetic field is set, 
the setup via \src{SYM} in \mline{SIAM:SYM}
switches to abelian U(1) spin symmetry. In this case,
as can be checked by inspection of \src{SS},
the impurity $S_z$ operator is returned as \src{SS(3) $\equiv$ 
SS(end)} in \mline{nrg:S3}.
The magnetic field is applied as $-B S_z^d$
at the impurity only [it is ignored for the bath, 
assuming $B\ll (D=1)$],
where the minus sign ensures that for $B>0$
one also obtains a positive magnetization
$\langle S_z^d\rangle$.

The above concludes the setup of the impurity.
Based on \src{A0}, \src{H0}, and \src{ff},
this could proceed with the MEX
routine \src{NRGWilsonQS} from here
[hard-coded \Cpp code, e.g., called via the
\src{rnrg} (`run NRG') script which for
the SIAM calls \src{setup/setupSIAM\_SU2x2.m}
as a more elaborate version of the setup above].
For the sake of transparency, however,
the iterative diagonalization is explicitly
transcribed into the Matlab script
\src{xnrgQS} discussed here.
\vspace{-1ex}

\paragraph{Iterative diagonalization}

To proceed with the iterative diagonalization,
the setup above is further adapted.
The couplings \src{ff} from \mline{nrg:ff} 
above
are in physical units of the half-bandwidth ($D=1$).
Since finite-size spectra
should be compared in rescaled units of order 1,
the exponentially decaying couplings
\src{ff}$\sim \Lambda^{-k/2}$ are also rescaled
to order 1,
\begin{minted}[escapeinside=??,firstnumber=last]{matlab}
% Rescale hoppings from global to iterative units of order 1 (see usage below)
  rL=sqrt(Lambda);  % rL = `root Lambda'
  ff=ff.* rL.^(1:N-1);
\end{minted}
Since at the first iteration \src{ff(1)} incorporates
the first Wilson site ${\bf f}_0$,
this leaves the Hamiltonian \src{H0} at iteration `0'
in its original units.

By default, the data from the iterative diagonalization
is stored to files (one file for each iteration,
starting from \src{*\_00.mat} for iteration `0'
which just diagonalizes the input \src{H0}),
\begin{minted}[escapeinside=??,firstnumber=last]{matlab}
% Save data by default to files in local Matlab data directory (LMA)
% as $\var{LMA}$/NRG/myNRG_##.mat with ## the Wilson shell index k
% where $\texttt{\idx[LMA]{LMA}}$ is expected to exist as environmental variable
  setdef('sflag',1); ?\mlbl{sflag}? % set sflag=0 to disable
  if sflag
     fout={'NRG' '/myNRG'};
     cto lma;  ?\mlbl{nrg:cto}?% `change to' directory $\var{LMA}$
     if ~isdir(['./' fout{1}]), system(['mkdir ' fout{1}]); end
     fout=[fout{:}];
     wblog('I/O','using %s_##.mat',fout); 
  end
\end{minted}
Storing the data to files allows one to use the data later,
e.g., for computing spectral functions via
\src{\idx[fdmNRG-0]{fdmNRG\_QS}} further below.
The code proceeds with initializing
variables and data spaces,
\begin{minted}[escapeinside=??,firstnumber=last]{matlab}
% Initialize iterative diagonalization
%    K  kept space      (AK $\equiv\ A_K$)
%    D  discarded space (AD $\equiv\ A_D$)
% scalar operators only have KK $\to$ K, DD $\to$ D (like HK $\equiv\ H_{K}$, HD $\equiv\ H_{D}$)
% general operators also have off-diagonal blocks KD, DK (cf. FKK $\equiv\ F_{KK}$)
  k=0;   % start at Wilson shell `0' (impurity only)
  AK=setitags(setitags(A0,'-A',k),1,'Lvac'); ?\mlbl{nrg:AK:itags}?  % sets itags {'Lvac','K00','s00'}
  HK=setitags(H0,'-op:K',k);                 ?\mlbl{nrg:HK:itags}?  % sets itags {'K00','K00*'}
  AD=QSpace;                                   % = QSpace(), inits to empty QSpace

% Fermionic operator F in KK space
  FKK=QSpace(1,nF);

% Keep track of general data along iterations
  EX=nan(N,2);     % 2 columns: heighest kept and lowest discarded energy
  E0=zeros(1,N);   % subtracted ground state energy for each iteration
  N4=[];           % number of kept and discarded multiplets/states

% Initialize structure to collect iterative diagonalization
  Inrg=struct('N',N,'Lambda',Lambda,'EK',[],'HK',QSpace(1,N));  ?\mlbl{Inrg:init}?

  fprintf(1,'\n'); disp(param); ?\mlbl{param:disp}?
  fprintf(1,'\n');
\end{minted}
By iteratively keeping all eigenstates below some
energy (\src{Etrunc}) or count (\src{Nkeep}) threshold
(cf. \mline{nrg:sdef}),
this partitions states into kept (\src{AK}$\equiv\, A_K$)
and discarded  (\src{AD}$\equiv\, A_D$) states
for any iteration. Ultimately, the collection of 
\src{AK} for all iterations together with the lowest
energy eigenstate for the last iteration $k=N$ constitutes
the ground state MPS for given Wilons chain of length $N$.
Expressing operator matrix elements in terms of
kept and discarded (based on \src{AK} and \src{AD},
respectively), leads to sectors \src{KK}, \src{KD},
\src{DK}, and \src{DD} (e.g. see operator \src{F}
$\to$ \src{FKK} $\equiv\, F_{KK}$). For scalar operators, there are 
no off-diagonal contributions, hence a single
label \src{K} or \src{D} suffices, such as for 
\src{HK} $\equiv\, H_K$. By using itags as in
\mlines{nrg:AK:itags}--\ref{nrg:HK:itags}
this will simplify the specification of contractions,
but also characterize \QSpace tensors more
generally, while making the output more readable 
and identifiable.
Basic NRG data will be collected into the info structure
\src{Inrg} initialized in \mline{Inrg:init}.
\mLine{param:disp}f summarizes the setup by displaying
the parameters used.

The following then
proceeds with the actual iterative diagonalization,

\begin{minted}[escapeinside=??,firstnumber=last]{matlab}
?\mlbl{NRG:iter:hdr}?% Iterative diagonalization based on couplings ff(k)
  for k=1:N  % Wilson shell index
     if k<N || ~sflag
          o={'Nkeep',Nkeep,'Etrunc',Etrunc}; ?\mlbl{iter:Nkeep}?
     else o={'Nkeep',0}; ?\mlbl{iter:Etrunc}? % by default, discard all states at last iteration
     end

   % Collect info  on total expanded dimension (K+D)
   % this copies 2 entries for non-abelian: [multiplet, state space] dimension
   % where q(1,*) are multiplet dimensions
     q=getDimQS(HK); N4(k,:,2)=q(:,2); ?\mlbl{nrg:Dtot}?% dimension of 2nd index of HK

   % Exact diagonalization of expanded space (may also use QSpace/eig wrapper here)
     [ee,I]=eigQS(HK,o{:}); ?\mlbl{nrg:eig}?

     ee=ee(:,1);      % $\texttt{\hyperref[I12.ee]{first column}}$ (relevant for non-abelian symmetries only)
     EK=QSpace(I.EK); % MEX files cannot return QSpace objects, only structures $\mlbl{I.EK}$
     ED=QSpace(I.ED); % hence the conversion to QSpace objects here          $\mlbl{I.ED}$

   % Subtract ground state energy [except for H0 (k=1), keep E0(1)=0]
     if k>1, E0(k)=min(ee); ee=ee-E0(k); end

     if EK % i.e., non-empty
      % EK is returned in compact diagonal format => take dimension on 2nd index
        Dk=getDimQS(EK); N4(k,:,1)=Dk(:,2); ?\mlbl{nrg:DK}?
        EK=EK-E0(k); q=EK.data;  % also subtract E0 in QSpace EK
        EX(k,1)=max([q{:}]);     % largest kept energy
     end
     if ED % repeat for ED
        ED=ED-E0(k); q=ED.data;
        EX(k,2)=min([q{:}]);     % smallest discarded energy
     end
     
     if sflag % obtain AD before AK is changed right next
     AD=contract(AK,I.AD,[1 3 2]); end
     AK=contract(AK,I.AK,[1 3 2]); % LRs index order convention
     if sflag
        q=struct('AK',AK,'AD',AD,'HK',EK,'HD',ED,'E0',E0(k));
        save(sprintf('%s_%02g.mat',fout,k-1),'-struct','q');
     end

   % Collect `finite size' energy spectra
   % (note that nrg_plot.m prefers Inrg.HK if present, plots EE otherwise)
     m=min(Nkeep,length(ee));
     EE(1:m,k)=ee(1:m); % plain energies without symmetry resolution $\mlbl{nrg:EE}$
     Inrg.HK(k)=EK;     % same energies but with symmetry resolution $\mlbl{nrg:HK}$

     if k==N, fprintf(1,'\n\n'); break; end

   % Wilson shall k actually starts here
   % compute matrix elements for next iteration (propagate FKK)
     for i=1:nF
      % operator index order KK[op] here by having FF to the right
        FKK(i)=contract(AK,'!2*',{AK,FF(i),'-op:^s'});
     end

   % Add new shell as described by `local space' in IS.E $\mlbl{nrg:add:site}$
     AK=getIdentity(I.AK,2,IS.E,[1 3 2]); % LRs convention
     AK=setitags(AK,'-A',k); % generates itags { K<k-1>, K<k>*, s<k>}

   % Rescale and propagate Hamiltonian
     HK=diag(EK*rL); % expand compact diagonal representation
     HK=contract(AK,'!2*',{HK,AK});

?\mlbl{Hcpl:hdr}?   % add hopping to newly added site
     for i=1:nF, q=ff(k)*Z*FF(i);  % include fermionic parity Z here!
        Q=contract(AK,'!2*',{FKK(i),'*',{AK,q,'-op:^s'}}); ?\mlbl{Hcpl:ctr}?
        HK=HK+Q+Q';  % +Q' adds hermitian conjugate (" + H.c.")
     end ?\mlbl{Hcpl:end}?

   % Make sure, zero-diagonal blocks are also included (important for k=1 only)
     HK=HK+0*getIdentity(AK,2); ?\mlbl{nrg:iter:done}?

   % Generic log output: current Wilson shell @ largest kept energy
   % (to be compared to chosen truncation energy Etrunc)
     fprintf(1,' %4d/%d (%g) EK=%.3g ... \r',k,N,max(Dk(1,:)),EX(k,1));

  end ?\mlbl{NRG:iter:end}?% of iteration (Wilson shell) k
\end{minted}
When adding a new Wilson site
in \mlines{nrg:add:site}--\ref{nrg:iter:done}
for the next iteration,
the extended state space is exactly diagonalized
in \mline{nrg:eig}.
This also carries out the truncation
into kept (\src{K}) and discarded~(\src{D}) states
[cf. \mlines{I.EK} and \ref{I.ED}].
Two parameters govern this truncation as
provided with the options \src{o\{:\}},
namely \src{Etrunc}\,$\equiv E_{\rm trunc}$ and 
\src{Nkeep}\,$\equiv N_{\rm keep}$.
This keeps at most $N_{\rm keep}$ states
(multiplets in the presence of non-abelian symmetries),
or fewer if the energy threshold $E_{\rm trunc}$
is reached earlier. Near-degenerate eigenstates
are never cut across by the truncation but kept
together (or truncated) in full to avoid
artificial symmetry breaking of symmetries that
are not explicitly enforced.
As per \mlines{iter:Nkeep}--\ref{iter:Etrunc},
both truncation parameters,
$N_{\rm keep}$ and  $E_{\rm trunc}$, are used
for all iterations except for the last one,
where all states are discarded. Hence with no
states kept, this way the Wilson chain is naturally
terminated \cite{Anders05}.

NRG converges exponentially with
increasing $E_{\rm trunc}$ \cite{Wb11_rho}.
On empirical grounds, typically a value of
\src{Etrunc}\,$\equiv E_{\rm trunc}
\gtrsim 7$ (in rescaled units as in the source above)
leads to good convergence. 
By truncating based on an energy threshold, this mimics
truncation at about a comparable discarded weight
and hence comparable accuracy along the Wilson
chain~\cite{Wb11_rho}.
Therefore it is typically recommended to truncate based on
$E_{\rm trunc}$, while keeping $N_{\rm keep}$ sufficiently
large, yet set nevertheless as a safeguard
on what is affordable numerically.
With this in mind, when increasing the
complexity of the impurity or number of (bath) flavors, 
this quickly also increases the number of states within the
energy interval $[0,E_{\rm trunc}]$ relative to the
ground state energy for any iteration.
In this case, one needs to work (exponentially) harder
by keeping more many-body states for comparable
accuracy, i.e., the same $E_{\rm trunc}$.

The iterative diagonalization is wrapped up 
again in the spirit of \src{NRGWilsonQS},
\begin{minted}[escapeinside=??,firstnumber=last]{matlab}
% Finalize Inrg info structure (used by nrg_plot.m)
  Inrg.E0=E0;
  Inrg.EK=EX;
  Inrg.EScale=Lambda.^(-(0:length(E0)-1)/2);
  Inrg.Itr=add2struct('-',Etrunc,Nkeep);  % Itr = info on truncation

% Cumulative `physical' value for E0
% i.e., global ground state energy in units of bandwidth
  Inrg.phE0=sum(E0.*Inrg.EScale); ?\mlbl{nrg:phE0}?

  if ndims(N4)==2
       Inrg.NK=N4;   % abelian symmetries only
  else Inrg.NK=reshape(N4,[],4); % non-abelian
  end

  if sflag
     f=sprintf('%s_info.mat',fout);
     save(f,'-struct','Inrg');
  end

% Finalize numeric array EE for finite-size spectra
% collected data sets have variable length => replace trailing zeros
% for shorter records by nan except so that these are ignored by plot()
  EE(~EE)=nan;            % same as EE(find(E==0))=nan;
  EE(1,isnan(EE(1,:)))=0; % restore zero for first value (ground state)
\end{minted}
The bare energy spectra collected in the numeric
array \src{EE} as finalized above may be plotted as is
or used for data analysis. 
The plot script \src{nrg\_plot.m} (see below), however, 
prefers the \QSpace array
\src{Inrg.HK} collected in \mline{nrg:HK}, as this contains
symmetry resolution. This is reflected then in the line
colors of the NRG `energy flow diagram' as generated
by \src{nrg\_plot}.

\begin{figure}[p]
\begin{center}
\includegraphics[width=1\linewidth,trim = 0.5ex 9ex 0.5ex 6ex, clip=true]{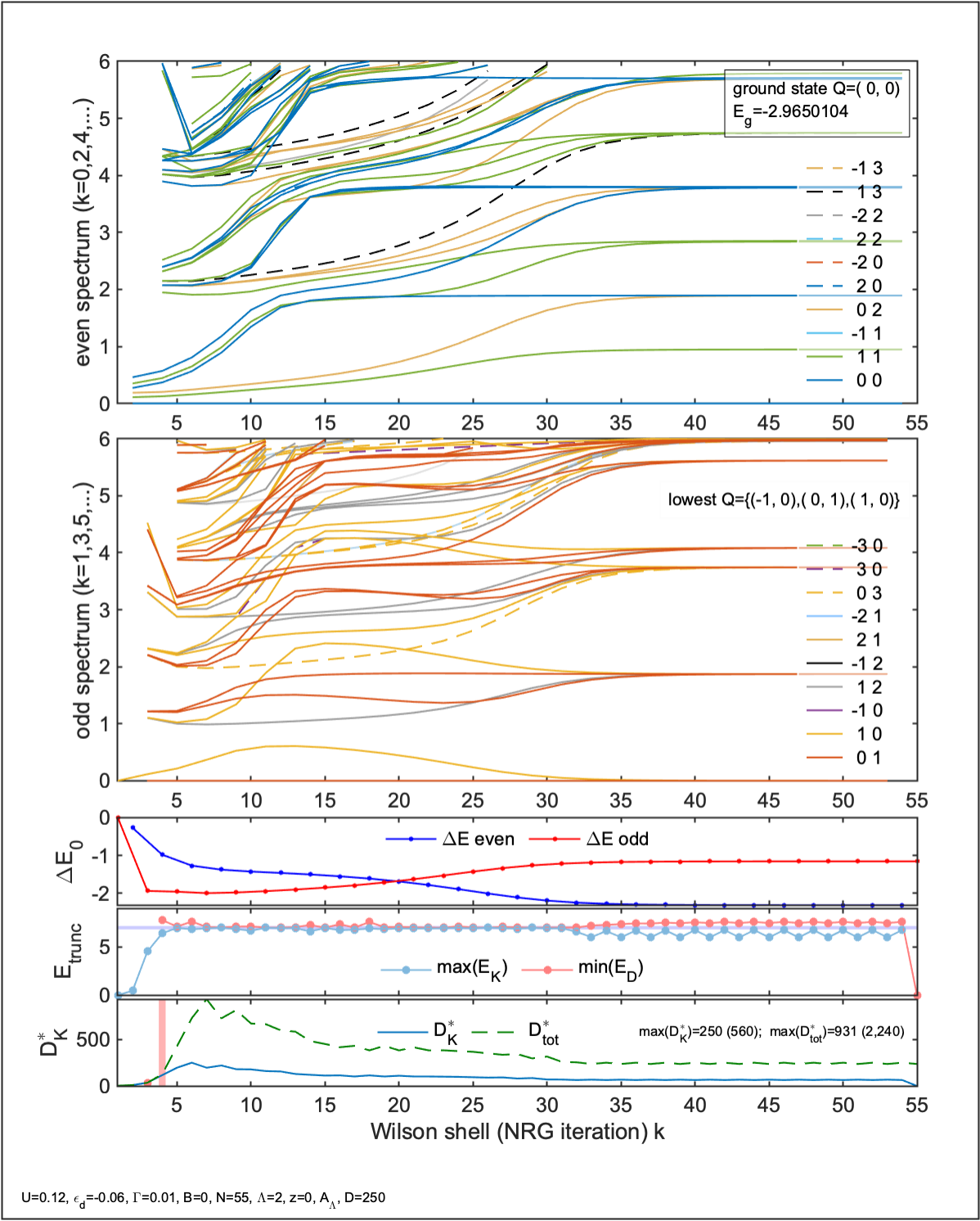}
\end{center}
\caption{Standerd NRG energy flow diagram (top two panels)
   for the collected finite-size spetra from the
   iterative diagonalization of the Wilson chain.
   This plot is generated by \src{nrg\_plot}
   (\mline{nrg:plot}),
   used with the default model parameters
   $U=0.12$, $\varepsilon_d=-0.06$, $\Gamma=0.01$,
   $B=0$ and hence
   ${\rm U}(1)_{\rm charge} \otimes {\rm SU}(2)_{\rm spin}$,
   as well as the default NRG-specific parameters
   $N=55$, $\Lambda=2$, $z=0$,
   \src{Etrunc}$=7$, and \src{Nkeep}$= 1024$.
   The lower three panels show 
   the subtracted ground state energy offset $\Delta E_0$
   for even and odd iterations,
   the truncation energy bounded by
   ${\rm max}(E_K) < {\rm min}(E_D)$
   relative to \src{Etrunc} (gray line),
   and the number of multiplets kept ($D_K^\ast$)
   together with the expanded Hilbert space dimension
   when adding a Wilson site ($D_{\rm tot}^\ast$).
   The maximum of $D_K^\ast\leq 250$ kept SU(2) multiplets
   corresponds to $D_K\leq 560$ states (reflecting
   the data in \src{Inrg.NK}).
   The dimensions of the expanded state space
   that had to be diagonalized exactly in \mline{nrg:eig} is also
   shown for reference, having $D_{\rm tot}^\ast \leq 931$
   multiplets corresponding to $D_{\rm tot} \leq 2\,240$ states.
   With $D_K^\ast\leq = 250 < (\src{Nkeep} = 1024)$,
   all truncation was performed based on energy,
   as also confirmed by the second to last panel.
} 
\label{fig:nrg-plot} 
\end{figure}%

\paragraph{NRG energy flow diagram}

The finite-size spectra collected from the iterative
diagonalization can be combined and plotted as standard NRG
energy flow diagrams \cite{Wilson75,Bulla08}.  \QSpace provides
the general plot script \src{nrg\_plot.m} for this purpose.
These energy flow diagrams give a concise detailed overview of
the relevant energy scales of the impurity problem down to the
lowest-energy fixed points.  For this, it is important to be
aware that the Wilson shell $k$ directly translates to an energy
scale $\Lambda^{-k/2}$ in units of the half-bandwidth $D$:
impurity models are 0+1 dimensional, with `0' the impurity, and
`+1' the direction of the chain indicating energy scale.  The
following thus generates the NRG energy flow diagram which
summarizes the collected finite-size spectra from the iterative
diagonalization,
\begin{minted}[escapeinside=??,firstnumber=last]{matlab}
% Summarize result by plotting NRG energy flow diagram
  if ~exist('plotflag','var') || plotflag  % disable by setting plotflag=0
     param.D=Nkeep;
     nrg_qdisp=10;  % turn on legend with symmetry labels (at most 10 entries)
     nrg_plot       % main plot script for NRG energy flow diagram $\mlbl{nrg:plot}$
  end
\end{minted}
The call to \src{nrg\_plot} in \mline{nrg:plot}
generates a plot with five panels as shown in \Fig{nrg-plot}.
All panels have the Wilson shell index on the horizontal axis,
which starts from high energies at the left
and progresses to exponentially small energy scales
towards the right. The top two panels
show standard NRG energy flow diagrams for even and
odd iterations.  Due to the open boundaries
of the Wilson chain at any shell (iteration) $k$,
intrinsic even-odd alternations arise in the energy spectra.
Hence to avoid zig-zag lines in plots, a smooth energy flow
is obtained by plotting even and odd iterations separately
(see labels for vertical axis).
While they provide qualitatively similar behavior in terms of 
flow across fixed points towards the low-energy
stable fixed point, they differ significantly in their details.
By convention, the impurity together with the
first Wilson site $\hat{f}_0$, i.e., $H_0$ in \Eq{Himp},
is considered an even iteration \cite{Wilson75,Bulla08}. 

For the default SIAM parameters
in \mlines{nrg:sdef:1}f,
the convergence in the energy flow diagram 
in \Fig{nrg-plot} for $k\gtrsim 35$ 
is attributed to having reached energies
sufficiently below the Kondo scale. 
Generally, the NRG finite-size spectra converge 
in the low-energy fixed point to the corresponding
problem-dependent stable fixed point spectrum.
Therefore once the spectra become self-repetitive
(straight lines for iterations $k\gtrsim 40$),
the iterative NRG diagonalization can be stopped
(here at iteration $k=N=55$).
The crossover around $k^\ast \sim 25$ towards
the low-energy fixed point
is in good agreement
with the analytical value of the Kondo temperature
\cite{Haldane78,Filippone18}, e.g., as returned
by the function \src{TKondo.m}
provided with the \QSpace repository: based on the parameters
in the global structure \src{param}, \src{TKondo}
without any further input parameters
returns $T_K^{\rm analytic} = 2.20\cdot 10^{-4}$
which compares well to $T_K^{\rm NRG} \sim
\Lambda^{-k^\ast/2} \sim 1.7\cdot 10^{-4}$
(bearing in mind that the Kondo scale is a crossover
scale, it is only
determined up to a prefactor of order one, depending
on the particular physical or numerical
prescription used to define $T_K$).

The third panel out of \src{nrg\_plot} in \Fig{nrg-plot}
shows the lowering of the ground state energy with
each iteration by adding another Wilson site.
This is based on the data in \src{Inrg.E0}
which is again plotted separately 
for even and odd iterations. 
One observes that below the Kondo scale $k \gtrsim k^\ast$,
even iterations lower the energy {\it more}.
Hence the `ground state' of the system is
better reflected by even iterations.
For this reason, \src{nrg\_plot} adds
the label `ground state' to the upper panel for even iterations, together with its symmetry sector $Q$
and the cumulative `ground state energy' $E_g$ taking
all NRG iterations into account (\src{Inrg.phE0}
in \mline{nrg:phE0}).
On physical grounds, the SIAM flows to strong coupling
in which case one may think of $H_0$ as already forming
a screened singlet which is also an even iteration.
Thus in the low-energy fixed point the impurity spin
has been absorbed into, and in this sense, has
become fully screened by the bath.

The fourth panel analyzes the truncation energy 
by plotting up to which energy states were kept
along each iterative diagonalization.
This also shows an additional line that corresponds to
the lowest discarded energy. 
When truncating based on energy,
the latter line lies above \src{Etrunc},
whereas the former line is upper bound by \src{Etrunc},
as seen in \Fig{nrg-plot}, having 
\src{Etrunc=7}.

The fifth (lowest) panel shows the number of states $D_K$
(or number of multiplets $D_K^\ast$ in the presence
of non-abelian symemetries) that were actually kept
along the iterative diagonalization (\mline{nrg:DK}),
together with the expanded state space $D_{\rm tot}$
(\mline{nrg:Dtot})
that had to be fully diagonalized (\mline{nrg:eig})
at each step before 
truncating, and thus splitting it into kept
and discarded states (\mlines{I.EK}--\ref{I.ED}).

When proceeding to lower energy scales, certain
degrees of freedom become frozen out. Generally, therefore,
the entanglement and hence also the number of required
states along the iterative diagonalization show 
a monotonically decreasing behavior.
This behavior is in agreement with Zamolodchikov's
c-theorem \cite{Zamolodchikov86}, and is typically well
reflected in the bottom panel of \Fig{nrg-plot}.
It is also reflected
in the upper two panels where levels tend to `dilute'
when moving to lower energy scales towards the right,
i.e., levels have an average tendency to move upward.
Of course, this excludes the start of the iterative
diagonalization before truncation sets in (vertical
marker line in the bottom panel) up until the maximum
of states is reached soon after.
The early iterations around when truncation sets in,
are thus the most expensive ones in an NRG simulation.
If these can be performed in a stable converged sense,
e.g., by keeping sufficiently many states up to
the required \src{Etrunc}, then arbitrary low-energy
scales can be reliably reached by NRG
at (typically considerably) lower numerical cost
per iteration.

\paragraph{fdm-NRG spectral functions}
\label{idx:fdmNRG-0}

The iterative diagonalization of the Wilson chain
has the unique and powerful property that it
builds an approximate, but complete many-body eigenbasis
that is fully tractable, in practice \cite{Anders05,Peters06}.
This permits one to obtain thermodynamical data in
a text-book like fashion by directly evaluating
Lehmann representations in the spirit of the
{\it full density matrix} (fdm-)NRG \cite{Wb07,Wb12_FDM}.
This includes thermal quantities both in the static
as well as the dynamical context.
Since the underlying procedure of fdm-NRG requires 
significantly more book-keeping as compared to
iterative diagonalization, e.g., when collecting spectral data
into energy bins, etc., this algorithm only exists hardcoded
as the MEX routine \src{fdmNRG\_QS}. Typically,
it is run after \src{NRGWilsonQS}, yet may also be run
after the iterative diagonalization based on
\src{xnrgQS} above under the condition that
the data was stored in a compatible file format
by not disabling \src{sflag} in \mlines{sflag}ff.
Generally, dynamic correlation functions
$\langle \mathcal{O}_1 \Vert \mathcal{O}_2^\dagger\rangle_\omega$
require the specification of a pair of operators
$\mathcal{O}_1$ and $\mathcal{O}_2$ which in the context
of the NRG are local operators. In \QSpace this
translates to operators that can be expressed within
the state space described by \src{A0}.
If $\mathcal{O}_1$ is not set, \QSpace assumes
$\mathcal{O}_1 = \mathcal{O}_2$, by default.
Then by choosing the fermionic operator
\src{$\mathcal{O}_2 \equiv $\,op2=F(1)} below
as the operator acting on the impurity,
this computes the impurity spectral function
$\langle \hat{d}\Vert \hat{d}^\dagger\rangle_\omega$
for the SIAM,
\begin{minted}[escapeinside=??,firstnumber=last]{matlab}
if sflag
 % Compute the dynamical correlation function $\langle\mathcal{O}_1|\mathcal{O}_2^\dagger\rangle$ where $\mathcal{O}_1=\,$op1 and $\mathcal{O}_2=\,$op2
 % are operators acting on the impurity (more precisely, acting within A0 here)
   op1=[];    % operator $\mathcal{O}_1$ (empty takes default: op1=op2)
   op2=FF(1); % operator $\mathcal{O}_2$: here fermionic annihilation operator
   zflags=1;  % whether to use fermionic parity Z with op[12] (here: yes)

   nostore=1; % do not store matrix elements of op[12] in NRG basis to files fout*
   locRho=1;  % use local thermal density matrix, i.e., do not store Rho(T) 
              % to files fout* either, but keep it in memory

 % Look for relevant options set in Matlab for fdmNRG_QS
 % where a trailing '?' indicates a non-mandatory option
 ?\mlbl{ofdm:1}?  ofdm=setopts('-','T?','zflags?','-nostore?','-locRho?','nlog?','emin?','emax?');?\mlbl{ofdm:2}?
 % main fdm-NRG routine
   [om,a0,Ifdm] = fdmNRG_QS(fout,op1,op2,Z, ofdm{:}); ?\mlbl{fdmNRG}?
   fdm_plot; ?\mlbl{fdm:plot}?  % plot script to summarize its output
end
\end{minted}
One may specify an arbitrary temperature \src{T}
in \mline{ofdm:1} (by default,
it is taken as the energy scale towards the end of,
but still within given Wilson chain;
in this sense, the default temperature is $T=0^+$;
the actual temperature used is returned in \src{Ifdm.T}).
The result out of \src{fdmNRG\_QS} in \mline{fdmNRG}
includes the discrete spectral data \src{a0}
[one column per requested operator pair in 
\src{op1(i)} and \src{op2(i)}, where $i\leq n$ 
correlation functions can computed in a single
call to \src{fdmNRG\_QS}].
The discrete spectral data \src{a0} is collected on a uniform
logarithmic frequency grid \src{om}\,$\equiv \omega$ with 
energy range [\src{emin}, \src{emax}] for both,
positive and negative energies and \src{nlog} bins per decade
(\mline{ofdm:2}). Any spectral weight outside this range
is added to the closest extremal bin, so no spectral
weight is lost. The returned frequency grid \src{om}
already catenates negative and positive frequencies
into a sorted vector that matches \src{a0} in length.
The spectral output is again simply analyzed within Matlab
based on the plot script \src{fdm\_plot} in \mline{fdm:plot}.
This also broadens the spectral data in a log-Gaussian fashion \cite{Wb07}.
A detailed discussion of this is outside
the scope of this documentation, and hence is left
for the future. In case of interest, nevertheless,
please feel free to explore and if there is need, to
\idx[contact]{contact support}.

\section{Conclusion}
\label{sec:conclusion}

This documentation of the \QSpace tensor library
focuses on its basic structure in the ecosystem
that it was developed in, namely \Cpp MEX embedded
into Matlab on Linux-like environments including macOS.
While the public repository with \QSpace v4 nevertheless
already also includes powerful applications
in the context of NRG and DMRG,
these are mostly only mentioned in passing here.
Further documentation in that regard is left for the future.

The strong link to Matlab is due to historical reasons.
Since the requirement of a standard Matlab license
represents some barrier for the usage of \QSpace
in the community, however, the longer-term perspective is
to replace the embedding into Matlab by other
open environments eventually, with Julia a
particularly attractive more recent candidate.
Yet the \QSpace core libraries
will continue in the \Cpp code base in any case. 
The environment, however, that these are embedded
into as an API library, may be subject to change
in the future.

\subsection{Acknowledgements}

I thank the following people for carefully looking
through this manuscript and providing valuable feedback:
Jan von Delft (Ludwig Maximilians University (LMU), Munich, Germany)
and his group members, as well as
Seung-Sup Lee (Seoul National University, South Korea),
Fabian Kugler (Flatiron Institute, New York, USA),
Matan Lotem (Tel Aviv University, Israel), 
Ji-Yao Chen (Sun Yat-sen University, China), and
Adrien Florio (Brookhaven National Laboratory, New York, USA).

I acknowledge long-standing support from the
condensed matter theory group of Prof. Jan von Delft (LMU)
until 2017 and the office space still provided
long after that during my frequent visits to Munich.
I very much enjoyed the many inspiring discussions
with students and postdocs who also provided valuable feedback 
on \QSpace all along. Since 2006 the students in alphabetic
order were: Arne Alex, Benedikt Bruognolo,
Andreas Gleis, Cheng Guo, Markus Hanl, Theresa Hecht,
Andreas Holzner, Jheng-Wei Li, Hamed Saberi, Frauke Schwarz,
Katharina Stadler, Elias Walter).
Former grad students outside or with a temporary link to the LMU:
Bin-Bin Chen (LMU; Beihang University, Beijing; University of Hong Kong, China),
Matan Lotem (Tel Aviv University, Israel), and
Anand Manaparambil (Adam Mickiewicz University, Poznan, Poland).
Former postdocs at the LMU, all of whom are faculty
elsewhere by now:
Wei Li (LMU; Institute of Theoretical Physics, CAS, Beijing, China),
Seung-Sup Lee (LMU; Seoul National University, South Korea),
Ji-Yao Chen (MPQ Garching; Sun Yat-sen University, China),
Ireneusz Weymann (LMU; Adam Mickiewicz University, Poznan, Poland),
Yilin Wang (BNL; Hefei National Laboratory, China).
Since my relocation as a staff scientist to the Brookhaven
National Lab (BNL, USA) in 2018,
I am very grateful also to Alexei Tsvelik and Robert Konik
to leave me with sufficient freedom to carry
on the \QSpace endeavor.

The \QSpace project was initiated and developed over many
years by myself while working at the LMU in Munich
at the crossroads of NRG and DMRG. A major incentive
thinking about a tensor library was an early DMRG
Matlab code kindly provided by Frank Verstraete early 2005
that demonstrated the power of simple transparent tensor
routines, even if no symmetries were exploited then.
The initial development (\QSpace v1 for abelian symmetries)
was funded under DFG-SFB631 and DFG-TR12. The later
development of \QSpace v2 onwards for non-abelian
symmetries was funded by 
DFG WE-4819/1-1 (Independent researcher),
DFG WE-4819/2-1 (Heisenberg fellowship),
and DFG WE-4819/3-1 (PhD student with a focus on applications).
Since 2018 the development of \QSpace
has been supported by the U.S. Department of Energy
with a major focus, also after repeated inquiries from
the physics community, to push toward open source
as of \QSpace v4.

\paragraph{Funding information}

While developing \QSpace v4 with the main incentive
to prepare the code for open source as well as working on
this documentation,
A.W. was supported by the U.S. Department of Energy, Office of Science, Basic Energy Sciences, Materials Sciences and Engineering Division.

\subsection{Feedback and support}
\label{idx:contact}

While this documentation provides the more general
background on the \QSpace tensor library, by the very volume
of the sources included in the public git repository,
this documentation is necessarily far from complete.
Therefore aside from this documentation,
interested users are encouraged to explore the git repository 
and specifically also the \src{Docu} folder there
for additional updated resources.
\vspace{-1ex}

\paragraph{Feedback and bug reports}

General feedback, as well as bug reports are welcome.
The primary contact in this regard should be via the
\href{https://bitbucket.org/qspace4u/qspace-v4-pub/}%
{public \QSpace repository}, e.g., by
\href{https://bitbucket.org/qspace4u/qspace-v4-pub/issues}%
{raising an issue} there.
The general contact will also be maintained 
and kept up to date with the
\href{https://bitbucket.org/qspace4u/qspace-v4-pub/wiki}{wiki pages}
also supported by the public repository.
As a last resort, an \href{mailto:weichselbaum@bnl.gov}{email}
may be sent directly to the author of this documentation
as provided on the front page.

When a reproducible bug
is encountered in \QSpace, be it on the Matlab side
or in the behavior of MEX files, a bug report is welcome.
It will be checked and fixed in due time.
Ideally, a bug report should include a minimal
Matlab script that contains the problematic command
that causes a particular error or unexpected behavior.
This should permit one to reproduce the error without
reference to other user-defined functions. This may be
complimented with a small mat-file that contains particular
instances of \QSpace tensors. All of this can be uploaded
with the git repository when
\href{https://bitbucket.org/qspace4u/qspace-v4-pub/issues}%
{raising an issue}.

\paragraph{Support}

The software in the public git 
repository for \QSpace is open source
under the Apache 2.0 license, hence free of charge,
and provided as is without any guarantee. In this sense
actual support beyond this documentation is subject
to the specific circumstances, the availability
of the author, or specific agreements.


\begin{appendix}
\newpage


\section{General Definitions}
\label{app:defs}

\subsection{Acronyms}
\label{app:acronyms}

The following list of acronyms
are central to \QSpace and this documentation.
It is grouped by their meaning rather than
sorted alphabetically, with more basic
definitions coming first.
The main text includes frequent hyperlinks
into this listing.

\paragraph{irep (multiplet)}
\label{idx:irep}

Irreducible representation --
Non-abelian symmetries organize state spaces into
multiplets that form irreducible units (ireps)
when acted upon with the generators of the symmetry.
That is, starting from any state in the multiplet
one can reach any other state by applying some
sequence of generators. In this sense, all states
within an irep are connected by symmetry.
For a particular
non-abelian symmetry, specific ireps include:

\subparagraph{Scalar representation}
\label{idx:scalar}

Trivial multiplet that consists of a single state
that is invariant under all symmetry operations.
It is also referred to as a singlet.
In \QSpace the scalar representation is always referred
to as $q=(0,\ldots,0_\Rsym) \equiv 0$,
for abelian and non-abelian symmetries alike,
with \Rsym defined in \Eq{Rsym}.
The apparent exception of parity symmetry with
$q \in \{+1,-1\}$ is equivalent to $\mathbb{Z}_2$
with $q \in \{0,1\}$, which then again also has $q=0$
as its scalar representation. In this sense, \QSpace
prefers $\mathbb{Z}_2$ to encode parity.

\subparagraph{Defining representation}
\label{idx:qdef}

The smallest non-trivial unit that by taking
tensor-products with itself can generate any
other multiplet for given symmetry.
This is also reflected in ireps represented by
Young tableaus for SU(N) where the defining irep
represents a single box, and the tensor product
of any irep (Young tableau) with the defining irep
adds a box.
This is also in the spirit of tensor networks
where many-body Hilbert spaces are generated
by iteratively adding sites that may be
given, e.g., in the defining irep.
For any non-abelian symmetry, similarly \QSpace always
also starts from the Lie algebra in the defining
representation. Its symmetry labels are always
$q_{\rm def} = (1,0,\ldots,0_\rsym)$ with $\rsym$
the \idx[rsym]{rank of the symmetry} [see also \Eq{qdef}].

\subparagraph{Adjoint representation and spin operator}
\label{idx:qadj}

For any multiplet there must exist a set of operators
that can carry out the symmetry transformations of
given non-abelian symmetry within the multiplet's
state space.
That is, there must exist an irreducible operator
for any multiplet that transforms according to
an irreducible representation that has as many
operators as there are generators in the Lie algebra,
i.e., has the dimension of the Lie algebra.
This irreducible representation is called the {\it adjoint}
($q_{\rm adj}$) which is always self-dual.
The respective irop is also referred to as 
`spin operator' with reference to SU(2).
From the above argument together with
\hsec[1j]{$1j$-symbols}, the adjoint must always appear
when fusing any multiplet $q$ with its \idx[dual]{dual}~$\bar{q}$
(see also construction of
\idx[op-basis]{complete operators basis}).
This permits a generic way to generate the adjoint,
namely fusing $q\otimes \bar{q}$ for any $q$. In the simplest
case when taking the defining irep $q_{\rm def}$,
the adjoint is guaranteed to occur without
outer multiplicity.
For example for SU(2), $\frac{1}{2} \otimes \frac{1}{2}
= 0 \oplus 1$, or ${\bf 1} + {\bf 3}$ in terms
of dimensionalities, or within \QSpace having
\idx[SU2:qlabels]{$q=2S$}, $1 \otimes 1 = 0 \oplus 2$.
Here the spin operator transforms like an $S=1$
multiplet, which is the adjoint representation
for SU(2), indeed, also referred to as triplet then.
For SU(3), this becomes $(10)\otimes(01) = (00) \oplus (11)
\equiv {\bf 1} + {\bf 8}$ with now the octet $(11)$
the adjoint.

\paragraph{irop} Irreducible operator --
Similar to states, in the presence of non-abelian
symmetries it also must be possible to organize all
operators into irreducible units that are connected
via commutators with the symmetry generators.
Hence an irop directly also reflects
the structure of a particular irep of a symmetry,
and is thus said to transform according to that irep.
A particularly simple example are \idx[scalar:op]{scalar operators}
(like Hamiltonians or density matrices) which
transform according to $q=0$, and thus commute
with all generators of the group.

\paragraph{CGC}
\label{idx:CGC}

refers to a standard textbook
Clebsch-Gordan coefficient space, i.e., a CGT of
rank $r=3$ with \idx[qdir]{$q$-directions} \qdir{++-}.
These arise out of tensor product decomposition
(fusion rules) of two multiplets $(q_1,q_2|q_{\rm tot})$.
Usually used with non-abelian symmetries, they
can also be used to (trivially) represent the fusion
rules for abelian symmetries or discrete symmetries.

\paragraph{CGT}
\label{idx:CGT}

refers to a {\it Clebsch Gordon tensor}
as a generalization of CGCs to tensors of arbitrary rank
\idx[rank01]{$r\geq 2$} via contractions.
Like CGCs, CGTs are also always real. In \QSpace,
they are stored in generalized sparse column-major order
and computed in high-precision format via MPFR
using quad precision at 128 bits.
For later retrieval, they are tabulated and stored
with \idx[CGT-CGR]{sorted $q$-labels in \RCS/CStore}.
This contains CGCs
as a subset in \src{CStore/\qdir{++-}/}
and \hsec[1j]{\onej{}}-symbols in \src{CStore/\qdir{++}/}.

\paragraph{CGR}
\label{idx:CGR}

CGT reference --
\QSpace tensors themselves no longer
carry explicit CGTs since these can be tabulated and
stored once and for all (as of \QSpace v3).
Therefore \QSpace tensors only carry references to 
\idx[CGT:sorted]{sorted CGTs}, referred to as CGRs.
As additional information, the CGRs carry
hence a permutation as well as the \idx[cgw]{$w$-matrices}.
In a \QSpace \src{X},
the CGRs are stored in \src{X.info.cgr}.
The \RCR{\RCS/CStore} only stores
\idx[CGT:sorted]{sorted CGTs}, meaning they all have
their \mbox{$q$-labels} lexicographically sorted w.r.t. to their legs
within the group of incoming and outgoing legs, 
with the incoming legs listed first
and at least as many incoming as there are outgoing legs
(for the reverse case, tensor conjugation is used).
For this reason, a CGR also implicitly encodes
an arbitrary permutation $p$ together with a potential
conjugate flag in addition to the CGT reference.
A CGR together with the stored
(and thus {\it sorted}) CGT can thus describe an {\it unconstrained}
CGT with respect to index order of incoming or outgoing legs.
This is central to the \hsec[QS:struct]{\QSpace data structure}.

\paragraph{RMT}
\label{idx:RMT}

refers to generalized reduced matrix elements tensor:
as known from the Wigner-Eckart theorem \cite{Sakurai94},
matrix elements
of operators factorize into reduced matrix elements
in a tensor product with the respective CGT
[cf. \Eq{tensor:decomp}]\cite{Singh10}.
In \QSpace,
generalized column-major order is assumed
for RMTs of any rank. The RMTs are always
stored in \src{double} precision format,
or interleaved \src{double} for complex-valued
tensors for real and imaginary parts.

\paragraph{MPS} Matrix product state
\cite{Dukelsky98,Schollwoeck11,Orus14tns} --
Tensor network state with a one-dimensional structure,
also referred to as {\it tensor train} \cite{Oseledets11},
typically with open (or in the case of translational
invariance, infinite) boundary conditions.

\paragraph{PEPS}
\label{idx:PEPS}

Projected entangled pair states --
A way to describe physical systems by tensor
networks in higher dimensions
\cite{Verstraete04_peps,Cirac09,Schuch10}
with extension to fermionic systems \cite{Corboz10}.

\paragraph{TNS}
\label{idx:TNS}

Tensor network state \cite{Cirac09,Eisert13,Bridgeman17}
-- general terminology for arbitrary
tensor network topographies, including MPS, PEPS, TTNs, etc.
From a physical perspective, it typically refers to
a single state, i.e, with no other external legs
than the physical site indices.

\paragraph{TTN}

Tree tensor network state \cite{Shi06,Murg10} --
arbitrary TNS in the absence of loops,
which includes MPS with open boundaries
as a trivial example.

\paragraph{LRs} 
\label{idx:LRs}

\idx[Atensor:LRs]{MPS index order convention} $(l,r,\sigma)$ with
$l\in$ \src{L}(eft) block, $r\in$ \src{R}(ight) block,
$\sigma \in$~\src{s} for local state space.
This keeps the typically large block dimensions for \src{L}
and \src{R} to the front, like a matrix, that
is then stacked with respect to the small state space $\sigma$.
Assuming column-major, the last index $\sigma$ is the slowest
in terms of data layout in computer memory. See also
\idx[irop]{operator index order} convention.


\paragraph{OC}
\label{idx:OC}

Orthogonality center \cite{Stoudenmire10} --
Tree tensor networks (TNSs), including the special case of an 
MPS with open boudary condition,
permit strict orthonormalization of all 
block state spaces w.r.t. any bond
since cutting any bond disects the TNS into
disconnected tensor network blocks. 
This also permits a clean implementation
of symmetries and well-conditioned local 
tensor environments \cite{Singh13,Lubasch14}.
In this case, tensors in a TTN can have multiple
indices that `enter', but at most one index that `leaves' it,
in the sense that it acts as an isometry that maps
all incoming state
spaces onto one fused state space that is typically
truncated.
Such a TTN can only have one tensor with all indices incoming,
which is referred to as the orthogonality center (OC).
All arrows in a TTN flow toward the OC then.
It ties together otherwise disconnected blocks of
a TTN into the wave function for the full system.
The OC may be moved iteratively throughout
the tensor network, a process that is referred to as
{\it sweeping} and also relates to
automatic differentiation.
\vspace{-1ex}

\paragraph{OM}

\idx[OM]{Outer multiplicity} -- only applies to CGTs,
and refers to the fact, from a tensor and symmetry point
of view that for fixed symmetry sectors on all legs,
the fusion of incoming legs can result in {\it multiple
orthogonal} combinations for the {\it same} fixed symmetry
sectors of the outgoing legs \cite{Wb12_SUN,Wb20}.
The presence of OM, $\mu=1,\ldots,M$ with $M>1$ thus translates
to an additional trailing OM index $\mu$ in CGTs
that links to an additional trailing OM index $\mu'$
in the respective RMTs via an auxiliary $w$-matrix
[cf. \Eq{tensor:decomp}].
\vspace{-1ex}

\paragraph{MEX files}
\label{idx:MEX}

\href{https://www.mathworks.com/help/matlab/call-mex-file-functions.html}{\it Matlab executables} (MEX, \cite{mw_mex})
are \Cpp binaries that are compiled via Matlab's \src{mex}
interface, with file extensions \src{*.mex*},
like \src{*.mexa64} (Linux) or
\src{*.mexmaci64} (macOS). They represent
functions that can be called directly from within Matlab
like regular Matlab functions, but are based on
externally compiled binaries that are interfaced
via Matlab's MEX API. 
This permits efficient complex
data structures with access to \Cpp coding standards
(\src{C++11}), highly optimized open-source
libraries such as the standard template libraries (STL),
as well as \Cpp style garbage collection.
The MEX compilation  requires
\src{\hyperref[mex-setup]{mex~setup~\Cpp}}.
\QSpace uses the option \str{-R2018a} with \src{mex} 
to enable standard interleaved complex data format.
\vspace{-1ex}

\paragraph{NB!} 
\label{idx:NB}

Nota bene -- Latin for `be aware' to indicate
something important. Frequently used in source code
comments, but also in \src{wblog} output.
\vspace{-1ex}

\paragraph{QS {\normalfont or} QSP} refers to \QSpace,
e.g., as trailing part of file names as a low-level
way to differentiate namespaces like \happ[MEX:list]{MEX functions}
\src{*QS}, or as prefix for \QSpace specific
\idx[QSP_NUM_THREADS]{environmental variables}
like \src{QS[P]\_*}.
\hspace{-1ex}

\paragraph{Wb} Alternative acronym
for the initials of the author (AW), frequently 
used in help or usage information,
and comments in the code.

\subsection{Terminology}
\label{app:TNsym}

The following glossary of terminology
is central to \QSpace and this documentation.
Like with the acronyms above,
again the order of entries is grouped
by their meaning rather than alphabetically sorted.
More elementary concepts come first,
so that reading through the entries
from top to bottom should be meaningful.
The main text includes frequent hyperlinks
into this listing.

\paragraph{Sites} are the smallest physical units in
a tensor network that are typically formulated
on a lattice
(a notable exception in the realm of physics
are continuous MPS \cite{Verstraete10}).
For a finite tensor network of $L$ sites then,
sites are indexed $i=1,\ldots,L$
in a well-defined order. Typically, sites are taken identical,
while nevertheless
the Hamiltonian may impose site-specific parameters.

\paragraph{Dimension} 
\label{idx:dimension}

refers to the index range of a
particular index (leg) of a tensor. It needs to be
differentiated from the number of legs of a tensor
which is referred to as its \idx[rank]{rank}.

\subparagraph{Local dimension $d$} refers to the Hilbert space
dimension $d$ of a site which typically is independent of site $i$. It is considered to be small and therefore
is denoted by lowercase $d_{(i)}$.

\subparagraph*{Bond dimension $D$} refers to the dimension
of the auxiliary or virtual state space on a bond
that connects two tensors within a TNS.
Since it is typically significantly larger than the local
dimension, it is denoted by a capital $D$,
and typically varies for different bonds across a TNS.

\subparagraph{Effective dimension $D^\ast$}
\label{idx:Deff}

When dealing with non-abelian symmetries with
multiplets that contain several states grouped as ireps,
the total number of states $D$ on any Hilbert space
can be {\it reduced} to an effective dimension $D^\ast$
which counts multiplets only. For larger-rank
symmetries, one quickly encounters $D^\ast \ll D$.
Empirically, one can gain a total dimensional reduction
on average by a factor of about $\sim 3^\rsym$
with $\rsym$~the \idx[rsym]{rank of the symmetry}
\cite{Wb12_SUN,Wb18_SUN},
e.g., $\sim 3^{N-1}$ for SU($N$).
This has dramatic effects on numerical efficiency.
Since the RMTs carry effective dimensions, e.g.,
the typical cost for matrix
operations of order $\mathcal{O}(D^3)$, 
when exploiting non-abelian symmetries
can be effectively reduced to $\mathcal{O}((D^*)^3)$.
For SU($N$), this implies a speedup of
$\mathcal{O}(3^{3(N-1)} \sim \mathcal{O}(30^{N-1})$ 
which makes a {\it significant} difference
for any step $N\to N+1$. 
For SU(2) this amounts to more than one order,
and for SU(3) about three orders of magnitude of speed up
for comparable accuracy of the simulation.

\subparagraph{Singleton dimension}
\label{idx:singleton}

refers to a leg in a tensor whose dimension is 1,
i.e., its index range is limited to a single value.
Consider, for example, a rank-3 tensor $X$
with dimensions $ D\times D \times 1$.
It has a trailing {\it singleton} dimension, i.e.,
an index space that has trivial range $X(i,j,1)$
on that last index. If in the presence of symmetries,
this index also carries the symmetry label $q=0$,
i.e., corresponds to the scalar irep like the vacuum state,
then this additional dimension is also irrelevant
from a symmetry point of view, and can be skipped
alltogether. In the case of trailing scalar singleton
dimensions beyond rank 2, the rank of a tensor thus can be
trivially reduced. This is partly used in \QSpace
with irops that represent a scalar operator which
reduces a rank-3 irop to a rank-2 scalar operator
[e.g., see \src{\AtQSpace/fixScalarSop()};
for the reverse process of reattaching a singleton
irop dimension, one may use \src{\AtQSpace/makeIrop()}].
Leading or intermediate singleton dimensions,
on the other hand, are never skipped.
 
\subparagraph{Multiplet dimension}

refers to the number of states in a specific multiplet.
It is denoted by $|q|$ for irep $q$.
See entry \idx[qdim]{multiplet / irep dimension} below
for typical dimensions to expect
depending on the non-abelian symmetry.

\paragraph{State spaces}

Every index (leg) in a tensor network represents
a state space. This can be either the physical
state space of a site, or the virtual or
auxiliary state space of a bond.

\subparagraph{Local state space} refers to the
state space of a site $i$ in the lattice model analyzed,
and is usually denoted by $\sigma_{(i)} \in s_{(i)}$
or $s$ itself, e.g., if $\sigma$ is already used to
specify spin. Typically, sites are considered identical,
thus skipping the subscript $i$ where unambiguous.

\subparagraph{Bond states} refer to the auxiliary
or virtual indices that connect tensors together.
They can be interpreted as many-body block states
if the tensor network has no loops,
such as a tree tensor network or an MPS.
In the presence of loops like in \idx[PEPS]{PEPS}
one may interpret the bond states as actual
fictitious auxiliary state spaces that live
on both sides of a bond and which thus
describe an orthonormal state space consistent
with symmetries {\it by fiat} in this ansatz.
The pair of auxiliary states
associated with a bond are then maximally entangled
into a singlet, followed by local projection to
the physical state space on every site. This
PEPS interpretation directly generalizes
the AKLT \cite{Affleck88vbs} construction.

\paragraph{Operators with and without hats (carets)}
\label{idx:Op:hats}

By definition, \idx[matel]{matrix or array coefficients}
represent operators or tensors in a particular basis
of a state space.
Therefore general operators, prior to casting them
into a particular basis, are shown with hats.
This is generalized to arbitrary tensors,
including \idx[Atensor]{$A$-tensors}
as in $A^{l\sigma}_r \equiv \langle l\sigma |\hat{A}| r\rangle$.
The hat is removed once specific matrix (or array) elements are addressed. In this sense, operators with hats can only occur
in the abstract, but not in the computational
context. Tensors expressed within \QSpace therefore
never carry hats in discussions.

\paragraph{(Generalized) matrix elements}
\label{idx:matel}

Matrix elements are coefficients of operators
in a particular basis, such as 
$S^{\sigma}_{\sigma'} \equiv
   S^{\sigma}_{\ \,\sigma'} \equiv
   \langle \sigma| \hat{S}| \sigma' \rangle
$ [cf. \Eq{S:indices}].
This can be generalized to arbitrary tensors, e.g.,
$T^{\sigma_1\sigma_2}_{\sigma_3 \sigma_4\sigma_5} \equiv
\langle \sigma_1 \sigma_2| \hat{T}|
\sigma_3 \sigma_4 \sigma_5\rangle$,
which are then referred to as coefficient spaces,
array elements, {\it generalized matrix elements},
or simply still also {\it matrix elements}.

\paragraph{Rank of a tensor}
\label{idx:rank}

The rank $r$ of a tensor, i.e., its {\it tensor rank},
refers to the number of
legs or indices associated with a tensor.
A tensor with $r$ legs is thus referred to as a tensor of rank $r$, or simply a rank-$r$ tensor.
With the terms tensor and array used interchangeably,
this also may be referred to as an array of rank $r$.
\\[2ex]
{\it Disambiguation:} The {\it tensor rank} is also known
as \Emph{order} or \Emph{degree} of a tensor. It
needs to be differentiated from the Schmidt or {\it matrix
rank}.\footnote{This semantics is also consistent,
for example, with Wolfram's
\href{https://mathworld.wolfram.com/TensorRank.html}{tensor rank} 
in Mathematica as obtained 
by the function \src{\href{https://reference.wolfram.com/language/ref/TensorRank.html}{TensorRank[]}} vs.
\src{\href{https://reference.wolfram.com/language/ref/MatrixRank.html}{MatrixRank[]}}. In Matlab, by contrast,
the tensor rank of an array is obtained
by \src{ndims()} which returns the {\it number of dimensions}
while skipping trailing singletons beyond tensor rank 2.}
The latter only applies to matrices and refers to the
number of non-zero Schmidt coefficients.
It is typically of lesser importance in tensor networks
in practice,
since the singular values from a singular value
decomposition for some arbitrary but fixed
bisection of a tensor are assumed to decay
without abruptly becoming exactly zero numerically.
A `low-Schmidt-rank' {\it approximation}
for tensors, in this sense, always translates
to truncation on an auxiliary intermediate index.
This comes with a truncation error, also referred
to as the discarded weight, that is
controlled by the number of states kept on that
intermediate index. 
The bisection of a tensor by organizing all its indices
into two non-empty groups, is also unique for matrices only.
This becomes ambiguous for tensors with three
or more indices. The particular generalization of
the matrix rank to tensors in terms of
the {\it canonical polyadic decomposition} (CPD)%
\footnote{Also referred to as PARAFAC decomposition \cite{parafac97}
which also may be seen as a generalization of a
principal component analysis (PCA) to higher dimensions.}
may be referred to as {\it CP rank}
to distinguish it from {\it tensor rank}.
Given a tensor of higher rank (many indices),
a {\it low-rank approximation} 
can also be obtained by a {\it tensor decomposition} 
\cite{Ritter24,Jeannin24}
described by some tensor network
of lower-rank tensors (i.e., tensors with fewer legs
and auxiliary intermediate indices).
A prototypical example of this is the decomposition
of a many-body wave function into an MPS.
In this sense, a low-rank approximation in the above
sense has the same objective as a low CP-rank
decomposition. Both refer to a tensor decomposition.

\paragraph{Rank of a symmetry}
\label{idx:rsym}

The rank $\rsym_{S}$ of a continuous non-abelian symmetry $S$
refers to the number of symmetry labels required
to specify an arbitrary multiplet.
This is derived from the number of generators that
can be simultaneously diagonalized. These are referred
to as the \idx[Cartan]{Cartan subalgebra} and generalize
the $S_z$ operator for SU(2).
For example, SU($N)$ is a symmetry of rank $\rsym=N-1$,
which implies a single label for SU(2), but two for SU(3), etc.
An abelian symmetry like U(1) has $\rsym=0$.
In that case the multiplet space becomes
trivial, as the multiplet and weight label
of its only state become identical.
Therefore $\rsym>0$ distinguishes non-abelian from abelian symmetries. 

\noindent
On practical grounds, it is convenient
to also define the adapted rank (note the marginally different font),
\begin{eqnarray}
   \Rsym \equiv {\rm max}(1,\rsym)
\label{eq:Rsym}
\end{eqnarray}
which then specifies the number of symmetry labels
required for any symmetry, abelian and
non-abelian alike.

\paragraph{Generators of Lie algebra}
\label{idx:Lie}

The space of generators in any Lie algebra can
be organized in a canonical way that generalizes
the combination of the diagonal operator $S_z$ and raising
and lowering operators $S_\pm$ known for SU(2).
This corresponds to and can also be
visualized by the \Emph{root diagram} for the
adjoint representation \cite{Cahn84}.

\subparagraph{Cartan subalgebra}
\label{idx:Cartan}

Maximal set of mutually commuting
generators in a Lie algebra. This generalizes
the single $S_z$ operator for SU(2) to a set of $\rsym$
operators, with $\rsym$ the
\idx[rsym]{symmetry rank}~\cite{Cahn84}.
Since mutually commuting, these
`$z$-operators' can be simultaneously diagonalized.
Their combined diagonal entries then define the
\heq[composite-index]{$q_z$ labels} in \Eq{composite-index}
for any irep ($\rsym$-tupel of numbers,
also referred to as weights).

\subparagraph{Simple roots}
\label{idx:roots}

The remainder of generators, aside from the 
\idx[Cartan]{Cartan subalgebra}, can be organized
into pairs of raising and lowering operators.
For any Lie algebra, there exists
a minimal set of so-called {\it simple} roots
which permits one to traverse any multiplet
(irep or irop)
in the spirit of raising/lowering operators~\cite{Cahn84}.
For a symmetry of rank $\rsym$, it therefore
suffices to constrain oneself to the $\rsym$
{\it simple positive roots} (`raising' operators),
together with their Hermitian conjugate,
the \rsym{} {\it simple negative roots}
(`lowering' operators).

\subparagraph{Maximum-weight state and $q$-labels}
\label{idx:max-weight}

The maximum weight state of a multiplet is
destroyed by any raising operator.
Then starting from the maximum-weight state,
one can sequence the entire multiplet
by applying the minimal set of lowering operators 
(simple negative roots) only. Eventually, one
reaches the complimentary {\it minimal weight}
state that is destroyed by any of the
lowering operators. The $q_z$-labels for the
maximum-weight state of an irep can be used to uniquely
identify the irep. \QSpace employs a
linear map that transforms the (non-standard,
but convenient) internal $q_z$ labels for the
maximum weight state
to standard multiplet labels consistent
with the literature \cite{Wb12_SUN},
e.g., consistent with labels
for the \heq[YTableau]{Young tableau} for SU($N$).
They are referred
to as $q$-labels [cf. \Eq{composite-index}]
and written in \hsec[compact:qs]{compact notation}.

\paragraph{Multiplet / irep dimension}
\label{idx:qdim}

A continuous non-abelian symmetry of \idx[rsym]{rank \rsym}
explores an $\rsym$-dimensional label space.
This consists of a non-negative dense set of
integers $q_i \in \mathbb{N}_0$ that are collected
into an \rsym-tupel
$q=(q_1,q_2,\ldots,q_{\rsym} \geq 0)$.
This not only
holds for the multiplet labels,
but also for the state space of each multiplet
individually, since the generalization of spin
\idx[states]{$S_z \to q_z$}
\cite{Wb12_SUN} also explores an \rsym-dimensional
weight space $q_z$.  Empirically, 
this gives rise to a typical range of the \Emph{multiplet dimension
$|q|$} encountered in tensor network simulations 
\begin{eqnarray}
   1 \leq |q| \lesssim 10^{\rsym}
\ .
\label{eq:qdim}
\end{eqnarray}
This assumes that the
entire physical state is globally `non-magnetic', i.e.,
in or close to the singlet symmetry sector $q_{\rm tot}=0$.
For example, abelian symmetries have $\rsym=0$.
Therefore their multiplet dimension is $|q|=1$ which is
consistent with \Eq{qdim}. In DMRG simulations
with non-spontaneously broken SU(2) spin symmetry,
the typical multiplet range reaches a dimension
$q+1 = \idx[SU2:qlabels]{2S}+1 \lesssim 10$, i.e., $S\lesssim 5$.
Since this range is there for every one of the $\rsym$
dimensions in the label space, this motivates
\Eq{qdim} as also confirmed in practice.

\EQ{qdim}  implies that the typical
multiplet dimension grows {\it exponentially} 
with the rank of the symmetry. For example, for 
tensor network simulations that truncate dynamically
on the bond state spaces while exploiting SU($N$) symmetry,
the SU($N$) multiplets quickly explore dimensions
from $1$ ($q=0$) up to $\sim 10^{N-1}$.
Therefore increasing the symmetry rank by one,
as in SU($N$) $\to$ SU($N+1$), always makes
a {\it significant} difference in numerical cost.
On the other hand, symmetries of the same rank have
comparable numerical cost.
Hence when using \QSpace it is important to be 
generally aware of the scaling of typical multiplet dimensions
with symmetry rank as indicated with \Eq{qdim}.
This strongly affects the numerical cost when generating the \RCS.
As compared to the cost of dealing with the RMTs,
however, the numerical cost for \RCS is typically
negligible for symmetries of rank $\rsym \leq 2$,
which thus includes SU(3).

Irep dimensions are known in \QSpace from the \RCS.
Generally, new ireps can only occur via decomposition
of tensor product spaces. This also
automatically provides their dimension
(the dimension of multiplets in any Lie algebra can also be
determined analytically via the Weyl character
formula \cite{Weyl25,Cahn84}).
Other than that, multiplet dimensions
are usually not particularly important from the user's perspective.
Nevertheless, the dimensions of the ireps are shown
for information purposes.
For any given \QSpace \src{X},
irep dimensions for multiplets are explicitly
stored with the CGRs in \src{X.info.cgr.size}.
These are used when displaying a \QSpace [e.g., see \Fig{QS:disp}].

\paragraph{Dual representation} 
\label{idx:dual}

For any symmetry sector $q$ there exists exactly
one unique dual representation denoted as $\bar{q}$.
It is the only irep that, when fused with $q$,
also generates the scalar irep,
$q \otimes \bar{q} = 0 \,\oplus\, q_{\rm adj} \,\oplus\, \ldots$
\  amongst other multiplets  for $q\neq 0$
(based on the construction of a \idx[op-basis]{complete
operator basis}, this always also includes the
\idx[qadj]{adjoint} $q_{\rm adj}$ at least once).
The scalar representation cannot
result out of any other fusion, i.e.,
$(q q'|0)=0$ for any $q' \neq \bar{q}$.
The defining property $(q \bar{q}|0)$ is 
essential to \hsec[1j]{\onej-symbols}.
One way to think about dual representations from
a physical point of view, is that if one creates
a particle with symmetry labels $q$, the respective hole
must transform in $\bar{q}$, such that they can annihilate each other.
That is, if a particle creation operator transforms
according to $q$, the annihilation operator must
transform according to $\bar{q}$.
In this sense, every irep $q$ has a unique dual $\bar{q}$ 
which represents its time-reversed partner.
Therefore the dual representation always also must
have the same multiplet dimension, i.e.,
$|q| = |\bar{q}|$.
For the case $q=\bar{q}$, the irep is referred to as self-dual.

For SU($N$), the dual for a particular irep $q$ is simply
given by $\bar{q}=\src{flip}(q)$, i.e., by
flipping the order of symmetry labels.
This can be easily motivated pictorially
with reference to \heq[YTableau]{Young tableaus}:
 $\bar{q}$ is
precisely the unique irep (Young tableau)
that when `flipped around'
(rotated by 180$^\circ$) perfectly complements $q$
in the sense that it can be combined with $q$
like two puzzle pieces that together result
in a completely filled rectangular block
of $(\rsym+1) \times (\sum_i q_i)$ boxes.
Being equivalent to the scalar $q=0$,
this $\bar{q}$ always permits the
fusion $(q\bar{q}|0)$.
It follows from the above argument
that a rank-1 non-abelian symmetry such as SU(2)
is necessarily self-dual, i.e., $q = \bar{q}$ for all $q$.
The symplectic symmetries Sp($2N$) as well as
SO($2n+1$) and SO($4n$) are self-dual as well,
whereas for SO($4n+2$) the last two values
in the $q$-label of an irep need to be swapped
to obtain the dual.

Many Abelian symmetries are also {\it not} self-dual.
For example, U(1) with $q + \bar{q} = 0$ and hence
$\bar{q}=-q$, this is clearly not self-dual.
It is instructive to compare to an SU(2) multiplet
with spin $S$ and spin-projection $S_z$:
the latter reflects
the abelian behavior, having $S_z \to -S_z$ under
time reversal, and thus `$\bar{S}_z = -S_z$' in case
of a U(1) spin symmetry. By contrast,
$S$ remains the same, $\bar{S}=S$,
since the weight diagram is symmetric under inversion.
\\[2ex]
{\it Disambiguation:} In \QSpace
the operator $^\ast$ (raised asterisk) does not denote `dual',
but is reserved for (tensor) conjugation.
The \idx[conj]{conjugation} of a tensor, by definition,
leaves symmetry labels invariant,
thus having  $q^\ast = q$, yet with reverted directions.

\paragraph{Symmetry labels (individual symmetry)}
\label{idx:qlabels}

Any given index or state space is 
\idx[states]{decomposed} into symmetry
sectors that are described by symmetry labels
[cf. \Eq{composite-index}].
For abelian symmetries, these can be chosen as plain
integers $q \in \mathbb{Z}$, such as U(1) or $\mathbb{Z}_n$.
For non-abelian symmetries, the symmetry labels specify
ireps, and are tuples of non-negative integers,
based on Dynkin labels which for SU($N$) directly
specify the respective Young tableau.%
\footnote{The emphasis on \hsec[QS:int]{integers} is also
convenient within the \QSpace core libraries since then
there cannot be any issues with rounding errors.
Initially, the irep labels for a non-abelian multiplet
are derived internally from a linear map
of the weights of the maximum weight state \cite{Wb12_SUN}.
The weights are chosen in a most convenient
non-standard way in double precision format.
The linear map on the maximum weight state also
includes non-integers. Yet the resulting $q$-labels
are standard, and thus must correspond to integers up to
rounding errors. This also serves as a simple internal
consistency check.}
The symmetry labels for 
SU(2) are chosen as \idx[SU2:qlabels]{$q=2S$},
and thus are also integers.
The length of the tuple is equal to the \idx[rsym]{rank of the symmetry}.
By having non-negative integers for non-abelian symmetries,
this permits a \idx[compact:qs]{compact notation} of ireps.
In \QSpace therefore all symmetry labels
are represented by \hsec[QS:int]{signed integers},
generally referred to as {\it $q$-labels}.
For an individual symmetry $s$, they are given
by $q\in \mathbb{Z}^{\Rsym_s}$, with $\Rsym_s \geq 1$
the number of labels [cf. \Eq{Rsym}].
The \idx[scalar]{scalar representation} is also
denoted by the shortcut notation $q=0$.

\subparagraph{Note on multiplet specification by dimension}
\label{idx:qlabel-dim}

The specification of an irep via the single label
of its state space dimension
only is generally insufficient for non-abelian
symmetries of rank $\rsym>1$, because it is not unique.
With reference to the literature,
the bold notation in terms of multiplet dimension
is used in parallel, nevertheless, if unambiguous,
like ${\bf 3} \equiv (10)$ for the defining representation
of SU(3). However, note for example, that
the ireps (40) and (21)
of SU(3) accidentally share the same multiplet
dimension $d=15$. Hence in that case the notation ${\bf 15}$
to specify these multiplets becomes ambiguous
[similarly so also for their respective duals
(04) or (12), for which $\bar{\bf 15}$ becomes ambiguous].

\paragraph{Symmetry labels (set of symmetries)} 
\label{idx:qlabels}

In the presence of multiple ($n_{\rm sym}$)
symmetries, for any given index or state space
the symmetry labels for all symmetries
are catenated into a single tuple. For generality,
also their combination is referred to as $q$-labels,
$q \equiv (q_1,\ldots,q_{n_{\rm sym}})$.
Here $q_s$ with $s = 1,\ldots n_{\rm sym}$  describes
the symmetry labels for symmetry $s$,
where $q_s$ itself may already contain multiple
labels for non-abelian symmetries depending
on the symmetry's rank.
For a \QSpace \src{X}, these combined $q$-labels
are collected as rows in the matrices $X.Q\{j\}$
for leg~$j$.
The vacuum state has the \idx[scalar]{scalar representation}
for all its symmetries, and is therefore also
denoted by the shortcut notation $q=0$,
meaning zero labels across all symmetries present.

\paragraph{Symmetry labels (multiple legs)} 
\label{idx:Qlabels}

For a tensor \src{X} of rank $r$  one may consider a 
particular symmetry sector with fixed $q$-labels 
for all legs. This corresponds to a
\idx[QS:recs]{\QSpace record} that via CGRs
references the respective CGTs (one for each symmetry).
Depending on the context,
the combined set of these $q$-labels is also 
collectively referred to as $q$-labels, for simplicity.

\subparagraph{Degenerate $q$-labels}
\label{idx:qdeg}

A CGT has well-defined $q$-labels on all its legs.
For a given CGT then, if multiple legs have the same $q$-labels
as well as the same directions, this is referred to
as {\it degenerate} $q$-labels in \QSpace \cite{Wb20}.
This is important for the interplay of 
\idx[CGT-CGR]{CGRs with sorted CGTs}.

\paragraph{Scalar operator}
\label{idx:scalar:op}

A scalar operator is an irop that transforms like the
\idx[scalar]{scalar representation} $q=0$. Therefore
it consists of a single operator where the trailing
singleton \idx[irop]{irop dimension} on the third index can be 
trivially skipped. Examples include Hamiltonians
or density matrices.
\\ 

\paragraph{$A$-tensor}
\label{idx:Atensor}

When fusing two (incoming) state spaces into the combined
(outgoing) state space, this is described by a
rank-3 tensor. The pairwise fusion of state spaces
is a generic operation for any type of tensor network.
When adhering to an \idx[LRs]{LRs}
index order as with an MPS, this is frequently referred
to as \idx[Atensor:LRs]{$A$-tensor} in \QSpace, and thus also
in this documentation.
A call to \src{getIdentity()} returns an {\it identity} $A$-tensor
which describes the full, i.e., untruncated combined state space
based on an identity matrix in multiplet space that is
sliced up into RMTs \cite{Wb12_SUN}. As such this represents
a map from input multiplets to fused multiplets.

\paragraph{\QSpace record}
\label{idx:QS:recs}

A \QSpace consists of a \hsec[QS:struct]{listing of non-zero blocks}
(\Aps{QS:struct}), each of which has well-defined symmetry
labels on all legs as discussed with \Eq{tensor:decomp}.
In a \QSpace tensor \src{X} these entries are referred to as {\it records},
indexed by $\irec = 1,\ldots,n_X$ with $n_X$ the total
number of blocks. Record $i$ then consists of the
RMT \src{X.data\{\irec{}\}} together 
with symmetry labels \src{X.Q\{l\}(\irec,:)} on leg $l$.
The \QSpace display as in \Fig{QS:disp} shows one line per record.

\paragraph{Vacuum state}
\label{idx:vacuum}

The vacuum state, like in the absence of any sites,
always carries the \idx[scalar]{scalar representation}
for all its symmetries. Hence it is described by
the trivial \idx[qlabels]{symmetry labels} $q=0$.
The identity operator for the vacuum state
based on the symmetry configuration specified
with reference to the symmetries of some existing
\QSpace tensor \src{X} can be obtained
via \Srt{getvac(X)}.

\paragraph{Zero blocks and scalar operators}
\label{idx:QS:zblocks}

When exploiting symmetries, tensors decompose into a
\heq[tensor:decomp]{block structure}
since many matrix elements are
zero because of symmetry. 
If entire blocks are zero, because forbidden
from a symmetry point of view, these are naturally
absent in a tensor library that exploits
symmetries. However, there can 
also be blocks that are permitted by symmetry,
but are zero nevertheless, in the sense that
the RMT's Frobenius norm is below a threshold
representing numerical noise.
Such `zero blocks' are generally also skipped,
with one notable exception: scalar operators.
Diagonal zero blocks in scalar operators
are kept when present, by default, to avoid confusion.
For example, a Hamiltonian may have zero
eigenvalues for an entire block, specifically
so it represents a small state space
such as single site. Such blocks with all-zero
eigenvalues contain relevant state space
information that generally is of interest,
and hence is not skipped.
This behavior is a safety measure to guard against
`missing' state space.
See also \src{\idx[skipZerosQS]{skipZerosQS}}.

\subsection{\QSpace data structure}
\label{sec:QS:struct}

The data structure of \QSpace within the MEX routines is
represented by an extensive \Cpp~class.
On the Matlab interface,
this is mapped onto a Matlab structure that, for convenience,
is cast into a Matlab wrapper class \AtQSpace. 
Assuming a \QSpace tensor \src{X} with $\irec=1,\ldots,n_X$
\idx[QS:recs]{records}, within Matlab its data is organized
as a structure object with the following fields
:
\begin{itemize}
\item \src{X.data} contains the RMTs: it is a cell array
   where \src{X.data\{\irec{}\}} contains the RMT
   for record \irec. It has 
   \src{double} floating point precision
   (interleaved double in case of complex matrix elements).
   With reference to symmetries, this corresponds to the
   block $\Vert X\Vert_{q}$ in \Eq{tensor:decomp}
   with $q \to \irec$.

\item \phantomsection\label{sec:QS:int}
   \src{X.Q} contains the symmetry labels ($q$-labels):
   \QSpace insists that all symmetry labels are plain integers
   ($q_i\in \mathbb{Z}$). In the \QSpace MEX core routines,
   \src{X.Q} is represented as a \Cpp signed \src{int} array
   (4 bytes), thus $|q_i|\lesssim 10^{9}$. 
   Like \src{X.data}, the field \src{X.Q} is also a cell array.
   In this case, however,
   \src{X.Q\{l\}(\irec,:)} contains the combined set of
   symmetry labels for record~\irec on leg~$l$.
\newpage
\item \src{X.info} represents a substructure with
   the following fields [e.g., see \mline{F.info} after
   \Eq{F:struct}]:
   \begin{itemize}
   \item \src{qtype} contains the symmetry setting in compact
       string notation (comma-separated list without
       white space, containing entries such as \str{SU2}
       for SU(2), or \str{A} for abelian
       which is taken synonymous for U(1), etc.).
       If \src{qtype} empty, the fully abelian
       setting of all-U(1) symmetries is assumed (see below).
       
   \item \src{otype} contains `operator type'
     (mostly a safeguard and this sense optional):
     values include \str{operator} to emphasize that given
     \QSpace is not any rank $r\leq 3$ tensor,
     but represents an
     operator, either scalar rank-2, or an
     \idx[irop]{irop} of rank 3 that adheres to
     index order conventions. In case a particular
     user-created tensor satisfies the definition
     of a standard operator of rank-2 or 3,
     the \src{operator} flag may be set manually
     via \src{X.info.otype=\str{operator}}.
     This field accepts one other much lesser
     used value, \str{A-matrix}. This serves
     to emphasize that given rank-3 \QSpace is
     an identity \idx[Atensor]{$A$-tensor}
     with an MPS in mind.
     
   \item \src{itags} (mandatory field)
     cell array of strings that contains the \idx[itags]{\itags}
     including \mbox{\idx[qdir]{$q$-directions}} via trailing 
     asterisks \str{*}. As of \QSpace v4,
     there must be one \itag for each leg
     in the order of legs as present. Hence
     the number of \itags, i.e., the length of the cell array,
     must match the number of entries in \src{X.Q}.
     This must hold even if the \itags per se are empty,
     because of the additional information of the
     \idx[itag:markers]{trailing markers} such as \src{*}.
     Itags are restricted in length to
     at most 8 characters (excluding the trailing \src{*})
     since within the \QSpace core library
     they are directly mapped into numerical IDs
     of type \src{unsigned long} for convenience.
     The sign bits are
     used for other internal purposes. Hence the character
     set is restricted to the
     \idx[itags]{ASCII range} 33-126 up to
     \idx[itag:markers]{marker characters}.
     As such this includes any case-sensitive alpha-numeric
     strings with simple separators, yet without white space.

   \item \src{cgr} is a structure array that contains
       the \idx[CGR]{CGRs} including
       the \idx[cgw]{$w$-matrix} as stored
       in the subfield \src{cgw}
       (always in \src{double} precision and real).
       The field \src{cgr} is only meaningful for non-abelian
       symmetries. Hence if is empty, the fully abelian
       setting of all-U(1) symmetries is assumed (see below).
       It is non-empty in the presence of (also)
       non-abelian symmetries.
       In this case the number of columns in 
       \src{cgr} equals the number of symmetries used
       in the order specified during setup
       (cf. \hsec[gls]{\Srt{getLocalSpace}}),
       i.e., the entry \src{X.info.cgr(\irec,s)}
       represents the CGR for record \irec and symmetry $s$.
       This also includes any abelian symmetries present
       then, with trivial CGRs generated on the fly for
       the sake of overall code homogeneity.

   \item \src{ctime} creation time stamp of \QSpace object
       (for tracking purposes only).

   \end{itemize}
   {\it Default U(1) symmetries} --
   If the field \src{info.cgr}
   together with \src{info.qtype} is empty, the abelian
   setting of all-U(1) symmetries is assumed.
   In this case, the symmetry labels in the matrix
   \src{X.Q\{l\}} for leg~$l$
   refer to a separate U(1) symmetry for each column then.
   This corresponded to the
   original setting in \QSpace v1.%
   \footnote{The field \src{info.cgr}
   was added with \QSpace v3, whereas \QSpace v2
   had the CGTs stored with the \QSpace tensors
   themselves still without any CGRs \cite{Wb12_SUN}.}
\end{itemize}
To reveal the bare data structure of a \QSpace \src{X},
one may call \src{struct(X)}, with an
example shown with \Eq{F:struct} and the subsequent lines.
Since this data structure is technical and therefore rather
unreadable in raw format, \QSpace provides the \src{display}
routine. It is invoked
automatically by Matlab whenever an object is displayed,
e.g., as part of an output or just by typing the name
of a \QSpace tensor. This conveniently extracts and displays
the most relevant information in the \QSpace data structure
in a formatted output.
See \Fig{QS:disp} for an example.

\paragraph{Interplay of \QSpace tensor and \RCS}
\label{idx:CGT-CGR}

A CGR contains a reference to a CGT in the \RCR{\RCS}.
However, since the latter only stores \idx[CGT:sorted]{sorted CGTs},
a CGR implicitly also stores a permutation
(potentially including a conjugate flag) to be
incorporated just in time when required with the sorted CGT.
In this sense, the CGR, while referencing a sorted CGT,
nevertheless can represent an unconstrained CGT
that is consistent with any index order of given
\QSpace tensor.
Note that this procedure also {\it defines} any unsorted CGT:
it equals a sorted CGT with the given permutation
(and possibly conjugation) applied. The presence
of degenerate $q$-labels (i.e., when exactly the same
symmetry label is present at multiple legs) leads
to subtleties that are properly dealt with
in \QSpace \cite{Wb20}.
The setting above, based on `reference plus permutation'
is explicitly used by the CGRs within the \Cpp core library.
This becomes somewhat hidden and this implicit
in the structure returned to Matlab: the 
$q$-labels and $q$-directions shown in \src{info.cgr}
{\it already} have the permutation (and conjugation flag,
if any) applied to them, while nevertheless also storing
IDs and high-resolution time stamps to the sorted CGT
in the field \src{cgr.cid}. By re-sorting the $q$-labels
in the CGR, the \Cpp code regenerates the relevant
permutation and identifies the underlying sorted CGT.
This permits \QSpace to translate the Matlab CGRs
in \src{info.cgr} into proper pointer references
in the MEX core libraries.
By also providing the IDs, this ensures and subsequently
enforces that the correct underlying \RCS is present
for consistency.


\section{\QSpace Installation and Environment} 
\label{sec:install}

\subsection{System requirements}
\label{app:requirements}

\subsubsection{Matlab and Unix-like environment}
\label{app:ENV:ML+unix}

The \QSpace core library is written in \Cpp. 
This is embedded into Matlab via the \idx[MEX]{MEX} API
which is included with the basic Matlab environment.
All of the configurations assume a Unix-like environment,
such as Linux (assumed, by default) or also macOS,
both within a bash shell environment and standard access to perl.
To be specific, \QSpace does not yet run on a
Windows operating system, unless Linux
is emulated.

Up to the current version \QSpace v4, 
one needs access to a fully functional Matlab environment. 
A basic Matlab license suffices for this, i.e.,
there is no need for any particular Matlab toolbox.
A potential exception is the Matlab compiler toolbox
for \idx[MCC]{MCC compilation}, if an application is intended
to be run on a high-performance computing (HPC) cluster environment.
If the Matlab license permits one to run arbitrarily many
Matlab instances, there is no need for the
MCC toolbox.
By the setup above, \QSpace requires an
environmental setup for Matlab, e.g., see
the \src{./system} folder provided with the git
repository. In particular, the Matlab script \src{system/ml.m}
shows how Matlab using \QSpace can be started
\idx[ml:term]{within a terminal} without having to load
the full Matlab desktop). 

The \QSpace functionality is fully packaged into 
hard-coded \idx[MEX]{MEX} routines as a wrapper
for Matlab intended for tensor network applications.
These binaries can be called directly from within Matlab
like any other Matlab command.
This was the philosophy of \QSpace from its very inception:
by embedding it into Matlab, this 
significantly enhances the user-friendliness of \QSpace.
This makes it efficient to test new
tensor network algorithms right on the Matlab prompt,
or run and debug Matlab scripts that use \QSpace.
This permits one to focus directly
on tensor network algorithms that fully exploit
complex symmetry settings, rather than having
to worry about the underlying implementation.
The inner
workings of the \Cpp core tensor library is mostly hidden
from the point of view of a user application within Matlab.

\paragraph{Matlab terminal mode}
\label{idx:ml:term}

Matlab can be started either with the full-fledged graphical
Matlab desktop, or within an existing terminal (referred to as
`terminal mode').
The latter still permits all graphical capabilities,
such as plotting figures which responds to the environmental
system variable \src{\idx[DISPLAY]{DISPLAY}}.
The embedding of \QSpace into Matlab was developed
in this slim mode.
Hence this is also the recommended way of using \QSpace with Matlab.
For one, this is light-weight and hence beneficial
when working remotely. On the other hand, by directly
using a regular terminal for the Matlab prompt,
this also permits \idx[QS:color]{color coded log output}
for the sake of readability (the latter is based
on printed escape sequences which are supported by
terminals, but not the Matlab desktop).

\paragraph{Matlab scoping}

The git repository provides a full Matlab environment
for \QSpace, including many convenient Matlab utility routines
in \src{./lib}. As this may interfere with other user-specific
environments within Matlab, the \QSpace Matlab environment
may be wrapped into a 
\href{https://www.mathworks.com/help/Matlab/Matlab_oop/scoping-classes-with-packages.html}{package as this supports namespaces}
to isolate project environments
(not included with the \QSpace git repository itself
at the current stage).
Using \src{import} statements should allow one to access to model
class or function names defined in other scopes (packages)
without having to use fully qualified references, throughout.
At a lower level, including or excluding the \QSpace
directories into the Matlab \src{path} 
as in \src{./startup.m} has a similar effect.

\subsubsection{\Cpp libraries required for MEX compilation}

Since \idx[MEX]{MEX} files are compiled for usage in Matlab,
it is advised to link against external libraries 
as provided with the Matlab install where present.
The main reason is that these libraries are modified such that
they can properly deal with error handling consistent
with the Matlab ecosystem (e.g., an error should not terminate
an entire Matlab session, but rather return to the Matlab 
prompt). The required dependencies are specified
in \src{Source/Makefile}. It consists of the following
three standard libraries:

\paragraph{GMP/MPFR multi-precision libraries}
\label{idx:MPFR:lib}

\QSpace uses a higher-precision data format
for all its \idx[CGT]{CGTs}
based on the GNU \idx[MPFR]{GMP/MPFR} multi-precision library
\cite{Fousse07,mpfr}.
All CGTs in the \RCR{\RCS} are stored in this
format, using quad precision at 128 bits.
The MPFR library is linked
from the location of the Matlab install
under \MLRh.
See \src{Source/Makefile} for more details.
By contrast, the GMP library is linked from the operating
system for technical reasons, hence must be installed
if not present (while the GMP is also present under
\srcMLR, it does not come with a header
file there, besides that it also appears to have
non-standard mathworks/Matlab adapted \str{mw\_*}
naming conventions).

\paragraph{LAPACK/BLAS libraries}
\label{idx:lapack}

\QSpace uses the standard LAPACK/BLAS library
\cite{lapack}
as provided with Matlab (\src{-lmwblas},
\src{-lmwlapack}; see \src{Source/Makefile}).
From a coding point of view, they behave
identically to standard LAPACK/BLAS routines.
It is {\it a priori} unclear, though, how Matlab itself
uses these libraries internally. Therefore
at times minor performance differences
may be observed when benchmarking with native
linear algebra routines at the Matlab prompt.
For example, there exist multiple LAPACK
routines with and without divide-and-conquer schemes,
such as \src{dgesvd} vs. \src{dgesdd} for
singular value decomposition.

\paragraph{OpenMP library}
\label{idx:omp}

\QSpace permits parallelization on top of standard
parallelization in library routines such
as LAPACK/BLAS, based on the {\it open 
multi-processing}
(\href{https://www.openmp.org/}{openMP} \cite{openMP})
framework to parallelize loops.
Together with the parallelization within LAPACK/BLAS
this results in nested parallelization.
The library \src{(lib)iomp5} is provided with Matlab
and hence also linked from there.
See \src{Source/Makefile} for more details.

\subsection{Download and install}
\label{app:download}

The public git repository for \QSpace includes
the \Cpp core sources, as well as the
extensive source for the Matlab wrapper environment.
As such there is no automated installation
process per se for \QSpace. The git repository
may be simply pulled into a dedicated \QSpace directory
in one's Matlab home directory, as specified by the
environmental variable \hMYM. 
References to directories or files in the git repository
therefore, by default, assume \srcMYMv
as the root directory. After setup of 
the system environment, only the MEX binaries need
to be compiled in \src{./Source}
$\equiv$ \srcMYMv{}\src{/Source/}
as discussed below.

Currently, there are two (nearly identical) public
git repos available under the \QSpace project
\cite{qspace4u} 
\href{https://bitbucket.org/qspace4u/}{https://bitbucket.org/qspace4u/}
\begin{itemize}

\item \src{qspace-v4-pub/}
-- main repository
developed on and intended for a Unix-like environment.
It contains the most recent and up-to-date \QSpace version,
yet without any MEX binaries. The MEX files thus need to
be (re)compiled by the user. This is also the setting
associated with this documentation.

\item \src{qspace-v4-osx12-monterey/}
-- duplicate
of the above intended for macOS, which then also includes
compiled MEX binaries for intel CPUs. These 
\src{*.mexmaci64} binaries
are compiled with OSX12/Monterey on CPU based macs.
Hence as of \QSpace v4.0, the ARM architectures (M1, M2, etc.)
are not yet supported. For updates, please consult
the \idx[Docu]{documentation} with the repository.
The advantage of this repository is that,
assuming that all required libraries are present,
\QSpace should work {\it as is} without the need for
any additional compilation. Yet because this git repository
also contains binaries, it is not updated as frequently.
In any case, because macOS is unix-like, one can always
also clone the repo \src{qspace-v4-pub} above instead,
and (re)compile for macOS.

\end{itemize}

\subsubsection{System environment setup}
\label{app:ENV:setup}

The setup of the \QSpace system environment
is centralized around
\src{system/matlab\_setup.sh} (bash shell script).
As this file is part of the git repository,
it should not be edited.
By default, it expects a bash script
\src{system/matlab\_setup\_user.sh} which is
not part of the git repository,
in order to avoid that it gets overwritten
during git updates. Hence this file needs to be
generated.
It is the only file that should be edited
to configure the environment
(see \happ[ENV]{environmental variables} for more).
A template is provided in 
\src{system/matlab\_setup\_user.sh-template}.
In summary, the following is the recommended procedure
to set up the \QSpace system environment
(command displays are shown with blue background here
to indicate \src{bash} script),
\begin{minted}[escapeinside=??]{bash}
  cd $MYMATLAB/system/
  cp matlab_setup_user.sh-template matlab_setup_user.sh
  edit matlab_setup_user.sh   # $\mlbl{mlsetup}$use your favorite text editor
\end{minted}
where the path \hMYMv specifies
the location of the \QSpace repository. This 
reflects the environmental variable of the same set in
\src{matlab\_setup\_user.sh} later.
The edit in \mline{mlsetup}
only requires very few lines to change
when using the specified template.
It can be done with any plain text editor
(in the case \src{edit} $\to$ \src{vim},
the repository also provides
the optional \src{system/.vim} and \src{.vimrc} 
to configure \src{vim};
for these files to take effect, one may \src{rsync}
them to one's home directory
via \src{rsync -irp ./system/.vim* \$HOME/}).
If one wants to use a different file other than
\src{matlab\_setup\_user.sh}, then the script
\src{system/matlab\_setup.sh}
permits one to define the environmental variable
\src{\idx[ML-CONFIG]{QS\_CONFIG\_ML\_SH}}, instead.
This then needs to specify the alternative user file
name with a fully qualified path if not located
within the \src{system/} folder of the repository.

The setup is finalized by the perl script
\src{system/matlab\_setup.pl} which is automatically
called at the end of \src{system/matlab\_setup.sh}.
Again since it is part of the repository,
it should be edited by the user.
This perl script cleans up the setup, in that
double-checks and completes the setup where possible.
For convenience,
one may define an alias (feel free to rename),
\begin{minted}[escapeinside=??,firstnumber=last]{bash}
  alias mlsetup="source $MYMATLAB/system/matlab_setup.sh"
\end{minted}
which may also be added to one's \src{\$HOME/.bashrc}
setup for later convenience.
Once the edit in \mline{mlsetup} is complete
and the alias above is defined, then the simple command
\begin{minted}[escapeinside=??,firstnumber=last]{bash}
  mlsetup
\end{minted}
should suffice to configure the required \QSpace Matlab
environment in the current shell.
Because the environmental setup is centralized 
around \src{system/matlab\_setup.sh},
the call to \src{mlsetup} is sufficient to setup
the Matlab environment whenever one deals with \QSpace.
This includes the regular usage of \QSpace
(e.g., see \src{system/ml}), but also compilation
(e.g., see \src{Source/Makefile} or also
\src{MCC/Makefile.template}).

\paragraph{Potential hickups}
\label{idx:mlsetup:caveats}

The environmental setup of \QSpace makes certains
assumptions that may lead to potential hickups
in the setup. These concern the system \src{PATH}:
\begin{itemize}

\item {\it Include current directoy} --
The environmental setup of \QSpace assumes
that the current directory \str{.} is included in the
system \src{\idx[PATH]{PATH}}.
This concerns, for example, the usage of auxiliary perl
scripts located in \src{./system} or \src{./Source}.
Hence the system \src{PATH} should include the
current directory \str{.},
e.g., with lower priority towards the end of \src{PATH}.

\item {\it Matlab programs} --
It is recommended to have \src{\idx[MLR]{\$MATLAB\_ROOT}/bin}
in the system's \src{\idx[PATH]{PATH}} with higher priority,
i.e., by listing it to the front of \var{PATH}
to avoid mixups with programs that accidentally
share the same command name, like \src{mex}, etc.

\end{itemize}

\subsubsection{MEX compilation}
\label{app:mex-compile}

Once the git repo is cloned and the environment 
set up, the last step is to (re)compile the
\idx[MEX]{MEX files}.
The recommended way for this is via
\src{./Source/Makefile} at the level of the
operating system, which calls Matlab's \src{mex}
command from there (in principle, \src{mex} also 
could be called from within Matlab).
In either case, this first requires the 
\happ[ENV:ML+unix]{Matlab environment}
to be configured as described above,
since the \src{Makefile} makes use of this.

The MEX compilation based on the \src{mex} command
requires a \Cpp compiler to be present
on the system. Matlab has constraints on which \Cpp compilers are
\href{https://www.mathworks.com/support/requirements/supported-compilers-linux.html}{supported and compatible}
with a given \href{https://www.mathworks.com/support/requirements/previous-releases.html}{Matlab release}.
Hence the recommended way to choose a compatible \Cpp
compiler is via Matlab's setup in this regard,
\begin{minted}[escapeinside=??,firstnumber=last]{bash}
  mex setup C++      # required once for a given Matlab release $\mlbl{mex-setup}$
\end{minted}
which may be run either from a terminal,
e.g., after a call to \src{mlsetup}, or from the Matlab prompt.
Note that the \QSpace \src{Makefile}s
assume \src{gcc} on linux and \src{Xcode/clang} on macOS.
The mex setup above looks
for compatible \Cpp compilers on the system.
If more than one compiler is found, this becomes
an interactive process where one needs to choose.
The information
on the chosen compiler is then stored
within one's \var{HOME} directory. Hence this only
needs to be run once for a given Matlab version.
However, since the selection of a compatible \Cpp
compiler is Matlab release specific,
the setup in \mline{mex-setup} needs to be repeated
whenever one switches to a new Matlab version. 

Once a compatible \Cpp compiler has been chosen,
one can proceed to compile the MEX files in a terminal.
A test compile (based on \src{helloworld.cc})
can be run first to double-check
whether the Matlab setup proceeds without error,
\begin{minted}[escapeinside=??,firstnumber=last]{bash}
  cd ./Source
  mlsetup && make test    # compiles helloworld.cc into MEX file
\end{minted}
When the above commands run successfully,
this can be followed by
the full compilation 
\begin{minted}[escapeinside=??,firstnumber=last,mathescape=false,texcomments=true]{bash}
  make -B all             # may add option \src{-j \$np} to parallelize compilation $\mlbl{make-j}$
\end{minted}
This compiles about 40 \idx[MEX]{MEX} files total.
About half of these are \QSpace
core routines that are compiled into the \src{./bin}
folder with file names ending in \src{QS},
thus also referred to as \src{QS}-routines.
The remaining utility routines are compiled
into the \src{./util} folder.
The full \mbox{(re-)}compilation can be rather
time consuming when run sequentially. However,
since there are no dependencies across the MEX files,
the compilation of the  MEX files can be trivially
parallelized  to speed up the process. This is achieved 
by specifying the option
\src{-j \var{np}} to \src{make} in \mline{make-j} 
where \src{np} specifies the number of compilation
targest to run in parallel.
A sensible choice for \src{np} will depend on the number
of available cores on one's system.

\paragraph{Safeguards and assertions}
\label{idx:safeguards}

The \Cpp core library of \QSpace includes many 
internal safeguards to ensure overall consistency.
These safeguards usually correspond to simple and thus fast
checks. In this sense, the overhead should be minimal.
Some of the more severe checks can be turned off,
however, by defining \src{\_\_WB\_SKIP\_ASSERT}
in \src{wblib.h}.

\subsubsection{Matlab startup}
\label{app:ml}

A typical Matlab startup proceeds as in the bash script
\src{system/ml} provided with the repository.
With the \QSpace repository located at \hMYMv,
starting Matlab from that directory 
(\src{cd~\srcMYMv}) automatically
sources \src{./}\Src{startup.m}
which is important to setup the \QSpace
environment {\it within} Matlab, e.g., by properly extending
the Matlab \Src{path}.
Since this script is part of the git repository, 
it should not be altered. For that purpose,
at the end of the script \src{./startup.m}, it looks
for another script \src{./startup\_loc.m} in the same
directory, and if present, also runs it. 
This is intended to add automated user-specific
configuration to one's Matlab environment,
by generating and editing that file (using
one's favorite text editor),
\begin{minted}[escapeinside=??,firstnumber=last]{bash}
  edit ./startup_loc.m       # optional Matlab specific user configuration
\end{minted}
As seen in \src{system/ml},
it first sources \src{system/matlab\_setup.sh}
(equivalent to \src{mlsetup} in lines 4-5 above),
switches to the directory \srcMYMv,
and then starts Matlab from there.
In the case of \src{system/ml}, this starts
Matlab into a terminal without the full desktop
environment (by specifying the option \src{-nodesktop})
and without the Matlab splash window during
startup (\src{-nosplash}). With the \src{\idx[DISPLAY]{DISPLAY}}
variable active,
i.e., not disabled via \src{-nodisplay},
the full graphical interface for figures
is available, nevertheless.
Running \src{system/ml}
without any errors or warnings is an essential first
assurance that \QSpace is set up properly.

\subsection{\QSpace Environment within Matlab}
\label{app:QS:env}

The extensive \Cpp code base provides the \QSpace core
functionality. This is packaged into elementary MEX routines
to run tensor network algorithms within Matlab.
By convention,
all MEX routines that deal with elementary \QSpace
tensor operations have a trailing \src{QS} to their
name, e.g., as in \src{eigQS()}.
They are referred to as \src{QS}-routines,
and are located in the \src{bin/} directory of the
\QSpace repository.
Other MEX utility routines are located in the
\src{util/} directory. These have no trailing \src{QS}
extension to their name. Both directories are
included in the Matlab path via the \Srt{startup}
script in \srcMYMv.

\paragraph{\atQSpace wrapper class}
\label{idx:QS:routines}

The \QSpace class in \HSec[atQSpace]{Class/@QSpace}
wraps the \Cpp \QSpace class object after being cast into a
Matlab \hsec[QS:struct]{data structure} by the MEX routines.
For the sake of the discussion in this documentation,
to differentiate it from the \Cpp counterpart,
this Matlab wrapper class is frequently referred
to as \AtQSpace or simply also \Srt{QSpace}
with differentiated \hsec[format]{color-coding}
to further emphasize the Matlab class context
where important.
 
By also overloading a range of Matlab functions,
the \atQSpace class simplifies operations in Matlab
and therefore makes \QSpace considerably more user-friendly.
Simple operations on \QSpace objects are directly
coded within Matlab in \atQSpace,
while it leaves the heavy-duty weight lifting to the MEX routines.
The class constructor  \Srt{QSpace()}
together with the more 
than a hundred other class routines
in \HSec[atQSpace]{Class/@QSpace} [cf. \Sec{atQSpace}]
ensure that the proper
\QSpace tensor structure is maintained
consistent with the MEX core routines.
Last but not least, this wrapper class also provides a
compact formatted 
\idx[QS:display]{display} of \QSpace objects that summarizes
the most important information about a \QSpace tensor
(e.g. see \Fig{QS:disp} in \Sec{QS:disp}).

For a \QSpace \Srt{X},
the command \src{X=struct(\Srt{X})}
strips it down to the underlying plain structure object in Matlab
[see \eqref{eq:F:struct} for an example].
Conversely, given such a structure,
it can be cast back to a class object via 
\Srt{X=QSpace(\src{X})}.
By construction, \atQSpace methods 
only apply to, and hence can only be called
on \QSpace objects. By Matlab convention, this implies
that an \atQSpace member function is called
if at least one \QSpace object is present in the input
argument list. It does not necessarily have to be
the first argument, though.

\subparagraph{MEX files return structures}

Matlab does not permit MEX routines
to return a user-defined Matlab class object
like \mQSpace. Hence \QSpace objects out of
MEX returns can only be returned as a structure object.%
\footnote{Effectively, the casting into a \mQSpace
Matlab class and the corresponding error handling
needs to occur within the Matlab
session, and cannot be outsourced to the MEX API.
The situation is different when MEX routines
store data into files, since Matlab does permit one 
to save a \QSpace data structure as a 
designated \mQSpace object into a \src{mat}-file.
The casting into the Matlab run-time object then
occurs when the \src{mat}-file is loaded.}
For that reason, very simple wrapper routines
exist for \src{QS}-routines,
that effectively behave like
\Srt{foo(X)} $\equiv$ \src{\mQSpace{}(fooQS(\Srt{X}))},
with \src{contractQS} an example.
The name for these routines is the
same as for the respective MEX routine
up to a trailing~\str{QS}, i.e.,
\src{fooQS()} becomes \Srt{foo()}. 
To locate uniquely which function gets called,
one may use Matlab's command \idx[which]{which}
in any context.

More generally, when an \atQSpace routine \Srt{foo()}
is called, it operates on an input that includes
at least one \QSpace object. It may include calls
to multiple MEX routines or none at all.
If \QSpace tensors are returned, they are usually
cast into \mQSpace objects if they exist as plain 
data structures, e.g., out of MEX routines.
The \atQSpace routines typically perform
additional tasks coded within Matlab.
Therefore the type and order of the return arguments are
not necessarily the same as compared to a respective
MEX routine called. For example, \Srt{eig()}
mimics the output of Matlab's native \src{eig()}
routine rather than that of \src{eigQS()}.

\subsection{MEX environment}

\idx[MEX]{MEX} routines are \Cpp compiled binaries that
can be called from
within Matlab like any other Matlab function.
From a \Cpp point of view, the standard entry point
\begin{minted}[escapeinside=??,firstnumber=0]{text}
   int main(int nargin, char **argin) { ... }
\end{minted}
is replaced by
\begin{minted}[escapeinside=??,firstnumber=last]{text}
void mexFunction(
    int nargout, mxArray *argout[],        ?\mlbl{nargout}?
    int nargin, const mxArray *argin[]) {  ?\mlbl{nargin}?
    ...
}
\end{minted}
where the Matlab arguments via the generic data
type \src{mxArray*} are handled via the Matlab API.
The input in \mline{nargin}
is by \src{const} pointer, and hence
{\it by reference} and read-only.
This way \QSpace can \idx[avoiding-copies]{avoid copies}
for input {\QSpace}s where read-only suffices.
All MEX routines in \QSpace are also effectively wrapper
routines that package a particular functionality
of the \Cpp \QSpace sources. This way, 
all MEX files are typically rather simple \Cpp
files themselves.

\paragraph{Memory management}

Once a function is called in Matlab, it stays resident
in memory so that later calls gain performance.
The same also holds for MEX routines,
with the effect that global variables in MEX routines
are {\it persistent} in memory, i.e.,
they keep their data from one MEX call to the next.
This is widely exploited by \QSpace
when dealing with the \RCR{\RCS}, \mbox{because}
data can be read on demand once and for all.
Nevertheless, this data cannot be shared across
different MEX files. 
Functions and their associated memory spaces
can be cleared in Matlab via
\src{clear functions}. This triggers standard
\Cpp garbage collection which ensures that the global data
with MEX routines is also properly released.
It is important to note, though, that \QSpace{} {\it does not}
use the memory allocation routine provided by the
Matlab API, but rather uses standard \Cpp memory management.
Hence while \Srt{clear functions} does release
all memory allocated by the MEX functions via the
underlying \Cpp class destructors, this is not necessarily
visible in the memory consumption of the 
Matlab process as a whole from the perspective
of the operating system. Subsequent MEX calls may reuse
freed up memory. The detailed behavior here, however,
very much depends on the particular operating system.

\paragraph{\QSpace avoids copies of arguments at MEX interface}
\label{idx:avoiding-copies}

The MEX API hands over input variables by constant reference
(\mline{nargin} above).
This is used by \QSpace to avoid generating explicit
copies of input \QSpace tensors where possible.
That is, in Matlab spirit, \QSpace copies
are generated {\it just in time} (JIT) only where required.
The data format of generalized column-major for tensors
is consistent across all Matlab and \QSpace arrays.
With Matlab's switch to interleaved complex data 
\idx[MEX]{since 2018}, this is also consistent with the data format
for complex numbers in standard LAPACK/BLAS routines.
Similarly, also the output in \mline{nargout} above
is handed over as a pointer.
Therefore to avoid copying output data back to Matlab,
output \QSpace tensors
are created in MEX compliant data spaces as soon as
possible within the \Cpp core routines.

\paragraph{Error handling}
\label{idx:ERR}

MEX routines temporarily switch the error handling
from Matlab to standard \Cpp
which also concerns the dealing with interrupts
such as \src{Ctrl-C}. Errors
are therefore caught within \Cpp. Before returning
to Matlab, however, the Matlab-specific
signal handlers are restored. This is ensured by an
outer try/catch block in every MEX routine.

When interrupting Matlab while inside a MEX routine,
the behavior is typically as follows:
When \src{Ctrl-C} is pressed the first time,
(1) this instantaneously prints a message
that confirms that the interrupt was received.
However, the MEX routine will continue and still try
to finish its current step until the next point
that is natural for the MEX routine to terminate,
e.g., after some finished iteration in a loop.
When pressing \src{Ctrl-C} additional times,
the MEX routine gives (2) a warning that the MEX routine
will be terminating immediately, and (3) actually
terminate immediately. Enforcing the termination is
considered unsafe, though, with potentially
erroneous behavior. This motivates the above behavior.

\paragraph{Warnings}
\label{idx:WRN}

Warnings in MEX routines are typically general
\src{\idx[wblog]{wblog}} messages to \src{stdout}
both within Matlab scripts, as well as in MEX files
with the tag \str{WRN} in them. In MEX routines,
this reflects standard output, though. Hence this
will typically not be recognized by the Matlab
debugger as a warning (this is in contrast to MEX
errors \str{ERR} which always are dealt with 
as an error also within Matlab by throwing
an exception).
Only in very few instances, does a warning
within a MEX function also escalate to the
Matlab environment as a proper Matlab warning that can be
caught via \src{dbstop if warning} with the Matlab debugger.
So far this only concerns \src{contractQS}. This MEX routine
issues a proper warning when it tries to automatically
fix missing \src{conj} flags when contracting tensors,
or when regular expressions operating on \itags within
\src{contractQS} have no effect because they do not match.
Since this should be checked and fixed in the Matlab
code right where it is called, it is useful to be able
to stop the Matlab execution there.

\subsection{HPC environment and parellization}

\paragraph{MCC compilation}
\label{idx:MCC}

The Matlab compiler (MCC) is a toolbox that can be acquired
with Matlab. It permits one to decouple an arbitrary
number of cluster jobs from Matlab license requirements
since the Matlab license is required only for
the MCC compilation itself. This collects all relevant
code from the Matlab path, including MEX binaries.
It builds a standalone application that can 
eventually be deployed without a Matlab license.
While there is no speedup in this process,
the output of the MCC compilation contains everything that
is required for a cluster job to run.

\paragraph{Parallelization (multiple jobs)}

When running multiple jobs, it is highly
recommended to have a mostly complete \RCR{\RCS} 
that is used in a centralized fashion, e.g., over
a network drive, in combination with a differential setup
with job-specific local storage (see more on this
with the environmental variable \idx[RCStore]{\RCS}).
If an \RCS needs
to be built from scratch, a good practice
is to run a test job upfront which generates
most of the symmetry-related data.
When post-analyzing job data where, rather than
just reading out data, one non-trivally operates
on \QSpace tensors still 
which may potentially add to an existing \RCS
like contractions,
then the (differential) \RCS that was used to
generate the data in the first place, must also be
accessible.

\paragraph{Parallelization (single job)}
\label{idx:OMP}

\QSpace parallelizes at the level of MEX routines
which is motivated by the following:
\QSpace tensors intrinsically consist of lists of
blocks (RMTs) each of which is stored in full, 
generalized column-major format. 
When working through such lists, e.g., in the context of
contraction, eigen- or singular-value decomposition,
it is desirable to parallelize the respective loops.
For the RMTs, this amounts to mostly trivial
parallelization.
For example, in the context of contractions
there are frequently a very large number of pairwise contractions
of RMTs to be performed which have matching symmetry sectors
in the contracted indices. In the presence of
multiple symmetries and rank $r>2$ tensors, this
can quickly reach hundreds or many thousands of individual
pairwise contractions \cite{Bruognolo21}. 

In this sense, {\it outer loops} are parallelized
in \QSpace using \texttt{openMP}.
This responds to the environmental variable
\src{\idx[QSP_NUM_THREADS]{QSP\_NUM\_THREADS}}.
At a lower level, also the elementary steps of operating
on individual (pairs of) RMTs are parallelized
simply by linking the standard
\idx[lapack]{\mbox{LAPACK/BLAS}} libraries.
These parallelize themselves, typically responding to
\src{OMP\_NUM\_THREADS} (since Matlab has a tendency
to overwrite or ignore those, one may rather use
\src{setNumThreads.m}, instead, as provided with 
the \QSpace repository).
Together with the \texttt{openMP} parallelization,
this gives rise to {\it nested} parallelization,
in that every worker in the loop itself can parallelize.

Parallelizing loops, ideally, relates to trivial and
thus efficient parallelization. This is mostly the
case for the RMTs, and hence also in the presence 
of only abelian symmetries, where all symmetry-related
constraints can be quickly obtained on the fly.
For non-abelian symmetries,
however, \QSpace makes use of an \RCR{\RCS} that
needs to be kept consistent across all workers.
Therefore, critical symmetry-related operations
that may lead to updates in the \RCS need to be
forced into serial mode via \src{critical} sections
or object-specific openMP locks (mutex and semaphores)
for synchronization.
Since symmetry-related operations can trigger a range 
of {\it a~priori} unexpected additional actions, this slows
down the parallelization of loops in \QSpace for larger
\src{QSP\_NUM\_THREADS}. Hence typically, one may use
\src{\idx[QSP_NUM_THREADS]{QSP\_NUM\_THREADS}}\,$\lesssim 8$.

From the above outset, the parallelization is optimized
by combining openMP with native parallelization
of LAPACK/BLAS routines. Much of this parallelization
is CPU-centered still. Hence as of now, \QSpace
does not support GPU parallelization. If in the future
openMP together with LAPACK/BLAS support GPUs in a seamless
fashion, this should enable access to GPUs for \QSpace.

\paragraph{Subtleties and potential issues with parallelization}

Tensor operations can trigger an {\it a priori} unexpected
set of additional actions that need to be properly 
synchronized across different threads when parallelizing.
The following, for example, shows how contractions
can trigger tensor product decompositions
which may result in proliferating requests for
CGT-specific locks, and worst case, in deadlocks.

To be specific, consider the example of contracting two tensors.
The contraction needs to be performed both, at the level
of RMTs, as well as at the level of CGTs for each non-abelian
symmetry involved. Because the latter are tabulated,
most of the CGTs may likely already be present from
earlier simulations.
But there is a chance that new symmetry-related data
is created.
Therefore when a particular CGT resulting out of the
contraction is requested as a reference (CGR) from the
internal buffer in memory,
there needs to be a lock on this particular object
in the buffer: after all, it may not be in memory yet
e.g., because a particular simulation just started.
The thread thus (i) issues an object-specific openMP lock
and then (ii) reads the object from the buffer. If present,
it can release the lock. If not yet present,
the thread keeps the lock active, and (iii) tries
to read the CGR from the \RCS.
If not present there either, the thread
(iv) will generate the requested CGT, e.g.,
by performing the contraction explicitly
and tabulating the result for the future,
before releasing the lock.

With the further assumption that the contraction 
results in a rank-3 tensor that has two incoming
and one outgoing index or vice versa,
this reflects a standard CGC that can also be obtained
from tensor product decomposition, instead.
If at least one input multiplet
is of smaller dimension, in the sense that
it may reasonably occur in a local state space of a site,
\QSpace does not obtain the CGT by contraction,
but (v) rather gives preference to the full tensor product
decomposition, as this automatically also generates full 
\idx[OM]{outer multiplicity} where present.
This is carried out by the present thread 
in addition to (iv) above.
The thread then (vi) projects the contracted CGT 
in (iv) onto the CGT out of the tensor product
decomposition in (v), before it finally releases the lock.

Now to further complicate things,
a tensor product decomposition not only generates
the particular CGT of interest, but typically also
additional other CGTs.
Accidentally, some of these may
coincide with CGTs that occur in the present tensor contraction
as encountered by {\it other} threads which simultaneously also
try to acquire locks themselves.
Worst case, such scenarios can
result in deadlocks if \src{\idx[QSP_NUM_THREADS]{QSP\_NUM\_THREADS}}
is chosen too large. This situation, however, can only occur
if a sufficiently large number of new objects needs to be generated
for the existing \RCS. Hence a pragmatic way to deal with competing
locks as described above, is to test run a particular
simulation first in serial mode by unsetting
\src{\idx[QSP_NUM_THREADS]{QSP\_NUM\_THREADS}}.
This will generate most of the required entries
in the \RCS, which then can be made accessible
globally in the spirit of a differential storage setup
via the environmental variable \RCR{\RCS}.

\subsection{Environmental variables}
\label{app:ENV}

The following environmental variables are specific
to \QSpace.
They are defined when setting up the
\happ[ENV:setup]{Matlab environment}
for \QSpace via \src{system/matlab\_setup.sh}.
The mandatory ones are indicated,
otherwise default values will be auto-determined
by \src{matlab\_setup.pl} which finalizes the setup.
As these variables are used in various places
either during MEX compilation, startup of Matlab,
or while using \QSpace within Matlab,
they are expected to be exported
to sub-shells, (e.g., using bash \src{export}).

\paragraph{\src{ARCH}} system architecture 
(auto-determined based on system call \src{uname~-sm},
by default) --
Used in \src{Source/Makefile}
to distinguish between system environments as specified
with Matlab in \src{\var{\srcMLR}/sys/os/\var{ARCH}/},
such as \src{glnxa64} for Linux or \src{maci64} for macOS. 

\paragraph{\src{LMA}}
\label{idx:LMA}

local Matlab data directory (path, mostly optional) --
In some applications provided with \QSpace,
this expects the designation of a (local)
Matlab directory intended for data storage. For example,
used with \str{cto lma} to change directories (`change to')
which then looks for the environmental variable
in upper case letters \src{LMA}. E.g., see \mline{nrg:cto}
with \Sec{NRG} above.

\paragraph{\srcMLR}
\label{idx:MLR}
Matlab root directory (path; mandatory) --
The location of the Matlab install: if multiple
Matlab versions are present, this typically
includes the Matlab version string.
The Matlab binary is expected to be located in
\src{\$(\srcMLR)/bin/[matlab]}.
The variable \srcMLR is important for the compilation via
the \src{Makefile}s since these need to link to headers
and libraries in the Matlab install.

\paragraph{\src{MCC\_TAG}}
system dependent subdirectory in \src{./MCC}
used with MCC compilation (string which
defaults to \str{bin*} based on the MEX file
extension \src{[mex]*}).

\paragraph{\srcMYM}
\label{idx:MYM}

location of the cloned \QSpace git repository (path; mandatory)
-- Since the repository also includes an extended Matlab
environment with wrapper and utility routines
from within which applications and simulations are run,
this is referred to as `my Matlab', typically
located somewhere within the HOME directory like
\src{$\sim$/Matlab/}.

\paragraph{\src{MEX}}
location of MEX / \Cpp source directory (path; defaults to \src{./Source}).

\paragraph{\src{QSP\_NUM\_THREADS}}
\label{idx:QSP_NUM_THREADS}

(non-negative integer) --
Number of threads to be used with \QSpace
on top of standard LAPACK/BLAS parallelization
(which itself parallelizes based on \src{OMP\_NUM\_THREADS}).
Combined, this gives rise to \idx[OMP]{nested parallelization}.
If \src{QSP\_NUM\_THREADS} is not set,
no such parallelization is performed.

\paragraph{\src{QS\_LOG\_COLOR}}
\label{idx:QS:color}

Enables/disables color coding
(intended for Matlab \idx[ml:term]{terminal mode} only, 
assuming a dark terminal background) --
By default, the color coding is turned on in interactive
terminal mode where it may be disabled by\,
\src{export QS\_LOG\_COLOR=0}. The color setting 
affects both, the output of MEX routines, as well as
log entries purely within Matlab, e.g., based on the
\src{lib/wblog.m} logging routine.

\paragraph{\src{QS\_CONFIG\_ML\_SH}}
\label{idx:ML-CONFIG}

Optional specification of bash configuration file
for Matlab \happ[ENV:setup]{system environment}
(used by \src{system/matlab\_setup.sh};
defaults to \src{system/matlab\_setup\_user.sh}).

\paragraph{\src{\RCS}}
\label{idx:RCStore}

The environmental variable
\[
     \src{{\RCS}=path1:$\ldots$:pathN}
\]
points to the location where symmetry-related
data is written or, subsequently, read. The variable \src{\RCS}
contains one or multiple paths separated by a colon
\str{:} for the sake of a {\it differential} storage setup,
where the last path is assumed to contain the newest
symmetry-related data if any. To be specific, 
the lookup of symmetry-related data proceeds as follows,{\color{csrc}{
\[
     \underbrace{
     \src{pathN} \ \to\ \src{path1} \ \to\ 
     \ldots\ \to\ \src{pathN-1}}_{\text{read until found}}
     \quad\overset{\text{not found}}{\longrightarrow}\quad
     \underbrace{\src{pathN}}_{\text{write}}
\]}}%
When \QSpace requires a particular symmetry
related object, this starts by looking into the {\it last}
(because newest) path, \src{pathN}.
If the object looked for is found there,
it is read and returned. If the object is not found,
\QSpace proceeds to the {\it first} and then the subsequent
directories from there, until the object is found.
If nowhere found, then the desired symmetry-related object
is computed, where by the iterative nature of \QSpace,
all required input for this is assumed to be present from
the earlier \RCS history. The result is then stored
in the {\it last} path, \src{pathN}, in \RCS
and, finally, also returned.

The above procedure permits a differential setup
where all paths
except the last one (\src{pathN}) are read-only.
New or updated data is written to the last
path only. For an interactive Matlab session,
a single path is recommended. For cluster jobs,
the recommended setting is two paths:
\src{{\RCS}=(global\_path):(local\_path)},
where \src{global\_path} points to an \RCS
that is mostly complete in the required symmetry-related data
based on earlier calculations. This may be available
from a network drive, where required entries are typically
read once and for all, and stored in memory (larger CGT
objects are frequently purged from memory, though,
since CGRs are typically all that is needed
once no more new OM
components are generated for the CGTs under consideration).
By contrast,
the \src{local\_path} is meant to be a local
directory with the cluster job. The major reason for 
this differential setup above is to avoid race conditions
and thus clashes across  different cluster jobs 
operating on the same \RCS 
[newly generated \hsec[RCStore]{CGTs get assigned IDs}
which are referenced elsewhere, such as in \Xsymbols;
therefore these must be consistent across the entire \RCS;
they may get corrupted in case of race conditions
from parallel executions of different simulations,
e.g., cluster jobs; by contrast,
the parallel execution within threads of the
{\it same} simulation
uses OMP critical sections, as well as other
OMP locks based on mutex and semaphores
for synchronization].

Once all jobs are finished, their \RCS 
have diverged from each other if \src{local\_path}, 
is non-empty. This way they have become incompatible 
with each other.
However, for future purposes, one may choose
one particular job, e.g., one that generated the most new
symmetry-related data in terms of the number of file entries
or disk usage in \src{local\_path},
and simply \src{rsync} all content in its
\src{local\_path} on top of \src{global\_path}
(assuming there are no more active jobs
that still access \src{global\_path}).

\paragraph{\RCSy}
\label{idx:RCSync}

(optional path)
The environmental variable \RCSy
is complimentary to \RCS. For historical reasons, it is
intended for file locks only to synchronize
symmetry-related operations across parallel jobs.
However, since file locks are
not entirely reliable over network drives,
if configured at all due to their impact on performance,
the initial concept of file locks has been mostly
abandoned in \QSpace, such that setting the
variable \RCSy is optional. For cluster jobs
the recommended setting, instead, is to use a differential
\RCR{\RCS}.

\newpage

\subsubsection*{Other relevant external environmental variables}

\paragraph{\src{DISPLAY}} 
\label{idx:DISPLAY}

(system variable)
handles the graphical display of graphical output
from Matlab when run in terminal mode (assuming Matlab
is not run with the \src{-nodisplay} option). The 
variable DISPLAY belongs to the X11 window system
(e.g., \src{XQuartz} on macOS) and is usually
automatically set if X11 is present on one's system.
\vspace{-1ex}

\paragraph{\src{PATH}} 
\label{idx:PATH}

(system variable) contains system path  --
Matlab and the \Cpp compiler must be visible on the system
PATH. Therefore it needs to be properly adapted. Furthermore,
the current directory~\str{.} is 
\idx[mlsetup:caveats]{expected} to be included
in the system \src{PATH}, such that \QSpace specific
local bash or perl scripts, e.g., in \src{./system}
or \src{./Source} are found.

\subsection{{\tt RC\_STORE} database}
\label{sec:RCStore}

The \RCS is a file-based database that stores all
symmetry-related content for non-abelian
symmetries \cite{Wb12_SUN,Wb20}.
All data is computed on demand once and for all, and then
stored for later usage.
The root path of this database is specified via the
environmental variable \idx[RCStore]{\RCS}.
In order to avoid race conditions across parallel jobs,
this supports a \idx[RCStore]{differential setup}
via the specification of multiple paths.
Nevertheless, for the sake of the argument here,
\RCS is assumed to specify a single path.

For each symmetry \var{sym}, \QSpace generates a sub-directory
\src{\var{\RCS}/\var{sym}/}, e.g., with \src{sym=SU2}, etc.
Referring to this as the current directory \str{.}
in what follows,
\QSpace maintains a detailed log file \src{./\var{sym}.log}
for all operations carried out for symmetry \var{sym}.
Furthermore,
\QSpace stores and maintains three database directories,
R-, C-, and X-stores (for historical reasons, this is only
partly reflected in the name {\tt `RC'\!\!\_STORE},
since \src{X} was added with \QSpace v3).
All data files are stored in Matlab binary format
(\src{mat}-files), even if all file extensions have been
altered by internal \QSpace conventions,
simply to reflect their content.
\begin{itemize}
  \setlength{\itemsep}{-0.1\baselineskip}

\item \src{./RStore} -- 
stores all irreducible representations $R_q$
for the ireps $q$ encountered during given
\RCS history for symmetry \src{sym}. The respective
data is stored in files \src{RStore/(q).rep}
where the extension indicates `representation'.
Here $R_q$ includes a full basis
decomposition in terms of weight labels which may 
include degeneracy, i.e., inner multiplicity \cite{Alex11},
as well as a sparse representation of the diagonal
\idx[Cartan]{Cartan subalgebra} (generalized `$z$-operators'
\src{Sz}$\equiv S_z$)
and of the \idx[roots]{simple roots}
(generalized minimal set of raising operators
\src{Sp}$\equiv S_+$)
of the Lie algebra \cite{Cahn84,Wb12_SUN}.

\item \src{./CStore} -- 
stores all \idx[CGT:sorted]{sorted CGTs} $C_q$ as files \src{CStore/\var{qdir}/(q).cgd}
where the extension indicates generalized
{\it Clebsch-Gordon data}. All CGTs are stored 
in sparse format generalized to arbitrary rank
and in roughly quad-precision (GMP/MPFR).
The latter ensures that all CGTs are exact
when converted to double precision.
The CGTs are grouped in
subdirectories based on their sorted \idx[qdir]{$q$-directions}
\var{qdir}
[incoming (\qdir{+}) before outgoing (\qdir{-})]
which in their combination thus also represents the tensor rank.
For example, standard CGCs are stored in the subdirectory
\src{CStore/\qdir{++-}/}. 
All \hsec[1j]{\onej symbols} are stored
in \src{CStore/\qdir{++}/}.
These duplicate the entries of the type $(q\bar{q}|0)$ in 
\src{CStore/\qdir{++-}/} if present \cite{Wb20}.
The latter directory also stores the fusion rules
as obtained from full tensor-product decompositions.
As these represent {\it maps} of $2$ in- into $1$
out-going leg, i.e., operates on a total of {\it 3} legs,
this is reflected in the file extension \src{*.mp3}).

\item \src{./XStore} -- 
stores the \Xsymbols \cite{Wb20}
derived from all encountered pairwise CGT contractions
$C \equiv A*B$ that are not trivially zero due to non-permissible%
\footnote{The check on zero-contraction of a pair of CGTs
is based on simple
checks concerning the resulting rank-2 or rank-3 CGTs $C$ only.
Therefore many contractions that resulted in zero for rank $r>3$
are stored, nevertheless. While such contractions may be zero
during the initial build of the \RCS
depending on the particular contraction performed,
the resulting CGT can be permissible, nevertheless,
and hence non-zero eventually.}
combinations of symmetry labels in the uncontracted
indices present in $C$.
With this the \src{XStore} typically
contains by far the largest number of files. Most of these
are small, because \Xsymbols are small objects in general.
The \Xsymbols are stored in files like
\src{XStore/\var{qdir}/(A)\_ia (B)\_ib.x3d} 
where the extension indicates \Xsymbol{} {\it data}
which always operates on {\it 3} outer multiplicity indices.
The file name itself shows the contraction performed
with reference to the input tensors \src{A} and \src{B}
(specified by their compact $q$-labels), as well as
the respective contracted indices \src{ia} and \src{ib}
(also in compact format if multiple indices are contracted
simultaneously).
The files are grouped into subdirectories based
on the \idx[qdir]{$q$-direction}
\var{qdir} of the {\it resulting} tensor $C$.

\end{itemize}

\paragraph{Sorted CGTs}
\label{idx:CGT:sorted}

The \src{./CStore} stores {\it sorted} CGTs only:
For one, this includes the more trivial ordering that
all incoming ({\tt +}) legs are listed before
all outgoing ({\tt -}) ones.
Yet this also includes the considerably more subtle ordering
within each group of directions where the legs are sorted
w.r.t. their $q$-labels in lexicographic order.
If there are more outgoing than incoming indices,
then the conjugate CGT is stored.
Therefore all CGTs in the \RCS have at least as many
incoming indices as there are outgoing ones.
These \idx[qdir]{$q$-directions} are then used as
subfolder names to organize the CGTs in the \RCS,
like \src{CStore/\qdir{++-}/} for standard rank-3 CGCs.
By only storing sorted CGTs in the above sense,
this avoids the proliferation of closely related entries
in the \RCS. A general non-sorted, and in this sense
unconstrained CGT can be obtained, nevertheless,
\idx[CGT-CGR]{in combination with a CGR}.

\paragraph{{\tt RC\_STORE} initialization}

Whenever a symmetry \var{sym} is requested for the first time,
\QSpace creates and initializes the above \hsec[RCStore]{directory
structure} in \src{\var{\RCS}/\var{sym}/}.
It creates the representation $R_{\rm def}$
for the defining irep in the \src{RStore}, as well
as its dual if different. \QSpace then loops
several times by generating tensor-product decompositions
of the existing content in the \src{RStore} with itself
while putting an upper threshold on the product dimension
of the input for the case of larger-rank symmetries.
This is intended at the very least to generate all smaller
ireps  that typically are required in the description
of a local site prior to performing any contraction.

The initialization process is very fast for small symmetries
such as seconds for SU(2), but becomes quickly more elaborate
for larger symmetries [the cost grows exponentially with the
\idx[rsym]{rank} of the symmetry; cf. \Eq{qdim}].
For rank-2 symmetries like SU(3), the initialization
process should be completed within minutes.
But for SU(4) one should be already prepared to wait
several hours while the \RCS is built the first time.
The initial process of generating lower-dimensional ireps
may be interrupted, though,
in the worst case by killing the process
(\idx[ERR]{Ctrl-C} while still waiting for the present
generation cycle to finish).
The next time the \RCS is accessed for a given symmetry,
it already exists.
Still, subsequent computations for larger-rank symmetries
may take considerable time in any case, since the \RCS
still needs to be extended in the course of
building a many-body Hilbert space.

Once a tensor network simulation is finished,
though, by construction all relevant symmetry related
data is present. That is, repeating the same or similar
simulations will have access to a (near) complete \RCS.
At this point the full speedup of exploiting non-abelian
symmetries is achieved. This is also a major reason
why a \idx[RCStore]{centrally maintained} \RCS is important
for larger symmetries, like beyond SU(3),
as this can make use of existing symmetry-related data.
Besides, later access to an \RCS is also important
if one wants to analyze and still operate non-trivially
on \QSpace tensors at a later time in a new Matlab
session or with memory cleared, since \QSpace tensors
are tied to a \idx[RCS:hist]{particular} \RCS.

\subsection{Logging}
\label{idx:wblog}

\QSpace uses a streamlined format when logging data
to standard output (\src{stdout}). This is based on
the routine \src{wblog()} that behaves similarly in
Matlab scripts (which call \src{lib/wblog.m}),
as well as with a \Cpp analog in MEX routines.
It represents a tweaked formatted output to the \src{stdout} stream,
like \src{fprintf(stdout,fmt,arg1,arg2,...)}
with \src{fmt} some format string. This translates
to \src{wblog(tag,fmt,arg1,arg2,...)}.
For example,
\begin{minted}{matlab}
   wblog('TAG','some info (value of pi=\%g)\nand more\N-> hint: ...',pi)
\end{minted}
generates output with the format
\begin{minted}[firstnumber=last]{text}
   [file-name:lineno]   HH:MM:SS  TAG some info (value of pi=3.14159)
   [file-name:lineno]   HH:MM:SS  TAG and more
   -> hint: ...
\end{minted}
It automatically prepends a header portion of about
$36$ characters that includes an automatically determined
pointer into the source code at the location of the
\src{wblog} command for reference
[file name (possibly abbreviated) and line number (lineno)],
a time stamp in 24H format,
and a 3-character tag
(the first input argument) as described below.
The remainder of the \mbox{input} to \src{wblog}
is mostly forwarded to \src{printf()} and hence accepts
the same syntax with minor additional tweaks
concerning the format string.
Newlines are permitted, as in the
example above, which then repeat the header portion.
Tweaks include, e.g., the non-standard
\src{{\textbackslash}N} which inserts a newline
{\it without} the header portion.
Every \src{wblog} entry is terminated with a newline.
This can be prevented by a trailing \str{\textbackslash\textbackslash} in the format string.

The 3-character TAG is used for a rough categorization
of log entries via tags. It can be chosen arbitrarily,
but is typically within the following set that has
special meanings assigned:\\[2ex]
\mbox{\hspace{.3in}}\begin{tabular}{p{.5in}l}
  \str{ERR} & for errors\footnotemark \\
  \str{WRN} & for warnings\footnotemark \\
  \str{<i>} & for more important infos \\
  \str{I/O} & for I/O related messages \\
  \str{NB!} & for critical things to be aware of (\idx[NB]{`nota bene'}), \\
  \str{SUC} & succeeded with a certain operation  \\
  \str{ok.} & some successful check performed \\
  \str{*{ } } & log-level 1 \\
  \str{ * }   & log-level 2 \\
  \str{ { }*} & log-level 3 \\
  \ \ \src{...} & etc.
\end{tabular} \\
\addtocounter{footnote}{-1}
\footnotetext{The \str{ERR} tag throws an error in the MEX files,
but continues in Matlab code. To also throw an
error in the latter case, \QSpace uses \src{wbdie()}.
For the debugger to also stop with \src{ERR} in Matlab code,
\QSpace provides the command \src{dberr}.}
\addtocounter{footnote}{1}
\footnotetext{
The \str{\idx[WNR]{WRN}} tag in the MEX files and also
Matlab code usually
does not raise any signal or issue, but simply continues.
It leaves it up to the user to check the log output for warnings.
Nevertheless, 
to stop the debugger in an interactive session
with \str{WRN} in Matlab code,
\QSpace provides the command \src{dbwrn}.
}

\noindent
When using \QSpace in \idx[ml:term]{interactive
Matlab terminal mode}, some log output is shown
in color for emphasis depending on their TAG
(like \src{ERR} messages in dark red, \src{WRN}s
in lighter red, \src{NB!} or \src{OK!} in dark green,
\src{ok.} in lighter green, etc.; the particular color
coding may change in future versions, though).

The leading pointer into the source code facilitates to 
track down where a particular log message originated from,
be it a \Cpp reference from within a MEX file,
or reference to a Matlab script.
The file name always reflects the bare file name,
i.e., always skips any path. It may be extended to include
program tags, current Matlab subroutine, etc. In order
to align log entries with respect to the subsequent
time stamp, 20 characters are reserved for \src{file-name:lineno}.
If it does not fit within that space, various schemes 
for abbreviating the file name are adopted
depending on the context, thus giving preference
to the readability of the log output by aligning entries.
Skipped string portions in the file name are indicated
by a prime (\src{'}).

\section{\QSpace applications in the repository} 
\label{app:QS:apps}

The \QSpace repository also includes
state-of-the-art implementations of
fdm-NRG \cite{Wb07} and DMRG.
The standard \hsec[NRG]{iterative diagonalization} of the NRG
was already discussed in more detail in \Sec{NRG}.
However, the full implementation of fdm-NRG
is considerably more involved and its documentation
beyond the scope of this documentation.
Similar so for the DMRG. Please consult
the \src{./Docu} folder in the git repository
for future updates. Nevertheless,
it is emphasized here in passing,
that these implementations already exist in the
public \QSpace repository. The interested
reader may find it rewarding to explore.
In case of more detailed interest, please
also feel free to \idx[contact]{contact the author}
in this regard.

\subsection{DMRG simulations}
\label{sec:DMRG}

By default, \src{runDMRG} located in the
\src{DMRG/} folder launches a DMRG
simulation for the ground state of the isotropic
Heisenberg model (\src{wsys=\str{Heisenberg}})
with nearest-neighbor interaction $J=1$
and next-nearest neighbor interactions set to $J_2=0.25$
exploiting SU(2) spin symmetry,
\begin{minted}[escapeinside=??]{matlab}
  runDMRG
\end{minted}
The Hamiltonian is encoded in \src{HAM}
which represents the \QSpace Matlab class
\Srt{@Hamilton1D}. Simply typing \src{HAM}
will show essential information concerning the
model Hamiltonian.
The model parameters are stored in the structure
\src{HAM.info.param}. The lattice structure
can be inspected via \src{plot(HAM)}.
A summary of the DMRG sweeps can be generated by the
plot script \src{plot\_Hamilton1D}.

\subsection{NRG impurity solver}

A typical NRG run involves two parts \cite{Wilson75,Bulla08}:
(i) \hsec[NRG]{Iterative diagonalization} (line 1 below which
by default selects the single impurity Anderson model).
This collects finite-size
spectra in kept and discarded resolution \cite{Anders05,Wb07}
as already explicitly demonstrated in \Sec{NRG}.
This then permits one to (ii) define a thermal state via
the \idx[fdmNRG-0]{full density matrix (fdm)}
for statistical properties but also spectral properties
(line 4 below)
that can be evaluated efficiently in textbook-like
fashion in Lehman representation \cite{Wb07,Wb12},
\begin{minted}[escapeinside=??]{matlab}
  rnrg;     % NRG iterative diaognalization
% nrg_plot  % generates NRG energy flow diagram

  rfdm      % computes fdm-NRG correlation functions
% fdm_plot  % summarizes computed fdm-NRG spectral data
\end{minted}
A graphical summary of the results 
can be generated by the respective plot scripts
in lines 2 and 5 (these scripts are automatically
called at the end of the respective previous commands
in lines 1 and 4, hence they are commented out).
Further hard-coded \QSpace MEX applications exist
in the fdm-NRG context in the repository, as listed 
in \Sec{NRG:apps} together with the MEX routines
underlying \src{rnrg} and \src{rfdm} above.


\section{Additional information and documentation}
\label{app:Xdocs}

\subsection{Additional usage and help info} 
\label{app:help}

All \QSpace routines, MEX as well as Matlab
functions, come with their own help and usage
information (like `man pages'). It can be looked up
with the standard Matlab \src{help} command.

\paragraph{Help and usage of MEX functions}
\label{idx:MEX:help}

For \idx[MEX]{MEX} routines denoted by \src{fooQS} here,
the following are all equivalent
\begin{itemize}
   \item \src{help fooQS}
   \item \src{fooQS -h}
   \item \src{fooQS -?}
\end{itemize}
where the last two are simple redirects to the first call.
That is, all help files to MEX files are stored as Matlab m-files.
For this reason, Matlab finds {\it two} entries for every MEX routine in its respective directory:
the actual MEX binary \src{fooQS.mex*},
but also \src{fooQS.m}. The latter contains usage information
based on an extended header comment only without any actual source.
It is recommended that both files are
located in the same folder in the \happ[QS:env]{Matlab path}.
Then when calling the function, this finds the \src{fooQS.mex*}
file first. When calling the help, Matlab ignores
the MEX file and thus finds the \src{fooQS.m} first, instead.
Further flags supported by the \QSpace MEX files are
\begin{itemize}

   \item \src{fooQS {-}-version} \quad
   shows version information of \QSpace and linked libraries 

   \item \src{fooQS {-}-ping\ } \qquad
   dummy call of a MEX routine with the effect \\
   \phantom{\src{fooQS {-}-ping\ }} \qquad 
   that it becomes resident in the Matlab workspace

\end{itemize}

\paragraph{Matlab debugger}
\label{idx:ML:debug}

The Matlab debugger offers a useful low-level tool that
permits one to gain better insight into the behavior
of particular Matlab sources. Matlab simulations can be stopped 
at any location in a Matlab function (at specified line \#)
by setting a breakpoint,
\begin{minted}[escapeinside=??]{console}
   dbstop in [class/]foo [at line #]
\end{minted}
This permits one to inspect and
alter variables in a particular context. Also in the absence
of any error whatsoever, this can be useful for analyzing
or a better understanding of the behavior of a code. The
commands \src{dbup} and \src{dbdown} traverse the caller stack.
See also the \QSpace related \src{dberr} and \src{dbwrn}
that respond to \src{\idx[wblog]{wblog}}.

\paragraph{Help and usage of Matlab (class) routines}
\label{idx:ML:help}

Similar to the help on MEX routines above,
also all Matlab routines have a help that can be looked up
in standard Matlab syntax.
For example, help on the Matlab \mQSpace class routines
(located under \src{Class/\atQSpace/}) based on the
placeholder routine \src{foo} can be obtained as follows:
\begin{itemize}
 \item\src{help foo} \quad
    display help for function \src{foo}
 \item\src{help QSpace/foo} \quad
    display help on a particular class method
 \item\src{methods QSpace} \quad
   listing of all existing methods in \src{Class/\atQSpace}
\end{itemize}

\paragraph{Identifying and locating functions}
\label{idx:which}

Given the many functions and methods in the \QSpace
environment within Matlab, one can identify and thus locate
the relevant functions 
at the Matlab prompt but also at any debug breakpoint,
with the native Matlab command
\begin{minted}[escapeinside=??]{console}
   which [path/]foo[QS] [-all]
\end{minted}
where terms in square brackets are optional.
The first line in the output then shows the routine
that has the highest precedence,
and hence would get called in given context.
The option \src{-all}
permits one to see all overloaded routines,
as well as class routines that share the same name.
With this,
all subsequent output lines, if any, show routines or
functions that are shadowed. A subsequent \idx[ML:help]{help}
on a particular function or MEX routine from that list
will provide more detailed usage information.

\paragraph{Matlab debugger}
\label{idx:ML:debug}

The Matlab debugger offers a useful low-level tool that
permits one to gain better insight into the behavior
of particular Matlab sources. Matlab simulations can be stopped 
at any location in a Matlab function (at specified line \#)
by setting a breakpoint,
\begin{minted}[escapeinside=??]{console}
   dbstop in [class/]foo [at line #]
\end{minted}
This permits one to inspect and
alter variables in a particular context. Also in the absence
of any error whatsoever, this can be useful for analyzing
or a better understanding of the behavior of a code. The
commands \src{dbup} and \src{dbdown} traverse the caller stack.
See also the \QSpace related \src{dberr} and \src{dbwrn}
that respond to \src{\idx[wblog]{wblog}}.

\subsection{\src{Docu/} folder}
\label{app:Docu}

The \QSpace repository also contains the documentation
folder \src{./Docu} that is maintained with the repository.
It may be looked up for comments on most recent changes,
such as \src{readme} files, or additional more recent
documentation. Other main entries are:

\paragraph{QSdocs.pdf}

This contains a complete listing of the \happ[help]{help} to all
MEX routines as well as \QSpace methods (member functions).
It has been auto-compiled into the single PDF file
\src{Docu/QSdocs.pdf}
(about 70 pages). This file may be consulted
in parallel to this documentation.

\paragraph{Doxygen}

Some general documentation of the \Cpp code
base that was auto-generated via \src{doxygen}.
This provides an overview of the general outline
of the \Cpp code structure. Its output is also located
in \src{Docu/html/} which can be viewed in a web browser
starting from the \src{Docu/html/index.html}.

\section{Listing of MEX Routines}
\label{app:MEX:list}

\subsection{\QSpace core MEX routines}

The following short description of the most relevant
MEX routines is sorted in alphabetic order. For a detailed
description of all MEX routines, please see
their respective \idx[MEX:help]{help}.
Note that the help output for
all MEX routines has also been automatically compiled
into the single PDF file \src{\happ[Docu]{Docu/QSdocs.pdf}}
for reference. The MEX routines are all coded in \Cpp,
with the sources located in \src{./Source}.
With about 100,000 lines of code, this 
far exceeds the present documentation which rather
focuses on \QSpace usage and applications.
Nevertheless, an auto-generated \happ[Docu]{doxygen}
documentation is included in the repository.

\paragraph{compactQS()}
\label{idx:compactQS}

Used by \Srt{\hsec[gls]{getLocalSpace()}} to bootstrap
\QSpace tensors from Fock space. This routine obtains reduced
matrix elements based on the Wigner-Eckart theorem
and existing rank-3 CGTs in the \RCS.
It ensures that all non-zero matrix elements
of the input were accounted for in magnitude and sign.
It is essential to \src{getLocalSpace},
but is typically never called explicitly or needed
elsewhere in an application.

\paragraph{contractQS()}
\label{idx:contractQS}

This routine performs (nested sets of)
pairwise contractions of \mbox{\QSpace tensors}.
It supports two modes: (i) An explicit mode for a single
pairwise contraction where contraction indices are fully
specified where $q$-directions together with \itags,
if set, need to match nevertheless.
(ii) \idx[ctr:semantics]{Auto-contraction}
based on \idx[itags]{\itags} with
options to set \itags on the fly: this mode
supports contractions of a set of tensors where 
the order of the contraction in terms of pair-wise
contractions is explicitly specified based on a nested
cell input data structure. In case of errors
in the input, the auto-contraction of nested
cell structures can be debugged by adding 
the verbose option \str{-v}.

\paragraph{diagQS()}
\label{idx:diagQS}

Returns all diagonal matrix elements of a scalar tensor as
a single column vector (hence for rank-2 tensors only);
In the presence of non-abelian symmetries,
this returns a second column that specifies
the degeneracies when converting multiplets to states.
See also \Srt{\QSpace/}\Idt[QS:diag]{diag.m}.

\paragraph{eigQS()}
\label{idx:eigQS}

Performs eigen-decomposition of a scalar tensor 
(hence for rank-2 tensors only). The first output argument
returns the full set of eigenvalues as column vector.
In the presence of non-abelian
symmetries, this returns a second column that specifies
the degeneracies when converting multiplets to states.
These degeneracy factors are important
as weight factors, e.g., for a normalized density
matrix spectrum.
This routine also implements various truncation schemes
with truncation based on combinations of
(i) maximal number of multiplets,
(ii) an energy threshold
(assuming the input is a Hamiltonian), or
(iii) weight threshold (assuming the input is a density matrix).
In this case the second output argument
returns a structure
with \QSpace tensors \src{AK}\,$\equiv A_K$,
\src{AD}\,$\equiv A_D$, \src{EK}\,$\equiv E_K$,
and \src{ED}\,$\equiv E_D$
for eigenstates (\src{A*)}
and eigenvalues (\src{E*})
split into kept (\src{K})
and discarded  (\src{D}) state spaces, respectively.
The eigenvectors in \src{A*} are stored 
as columns, and the eigenvalues in \src{E*}
in compact diagonal \QSpace representation,
meaning that this only stores the diagonals
as a vector in the RMTs. It may be expanded
using \idx[QS:diag]{\atQSpace/diag()}.

\paragraph{getDimQS()}
\label{idx:getDimQS}

(integer vector or matrix)
Get overall dimensions of a \QSpace tensor
for all legs based on all 
\idx[QS:recs]{\QSpace records}. For a tensor of rank $r$,
a $1\times r$ vector is returned for all-abelian symmetries.
In the presence of non-abelian symmetries,
a second row is returned. In this case,
the first row contains the effective
dimension (in terms of multiplets), and the second
row the corresponding full dimension based on states.
See also \idx[getQDimQS]{getQDimQS} for detailed
dimensions of symmetry sectors on a particular leg.

\paragraph{getIdentityQS()}
\label{idx:getIdentityQS}

Get identity tensor either (i) for the state
space with respect to a single \QSpace,
with an optional flag \str{-0} to obtain the
\hsec[1j]{\onej}-tensor, instead,
or (ii) get the identity \idx[Atensor]{$A$-tensor}
for the tensor product space of two state spaces
with reference to two input \QSpace tensors.
By default, as of \QSpace v4, the \onej symbol in case (i)
\idx[itag:markers]{marks} the dual space
on the second index with a trailing prime
(\src{'}) [for backward compatibility,
the same itag for the dual space is returned with
the option \str{-z}, standing for `zero'
as an alternative to \str{-0}.
However, the recommended way is to use markers
as with the updated behavior of \str{-0}].
\vspace{-1ex}

\paragraph{getQDimQS()}
\label{idx:getQDimQS}

Get detailed dimensions for all symmetry sectors on
a specified leg of a \QSpace tensor,
returning $q$-label resolved dimensions
both for the reduced multiplet level (RMTs),
as well as the combined multiplet dimensions.
See also \idx[getDimQS]{getDimQS} for total dimensions.
\vspace{-1ex}
 
\paragraph{getRC()}

Load \RCS specific data (for inspection, experimental, or debugging purposes).

\paragraph{getSymStates()}
\label{idx:getSymStates}

Group Fock space decomposition into symmetry spaces
prior to \idx[compactQS]{compactQS} 
in \src{getLocalSpace}. Like \src{compactQS},
this routine is never explicitly called or needed
elsewhere in an application.
\vspace{-1ex}

\paragraph{isIdentityCG()}

(boolean) Whether given \QSpace tensor has identity CGRs;
returns 1 for scalar operators or rank-2 tensors with
fully abelian symmetries. By contrast,
\onej tensors with non-abelian symmetries return 0.
\vspace{-1ex}

\paragraph{isIdentityQS()}

(boolean) Whether \QSpace represents an identity operator
as a whole which thus also checks the RMTs to be
identities within numerical (double precision) noise.
\vspace{-1ex}

\paragraph{isHConjQS()}

(boolean) Whether rank-2 \QSpace is Hermitian conjugate
within numerical noise. In the case of all-abelian symmetries,
the input tensor $H$ may represent a scalar operator
decomposed as a tensor of even rank $r=2n$, where a return
value of 1 requires, aside from additional checks on the RMTS,
that the first $n$ indices are
incoming, and the remaining $n$ indices outgoing.
\vspace{-1ex}

\paragraph{normQS()}
\label{idx:normQS}

(number) Computes \idx[QS:norm]{Frobenius norm}
$\Vert X \Vert \equiv \sqrt{{\rm tr}(X^\dagger X)}$
for \QSpace \src{X} as in \Eq{QS:norm}.

\paragraph{orthoQS()}
\label{idx:orthoQS}

Computes QR-like decomposition based on prior SVD
for given \QSpace tensor,
also allowing for truncation
[same as \src{\idx[svdQS]{svdQS}}, except that singular values
are already contracted onto one of the isometries out of
the SVD as specified by the input].
\vspace{-1ex}

\paragraph{permuteQS()}
\label{idx:permuteQS}

\idx[perm]{Permute} legs of \QSpace tensor
together with optional tensor 
\idx[conj]{conjugation}.
\vspace{-1ex}

\paragraph{plusQS()}
\label{idx:plusQS}

Add \QSpace tensors, as in \src{C = A+bfac*B}
where, by default, \src{bfac=1} results in plain addition.
By contrast, specifying \src{bfac=-1}
results in subtraction, etc.
\vspace{-1ex}

\paragraph{setupRCStore()}
\label{idx:setupRCStore}

Triggers initial setup and generation
of \RCS for specified symmetry (this is automatically
triggered anyways when calling \idx[compactQS]{compactQS}
with a new symmetry in \src{getLocalSpace},
so there is usually no need
to explicitly call \src{setupRCStore}).

\paragraph{skipZerosQS()}
\label{idx:skipZerosQS}

Skip \QSpace \idx[QS:recs]{records} that have all matrix
elements in their RMT below some numerical threshold
(default $10^{-14}$ for double precision noise).
These \idx[QS:zblocks]{zero blocks}
are allowed by symmetry, but are zero
due to application specific reasons.
By default, diagonal zero blocks for rank-2, i.e.,
scalar operators are kept, nevertheless.
Zero-blocks for rank-2 operators
can be forced to also be skipped with
the option \str{-{-}all}.

\paragraph{svdQS()}
\label{idx:svdQS}

Singular value decomposition (SVD) of some tensor
typically representing an orthogonality center (OC).
For non-abelian symmetries, the SVD is constrained
w.r.t. a particular leg since no automated fusion of legs
is performed in this routine.

\paragraph{traceQS()}
(number)
Full trace of rank-2 scalar \QSpace tensor,
properly including weights by degeneracies
within multiplets.

\subsection{NRG related hard-coded MEX applications}
\label{sec:NRG:apps}

The following MEX routines are also located in the
\src{./bin} folder. 
Most of these have already been applied in the past,
as indicated by the references below.
In this documentation, however,
these are mentioned in passing for completeness only.
In any case, a \idx[MEX:help]{help} is provided
with each of these routines.

\paragraph{NRGWilsonQS()}
\label{idx:NRGWilsonQS}

Iterative diagonalization for 
Wilson's Numerical Renormalization Group
(NRG, \cite{Wilson75})
for quantum impurity models \cite{Wb07,Wb12}.

\paragraph{fdmNRG\_QS()}
\label{idx:fdmNRG_QS}

Full density matrix (fdm-NRG) approach to computing
impurity spectral that conserve spectral sum rules
to within numerical precision for arbitrary
temperatures \cite{Wb07,Wb12}. 
Together with \src{NRGWilsonQS} this is a powerful
black-box quantum impurity solver.

\paragraph{fgrNRG()}
\label{idx:fgrNRG}

Fermi Golden rule adaptation of \src{fdmNRG\_QS} 
to compute absorption and emission spectra
on quantum impurities \cite{Tureci11,Wb12,Haupt13}.

\paragraph{tdmNRG()}
\label{idx:tdmNRG}

Time-dependent quench adaptation of fdm-NRG
starting from thermal equilibrium in the initial Hamiltonian
(closely related to fgr-NRG transformed to the real-time domain).

\paragraph{getSmoothSpec()}

Log-Gauss broadening of spectral data,
e.g., as obtained from fdm-NRG \cite{Bulla08,Wb07},
with smooth transition
to linear binning across zero energy.

\paragraph{getSmoothTDM()}

Broadening of spectral data for quantum quenches
to smoothen data in the real-time domain \cite{Wb12}.
Closely related to \src{getSmoothSpec()}.

\subsection{\QSpace MEX utility routines}

The \src{./util} folder contains hardcoded \Cpp utility 
MEX routines. The listing below only mentions the more
important ones. Additional MEX files that are present
in the public repository but not listed here,
were partly for testing purposes and hence
can be safely ignored. Yet as always,
the \idx[MEX:help]{help}
provided with them explains their purpose.

\paragraph{helloworld()}
\label{idx:helloworld}

Hello world routine to test MEX file setup.

\paragraph{kramers()}
\label{idx:kramers}

Kramers-Kronig transformation of spectral data
(back and forth from real to imaginary and vice versa;
e.g., used with NRG spectral data).

\paragraph{matchIndex()}
\label{idx:matchIndex}

matches two sets of indices that are stored as rows
in a matrix (typically containing integer values,
such as symmetry labels, to avoid issues with double
precision noise). It returns the indices of the
matching rows, similar to Matlab's
\src{intersect(A,B,\str{rows})}, yet with crucial 
practical differences.
If rows are not unique, in the sense that
certain index sets appear multiple times 
identically across different rows, then all-to-all
matches are included within such `degenerate' subspaces
as in a tensor product.
An optional additional info structure is returned
as third argument. This contains additional data
such as indices of non-matching entries, etc.
See also \idx[uniquerows]{uniquerows}.

\paragraph{cgs2double()}
\label{idx:cgs2double}

converts an \idx[MPFR]{MPFR-encoded} 
Clebsch-Gordan data space,
like the multidimensional arrays as stored in the \RCS,
to regular double-precision decimal format (mostly for
inspection and testing purposes only).
See also lower-level routine \idx[mpfr2dec]{mpfr2dec}
or \hsec[SymStore]{\atSymStore}.

\paragraph{mpfr2dec()}
\label{idx:mpfr2dec}

converts \idx[MPFR]{MPFR-encoded} numerical arrays
as stored it the \RCS
to regular double-precision decimal format (mostly for
inspection and testing purposes only).
See also higher-level routine \idx[cgs2double]{cgs2double}.

\paragraph{uniquerows()}
\label{idx:uniquerows}

group rows in a matrix (typically assuming integer
values to avoid issues with double precision noise),
returning a matrix with unique rows. Additional return
arguments provide detailed information on the dimension
of grouped blocks with identical row entries
together with indices to the original data.
See also \idx[matchIndex]{matchIndex}.

\paragraph{wbrat()}
\label{idx:wbrat}

Acquire a rational approximation for float numbers
based on continued fractions
(similar in functionality to Matlab's \src{rat.m}
function, but this routine automatically also checks
for possible representations in terms of the square root of
rationals, as they frequently occur in CGTs.
It picks the closest representation
(with or without square root)
in terms of numerical accuracy. 
This routine returns a string with syntax that can
be reevaluated within Matlab. An additional info
structure returned as second argument
provides the integers \src{P} and \src{Q} as obtained
for the rational approximation \src{P/Q}.
If this routine fails
a find a rational approximation with the requested
accuracy, the original value is returned, instead.

\section{\QSpace Related Matlab Environment}
\label{app:QS:env-ML}

The following gives a listing and brief description of the
more important Matlab routines and additional
class constructions.
Since there are plenty of helper routines, including
some 250+ m-files in \src{./lib}, these cannot be
all described in detail here.  Yet as always,
all routines are provided with a \idx[ML:help]{help}
that explains their purpose.

\subsection{Class/\atQSpace}
\label{sec:atQSpace}

The \Idx[QS:routines]{\atQSpace class}
is a Matlab wrapper environment for the \Cpp
\QSpace class used in MEX files.
Since a tensor \src{X} thus becomes \QSpace object \src{X},
it also simly referred to as \QSpace~\src{X}~then.
The~\atQSpace class contains more than a hundred class methods
which cannot be all discussed in detail here.
Only the most noteworthy ones
are listed in the following. A complete listing
including their \src{help} output has also been compiled
and included in the
PDF file \src{\happ[Docu]{Docu/QSdocs.pdf}}
in the \QSpace repository.
Rather than alphabetically sorted,
the following entries are mostly grouped by topic.

\paragraph{QSpace()}
\label{idx:QS:QSpace}

   constructor of the \AtQSpace wrapper class.

\paragraph{display()}
\label{idx:QS:display}

   prints a convenient formatted summary of a given \QSpace tensor.
   It is automatically invoked by Matlab when displaying
   a variable by simply typing its name at the Matlab prompt.
   It shows a 2-line header that is generated by {\bf info()},
   followed by a listing of records.   
   See \Fig{QS:disp} for an example, together with a detailed 
   explanation of the printed output.
   For larger {\QSpace}s with more than about 12 non-zero
   blocks (records),
   \src{display} truncates the listing, by default:
   it shows the first 5 entries,
   up to two intermediate entries if largest in RMT size 
   followed by the last two entries. The listing
   of all entries can be enforced, nevertheless, by explicitly
   invoking the \src{display} routine with the equivalent
   options \str{-f} or \str{-a}.

   Like logging with \src{\idx[wblog]{wblog()}}, the display of \QSpace
   tensors also supports color-highlighted output for readability
   when Matlab is run in interactive Matlab 
   \idx[ml:term]{terminal mode}.
   In this case, \itags are color-coded by
   interpreting trailing \idx[itag:markers]{marker characters}.

\paragraph{plus(), minus(), sum()}
\label{idx:QS:plus}

   Overload of simple algebraic operators
   \src{+} (\src{plus}) or \src{-} (\src{minus}).
   These are extended wrappers of the MEX routine
   \idx[plusQS]{plusQS}. 
   The routine \src{sum(A,dim)} permits to sum up a
   \QSpace array $A$
   along dimension \src{dim} (by default, \src{dim=1};
   this follows similar semantics as Matlab's native \src{sum}
   routine). This adds up \QSpace{}s and hence returns
   a \QSpace (vector).

   Simpler cases where for a rank-2 tensor a plain 
   numerical value $x$ is added or subtracted,
   are handled within the Matlab wrapper routine itself.
   For example, in \src{H+x} with \src{H} some rank-2 tensor,
   the scalar $x$ is interpreted as proportional to the
   identity matrix
   for all symmetry sectors that are present in \src{H}
   (be aware, though, that \src{H} may have missing 
   diagonal \idx[QS:zblocks]{all-zero} blocks;
   hence to ensure a full state space, one may
   rather use \src{H+x} $\to$ \src{H+x*Id}, instead,
   with \src{Id} representing a complete identity operator
   within given state space).

\paragraph{oplus()}
\label{idx:QS:oplus}

   The routine \src{oplus(A,B,d12)} permits the direct
   sum of two tensors along dimensions \src{d12}
   (default: \src{[1\,2]})
   while respecting the symmetry structure.
   Both \QSpaces $A$ and $B$ must have the same tensor structure
   (same rank, same $q$-directions, etc).
   If \src{d12} specifies two dimensions (indices),
   this catenates blocks in a `block-diagonal' 
   fashion along specified dimensions \src{d12}
   [the Matlab analogon for matrices with \src{d12=[1\,2]}
   is \src{blkdiag(A,B)}, except that \src{oplus} only
   combines blocks that also share the same symmetry sectors;
   other blocks unique to either \src{A} or \src{B}
   are included as is in the output].
   If \src{d12} specifies a single dimension (index),
   symmetry matching blocks are catenated along that
   dimension only [the Matlab analogon in this case
   would be \src{cat(d12,A,B)}].

\paragraph{uplus(), uminus()}
\label{idx:QS:uminus}

   are unary operators that implement the unary plus
   and minus, i.e., syntax such as \src{X=+Y} or \src{X=-Y}.

\paragraph{transpose(), ctranspose()}
\label{idx:QS:hconj}

   overload the operators
   \src{.\textquotesingle} and \src{'} 
   for plain transpose and Hermitian conjugate, respectively.

\paragraph{times(), mtimes(), mrdivide()}
\label{idx:QS:times:etc}

   overload the respective algebraic operators
   \src{.*}, \src{*},  or \src{/} (for division by number only).
   The routine \src{mtimes} is an extended
   wrapper routine to the MEX function \idx[contractQS]{contractQS}.
   Simple multiplications with scalar numeric values
   are handled within the wrapper routine itself.

\paragraph{comm(), acomm()} 
\label{idx:QS:comm}

   computes commutator or anticommutator, respectively,
   for a pair of operators.
   For \idx[irop]{irops}, the irop index will necessarily
   also be contracted in this process.
   These routines are based on  \idx[contractQS]{contractQS}.

\paragraph{real(), imag(), isreal()} 
\label{idx:QS:real}

   extract real and imaginary parts of a \QSpace
   (simple Matlab functions that operate on the RMTs
   (i.e., on \src{X.data} for a \QSpace \src{X}),
   or check whether \QSpace is complex
   [\src{\textasciitilde{}isreal()}].

\paragraph{sqrt()} takes square root of rank-2
\QSpace that are diagonal (and diagonal only; this is enforced).
Warnings are issued if input data is negative or complex.

\paragraph{logical(), not(), isempty()}
\label{idx:QS:bool}

boolean operators that check whether a \QSpace is empty.
The method \src{not()} is the inverse of \src{logical()}.
For example, for {\QSpace}s \src{X} and \src{Y},
this permits simple syntax such as
\src{if X \&\& \textasciitilde{}Y, (do something); end}.
The method \src{isempty()} is somewhat more elaborate
than \src{logical()}, as it includes further structural checks.

\paragraph{eq(), ne(), isequal(), sameas()}
\label{idx:QS:equal}

Overload of Matlab's \src{==} and \src{\textasciitilde{}=} operators
(wrappers to \src{\atQSpace/isequal}).
This permits the syntax \src{X==Y} for {\QSpace}s \src{X} and \src{Y}.
The method \src{sameas()} permits difference in data up to some
epsilon (default: $10^{-10}$ on a relative scale).

\paragraph{getsub()}
\label{idx:QS:getsub}

selects sub-list of \idx[QS:recs]{\QSpace records}
while maintaining the \QSpace structure, otherwise. 
The index of records to select may be specified
explicitly, as in \src{getsub(X,I)} with record index \src{I},
or implicitly by selecting a particular symmetry sector,
as in \src{getsub(X,[qlabels],dim)} which only
selects the records that have specified \src{qlabels}
(set specified as rows in a matrix)
on leg \src{dim}.

\paragraph{subsref(), subsasgn()}
\label{idx:QS:subsref}

overloads indexing and access to fields in a \QSpace object.
As a technical subtlety concerning Matlab,
note that for a \QSpace \src{X},
typing \src{X.Q\{:\}} on the Matlab prompt does not
work as expected for a plain Matlab structure.
Since the operation \src{X.Q} needs to be rerouted via the
overloaded method \src{subsref} here, this itself operates
like a function call. Therefore typing \src{X.Q\{:\}}
on the Matlab prompt only displays \src{X.Q\{1\}}.
To display all entries in \src{X.Q}, this
requires, e.g., a temporary assignment followed
by its display, like \src{Q=X.Q; Q\{:\}}.
Similarly, catenating symmetry labels across all legs
as in \src{Q=X.Q; Q=[Q\{:\}]} is not equivalent to
\src{Q=[X.Q\{:\}]}. The latter rather corresponds to
\src{X.Q\{1\}} only.
This is a Matlab peculiarity, which returns a comma-separated
list for \src{Q\{:\}}. For a function call,
this would require the rather awkward syntax
\src{Q=cell(1,numel(X.Q)); [Q\{:\}]=X.Q\{:\}}
which is simply equivalent to \src{Q=X.Q}
in the first place.

\paragraph{diag()} 
\label{idx:QS:diag}

toggle rank-2 \QSpace tensor between diagonal and
non-diagonal representation. This mimics the behavior
of Matlab's \src{diag()} function. This function is useful
to deal with the compact diagonal behavior, e.g.,
as returned by \idx[eigQS]{eigQS} or \idx[svdQS]{svdQS}.
With the option
\str{-d}, this routine returns a plain numerical
vector of all diagonal entries
(see also \idx[diagQS]{diagQS} in this regard).

\paragraph{getitags()}
\label{idx:QS:getitags}

gets the full set of \itags for given \QSpace tensor.

\paragraph{setitags()}
\label{idx:QS:setitags}

Set \idx[itags]{\itags} of \QSpace
while preserving \idx[qdir]{$q$-directions}
(trailing conjugate flags \str{*}).
Any trailing \str{*} specified
with the input strings will be ignored.
This routine accepts a range of specialized settings.
For example, for \idx[Atensor]{$A$-tensors},
the following are equivalent,
\begin{minted}[escapeinside=??]{text}
   A=setitags(A,'-A',4)
   setitags(A,{'K03','K04', 's04'})
   setitags(A,{'K03','K04*','s04'})
\end{minted}
as may be used, e.g., for site $4$ in an MPS
(e.g., see \Fig{arrows}).
Conjugate flags, as in line 3, are ignored.
Since line 2-3 specify \src{A} by name, they will also assign
the changes to \src{A} in the workspace same as line 1.
For \idx[irop]{operators}, \src{S=setitags(S,\str{-op:s},4)}
sets itags \src{\{\str{s04},\str{s04} [,\str{op}]\}},
with the last `operator'
index skipped if \src{S} is of rank $2$.
See \idx[ML:help]{help} on \src{setitags} for more details.

\paragraph{itagrep()}
\label{idx:QS:itagrep}

Change \itags based on regular epxressions
(uses Matlab's \src{regexprep}).

\paragraph{untag()}
\label{idx:QS:untag}

Remove \itags for specified \QSpace tensors.
For convenience, if no return argument is requested and
all tensors are specified by name, as in \src{untag(A,B,$\ldots$)},
all objects will get updated in the caller workspace,
nevertheless, using Matlab's \src{assignin(\str{caller},...)}.
Therefore \src{[A,B,...] = untag(A,B,...);} and \src{untag(A,B,...)}
are equivalent.

\paragraph{isscalarop()}
\label{idx:QS:isscalarop}

(boolean) whether input represents scalar operator,
i.e., is of rank 2 and is block-diagonal for all symmetries.

\paragraph{fixScalarOp()}
\label{idx:QS:fixScalarOp}

for scalar operators with trailing \idx[irop]{irop index}
having $q_{\rm irop}=0$, this trailing singleton irop index
(third leg) is skipped. See also reverse operation 
\src{makeIrop()}.

\paragraph{makeIrop()}
\label{idx:QS:makeIrop}

transform scalar rank-2 operator back to an irop
by explicitly adding a trivial \idx[irop]{irop index} (third leg)
with scalar symmetry labels $q=0$.
This is the reverse operation
to \src{fixScalarOp()}. 

\paragraph{appendSingletons()}
\label{idx:QS:appendSingletons}

Append indices in the scalar symmetry sector $q=0$
for all symmetries
and thus of dimensions $1$ (singletons) to given \QSpace tensor.

\paragraph{addSymmetry()}
\label{idx:QS:addSymmetry}

Add new symmetry to specified \QSpace tensor.
By default, this appends the symmetry in the irep of the
scalar representation $q=0$ as an additional symmetry.
This routine is intended for specialized model setups to tweak
existing output of \src{getLocalSpace()}.
Like \src{getLocalSpace},
there should be no need to ever call this routine
after the model setup.

If the symmetry label $q$ for the new symmetry is explicitly
specified and non-scalar with  multiplet
dimension $|q|>1$, then a $\sqrt{|q|}$ factor
is applied to given \QSpace \src{X}. This mimics having
added $|q|$ copies, such that the \idx[QS:norm]{norm}
$\Vert X\Vert^ 2 =$\src{norm(X)\textasciicircum2}
will change by a factor $|q|$.

\paragraph{getvac()}
\label{idx:QS:getvac}

Get identity operator (`\idx[nolegs]{state space}')
in the vacuum state for the symmetry setting
of the input \QSpace. Hence this has dimensions
$1\times 1$ and all-zero $q$-labels, yet is aware
of all symmetries and their order.
With the option \str{-1d}, this returns a vector of
length $d=1$, with an example shown in \eqref{eq:rank1}.

\paragraph{SEntropy()} 
\label{idx:QS:SEntropy}

compute von-Neumann entropy
assuming that the input \QSpace represents a scalar
density matrix.

\paragraph{plotQSpectra()} 
\label{idx:QS:plotQSpectra}

Plot eigenspectrum of scalar input operator
in a symmetry-resolved manner. With option \str{-ES},
this plots the entanglement spectrum, instead (the latter is
intended for properly normalized density matrices only).

\paragraph{contract()}
simple \idx[QS:routines]{wrapper} for \src{contractQS}.

\paragraph{dim()} simple wrapper for \src{getDimQS}.

\paragraph{eig()} simple wrapper for \src{eigQS}.

\paragraph{getIdentity()} simple wrapper
for \src{getIdentityQS}.

\paragraph{permute()} simple wrapper for \src{permuteQS}.

\paragraph{skipzeros()} simple wrapper for \src{skipzerosQS}.

\subsection{Class/\atSymOp}
\label{sec:SymOp}

Utility class to generate a minimal set
of generators (symmetry operators, or `\src{SymOp}' for short)
in terms of raising and $z$-operators
(simple positive \idx[roots]{roots} \src{Sp} and
\idx[Cartan]{Cartan subalgebra} \src{Sz}, respectively)
when setting up symmetries in models.
It is used in \src{\hsec[gls]{getLocalSpace}} only,
e.g., with objects of this type returned in \src{IS.SOP}
$\to$ \src{Sp}, \src{Sz} together with 
the MEX routine \src{\idx[getSymStates]{getSymStates}}
to bootstrap
symmetry spaces. This class should
not be relevant otherwise, hence can be safely 
ignored from an application point of view.

\subsection{Class/\atSymStore}
\label{sec:SymStore}

The \atSymStore class offers basic utility routines
to access the \RCS. It is intended mostly for 
information, testing, and debugging purposes.
It does not try to generate any symmetry
related data. Therefore if the requested data is
not yet present in the \RCS, this class simply returns
an error.

For example, the minimal set of generators
(simple roots and Cartan subalgebra, or conversely,
raising and $z$-operators) for SU(2) for the
defining representation $\idx[SU2:qlabels]{q=2S}=1$
can be loaded from the \RCS
via \src{I=SymStore(\str{SU2},\str{-R},\str{(1)})}
with the option \str{-R}
which thus looks up the folder \src{\var{\RCS}/SU2/RStore/}.
The round brackets with the string of the $q$-labels
in the last input argument are optional.
The returned generators are encoded as sparse
\idx[MPFR]{MPFR} arrays.
These can be converted to regular double precision arrays
via the MEX utility routine \src{cgs2double(I.Sz)}.
Since the output is rank-2 in the present context,
this permits and thus also returns a regular sparse
double array.

Standard Clebsch Gordan coefficients (CGC) can be 
loaded with the option \str{-C}
which looks up the \src{CStore} folder.
For example,
\src{I=SymStore(\str{SU2},\str{-C},\str{(2,2;2)},\str{-f})}
loads the CGC for the spin tensor product
$(S_1=1) \otimes (S_2=1) \to (S_{\rm tot}=1)$
bearing in mind the labeling convention with \Eq{qlabels:tensor}
and that within \QSpace, \idx[SU2:qlabels]{$q=2S$},
thus resulting in \src{(2,2;2)}.
Since CGTs may get large, the data returned with 
the \src{-C} option effectively contains a reference,
by default. The option \str{-f} then enforces that
the full CGT is loaded. With this,
\src{I.cdata} contains the CGT 
in MPFR format, while \src{I.CData} $\equiv$
\src{cgs2double(I.cdata)} contains the same
data but already converted to a double-precision
multidimensional array.

Lastly, the fusion map out of two multiplets
can be accessed via the option \str{-M}
which looks up the currently generated
\src{CStore/++-/*.\hsec[RCStore]{mp3}} files.
For example, with reference to the previous
example, the command
\src{M = SymStore(\str{SU2},\str{-M},\str{(2,2)})}
returns all outcomes of the fusion of the
two SU(2) multiplets $(S_1=1) \otimes (S_2=1)$,
with \src{M.J12} reconfirming the input,
and the fused multiplets shown with \src{M.J}
as $q \in \{0,2,4\}$, i.e., $S \in \{ 0,1,2\}$,
as expected.

\subsection{Class/\atHamilton}

An application class that underlies
$\src{DMRG/\hsec[DMRG]{runDMRG}.m}$.
This permits to run DMRG simulations on simple pre-setup
models as defined in \src{runDMRG} with the option
to adapt to one's own needs. Since this class is rather
extensive on its own, this will be described in more
detail elsewhere. Yet please feel free to explore.
In case of issues or more detailed questions,
please also feel free to  \idx[contact]{contact the author}
of this documentation in this regard.

  \end{appendix}

  \newpage

  \bibliographystyle{SciPost_bibstyle}
  \bibliography{mybib.bib,refs.bib}

\end{document}